\documentclass[12pt]{article}
\usepackage{amssymb,amsmath,amsthm}
\usepackage{hyperref}
\usepackage{array,longtable}

\usepackage[dvips]{graphicx}
\usepackage[dvips]{color}

\usepackage{wrapfig}
\usepackage{epsfig}
\makeatletter
   \newcommand\tabcaption{\def\@captype{table}\caption}
\makeatother

\usepackage[mathscr]{eucal}

\newcommand{\be}{\begin{equation}}

\newcolumntype{V}{>{$}m{4cm}<{$}}
\newcolumntype{C}{>{$}c<{$}}
\newcolumntype{L}{>{$}l<{$}}
\newcolumntype{R}{>{$}r<{$}}
\newcommand{\Zh}{{\mathbb Z}}
\newcommand{\Rh}{{\mathbb R}}
\newcommand{\Ch}{{\mathbb C}}
\newcommand{\Dd}{{\mathcal D}}
\newcommand{\Pp}{{\mathcal P}}

\newcommand{\ssim}{\mathop{\sim}\limits}

\newcommand{\Nc}{\mathcal{N}}

\newcommand{\ra}{\rightarrow}

\newcommand{\Ac}{\mathcal{A}}

\newcommand{\Hc}{\mathcal{H}}
\newcommand{\Oc}{\mathcal{O}}
\newcommand{\pd}{\partial}
\newcommand{\opd}{\overline{\partial}}
\newcommand{\oz}{\overline{z}}
\newcommand{\bb}{\mathbf{b}}
\newcommand{\cb}{\mathbf{c}}
\newcommand{\eps}{\varepsilon}
\newcommand{\Tr}{\mathop{\mathrm{Tr}}\nolimits}
\newcommand{\la}{\langle\!\langle}
\renewcommand{\ra}{\rangle\!\rangle}
\newcommand{\Res}{\mathop{\mathrm{Res}}\limits}

\newcommand{\ap}{\alpha^{\,\prime}}

\newcommand{\vac}{|0\rangle}

\newcommand{\df}{{f^{\prime}}}
\newcommand{\ddf}{{f^{\prime\prime}}}
\newcommand{\dddf}{{f^{\prime\prime\prime}}}

\allowdisplaybreaks[3]

\begin{document}
\thispagestyle{empty}

\vspace*{2.0ex}

\centerline{\huge \bf Noncommutative Field Theories }

\medskip

\centerline{\huge \bf and }

\medskip

\centerline{\huge \bf (Super)String Field Theories }


\vspace*{4.0ex}

\centerline{\large\rm
I.Ya.~Aref'eva$^{a,\,}$\footnote{E-mail: \texttt{arefeva@mi.ras.ru}},
D.M.~Belov$^{b,\,a,\,}$\footnote{E-mail: \texttt{belov@physics.rutgers.edu}},
A.A.~Giryavets$^{c,\,}$\footnote{E-mail: \texttt{alexgir@mail.ru}},}
\centerline{\large\rm
A.S.~Koshelev$^{a,\,d,\,}$\footnote{E-mail: \texttt{koshel@orc.ru}} and
P.B.~Medvedev$^{e,\,}$\footnote{E-mail: \texttt{medvedev@heron.itep.ru}}}

\medskip
\centerline{\it ~$^a$Steklov Mathematical Institute,
Russian Academy of Sciences }

\centerline{\it ~$^b$Department of Physics, Rutgers
University, Piscataway, NJ 08855}

\centerline{\it ~$^c$Faculty of Physics, Moscow State
University}

\centerline{\it ~$^d$Physics department, University of
Crete, Greece}

\centerline{\it ~$^e$Institute of Theoretical and
Experimental Physics}

\begin{center}
\it{Based on lectures given by I.Ya.~Aref'eva \\ at the Swieca Summer
School,
Brazil, January 2001; \\
\it{Summer School in Modern Mathematical Physics, Sokobanja,}\\
\it{ Serbia,           }
Yugoslavia,
 13-25 August 2001;}\\
\it{ Max Born Symposium, Karpacz, Poland  21-24 September, 2001;}\\
\it{ Workshop "Noncommutative Geometry,
Strings and Renormalization",} \\ \it{ Leipzig, Germany, September 2001}
\end{center}

\centerline{\bf Abstract}
In this lecture notes we explain and discuss some ideas
concerning noncommutative geometry in general, as well as
noncommutative field theories and string field theories.
We consider noncommutative quantum field theories emphasizing
an issue of their renormalizability and the UV/IR mixing.
Sen's conjectures on open string tachyon condensation
and their application to the D-brane physics have led to wide
investigations of the covariant string field theory proposed
by Witten about 15 years ago. We review main ingredients
of cubic (super)string field theories using various formulations:
functional, operator, conformal and the half string formalisms.
The main technical tools that are used to study conjectured
D-brane decay into closed string vacuum through the tachyon
condensation are presented. We describe also methods which are
used to study the cubic  open string field theory around
the tachyon vacuum: construction of the sliver state,
``comma'' and matrix representations of vertices.

\vfill \eject

\baselineskip=16pt

\pagenumbering{arabic}
\newpage
\tableofcontents

\newpage
\section{Introduction.}
\setcounter{equation}{0}

It is well known that the gauge invariance principle is a basic
principle in construction of modern field theories. Both general relativity
and Yang-Mills theory use the gauge invariance. Another example of gauge
invariant theory is string field theory
\cite{Witten}. To formulate this theory it is
useful to employ notions of noncommutative differential geometry.

Noncommutative geometry has appeared in physics in the works of the founders
of quantum mechanics. Heisenberg and Dirac have argued that
the phase space of quantum mechanics must be noncommutative
and it should be described by quantum algebra. After works of von Neumann
and more recently by Connes mathematical and physical investigations
in noncommutative geometry became very intensive.

Noncommutative geometry uses a generalization of the known duality
between a
space and the algebra of functions on it \cite{book,Mad}.
One can take ${\cal A}$ to
be an abstract noncommutative associative algebra ${\cal A}$ with a
multiplication $\star$
and
interpret  ${\cal A}$ as an
"algebra
of functions" on a (nonexisting) noncommutative space.
One can introduce many geometrical notions in this setting.
A basic set  consists  of an
external  derivative $Q$ (or an algebra ${\cal G}$ of derivatives)
 and an integral, which satisfy ordinary axioms.
 In Section~\ref{sec:NC} we will present several examples
 of noncommutative spaces.

 Let  us sketch motivations/applications of models of
 noncommutative differential geometry (NCDG) in physics:
\begin{itemize}
\item Noncommutative space-time;
\item Large $N$ reduced models (quenched Yang-Mills theory);
\item D-brane physics, (M)atrix theory;
\item Description of the behavior of models in external fields;
\item Noncommutative Solitons and Instantons;
\item Smoothing of singularities;
\item String Field Theory (SFT).
\end{itemize}

\bigskip

\noindent{\bf Noncommutative Space-Time}
\medskip

\begin{itemize}
\item[$\blacktriangleright$] Uncertainty principle in gravity
\end{itemize}

There is an absolute limitation on length measurements in quantum gravity
\begin{equation}
\label{lim}
\ell\geqslant \ell_{Pl}.
\end{equation}
The reason is the following. If we want to locate a particle
we need to have energy greater than the Planck mass
$m_{Pl}=(\hbar c G^{-1})^{\frac12}$
($G$ is Newton's constant).
The corresponding gravitational field will have a
horizon at $r=2Gm_{Pl}c^{-2}=2\ell_{Pl}$, shielding
whatever happens inside the  Schwarzschild radius and therefore
the smallest ``quanta''  of space and time exists.
It is  tempting to relate the uncertainty $\delta x$
in the length measurement (\ref{lim}) to the noncommutativity
of coordinates
\begin{equation}
\label{ncc}
[x^\mu,x^\nu]=i\hbar \theta ^{\mu\nu}.
\end{equation}
The idea of quantization of space-time using noncommutative
coordinates $x_\mu$ is an old one \cite{Markov}-\cite{Fin}.
Snyder \cite{Sny47a,Sny47b} has proposed to use the noncommutative space to improve
the UV behavior in quantum field theory.
An extension of the Snyder's idea was considered in \cite{Golfand,Kad} where
the field theory with the momentum in the
de Sitter space  was considered.
A proposal for a general (pseudo)Riemannian geometry
in the momentum space was discussed in \cite{Tamm}.

Deriving relation
(\ref{lim}) from (\ref{ncc}) by analogy with the derivation of
the Heisenberg uncertainty
inequality one gets the problem because (\ref{lim}) has
to hold even for one-dimensional space. In \cite{VVZ}
it was suggested that (\ref{lim}) indicates
that at the Planck scale one should use non-Archimedean geometry.
For recent discussion of space-time noncommutativity see
\cite{WZ}-\cite{Luk}. The space-time stringy uncertainty has been
discussed in \cite{AVqp,Yoneya}.

\begin{itemize}
\item[$\blacktriangleright$]  UV divergences in QFT
\end{itemize}

One of earlier
motivations to consider  field  theories on
noncommutative space-time (noncommutative field  theories) was a hope
that it would be possible to avoid  quantum field theory
divergences \cite{WZ,AVqp,Mad,filk,Kempf,CDP,AV}. Now a commonly
accepted belief is that a theory on a noncommutative space is
renormalizable if (and only if) the corresponding commutative theory is
renormalizable. Results on the one-loop renormalizability of
noncommutative gauge theory \cite{ren1}-\cite{ren3} as well as the two-loop
renormalizability of the noncommutative scalar $\phi _4^4$ theory
\cite{ABK1}   support
this belief. However the renormalizability \cite{Ch} does not guarantee that the
theory is well-defined. There is a mixing of the UV and the IR
divergences \cite{MRS,ABK1}.
 For further developments  see
\cite{Hay}-\cite{2loop2} and refs. therein.

\bigskip
\noindent{\bf Large $N$ reduced models}
\bigskip

It has been found  that the large $N$ limit in QCD can be described
by a reduced quenched model \cite{EguKaw,Gonzalez}.
This model is equivalent to a
noncommutative
gauge model. A special large $N$ matrix model was proposed to describe
strings \cite{9612115,9910004} (see \cite{AMNS} in the context
of noncommutative
gauge models). It occurs that the master fields in the large $N$
matrix models
are the subject
of the noncommutative plane \cite{AV-largeN}.

\bigskip
\noindent{\bf D-branes and M(atrix) theory}
\bigskip

Connes, Douglas and Schwartz \cite{CDS} (see also \cite{DoHu}) have shown
that supersymmetric gauge theory on a noncommutative torus is naturally
related to a compactification of the M(atrix) theory \cite{BFSS,bilal}.
M(atrix) theory
is a maximally
supersymmetric quantum mechanics with the action:
\begin{equation}
\label{BFSS-act}
S = \int dt\,\Tr \sum_{i=1}^9 (D_t X)^2 - \sum_{i<j} [X^i,X^j]^2 +
\chi^\dag (D_t + \Gamma_i X^i) \chi .
\end{equation}
Here $D_t = \partial/\partial t + iA_0$, and variation over $A_0$ yields a Gauss law.
This action is also known as a regularized form of the
supermembrane action \cite{deWit}.
Equations of motion for the action (\ref{BFSS-act})
admit the following static solutions
\begin{equation}
\label{Ncc}
[X^\mu,X^\nu]=i\hbar \theta ^{\mu\nu}
\end{equation}
that are nothing but commutation relations (\ref{ncc})
defining noncommutative space-time.
About spectral properties of the corresponding Hamiltonian
see \cite{Hope} and \cite{AKM} and references therein.

Compactification on torus \cite{TaylorD}
$$
U_i^{-1} X^j U_i = X^j + \delta_i^j 2\pi R_i .
$$
leads to the relations
$$U_i U_j = e^{i\theta_{ij}} U_j U_i,$$
which define the algebra of functions
on a noncommutative torus with the noncommutativity parameter
$\theta$.
For reviews and further developments see \cite{Konechny}-\cite{ArArSh}.

One of the natural questions is
whether noncommutative gauge theory can be used to describe more general
compactifications in string theory. In order to have progress in this
direction  one
has to develop the theory of noncommutative quantum gauge fields not only
on the torus, but also on  more general manifolds \cite{S2}-\cite{NCkel}.
A framework for a noncommutative  gauge theory on  Poisson
manifolds by using recently developed deformation quantization
\cite{BFD,BerS} has been discussed in \cite{AV}.

\bigskip

\noindent{\bf Models in external fields}

\begin{itemize}
\item[$\blacktriangleright$] Dipole in a magnetic field
\end{itemize}
It is well known that the position operators in quantum mechanics in
the presence of a magnetic field
does not commute.
A model consisting   of a pair of opposite charges moving in a
strong magnetic field can be described as a particle moving on the
noncommutative 2-plane \cite{SB}.
\begin{itemize}
\item[$\blacktriangleright$] Quantum Hall effect
\end{itemize}
The model describing electrons in a magnetic field
projected to the lowest Landau level can be  naturally treated  as
a noncommutative field theory.  Thus noncommutative field theory
is relevant to the theory of
the quantum Hall effect \cite{SB,Jac}. It has been proposed that a
good description of the
fractional quantum Hall effect  can be obtained using
noncommutative  Chern-Simons theory
\cite{Susqh,Polych}.
\begin{itemize}
\item[$\blacktriangleright$] Strings in a magnetic field
\end{itemize}

Noncommutative geometry appears
in string theory with a
nonzero B-field \cite{SW} (for earlier considerations see
\cite{ChKr}-\cite{Som})
\begin{equation}
\label{strB}
S = \frac{1}{4\pi \alpha^{\prime}} \int_{\Sigma}d^2z
\left( g_{ij} {\partial}_{a} x^i {\partial}^a x^j -
2\pi i \alpha^{\prime} B_{ij}
\epsilon^{ab} {\partial}_a x^i {\partial}_b x^j \right).
\end{equation}

Seiberg and Witten have found
a limit in which
the entire string dynamics is described by a supersymmetric gauge
theory on a noncommutative space and shown an
equivalence between
ordinary gauge fields and noncommutative gauge fields.
See \cite{Oku}-\cite{0104139} for
further developments of these ideas.

\bigskip

\noindent{\bf Noncommutative Solitons and Instantons}
\bigskip

In commutative scalar field theories
there is Derrick's theorem, which prohibits the existence of
finite energy
 solitons in two and more spatial dimensions.
 The proof is based on a simple scaling argument
 according to which no finite size minimum can exist, since
  upon shrinking all length
scales  both kinetic and potential energies
decrease.
This argument does not work in the presence of a
length scale $\sqrt{\theta}$. It has been shown in \cite{GMS},
 that for sufficiently
large $\theta$  stable solitons  exist in the noncommutative
 scalar field theory.
For example, for a cubic potential
solitons are defined as solutions of the following equation
\begin{equation}
\label{ncsol}
  \phi =\phi\star \phi .
\end{equation}
While in a commutative theory this equation would admit only constant
solutions, in  a noncommutative theory
there are countable many solutions, see \cite{Minw,Harvey}
for more details.

As for noncommutative instantons it  was found that the ADHM construction admits
a generalization to the
noncommutative case
\cite{NekSch}.  Noncommutative instantons in the limit
$\theta \to 0$ are pure $U(1)$ gauge  configurations with
a singularity at the origin. For nonzero  $\theta$
instantons  are  nonsingular objects with a size of order $\sqrt{\theta}$.
\bigskip

\noindent{\bf Smoothing of singularities }

\bigskip

Solutions of a local field theory
usually can be extended  to the solutions
of noncommutative theory.
 As a rule in such extensions
 any singularities disappear
due to the position-space uncertainty.
In particular, the noncommutative rank $1$ theory has
non-singular instanton, monopole and vortex solutions.

Solitons in noncommutative theories can be stable even when their
commutative counterparts are
not, so noncommutativity provides a natural
mechanism of stabilization for objects of the size $\sqrt{\theta}$.
This behavior is naturally related to nonlocality. In other words,
noncommutativity inevitably
leads to  nonlocality.

\bigskip

\noindent{\bf String Field Theory}

\bigskip

One of the main motivations to construct
string field theory (SFT) --- an off-shell
formulation of a string theory ---
was a hope to study non-perturbative
phenomena in string theory \cite{Witten}.
The Witten SFT action for open bosonic  string is
\begin{equation}
S=\frac12 \int \Phi \star Q\Phi+
\frac13 \int \Phi \star \Phi \star\Phi .
\label{WA}
\end{equation}
Open bosonic  string has a tachyon that leads to a classical instability
of the perturbative vacuum. In the early works by Kostelecky and
Samuel \cite{KS} it was proposed to use SFT to describe
condensation of the tachyon to a stable vacuum.
By using the level truncation scheme
they have shown that the tachyon
potential in open bosonic string has a nontrivial minimum.
Further \cite{AMZ3}, the level truncation method has been applied
to study an effective potential of auxiliary fields in the cubic
superstring field theory (SSFT) \cite{AMZ1,PTY,AMZ2}. It was found
that some of the low-lying auxiliary scalar fields acquire
non-zero vacuum expectation values
that was considered  as an indication of
supersymmetry breaking in this vacuum.

Later on, Sen has proposed  \cite{sen-con} to interpret the
tachyon condensation  as a decay of unstable D-brane. In the
framework of this interpretation the vacuum energy of the open
bosonic string in the  Kostelecky and Samuel vacuum cancels the
tension of the unstable
space-filling 25-brane.
This cancellation has been checked with high accuracy
\cite{9912249,0002237}.

The cubic  open string field theory
around the tachyon vacuum,
or the vacuum string field theory (VSFT) was proposed in \cite{F2}
and it is investigated  now very intensively \cite{RSZF}-\cite{rsz-7}.
This VSFT has the same form
as (\ref{WA}) but with a new differential operator, which
after restriction to the space of string fields
can be written in the following form
$$
Q_{new}=Q+\{\Phi _{vac},~\cdot~ \}.
$$

 The triviality of the  cohomology of the new
BRST operator $Q_{new}$ around the tachyon vacuum is expected
\cite{sen-con}. This means that this theory should not contain
physical open string excitations. The evidence in favor of this
conjecture was also found by demonstrating the absence
of kinetic term for the tachyon field in nonperturbative vacuum
\cite{HATA,Taylor}. VSFT is conjectured to describe the appearance
of closed strings after the open string tachyon condensation.
 Moreover, it was suggested
 \cite{F2} that after some (maybe singular) field redefinition
 $Q_{new}$ can be written as a
pure ghost  operator ${\cal Q}$.
If this conjecture is true it is instructive to search solutions of
VSFT equation of motion
\begin{equation}
\label{SEQ}
{\cal Q}\Phi +\Phi\star\Phi =0
\end{equation}
in the factorized form $\Phi =\Phi_{matter}\otimes \Phi_{ghost}$,
 thus obtaining a
projector-like equation for matter sector:
\begin{equation}
\Phi_{matter}=\Phi_{matter}\star~\Phi_{matter}.
\label{proj}
\end{equation}
An equation similar to (\ref{proj}) has appeared in construction of
solitonic
solutions in noncommutative field theories in the large coupling
limit \cite{GMS}.

As an example of the projector-like solution,
there is a state called sliver, $|\Xi\rangle$.
This projector was constructed
in the oscillator formalism by Kostelecky and
Potting \cite{Kostelecky-Potting}.
 In fact a wedge state as a candidate to solve (\ref{proj})
was appeared first in the CFT framework  \cite{zwiebach}.
Then it was at first identified with the  sliver  numerically
\cite{RSZF} and later on by direct calculations \cite{Japan2}.
The sliver state is special in a sense
that it can be defined in an arbitrary boundary CFT \cite{zwiebach}.

The projector-like form of eq. (\ref{proj}) gave a new impulse for the
development of the half-string formalism
\cite{Witten},\cite{AV-M}-\cite{9902176},
 which drastically
simplifies Witten's $\star$-product.
To construct projections more systematically,
it is useful to find a matrix
representation of open string fields.
To  see matrix representation in a more transparent way it is
fruitful to use the so-called ``comma'' form of the overlap vertex
\cite{bcnt,Abdurrahman-Bordes},\cite{Okuyama},\cite{Japan2}.
The Bogoliubov
transformation that  makes the sliver
to be a new  vacuum requires the 3-string vertex to
take a comma form \cite{Kostelecky-Potting}.
The Bogoliubov
transformation  relating  the sliver and initial Fock vacuum cannot be
defined as an operator in the Fock space \cite{MooreTaylor}.
This is not surprising since we deal
with infinite number of degrees of freedom.
One can use the dressing
transformation technique \cite{AKulish} to give to the sliver a meaning
of a state in the
Hilbert space. Just due to the infinite dimensional nature of
the  star product
appear associativity anomaly \cite{Horowitz1,Horowitz2,AM} and twist
anomaly \cite{HataSliver2}.

 Several tools are used to study VSFT:
\begin{enumerate}
\item the functional approach, where string fields are discussed in terms of
left-right modes functionals, is developed in \cite{GT01},\cite{GT02},

\item the operator approach, where explicit Fock-space operators for
left-right parts of the string are explored, is presented in
\cite{PR99}.
\end{enumerate}

Using these two descriptions the sliver states and their
generalizations are studied in \cite{rsz-3,0105184,0106242,Matsuo,
0110204}. In these lectures (Section~\ref{sec:SFTM})
we will discuss the half-string formalism
with main focus on the operator description of the matter sector.

In \cite{F2}
a  simple linear form for ${\cal Q}=c_o+\sum f_n(c_n+(-1)^nc^{\dag}_n)$
was proposed.
Under this assumption
the ghost part of the equation  admits an analytical investigation
\cite{0110124,HataSliver2,MooreTaylor,0111087,rsz-6,rsz-7}.
Since an analytic tachyon vacuum solution is unknown, it is impossible
to derive this form of VSFT action \cite{F2} from the initial Witten
action (\ref{WA}). However, one can check if the proposed VSFT action passes some
tests to be acceptable. In particular,
in \cite{0108150} a classical solutions of VSFT that reproduces the
perturbative open string vacuum is  constructed
(see \cite{HataSliver2,rsz-7}). Several attempts of
getting closed strings from VSFT have been recently performed
\cite{0111092,rsz-6}.

 This consideration concerns
only bosonic cubic SFT.
To study the same questions in the superstring case it
is worth to work with a formalism
which deals with equations of motions in the simple form
(\ref{SEQ}) without any modifications. String fields in
$-1$ picture   \cite{AMZ1},
\cite{PTY},\cite{AMZ2} (see also \cite{Thorn}) give us a suitable
 formalism. We will discuss this formalism in Section 6.

Let us mention that there are several covariant string field theories.
We collect them in the Table below.
\bigskip
\begin{center}
\renewcommand{\arraystretch}{1.2}
\begin{tabular}{||l|l||}
\hline
$~~~~~~$Bosonic strings & $~~~~~$Superstrings
\\
\hline
1. Witten's Open Cubic SFT$^*$ & 1. Modified Cubic SFT$^*$
\\
&$\quad$ a) 10D SUSY, GSO$+$ sector
\\
&$\quad$ b) no SUSY, GSO$+$ and GSO$-$ sectors
\\
\hline
2. Nonpolynomial Closed SFT & 2. Nonpolynomial Super SFT
\\
\hline
3. Background independent OSFT & 3. Background independent OSSFT.
\\
\hline
\end{tabular}
\end{center}
\bigskip

By $^*$
we mark the theories which we are going to discuss in details in
these lectures
(see also a review paper \cite{0102085} and refs. therein about the others).

In this lecture notes we would like to explain and discuss some ideas
concerning noncommutative
geometry in general, as well as noncommutative field theories and string field theories.
The material of the lectures is organized as follows.
In Section 2, we explain basic mathematical notions of noncommutative
differential geometry (NCDG) and give the most important examples of
noncommutative
spaces used in physical applications.
In Section 3, we consider noncommutative quantum field theories emphasizing
an issue of their renormalizability. Also we pay a special attention to the
problem of the UV/IR mixing.
In Sections 4-8, we  expose  cubic (super)string field theories
focusing on the recent developments in this area.

One can read the text using the following logic scheme.
\begin{figure}[h!]
\centering
\includegraphics[width=230pt]{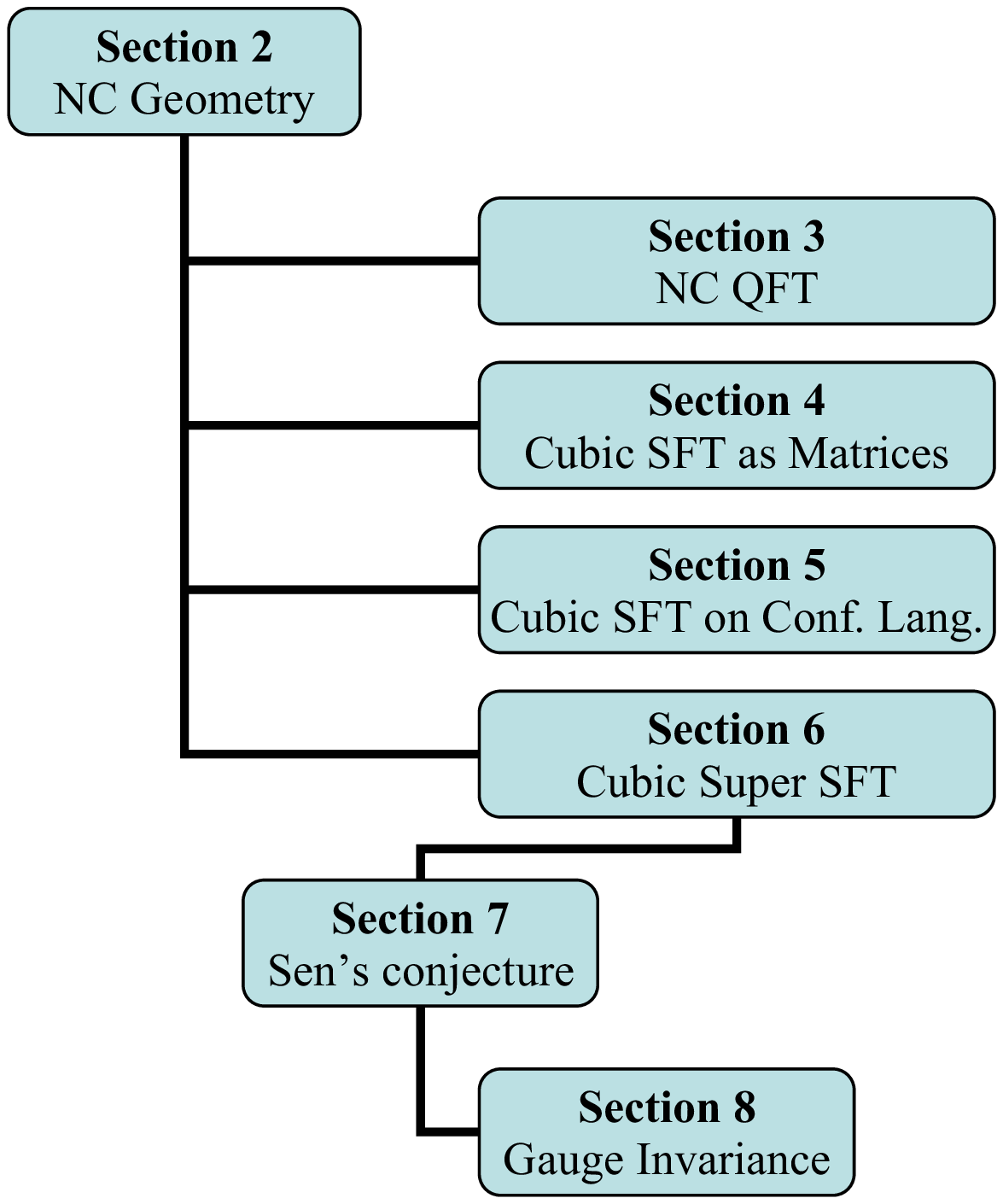}
\end{figure}
\newpage
Let us say a few words about the style
which we tried to follow in
these lecture notes.
We do not pretend to write a textbook
on noncommutative quantum filed theories and string field theories,
but would like to present key ideas about these subjects
and show explicitly how to perform the calculations.

Sometimes the reader may find our exposition very technical but we
try to be
explicit and to keep track  of  instructive computations.
However, not  to make a text extremely cumbersome
 we will  present in several cases
just the results.

We do not pretend to put the complete list of the references on the subject,
and apologize to those authors whose papers are not in the list.

\newpage
\section{Noncommutative Geometry. Examples.}
\label{sec:NC}
\setcounter{equation}{0}

In this section we consider the basic ideas of the noncommutative
geometry \cite{book} and give several examples of noncommutative
spaces.

Noncommutative geometry uses a generalization of the known duality
between a space and its algebra of functions, see
\cite{book,Mad,CDS}. If one knows the associative commutative
algebra ${\cal A}(M)$ of complex-valued functions on topological
space $M$ then one can restore the space $M$. Therefore, all
topological notions can be expressed in terms of algebraic
properties of ${\cal A}(M)$. For example, the space of continuous
sections of a vector bundle over $M$ can be regarded as a projective
${\cal A}(M)$-module. (We are speaking about left modules. A
projective module is a module that can be embedded into a free
module as a direct summand). So, the vector bundle over a compact space
$M$ can be identified with projective modules over ${\cal A}(M)$.

\subsection{Axioms of Noncommutative
Differential Geometry}
 Let ${\cal A}$
be an abstract  associative algebra with the multiplication
$\star$
\begin{equation}
\star :\; {\cal A}\otimes {\cal A}\to {\cal A}
\label{star2}
\end{equation}
One can
interpret  ${\cal A}$ as an
"algebra
of functions" on a (nonexisting) noncommutative space.

Let us introduce some basic geometrical notions on this algebra $\Ac$:
external  derivative $Q$ (or an algebra ${\cal G}$ of derivatives)
\begin{subequations}
\label{axiom-set1}
\begin{equation}
Q: {\cal A}\to {\cal A}
\label{Q}
\end{equation}
and integral (linear map)
\begin{equation}
\int: {\cal A}\to \Ch.
\label{integral}
\end{equation}
 Noncommutative space is the  set
\begin{equation}
\left(\Ac,Q,\int\right),
\label{triplet}
\end{equation}
which satisfies the following axioms:
\begin{itemize}
\item associativity
\begin{equation}
A\star (B\star C)=(A\star B)\star C,~~A,B,C \in {\cal A};
\label{assoc}
\end{equation}
\item nilpotency
\begin{equation}
Q^2=0;
\label{nilp}
\end{equation}
\item Leibnitz rule
\begin{equation}
Q(A\star B)=(QA)\star B)+A\star (Q B),~~A,B \in {\cal A};
\label{Leibnitz}
\end{equation}
\item cyclicity property
\begin{equation}
\int (A\star B)=\int (B\star A);
\label{tr}
\end{equation}
\item ``integration by part''
\begin{equation}
\int (QA)=0.
\label{int-by-part}
\end{equation}
\end{itemize}
\end{subequations}

In the most applications the algebra $\Ac$ is also
equipped with a natural $\Zh_2$-grading: $|A\star B|=|A|+|B|$.
For example, in Section~\ref{sec:NCFT} $\Zh_2$-grading
is usual boson/ferimon super-grading.
One has to modify Leibnitz rule (\ref{Leibnitz}) and
cyclicity property (\ref{tr})
to be in agreement with  $\Zh_2$-grading:
\begin{equation}
Q(A\star B)=(QA)\star B+(-1)^{|A|}A
\star (Q B),~~A,B \in {\cal A};
\tag{\ref{Leibnitz}$'$}
\label{Leibnitz-gr}
\end{equation}
\begin{equation}
\int (A\star B)=(-1)^{|A||B|}\int (B\star A),~~A,B \in {\cal A}
\tag{\ref{tr}$'$} \label{tr-gr}
\end{equation}
where $(-1)^{|A||B|}=-1$ if both $A$ and $B$ are odd elements
and equal 1 otherwise.

If ${\cal A}$ admits a Lie algebra of derivatives  one can develop
a noncommutative bundles theory  generalizing the
standard theory.  Let ${\cal G}$  be a Lie algebra of derivatives on
 ${\cal A}$ and $\alpha _1,\dots,\alpha _d$ be generators of  ${\cal G}$.
Then one can generalize the notion of non-commutative space
\eqref{triplet} to the set
 \begin{subequations}
 \label{axiom-set2}
\begin{equation}
\left({\cal A},{\cal G},\int\right),
\label{triplet-G}
\end{equation}
which satisfies axioms (\ref{assoc}) and (\ref{tr})
as well as
\begin{equation}
\alpha _i(A\star B)=(\alpha _iA)\star B+A\star (\alpha _i B),~~
A,B \in {\cal A},~~\alpha _i\in {\cal G};
\label{Leibnitz-G}
\end{equation}
and
\begin{equation}
\int \alpha _i ~A=0.
\label{int-by-part2}
\end{equation}
\end{subequations}
If $V$ is a projective module over  ${\cal A}$ (i.e.
 a ``vector bundle over
 ${\cal A}$'') one defines a connection in $V$ as the set of linear operators
 $\nabla_1,\dots,\nabla_d$ on $V$ satisfying
 \begin{equation}
\label{n.1}
 \nabla_i(a\phi)=a\nabla_i(\phi)+\alpha _i(a)\phi
\end{equation}
for all $a\in  {\cal A}$ and $\phi \in V,~~i=1,\dots ,d.$
The curvature of connection $\nabla_i$ is
\begin{equation}
\label{n.2}
 F_{ij}=\nabla_i\nabla_j-\nabla_j\nabla_i
 -f_{ij}^k\nabla_k
 \end{equation}
belongs to the algebra  of
endomorphisms of the ${\cal A}$-module $V$.

 A profit from such kinds of constructions is that we can immediately
write an action that is invariant under gauge transformations.
If we deal with the set of axioms (\ref{nilp})-(\ref{Leibnitz-gr})
we can define the gauge transformation as
\begin{equation}
\delta A=Q\Lambda+A\star \Lambda -\Lambda\star A,~~A,\Lambda \in {\cal A}.
\label{gaugetr}
\end{equation}
The gauge invariant action is \cite{Witten}
\begin{equation}
S[A]=\frac12\int A\star QA+\frac13\int A\star A\star A.
\label{action}
\end{equation}
This is a generalization of the Chern-Simons action.

In the case of axioms (\ref{assoc}), (\ref{tr}), (\ref{n.1}),
 (\ref{Leibnitz-G}) and (\ref{int-by-part2}) one
has a generalization of the Yang-Mills action
\begin{equation}
\label{g.3}
S=-\frac14 \int  (F_{ij}F^{ij})
\end{equation}
 with
 \begin{equation}
\label{g.32}
F_{ij}=\alpha _i A_j-\alpha _j A_i+A_i\star A_j- A_j\star A_i.
\end{equation}
The action is invariant under the following gauge transformations
\begin{equation}
\label{g.4}
\delta (A_i)=\alpha _i(\omega)+ A_i\star  \omega-\omega\star A_i,
\end{equation}
where $\omega\in {\cal A}$ is a gauge parameter.

There are several examples of realization of axioms (2.2)
and (2.3).
We will  discuss the following realizations of axioms
\eqref{axiom-set2}:
\begin{itemize}
\item
The d-dimensional noncommutative torus $\mathbb{T}_{\theta}^d$.
\item
The d-dimensional noncommutative Euclidian space $\Rh_{\theta}$.
\end{itemize}

String field theory (SFT) presents an example of axioms (2.2).
There is also a realization of axioms (2.2) on matrices \cite{0109182}.

\subsection{Examples.}

\subsubsection{Noncommutative torus $\mathbb{T}_{\theta}^d$.}

The d-dimensional noncommutative torus is defined by its algebra
 ${\cal A}_\theta$ with generators $U_1,\dots,U_d$
 satisfying the relations
$$
 U_iU_j=e^{2i\pi \theta _{ij}}U_jU_i
$$
 where $i,j=1,\dots,d$ and $\theta =(\theta _{ij})$
 ia a real antisymmetric matrix.
 The algebra ${\cal A}_\theta$ is equipped with an antilinear involution $*$
 obeying $U_i^*=U_i^{-1}$ (i.e. ${\cal A}_\theta$ is a *-algebra).
 An element of ${\cal A}_\theta$ is a power series
$$
 f=\sum f(p_1,...,p_d)U_1^{p_1}...U_d^{p_d}
$$
 where $p=(p_1,...,p_d)\in Z^d$ and the sequence of complex coefficients
 $f(p_1,...,p_d)$ decreases faster than any power of
 $|p|=|p_1|+...+|p_d|$ when $|p| \to \infty$.
 The function $f(p)$ is called the symbol of the element $f$.
 We denote by $U^p$ the product $U_1^{p_1}...U_d^{p_d}$.
 Then one has
 $U^pU^k=e^{2i\pi \varphi (p,k)}U^{p+k},$
 where $\varphi (p,k)=\sum \varphi_{ij}p_ip_j$
 and $\varphi_{ij}$ is a matrix obtained from $\theta$
after deleting all its elements below the diagonal.
To simplify the product rule we replace $U^p$
by $e^{i\pi \varphi (p,q)}U^p$ so that we have
$$
 U^pU^k=e^{i\pi \theta (p,k)}U^{p+k}.
$$

If $f$ and $g$  are two elements of ${\cal A}_\theta$,
$$
 f=\sum _pf(p)U^{p},~~~ g=\sum _k g(k)U^{k}
$$
then the product
$$
 fg=\sum _{p,k}f(p)g(k)U^{p}U^{k}=
$$
$$
 =\sum _{p,k}f(p)g(k)e^{2i\pi \theta (p,k)}U^{p+k}=
 \sum _q(f\star g)(q)U^{q},
$$
where the star-product $(f\star g)(q)$ of symbols $f(p)$ and
$g(k)$ is
$$
 (f\star g)(q)=\sum _{p}f(p)g(q-p)e^{i\pi \theta (p,q-p)}.
$$

The differential calculus on the noncommutative torus is introduced by means
of the derivations $\partial _j$ defined as
$$
 \partial _j U^p=ip_jU^p,~~j=1,\dots,d.
$$
They satisfy the Leibnitz rule $\partial _j(fg)=
\partial _jf \cdot g+ f\cdot \partial _j g$
for any $f,g$ $\in {\cal A}_\theta$.

The integral of
$f=\sum f(p)U^p$ is defined as
$\int f=f(0)$, which is in correspondence with the commutative case.
The integral has the property of being the trace on the algebra
${\cal A}_\theta$, i.e. $\int fg=\int gf$
for any $f,g\in {\cal A}_\theta$. Moreover one has
$$
 \int \partial _j f\cdot g=-\int \partial _j g\cdot f.
$$

The gauge field $A_i$ on the noncommutative torus is defined as
$$
A_i=\sum _{p\in Z^d}A_i(p)U^p,~~i=1,\dots,d.
$$
Here $A_i(p)$ is a sequence of $N\times N$ complex matrices
indexed by a space-time index. It corresponds to the Fourier
representation of the ordinary gauge theory on commutative torus.
The gauge field is antihermitian, $A_i^*=-A_i$, or
$A_i(p)^*=-A_i(-p)$. $A_i$ is an element of a matrix algebra with
coefficients in ${\cal A}_\theta$ and its curvature is defined as
$$
F_{ij}=\pd_i A_j-\pd_j A_i+[A_i,A_j],
$$
where $\partial _i$ is a derivative on ${\cal A}_\theta$ defined
above. The Yang-Mills action is
$$
S=-\frac14 \int \mbox{tr} (F_{ij}F^{ij}).
$$
This action is invariant under gauge transformations
$$
A_i\to \Omega A_i \Omega ^{-1}+\Omega \partial _i \Omega ^{-1},~~
F_{ij}\to \Omega F_{ij}\Omega ^{-1},
$$
where $\Omega$ is a unitary element of the algebra of matrices
over ${\cal A}_\theta$.

\subsubsection{Noncommutative flat space $\Rh_{\theta}$.}

The main property of the noncommutative flat space $R_{\theta}^d$
space is the non-zero commutator of the coordinates
$$
[x_i,x_j]=i\theta_{ij}
$$
where $\theta$ is called the parameter of noncommutativity. We
begin with the formulation of the Moyal product as one of the
important realization of such multiplication law.

Let us consider one-dimensional quantum mechanics in the Hilbert
space of square integrable functions on the real line $L^2({\bf
R})$ with ordinary canonical operators of position $\hat{q}$ and
momentum $\hat{p}$,
$$
[\hat{p},\hat{q}]=-i\hbar,
$$
acting as $\hat{q}\psi(x)=x\psi(x)$, $\hat{p}\psi(x)=-i \hbar
d\psi(x)/dx$. If a function of two real variables $f(q,p)$ is
given in terms of its Fourier transform
$$
f(q,p)=\int e^{i(rq+sp)}\tilde{f}(r,s)dr ds,
$$
then one can associate with it an operator $\hat{f}$ in $L^2({\bf
R})$ by the following formula
$$
\hat{f}=\int e^{i(r\hat{q}+s\hat{p})}\tilde{f}(r,s)dr ds.
$$

This procedure is called the Weyl quantization and the function
$f(q,p)$ is called the {\textit symbol} of the operator $\hat{f}$
\cite{BFD,BerS}. One has the correspondence
$$
\hat{f} \longleftrightarrow f=f(q,p).
$$
If $\hat{f}^{*}$ is the Hermitian adjoint to $\hat{f}$ then its
symbol is $f^{*}=f^{*}(q,p)$
$$
\hat{f}^{*} \longleftrightarrow \hat{f}^{*}(q,p)=\bar{f}(q,p).
$$
If two operators $\hat{f}_1$ and $\hat{f}_2$ are given with
symbols $f_1(q,p)$ and $f_2(q,p)$ then the symbol of product
$\hat{f}_1\hat{f}_2$ is given by the Moyal product $f_1\star
f_2=(f_1\star f_2)(p,q)$ as \cite{BFD,BerS}
$$
(f_1\star f_2)(p,q)= \sum _{\alpha, \beta}
\frac{(-1)^{\beta}}{\alpha ! \beta
!}(\frac{i\hbar}{2})^{\alpha+\beta} (\partial ^\alpha _q \partial
^\beta _p f_1 (p,q)) \cdot (\partial ^\beta _q \partial ^\alpha _p
f_2 (p,q))=
$$
$$
=e^{i\hbar
L}(f_1(q_1,p_1)f_2(q_2,p_2))|_{q_1=q_2=q,~~p_1=p_2=p},$$ where
$$
L=\frac{1}{2}(\frac{\partial ^2}{\partial q_1 \partial p_2}-
\frac{\partial ^2}{\partial q_2 \partial p_1}).
$$
This is also can be written as (we assume $\hbar =1$)
$$
(f_1\star f_2)(p,q)= \frac{1}{(2\pi )^2} \int
e^{2i[(q-q_2)p_1+(q_1-q)p_2+(q_2-q_1)p]}f_1(q_1,p_1)f_2(q_2,p_2)
dq_1dq_2dp_1dp_2.
$$
Introducing the constant Poisson structure on the plain $\omega
^{\mu\nu}=-\omega ^{\nu \mu}$, $\omega ^{12}=1$ and notations
$x=(q,p)$, $x_i=(q_i,p_i)$, $i=1,2$ , $x_1\omega
x_2=q_1p_2-q_2p_1$ one obtains
$$
(f_1\star f_2)(x)=\frac{1}{(2\pi)^2}\int f_1 (x_1)f_2(x_2)
e^{2i(x\omega x_1+x_1\omega x_2+x_2\omega x)}dx_1dx_2.
$$
On the other hand the exponential factor can be rewritten as
$$
2i(x\omega x_1+x_1\omega x_2+x_2\omega x)=2i\int _\triangle pdq,
$$
where $\triangle$ is the triangle on the plane with vertices
$x=(p,q)$, $x_1=(p_1,q_1)$ and $x_2=(p_2,q_2)$ and there is the
path integral representation
$$
e^{2i\int _{\triangle} pdq}=\int e^{2i\int _\triangle pdq}\prod
dx(\tau).
$$
One integrates over trajectories $x=x(\tau)$, $0\leqslant \tau
\leqslant 1$ in the phase plane with the boundary conditions
$$
x(0)=x(1)=x,~~ x(1/3)=x_1,~~x(2/3)=x_2.
$$

Therefore, we conclude that the Moyal product is represented in
the form
$$
(f\star g )(x)=\int K(x,x_1,x_2)f(x_1)g(x_2)dx_1dx_2
$$
and the kernel $K(x,x_1,x_2)$ has a path integral representation
$$
K(x,x_1,x_2)=\int e^{iS}\prod dx(\tau),
$$
where $S=2\int pdq$ and the path integral is taken over the
trajectories $x=x(\tau), $ $0\leqslant \tau \leqslant 1$ subject
to the boundary conditions specified above. About a connection with
 deformation quantization see \cite{AV} and refs. therein.

\subsubsection{Noncommutative sphere $S^2_\theta$.}

The last example of the noncommutative space we would like to
describe in this section is the noncommutative sphere $S^2_\theta$
introduced by Madore \cite{Mad} (see also \cite{S2}).
This space is
associated with an algebra of operators $x^i$, $i=1,2,3$ satisfying the
relations

$$
[ x^i, x^j ] = i\theta\epsilon^{ijk} x^k
$$
and
$$
\sum_{i=1}^{3}(x^i)^2=R^2.
$$

 The derivative and curvature can
be defined in the following form:
$$
\partial_i f = [X^i, f]
$$
and
$$
F_{ij} = i[\partial_i + A_i,\partial_j + A_j] -
i[\partial_i,\partial_j].
$$

\newpage
\section{Noncommutative Field Theory Models.}
\label{sec:NCFT}
\setcounter{equation}{0}

In this section we  discuss main features of
field theory models on $R_\theta$. We pay the most attention to the
problem of the renormalizability of these theories
and the UV/IR mixing.

\subsection{Noncommutative $\varphi ^4$-model.}
\label{sec:ncft.phi4}

The simplest nontrivial model
which can demonstrate the most important  properties of
the noncommutative field theories is a non-commutative $\varphi ^4$-model
\cite{filk}.

The theory is defined by the action
\begin{equation}
\label{nc.phi4.action} S=S_0+S_{int}=\int d^dx\,[
\frac{1}{2}(\pd_{\mu}\varphi)^2 + \frac{1}{2}m^2\varphi^2 + \frac
g{4!}(\varphi \star\varphi \star\varphi \star\varphi)(x)],
\end{equation}
where $\star$ is a Moyal product
$$
(f\star
g)(x)=e^{i\xi\theta^{\mu\nu}\pd_{\mu}\otimes\pd_{\nu}}f(x)\otimes
g(x),
$$
$\xi$ is a deformation parameter, $\theta^{\mu\nu}$ is a
non-degenerate skew-symmetric real constant matrix, $\theta^2=-1$,
$d$ is even.
In this section  we deal with
four dimensional Euclidean space.
Also, we introduce the convenient notation
$$p_1\wedge p_2\equiv\xi p_1\theta p_2.$$
Let us
rewrite the interaction term in the Fourier representation
\begin{equation}\label{nc.phi4.int}
S_{int}=\frac{g}{4!(2\pi)^d}\int dp_1dp_2dp_3dp_4\, e^{-i p_1\wedge
 p_2-i p_3\wedge p_4}
\varphi(p_1)\varphi(p_2)\varphi(p_3)\varphi(p_4)\delta(p_1+p_2+p_3+p_4).
\end{equation}
There are  the following distinguished properties of the deformed theory
as compared with standard local $\varphi^4_d$ model:
\begin{itemize}
\item
There are non-local phase factors in the vertex.
\item
These factors provide regularization for some loop integrals but
not for all \cite{ren1,ren2,ren3,AV,SB}.
\item
To have renormalizability  the sum of divergences in each order of
perturbation theory must have a phase factor already present in
the action.
\end{itemize}

To single out phase factors it is convenient  to use the 't Hooft
double-line graphs and a notion of planar graphs. For planar
graphs the  phase factors do not affect the Feynman integrations
at all (see \cite{filk} for details).
In particular the planar graphs have exactly
the same divergences as in the commutative theory \cite{SB}. There
are no superficial divergences in nonplanar graphs since they are
regulated by the phase factors \cite{ren1,ren2,ren3,AV,SB,Ch}. Moreover,
oscillating phases regulate also divergent subgraphs, unless they
are not divergent planar subgraphs.

\begin{figure}[h]
\begin{center}
\epsfig{file=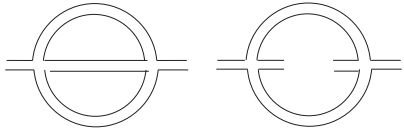,
   width=150pt,
  angle=0,
 }
\caption{Planar graph with a nonplanar subgraph.}
\label{nc.phi4.Fnp}
\end{center}
\end{figure}

So, at first sight  it seems that the proof of renormalizability
is rather trivial. One has divergences only in planar graphs, so
one can use the fact that planar theory is renormalizable if its
scalar counterpart does \cite{Hooft}. In other words, one can
expect that it is enough to take the planar approximation of an ordinary theory,
find divergences
within this approximation and write  counterterms in
the noncommutative theory as  divergent parts of planar graphs
multiplied  on the phase factors. However, these arguments work
only for superficial divergences. The situation is more subtle for
divergent subgraphs. The reason is that a planar graph can contain
nonplanar subgraphs (see  simple example on Fig.\ref{nc.phi4.Fnp})
and these divergences should be also removed. Therefore,
renormalizability of the theory (\ref{nc.phi4.action}) is not
obvious. This gave us \cite{ABK1} a reason to consider explicitly the two-loop
renormalization of the theory (\ref{nc.phi4.action})
and the authors of
\cite{Ch} to study the renormalizability  of
(\ref{nc.phi4.action}) up to an arbitrary
order.

\begin{figure}[h]
\begin{center}
\epsfig{file=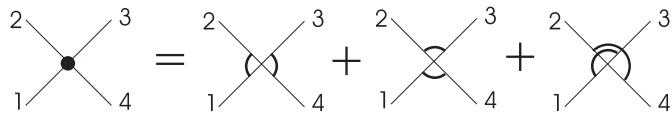,
   width=250pt,
  angle=0,
 }
\caption{The symmetric vertex.} \label{nc.phi4.Fint}
\end{center}
\end{figure}

In explicit calculations we will use single-line graphs and
symmetric vertices. After the symmetrization of (\ref{nc.phi4.int}) we
get
\begin{eqnarray}
S_{int}=\frac{g}{3\cdot 4!}\frac{1}{(2\pi)^d}\int
dp_1dp_2dp_3dp_4\,
\varphi(p_1)\varphi(p_2)\varphi(p_3)\varphi(p_4)\delta(p_1+p_2+p_3+p_4)
\nonumber\\ \times\left[\cos(p_1\wedge p_2)\cos(p_3\wedge
p_4)+\cos(p_1\wedge p_3) \cos(p_2\wedge p_4) +\cos(p_1\wedge
p_4)\cos(p_2\wedge p_3)\right]\nonumber
\end{eqnarray}
and the vertex is a sum of three terms (see
Fig.\ref{nc.phi4.Fint}).

Here we would like to say a few words about  a general technique of
calculations in the presence of non-local phase factors in the
vertex. In our calculations we always denote external momenta by
$p_i$ and loop momentum by $k$.
In order to perform an integration over loop momenta the
$\alpha$-representation and Feynman parametrization are used. The
useful formulae are the following \cite{BogSh}
$$
\frac 1{F_1F_2\dots
F_n}=(n-1)!\int_0^1\frac{\delta(x_1+x_2+\dots+x_n-1)dx_1dx_2\dots
dx_n}{(x_1F_1+x_2F_2+\dots+x_nF_n)^n},
$$
$$
\frac 1{F^s}=\frac
1{\Gamma(s)}\int_0^{\infty}\alpha^{s-1}e^{-\alpha F}d\alpha.
$$
These expressions are used to pass to the Gauss type integral over
internal momentum
\begin{equation}
\int e^{-\omega {\cal A}
\omega+l\omega}d^D\omega=\frac{\pi^{D/2}}{\sqrt{\det {\cal
A}}}e^{\frac 14l{\cal A}^{-1}l} \label{nc.phi4.gauss}.
\end{equation}
Namely, an expression for an arbitrary graph is a product of $n$
propagators with exponential factor coming from trigonometric
structure of the vertex. We put the parameter $\alpha_i$ in
correspondence to the $i$-th propagator.

The most general expression for a one loop graph can be written
schematically as
$$
L_1=\int d^dk \exp\left[-(ak^2+lk+M^2)+ib\theta k\right],
$$
where $a$ depends on $\alpha_i$, $l$ and $M$ depend on $\alpha_i$
and $p_j$, $b$ depends on $p_j$. Using (\ref{nc.phi4.gauss}) the
integration over $k$ is evident and the result is
$$
L_1=\frac{\pi^{d/2}}{a^{d/2}}e^{-M^2+\frac
1{4a}(l^2-b^2+2il\theta b)}
$$

The next step is an integration over parameters $\alpha_i$. There
are two formulae we work with
\begin{equation}
\int_0^{\infty}\alpha ^{\nu-1}e^{-\mu \alpha }d\alpha =\frac
1{\mu^{\nu}}\Gamma(\nu)\label{alphaa},
\end{equation}
\begin{equation}
\int_0^{\infty}\alpha ^{\nu-1}e^{-\gamma
\alpha -\frac{\beta}{\alpha }}d\alpha =2\left(\frac{\beta}{\gamma}\right)^{\nu/2}
K_{\nu}(2\sqrt{\beta\gamma})
\label{alphab}.
\end{equation}
Note that
function $\Gamma(\nu)$ has simple poles at points $\nu=0, -1, -2,\dots$,
which correspond to UV divergencies
(the momentum space ultraviolet divergencies become
 a small  $\alpha$ divergence). Meanwhile the RHS of (\ref{alphab})
for $\beta \neq 0$ has no singularities  at points $\nu=0, -1, -2,\dots$,
since
 the factor
$e^{-\frac{\beta}{\alpha}}$ produces a regularization
for small $\alpha$. It is easy to
show that the integral
$$
\int L_1d\alpha_1\dots d\alpha_n
$$
also is well defined if $b\neq 0$.

Now we are going to compute explicitly one-loop counterterms using
dimensional regularization $d=4-2\epsilon$.
We will also present
the explicit form of finite part for two point and four point 1PI
functions $\Gamma ^{(2)}$
and $\Gamma ^{(4)}$ in the one loop
approximation. We use the standard notations for perturbation
expansion of 1PI-functions $$\Gamma ^{(i)}=\sum _n g^n\Gamma
_{n}^{(i)},$$ $$\Gamma ^{(2)}=\Gamma _{f.p.}^{(2)}+\Delta\Gamma
^{(2)} $$ and $$\Gamma ^{(4)}=\Gamma _{f.p.}^{(4)}+\Delta \Gamma
^{(4)}.$$
\begin{figure}[h]
\begin{center}
\epsfig{file=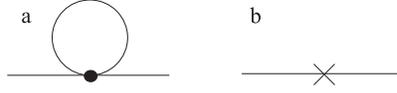,
   width=150pt,
  angle=0,
 }
\caption{$\Gamma _{1}^{(2)}-\Delta\Gamma _{1}^{(2)}$.}
\label{nc.phi4.F1PI2}
\end{center}
\end{figure}
The only graph \ref{nc.phi4.F1PI2}a contributes to $\Gamma
_{1}^{(2)}$ and
$$
\Gamma _{1}^{(2)}=-\frac{g(\mu^2)^{\epsilon}}{6(2\pi)^d} \int
dk\frac{2+\cos 2p\wedge k}{k^2+m^2}.
$$
Here we have introduced a new parameter $\mu$ of dimension of mass to
leave the action dimensionless in $4-2\epsilon$ dimensions.

We present this expression as a sum of planar and nonplanar parts
$$
\Gamma _{1}^{(2)}=\Gamma _{1,pl}^{(2)}+\Gamma _{1,npl}^{(2)}
$$
where
$$
\Gamma _{1,pl}^{(2)}=-\frac{g(\mu^2)^{\epsilon}}{6(2\pi)^d} \int
dk\frac{2}{k^2+m^2},
$$
$$
\Gamma _{1,npl}^{(2)}=-\frac{g(\mu^2)^{\epsilon}}{6(2\pi)^d} \int
dk\frac{\cos 2p\wedge k}{k^2+m^2}.
$$
We see that only the planar part has a divergence
\begin{equation}
\Gamma _{1,pl}^{(2)}=\frac{g}{32\pi^2}\frac23
m^2\left(\frac{1}{\epsilon}+\psi(2)-\ln\frac{m^2}{4\pi\mu^2}+O(\epsilon)\right)
\label{nc.phi4.1PI2'}
\end{equation}
and this divergent part is subtracted by the counterterm
\ref{nc.phi4.F1PI2}b.

The nonplanar part can be calculated explicitly using
$\alpha$-representation
\be
\Gamma _{1,npl}^{(2)}=-\frac{g}{32\pi^2}\frac23
\sqrt{\frac{m^2}{\xi^2p^2}}K_1(2m\xi |p|)
\label{nc.phi4.1PI2}
\end{equation}
and as has been mentioned above has no divergencies.
 Note that this representation takes place
only for $p\neq 0$.

\begin{figure}[h]
\begin{center}
\epsfig{file=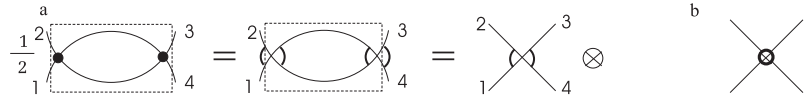,
   width=300pt,
  angle=0,
 }
\caption{a) $\Delta\Gamma _{2,s}^{(4)}$, $\quad$ b) Cross denotes
the 4-vertex counterterm.} \label{nc.phi4.F1PI4}
\end{center}
\end{figure}

$\Gamma _{2}^{(4)}$ is a  sum of s-, t- and u-channel graphs,
$\Gamma _{2}^{(4)}=\Gamma _{2,s}^{(4)}+ \Gamma _{2,t}^{(4)}+\Gamma
_{2,u}^{(4)}$. The explicit form of $\Gamma _{2,s}^{(4)}$ is
$$
\Gamma_{2,s}^{(4)}=\frac{g^2(\mu^2)^{2\epsilon}}{18(2\pi)^{d}}\int
\frac{\Pp_{\ref{nc.phi4.F1PI4}a}(\{p\},k)}{(k^2+m^2)((P+k)^2+m^2)}
dk
$$
where $P=p_1+p_2$,
$$
\Pp_{\ref{nc.phi4.F1PI4}a}(\{p\},k)=[2\cos(p_1\wedge
p_2)\cos(k\wedge P)+\cos(p_1\wedge p_2+(p_1-p_2)\wedge k)]\times
$$
$$
[2\cos(p_3\wedge p_4)\cos(k\wedge P)+\cos(p_3\wedge
p_4+(p_4-p_3)\wedge k)].
$$
The trigonometric polynomial
$\Pp_{\ref{nc.phi4.F1PI4}a}$  can be rewritten in the form
\begin{equation}           \label{nc.phi4.tp1PI4}
\Pp_{\ref{nc.phi4.F1PI4}a}=2\cos(p_1\wedge p_2)\cos(p_3\wedge
p_4)+ \sum _{j} {}' c_j e^{i\Phi^j(p)+ib^j(p)\wedge k}
\end{equation}
where the sum $\sum '$ goes over all $j$ for which linear
functions $b^j(p)$ are nonzero for almost all $\{p\}$. This
representation is useful to separate the terms which do not depend
on internal momentum and, therefore, produce divergencies. Other
terms have phase factors which contain $k$ and provide a finite
answer. Thus, the only first term in (\ref{nc.phi4.tp1PI4})
contributes to a pole part (see Fig. \ref{nc.phi4.F1PI4}a) and we
have
$$
\Gamma _{2,s}^{(4)}=\frac{g^2}{32\pi^2}\frac29
\left[\cos(p_1\wedge p_2)\cos(p_3\wedge p_4)\left(\frac
1{\epsilon}+\psi(1)-\int_0^1dx\ln\frac{m^2+x(1-x)P^2}{4\pi\mu^2}
\right) +\right.$$
$$
+\left.\sum_j {}'c_j e^{i\Phi^j(p)}
\int_0^1dxK_0(\sqrt{\xi^2{b^{j}}^2(m^2+P^2x(1-x))}) e^{ixb^j\wedge
P}\right].
$$
This representation is well defined only if all $b^{j}$ are
nonzero, i.e. for non exceptional momenta. It is the matter of
a simple algebra to sum up the divergent parts of $\Gamma
_{2,s}^{(4)}$,  $\Gamma _{2,t}^{(4)}$ and $\Gamma _{2,u}^{(4)}$ to
obtain a graph with the symmetric vertex.

Therefore, at the one-loop we have
$$
\Delta \Gamma_{1l}^{(2)}= \frac{g}{48\pi^2}\frac{m^2}{\epsilon},
$$
$$
\Delta \Gamma_{1l}^{(4)} =\frac{g^2}{16\pi^2}
\frac{1}{9\epsilon}[\cos(p_1\wedge p_2)\cos(p_3\wedge p_4)+ \cos(
p_1\wedge p_3)\cos(p_2\wedge p_4) +\cos(p_1\wedge p_4)\cos(
p_2\wedge p_3)].
$$

\subsection{Noncommutative Complex $\varphi ^4$-model.}

Here  we consider noncommutative quantum field theories
 of complex scalar field
\cite{AV} whose commutative analogue  $(\phi^*\phi)^2$ is
renormalizable in four-dimensional case.
There are only two noncommutative structures that generalize a
commutative quartic interaction $(\phi^*\phi)^2$:
\begin{description}
  \item[(a)] $~~~~~\phi^*\star\phi\star\phi^*\star\phi$,
  \item[(b)] $~~~~~\phi^*\star\phi^*\star\phi\star\phi$.
\end{description}

In the commutative case the quartic interaction $(\phi^*\phi)^2$
is invariant under  local $U(1)$-transformations. In the
noncommutative theory we can consider a "deformed" $U(1)$-symmetry
($U\star U^*=1$).
 One sees that only the structure (a)
is invariant under these transformations. Using (a) and (b) we can
construct an interaction
$$
V(\phi^*,\phi)=A\phi^*\star\phi\star\phi^*\star\phi + B
\phi^*\star\phi^*\star\phi\star\phi=
$$
$$
(A-B)\phi^*\star\phi\star\phi^*\star\phi+
\frac{B}{2}(\{\phi^*,\phi\}\star\{\phi^*,\phi\}),
$$
where $\{,\}$ is the Moyal anticommutator $\{f,g\}=f\star g+g\star
f$. The action of the theory is
\begin{equation}
S=\int d^dx\left(\pd_{\mu}\phi^*\pd_{\mu}\phi+m^2\phi^*\phi+
\lambda V(\phi^*,\phi)\right). \label{nc.c.action}
\end{equation}
Let us rewrite the interaction term in the Fourier components and
symmetrize it, i.e.
$$
V(\phi^*,\phi)=\frac{1}{(2\pi)^4}\int dp_1\dots dp_4\delta(\sum
p_i)\times
$$
$$
\times[A\cos(p_1\wedge p_2+p_3\wedge p_4)+B\cos(p_1\wedge
p_3)\cos(p_2\wedge p_4)] \phi^*(p_1)\phi(p_2)\phi^*(p_3)\phi(p_4).
$$

\begin{figure}[h]
\begin{center}
\epsfig{file=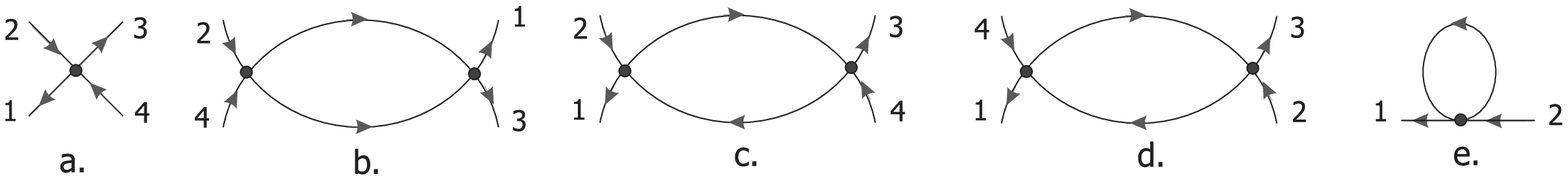,
   width=400pt,
  angle=0,
 }
\caption{The vertex and one-loop graphs.} \label{nc.c.F1PI}
\end{center}
\end{figure}

We would like to analyze counterterms to one loop Feynman graphs
in the theory (\ref{nc.c.action}) and find conditions when this
theory is renormalizable by renormalization of one
coupling constant $\lambda$.
We will show that this takes place only
 in two special cases: $B=0$ and $A=B$. Moreover, in
the case $B=0$ the model does not suffer from IR divergencies at
least at one-loop insertions level.
All one-loop corrections to the
interaction are presented on Fig.\ref{nc.c.F1PI}:b,c,d. "In"
arrows are the fields "$\phi$" and "out" arrows are the fields
"$\phi^*$".

The following analytic expression corresponds to the graph on
Fig.\ref{nc.c.F1PI}:b
\begin{equation}
\Gamma_{\ref{nc.c.F1PI}b}=\frac{N_b(\mu^2)^{2\epsilon}}{(2\pi)^d}\int
d^dk \frac{\Pp_{\ref{nc.c.F1PI}b}(p,k)}{(k^2+m^2)((k+P)^2+m^2)},
\label{nc.c.1PI.b}
\end{equation}
where $N_b$ is a number of graphs ($N_b$=8), $P=p_2+p_4=-p_1-p_3$
and $\Pp_{\ref{nc.c.F1PI}b}(p,k)$ is the trigonometric polynomial
$$
\Pp_{\ref{nc.c.F1PI}b}(p,k)=[A\cos(k\wedge p_2+(-k-p_2)\wedge
p_4)+B\cos(p_2\wedge p_4)\cos(k\wedge P)]
$$
$$
\times [A\cos(p_1\wedge(-k)+p_3\wedge (k-p_1))+B\cos(p_1\wedge
p_3)\cos(k\wedge P)].
$$
The terms containing $\exp[(\dots)\wedge k]$ give a finite
contribution to (\ref{nc.c.1PI.b}). Divergencies come from the
terms $\Delta\Pp_{\ref{nc.c.F1PI}b}$ of the polynomial
$\Pp_{\ref{nc.c.F1PI}b}$
$$
\Delta\Pp_{\ref{nc.c.F1PI}b}=\frac{B^2}{2}\cos(p_1\wedge p_3)
\cos(p_2\wedge p_4).
$$

The graphs Fig.\ref{nc.c.F1PI}:c and \ref{nc.c.F1PI}:d mutually
differ by permutation of momenta $1\leftrightarrow 3$ only and the
analytic expressions for these graphs coincide. For the graph
Fig.\ref{nc.c.F1PI}:c we have
$$
\Gamma_{\ref{nc.c.F1PI}c}=\frac{N_c(\mu^2)^{2\epsilon}}{(2\pi)^d}\int
d^dk \frac{\Pp_{\ref{nc.c.F1PI}c}(p,k)}{(k^2+m^2)((k+P)^2+m^2)},
$$
where $N_c$ is a number of graphs ($N_c=16$), $P=p_1+p_2=-p_3-p_4$
and $\Pp_{\ref{nc.c.F1PI}c}(p,k)$ is the trigonometric polynomial
$$
\Pp_{\ref{nc.c.F1PI}c}(p,k)=[A\cos(p_1\wedge p_2+(-k-P)\wedge
k)+B\cos(p_1\wedge (k+P))\cos(p_2\wedge k)]
$$
$$
\times [A\cos(p_3\wedge p_4+(-k)\wedge (k+P))+B\cos(p_3\wedge
k)\cos(4\wedge (k+P))].
$$
The polynomial $\Delta\Pp_{\ref{nc.c.F1PI}c}$ that gives
contribution to a divergent part of this graph is equal (after
symmetrization $p_2\leftrightarrow p_4$) to
$$
\Delta \Pp_{\ref{nc.c.F1PI}c}=\cos(p_1\wedge p_2+p_3\wedge p_4)
\left[\frac{A^2}{2}+\frac{B^2}{8}\right]+
\frac{AB}{2}\cos(p_1\wedge p_3) \cos(p_2\wedge p_4).
$$
We obtain the same answer for the graph on Fig.\ref{nc.c.F1PI}:d,
i.e.
$$
N_d=N_c,\qquad
\Delta\Pp_{\ref{nc.c.F1PI}c}=\Delta\Pp_{\ref{nc.c.F1PI}d}.
$$

The following condition is equal to one coupling constant
 renormalizability of the theory (\ref{nc.c.action}) at one-loop
 level
$$
N_b\Delta\Pp_{\ref{nc.c.F1PI}b}+2N_c\Delta\Pp_{\ref{nc.c.F1PI}c}=C
[A\cos(p_1\wedge p_2+p_3\wedge p_4)+B\cos(p_1\wedge
p_3)\cos(p_2\wedge p_4)],
$$
where $C$ is a constant. This condition yields two algebraic
equations:
\begin{eqnarray}
N_c\left[A^2+\frac{B^2}{4}\right]=AC,\nonumber\\
N_b\frac{B^2}{2}+N_cAB=BC.\nonumber
\end{eqnarray}
This system is self consistent if
$$
B(BN_c-2AN_b)=0.
$$
The latter equation has two solutions: $B=0$ and $A=B$. Therefore,
one coupling constant renormalizability takes place
at one-loop only in two cases
\begin{eqnarray}
B=0 & \mbox{and} & V(\phi^*,\phi)=A(\phi^*\star\phi)^2,\nonumber\\
A=B & \mbox{and} &
V(\phi^*,\phi)=\frac{B}{2}(\{\phi^*,\phi\})^2.\nonumber
\end{eqnarray}

It is straightforward to check the renormalizability of the
propagator. We do not perform here explicit calculations of the
one-loop correction to the propagator.

Here and in three next subsections we consider the corrections to
the interaction only. Corrections to the propagator are more
simple and we do not write down explicit calculations. The result
is that these corrections do not impose any new restrictions on
the coupling constants in the case of the complex scalar field,
noncommutative $U(1)$ Yang-Mills theory, noncommutative scalar
electrodynamics or $\Nc=2$ noncommutative super Yang-Mills theory.
Nevertheless, we consider one-loop corrections to the propagator
describing the UV/IR mixing in the last subsection of this
section.

\subsection{Noncommutative Yang-Mills Theory.}

In this subsection we explain the very important example of
noncommutative theories: the noncommutative Yang-Mills Theory. But
in contrast to the ordinary commutative theories we have
a nontrivial interaction already with $U(1)$ group. Roughly
speaking, this is noncommutative electrodynamics but it exhibits
properties of the Yang-Mills theory with non-abelian gauge group.
This is because of the deformation of the abelian gauge group.
This deformation was described in \cite{FL}. We have to replace
the ordinary product by the Moyal one. Such a replacement yields
the following form of the deformed group element:
$$
U=e^{\star
i\lambda}=\sum_{n=0}^{\infty}\frac1{n!}(i\lambda)^{\star n}
$$
where the $\star$ denotes the Moyal multiplication.

The $U(1)$ noncommutative connection $\Dd_{\mu}$ is the following:
$$
\Dd_{\mu}=\partial_{\mu}-ig[A_{\mu},\cdot]_{\star},
$$
where $g$ is a coupling constant and
$$
[f,g]_{\star}(x)=(f\star g)(x)-(g\star f)(x) =2i
\sin{(\theta^{\mu\nu}{\partial_\mu}\otimes {\partial_\nu})}\, f(x)\otimes
g(x).
$$
The expressions for the curvature is the following
$$
F_{\mu\nu}=\partial_{\mu}A_{\nu} -\partial_{\nu}A_{\mu}
-ig[A_{\mu},A_{\nu}]_{\star}.
$$
Indeed, the commutator does not vanish and it produces three- and
four-point vertices for the gauge field and a nontrivial interaction
with the Faddeev-Popov ghosts to be introduced below.

The noncommutative $U(1)$ Yang-Mills (NCYM) action is given by
\be
S=-\frac{1} {4}\int d^d x F_{\mu\nu}F^{\mu\nu},
\label{nc.U1.action}
\end{equation}
where we have dropped the $\star$ because of the property of the
integral. This action is invariant under the following
transformation:
$$
A_{\mu}\to U A_{\mu} U^{\dag} +\frac{i}{g} U\pd_{\mu}U^{\dag}
$$

A gauge fixed generating functional is of the form
$$
{\cal Z}[J,\eta,\bar{\eta}]=\int [dA] [dX] [d\Theta][d\bar{C}][dC]
e^{-S[A,X,\Theta]+S_{GF}[A,X,\Theta]+S_{FP}[A,C,\bar{C}]+\text{sources}},
$$
where $C$ and $\bar{C}$ are Faddeev-Popov ghosts, $S_{GF}$ is the
gauge fixing term
$$
S_{GF}[A_{\mu}]=-\frac{1}{2\alpha}\int (\partial_{\mu}A_{\mu})^{2}
$$
and $S_{FP}$ is Faddeev-Popov term
$$
S_{FP}[A_{\mu},C,\bar{C}]=\int
\partial_{\mu}\bar{C}(\partial_{\mu}C-ig[ A_{\mu},C]).
$$

The  generating functional can be computed perturbatively using
Feynman graphs \cite{ren1}-\cite{ren3}. The quadratic terms
(propogators) are identical to the ones appearing in non abelian
gauge theories. However, interaction vertices involve non
polynomial functions of the momenta (see explicit definition of
the Feynman rules in the next subsections).

The generalization of this action to the other groups $U(N)$ is
possible. However, Lie algebras of  $SU(N)$, $SO(N)$ or $Sp(N)$
are not closed under the Moyal commutator and, therefore, cannot
be considered. This fact was shown in \cite{Sus}. Also this
action can be made supersymmetric by adding the correct fermionic
and scalar degrees of freedom \cite{FL,ren2}.

Explicit calculations  \cite{ren1}-\cite{ren3} show that noncommutative U(1)
 gauge theory is renormalizable at one-loop.

\subsection{Noncommutative Scalar Electrodynamics.}

Next, we proceed with a consideration of noncommutative scalar
electrodynamics in Euclidean space $\Rh^4$. The classical action
is given by
\begin{eqnarray}
&S=\int d^4x\left(-\frac{1}{4} F_{\mu\nu}\star F^{\mu\nu}
+(\Dd_{\mu}\phi^*)\star
(\Dd^{\mu}\phi)+V[\phi^*,\phi]\right),&\nonumber
\\
&V[\phi^*,\phi]=\lambda^2(a\phi^*\star\phi\star\phi^*\star\phi+
b\phi^*\star\phi^*\star\phi\star\phi),&\label{nc.sed.action}
\end{eqnarray}
where the covariant derivative is defined by
$\Dd_{\mu}\phi=\pd_{\mu}\phi- ig[A_{\mu},\phi]_{\star}$. $g$ and
$\lambda$ are coupling constants and $a$ and $b$ are fixed real
numbers. It has been shown in \cite{ABK2}, that the pure complex
scalar field theory is one-loop renormalizable only if $a=b$ or
$b=0$. The purpose of the present analysis is to find the
analogous restrictions on $a$ and $b$ in the case of scalar
electrodynamics.

The action (\ref{nc.sed.action}) is invariant under the following
gauge transformations
$$
\phi\mapsto U\star\phi\star U^{\dag},\quad \phi^*\mapsto
U\star\phi^*\star U^{\dag},\quad A_{\mu}\mapsto U\star
A_{\mu}\star U^{\dag}+\frac{i}{g} U\pd_{\mu}U^{\dag},
$$
where $U$ is an element of the noncommutative $U(1)$ group
\cite{FL}. Note that since our fields are in adjoint
representation we could consider the theory with two real scalar
fields instead of one complex.

The Feynman rules for the theory (\ref{nc.sed.action}) are
presented in Table 1. Solid lines denote the scalar fields, "in"
arrows stand for the field $\phi$ and "out" arrows stand for the
field $\phi^*$.

\begin{center}
\renewcommand{\arraystretch}{2.5}
\begin{tabular}{cl}
\epsfig{file=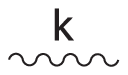,
   width=30pt,
  angle=0,
 }
 & $~~~~~D_{\mu\nu}(k)=\frac{1}{k^2}\left[\delta_{\mu\nu}-
 (1-\alpha)\frac{k_{\mu}k_{\nu}}{k^2}\right]$ \\
\epsfig{file=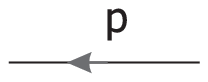,
   width=40pt,
  angle=0,
 }
 & $~~~~~D(p)=\frac{1}{p^2}$ \\
\epsfig{file=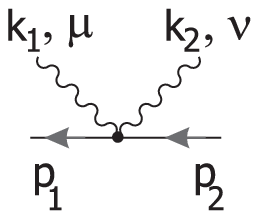,
   width=50pt,
  angle=0,
 }
 & $~~~~~4g^2\delta^{\mu\nu}
 [\cos(k_1\wedge p_1+k_2\wedge p_2)-\cos(p_1\wedge p_2)\cos(k_1\wedge k_2)]$ \\
\epsfig{file=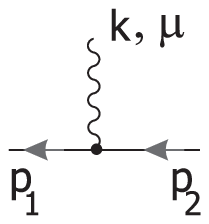,
   width=40pt,
  angle=0,
 }
 & $~~~~~2ig(p_1-p_2)_{\mu}\sin(p_1\wedge p_2)$ \\
\epsfig{file=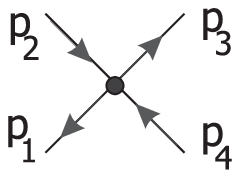,
   width=40pt,
  angle=0,
 }
 & $~~~~~-4\lambda^2[a\cos(p_1\wedge p_2+p_3\wedge p_4)+b\cos(p_1\wedge p_3)
 \cos(p_2\wedge p_4)]$ \\
\epsfig{file=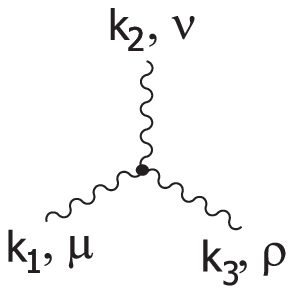,
   width=50pt,
  angle=0,
 }
 & $~~~~~-2ig\sin(k_1\wedge k_2)[
 (k_1-k_2)_{\rho}\delta_{\mu\nu}+
 (k_2-k_3)_{\mu}\delta_{\nu\rho}+(k_3-k_1)_{\nu}\delta_{\mu\rho}]$ \\
\end{tabular}

\begin{minipage}[b]{54pt}
\centering \epsfig{file=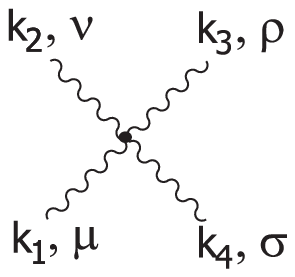, width=50pt, angle=0,}
\end{minipage}
\begin{minipage}[b]{.75\textwidth}
\raggedright
\begin{eqnarray}
4g^2[(\delta_{\mu\rho}\delta_{\nu\sigma}-\delta_{\mu\sigma}\delta_{\nu\rho})
\sin (k_1\wedge k_2)\sin (k_3\wedge k_4)\nonumber\\
+(\delta_{\mu\nu}\delta_{\rho\sigma}-\delta_{\mu\sigma}\delta_{\nu\rho})
\sin (k_1\wedge k_3)\sin (k_2\wedge k_4)\nonumber\\
+(\delta_{\mu\nu}\delta_{\rho\sigma}-\delta_{\mu\rho}\delta_{\nu\sigma})
\sin (k_1\wedge k_4)\sin (k_2\wedge k_3)]\nonumber
\end{eqnarray}
\end{minipage}
\tabcaption{Feynman rules for scalar electrodynamics.}
\renewcommand{\arraystretch}{1}
\end{center}

We do not specify the Feynman rules for ghosts, since they do not
contribute to one-loop graphs with the external matter lines.

All calculations are performed in Landau gauge $\alpha=0$. This is
a convenient choice, since in this gauge a great number of graphs
do not have divergent parts. One can prove that the theory is
gauge invariant on quantum level (cf. \cite{ren1,ren2,ren3,Sus}), so our
results and conclusions are valid for an arbitrary value of
$\alpha$.

\begin{figure}[h]
\begin{center}
\epsfig{file=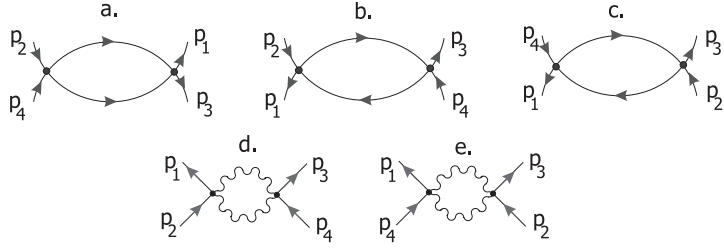, width=270pt, angle=0, } \caption{One-loop
corrections to the $4$-point scalar vertex.}
\label{nc.sed.F1PI4.1}
\end{center}
\end{figure}

The above mentioned restriction on $a$ and $b$ can come from
one-loop corrections to the 4-scalar and 2-scalar-2-gluon
vertices. First we consider one-loop corrections to the $4$-point
scalar vertex. The graphs that have non zero divergent parts in
Landau gauge are presented in Figure \ref{nc.sed.F1PI4.1}.
Using the dimensional regularization ($d=4-2\epsilon$) we find
that the sum of divergent parts of these graphs is equal to
\be
\frac{4}{(4\pi)^2\epsilon} [(3g^4+4\lambda^4a^2+\lambda^4b^2)
\cos(p_1\wedge p_2+p_3\wedge p_4)+
(3g^4+4\lambda^4ab+\lambda^4b^2) \cos(p_1\wedge p_3)\cos(p_2\wedge
p_4)]. \label{nc.sed.1PI4.1}
\end{equation}
The condition of one-loop renormalizability yields a system of two
algebraic equations on $a$ and $b$
\begin{eqnarray}
3g^4+4\lambda^4a^2+\lambda^4b^2&=&ca,\nonumber\\
3g^4+4\lambda^4ab+\lambda^4b^2&=&cb,\nonumber
\end{eqnarray}
where $c$ is a constant. These equations are self-consistent only
in the case $a=b$. Therefore the renormalizable potential for
scalar electrodynamics has the form
\begin{equation}
V[\phi^*,\phi]=a\frac{\lambda^2}{2}(\{\phi^*,\,\phi\}_{\star})^2,
\label{nc.sed.potential}
\end{equation}
where $\{f,\,g\}_{\star}=f\star g+g\star f$. Note that in contrast
to the pure noncommutative complex scalar field theory \cite{ABK2}
we do not have the solution $b=0$.

\begin{figure}[h]
\begin{center}
\epsfig{file=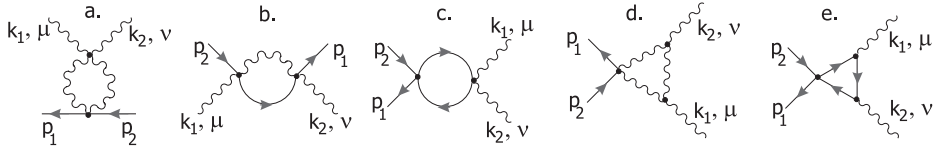,width=350pt, angle=0,} \caption{One-loop
corrections to the $2$-scalar-$2$-gluon vertex.}
\label{nc.sed.F1PI4.2}
\end{center}
\end{figure}

Let us turn to an analysis of one-loop corrections to the
$2$-scalar-$2$-gluon vertex. The graphs that have non zero
divergent parts in the $\alpha=0$ gauge are presented in Figure
\ref{nc.sed.F1PI4.2}.

The sum of the divergent parts of these graphs is
\begin{eqnarray}
\frac{12}{(4\pi)^2\epsilon} g^4\delta_{\mu\nu}[\cos(p_1\wedge
k_1+p_2\wedge k_2)-\cos(p_1\wedge p_2)\cos(k_1\wedge k_2)].
\label{nc.sed.1PI4.2}
\end{eqnarray}
Note that the graphs \ref{nc.sed.F1PI4.2}c and
\ref{nc.sed.F1PI4.2}e do contain $4$-point scalar vertex and there
are terms depending on $a$ and $b$. However the contributions of
these graphs mutually cancel and the sum does not depend on $a$
and $b$. Therefore, there are no new restrictions on these
constants. We see that a counterterm requiring for a cancellation
of (\ref{nc.sed.1PI4.2}) has just the same trigonometric structure
as the initial $2$-scalar-$2$-gluon vertex in the action, i.e.
this vertex is one-loop renormalizable.

Thus the above analysis leads to the conclusion that
noncommutative scalar electrodynamics (\ref{nc.sed.action}) is
one-loop renormalizable only if the scalar potential has the
anticommutator form (\ref{nc.sed.potential}).


\subsection{Noncommutative $\Nc=2$ Super Yang-Mills Theory.}

The action for the Euclidean noncommutative $\Nc=2$ SUSY
Yang-Mills theory reads \cite{additive_zum,additive_nie}
\begin{eqnarray}
&S=\int d^4x\left(-\frac{1}{4} F_{\mu\nu}\star F^{\mu\nu}
+(\Dd_{\mu}\phi_-)\star (\Dd_{\mu}\phi_+)
-i\chi^*\star\widehat{\Dd}\chi\right.&\nonumber\\
&\left.-g\sqrt{2}\chi^*\star(R[\chi,\phi_+]_{\star}+L[\chi,\phi_-]_{\star})
-\frac{g^2}{2}([\phi_-,\phi_+]_{\star})^2\right),&
\label{nc.N2.action}
\end{eqnarray} where
$\Dd_{\mu}=\pd_{\mu}-ig[A_{\mu},\cdot]_{\star}$,
$L,R=\frac12(1\pm\Gamma^5)$, $\phi_{\pm}$ are real scalar fields,
$\chi$ is a complex four-component spinor\footnote{Note that
$\Nc=2$ SYM in Minkowski space contains a complex scalar field.}.
The action (\ref{nc.N2.action}) is a noncommutative generalization
of Euclidean $\Nc=2$ SYM theory. A formulation of
noncommutative $\Nc=2$ supersymmetric theories in terms of
superfields was given in \cite{FL}.

Note that the scalar electrodynamics examined in the previous
section can be considered as a bosonic part of $\Nc=2$ NCSYM. The
identification is evident: $\phi$ and $\phi^*$ corresponds to
$\phi_+$ and $\phi_-$, respectively. Also we should replace
$\lambda$ by $g$ and take $a=-b=-1$. The fact that fields
$\phi_{\pm}$ are not complex conjugate does not affect the
performed calculations. The Feynman rules for the bosonic part of
the action (\ref{nc.N2.action}) can be easily obtained from the
Feynman rules for scalar electrodynamics (see Table 1) using the
above mentioned identification. The Feynman rules for fermion
fields are presented in Table 2.

\renewcommand{\arraystretch}{2.5}
\begin{minipage}{0.5\textwidth}
\begin{tabular}{cl}
\epsfig{file=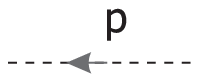,
   width=40pt,
  angle=0,
 }
 & $~~~~~S(p)=\frac{\Gamma^{\mu}p_{\mu}}{p^2}$ \\
\epsfig{file=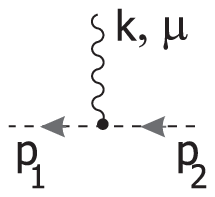,
   width=40pt,
  angle=0,
 }
 & $~~~~~-2ig\Gamma^{\mu}\sin(p_1\wedge p_2)$ \\
\end{tabular}
\end{minipage}
\begin{minipage}{0.5\textwidth}
\begin{tabular}{cl}
\epsfig{file=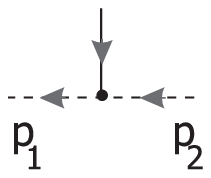,
   width=40pt,
  angle=0,
 }
 & $~~~~~2ig\sqrt{2}\sin(p_1\wedge p_2) R$ \\
\epsfig{file=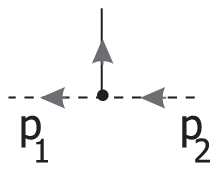,
   width=40pt,
  angle=0,
 }
 & $~~~~~2ig\sqrt{2}\sin(p_1\wedge p_2) L$ \\
\end{tabular}
\end{minipage}
\tabcaption{Feynman rules for fermion fields.}
\renewcommand{\arraystretch}{1}

\begin{figure}[t]
\begin{center}
\epsfig{file=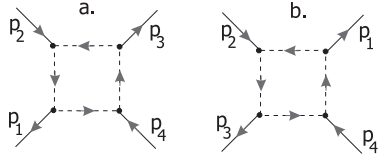, width=140pt, angle=0, } \caption{One-loop
fermion corrections to the $4$-point scalar vertex.}
\label{nc.N2.F1PI4.1}
\end{center}
\end{figure}

As in the case of scalar electrodynamics we start with the
examination of the one-loop corrections to the $4$-point scalar
vertex. The graphs with fermion loops are presented in Figure
\ref{nc.N2.F1PI4.1}.
The divergencies coming from these graphs are
\begin{equation}
-\frac{32}{(4\pi)^2\epsilon} g^4\cos(p_1\wedge p_2+p_3\wedge p_4).
\label{nc.N2.1PI4.1}
\end{equation}
Thus, taking into account the contribution (\ref{nc.sed.1PI4.1})
of the boson graphs we find that the boson and fermion
divergencies mutually cancel. The similar result is valid for
ordinary $\Nc=2$ SUSY Yang-Mills theory where the 4-point scalar
vertex is finite at one-loop \cite{booksuper}.

\begin{figure}[t]
\begin{center}
\epsfig{file=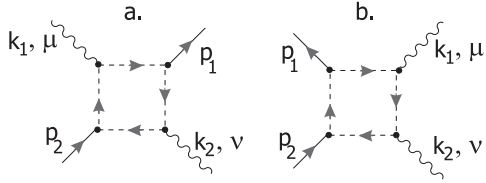,width=180pt, angle=0,} \caption{One-loop
fermion corrections to the $2$-scalar-$2$-gluon vertex.}
\label{nc.N2.F1PI4.2}
\end{center}
\end{figure}

Next we calculate one-loop corrections to the $2$-scalar-$2$-gluon
vertex. The graphs with fermion loops are presented in Figure
\ref{nc.N2.F1PI4.2}.
The sum of divergent parts of these graphs is
\begin{equation}
-\frac{16}{(4\pi)^2\epsilon} g^4\delta_{\mu\nu}[\cos(p_1\wedge
\label{nc.N2.1PI4.2}
\end{equation}
Summing the contributions of the boson (\ref{nc.sed.1PI4.2}) and
fermion (\ref{nc.N2.1PI4.2}) graphs we get
$$
-\frac{4}{(4\pi)^2\epsilon} g^4\delta_{\mu\nu}[\cos(p_1\wedge
k_1+p_2\wedge k_2)-\cos(p_1\wedge p_2)\cos(k_1\wedge k_2)].
$$
Note that one-loop fermion corrections as well as boson ones
restore the trigonometric structure of the initial vertex. So, we
conclude that the noncommutative $\Nc=2$ $D=4$ SYM with the action
(\ref{nc.N2.action}) is one loop renormalizable.

\subsection{UV/IR Problem.}

It is evident that our renormalized 1PI functions do not have the
limit $\xi \to 0$. Due to this there is a nontrivial mixing of UV
and IR divergencies (see \cite{MRS,ABK1} for details). It turns out
that a type of the UV divergency of the integral corresponding to
the Feynman graph in a local model
$$
J_{UV}( p)=\int f(k,p)dk
$$
is the same as a type of the IR behavior of the integral
corresponding to the Feynman graph in a noncommutative model
$$
J_{IR}(\xi p)=\int e^{i\xi k\theta p}f(k,p)dk.
$$
For example, the logarithmically divergent integral $J_{UV}\sim \log
\Lambda$
corresponds to  a logarithmic singularity $\log(\xi|\theta p|)$ in
$J_{IR}$, the quadratically divergent integral $J_{UV}$ corresponds to
a quadratic singularity $(\xi|\theta p|)^{-2}$ in $J_{IR}$, and so
on.

Indeed, the UV behavior of the Feynman graph is defined by its index of
divergence $\omega$.
For $\omega >0$ one has the asymptotic UV behaviour
as $\Lambda \to \infty$
\begin{equation}
J^{as}_{UV}=\int _0^\Lambda k^{\omega -1}dk=
\frac{1}{\omega}\Lambda^\omega.
\label{UV}
\end{equation}
The IR behavior is described by integral
\begin{equation}
J^{as}_{IR}=\int _0^\infty k^{\omega -1}e^{i\xi k}dk
\end{equation}
as $\xi \to 0$. This integral is  the Fourier transform of
the distribution $k_+^{\omega -1}$
\begin{equation}
J^{as}_{IR}=ie^{i\frac{\pi}{2}(\omega -1)}\frac{\Gamma (\omega)}{(\xi +i0)^\omega},
\end{equation}
i.e. for $\omega >0$ one has the  correspondence with UV behaviour (\ref{UV}).

To see logarithmic singularities on $\xi$
let us note that
\begin{equation}
\int _m^\Lambda k^{-1}dk\sim
C \int _{-\Lambda}^\Lambda \frac{d^2k}{k^2+m^2}\sim \log \Lambda
\end{equation}
and the corresponding non local integral is
\begin{equation}
J^{as,0}_{IR}(\xi p)=  \int \frac{e^{i\xi p\wedge k}}{k^2+m^2}d^2k.
\end{equation}
The latter integral can be calculated explicitly and the result is
\begin{equation}
J^{as,0}_{IR}=2\pi K_0(m\xi |p|).
\end{equation}
The modified Bessel function $K_{0}(z)$ has an expansion
$$
K_0(z)(\xi p)=-\ln(z)+\ln(2)-C+O(z^2)
$$
where $C$ is Euler constant, i.e. $J^{as,0}_{IR}(\xi p)$
has the logarithmic dependence on $\xi$.

IR poles appear in the corrections to propagators and can produce
IR divergencies in multi-loop graphs even in the massive theories.
One has not to worry  about logarithms, since in the origin
the logarithm is an integrable function. Let us consider several
examples.

{\bf Example 1}. One can calculate explicitly the finite part of
the tadpole graph in the noncommutative $\varphi^4$ theory. The
answer is given by the sum of (\ref{nc.phi4.1PI2}) and
 (\ref{nc.phi4.1PI2'}).
The
behavior of (\ref{nc.phi4.1PI2}) in the limit $p^2 \to 0$
(the same as the limit $\xi \to
0$) is the following
$$
\Gamma ^{(2)}_{1,f.p.} \ssim_{p^2 \to 0} \frac{c}{(\xi |p|)^2}.
$$
This result can be easily obtained from the series expansion of
the modified Bessel function $K_{1}(z)$
$$
K_1(z)=1/z+\frac12z\ln(z)+(\frac12C-\frac14-\frac12\ln(2))z+O(z^3)
$$
where $C$ is Euler constant. Caused
by this asymptotic there are problems with an IR behavior of
graphs with tadpoles. They produce divergence in the IR region if
the number of insertions is three or more (compare with an example
of ref. [21] in \cite{Ch}). Thus, theories with a real scalar field
have problems with infrared behavior \cite{MRS,ABK1} originated in
multi one-loop insertions.

{\bf Example 2}. Another example is a tadpole
Fig.\ref{nc.c.F1PI}:e in the case of complex scalar field. The
analytic expression for this graph is the following
$$
\Gamma(p)=\frac{(\mu^2)^{\epsilon}}{(2\pi)^d}\int
d^dk\frac{A+B\cos^2(k\wedge p)}{k^2+m^2}
$$
\be=
\frac{(A+\frac{B}{2})(\mu^2)^{\epsilon}}{(2\pi)^d}\int
d^dk\frac{1}{k^2+m^2}+\frac{B(\mu^2)^{\epsilon}}{2(2\pi)^d} \int
d^dk\frac{e^{i2k\wedge p}}{k^2+m^2}. \label{nc.uvir.ex}
\end{equation}
Integrating this expression over momentum $k$ we obtain
$$
\Gamma(p)=\frac{m^{d-2}(\mu^2)^{\epsilon}}{(4\pi)^{d/2}}(A+\frac{B}{2})\Gamma(1-d/2)+
\frac{B}{(4\pi)^{d/2}}\left[\frac{m}{\xi|\theta p|}\right]^{d/2-1}
K_{d/2-1}(2m\xi|\theta p|).
$$

For $d=4$ the second term is singular when $p\to 0$. But in the
case $B=0$ (one of the possible solution of the UV
renormalizability) this term disappears and hence there is no IR
problem at least at one-loop level.

\begin{figure}[h]
\begin{center}
\epsfig{file=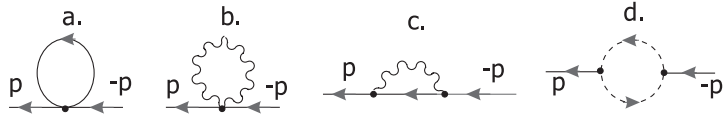,
   width=270pt,
  angle=0,
 }
\caption{One-loop corrections to the scalar field propagator.}
\label{nc.uvir.F1PI}
\end{center}
\end{figure}

{\bf Example 3}. In the case of scalar electrodynamics all
corrections to the scalar field propagator are presented in Figure
\ref{nc.uvir.F1PI}a,b,c.
For $a=b$ (see notations in the subsection about scalar
electrodynamics) the sum of the divergent parts is:
$$
\frac{6g^2(\mu^2)^{2\epsilon}}{(2\pi)^4}\int\frac{d^4k}{k^2}\left[1-\frac{a\lambda^2}{g^2}-
\left(1+\frac{a\lambda^2}{3g^2}\right)\cos(2k\wedge p)\right]$$
$$-
\frac{8g^2(\mu^2)^{2\epsilon}}{(2\pi)^4}\int
d^4k\frac{1-\cos(2k\wedge p)}{k^2(p+k)^2}
\left[p^2-\frac{(pk)^2}{k^2}\right]
$$
where $p$ is an external momentum. A quadratic UV divergence here
is removed by a mass renormalization. To remove IR poles one has
to impose the condition $\lambda^2a=-3g^2$.

Computing a one-loop correction to the scalar field propagator in
the $\Nc=2$ $d=4$ NC SYM (\ref{nc.N2.action}) we have one more
graph presented in Figure \ref{nc.uvir.F1PI}d. Summing all
contributions we have
$$
-4g^2(\mu^2)^{2\epsilon}\int\frac{d^4k}{(2\pi)^4}\frac{1-\cos(2k\wedge
p)}{k^2(p+k)^2} \left[p^2+2\frac{(pk)^2}{k^2}\right].
$$
Since all UV divergencies are logarithmic there is no IR poles in
the case of $\Nc=2$ $D=4$ NC SYM theory.

All calculations of the one-loop corrections was performed using a
dimensional regularization. Now we would like to show that
appearance of the IR singularities does not depend on
a regularization. Namely, we compute an answer for the tadpole graph
in the noncommutative complex scalar field theory using another
methods. The expression for this graph is given by
(\ref{nc.uvir.ex}).

 First, we perform calculations using Pauli-Villars
 regularization. To do this we rewrite (\ref{nc.uvir.ex}) in the
 following form
$$
\Gamma_{reg,PV}(p)=\frac{1}{(2\pi)^d}\int
d^4k\left[A+B\cos^2(k\wedge
p)\right]\left[\frac{1}{k^2+m^2}-\frac{C_1}{k^2+M_1^2}-\frac{C_2}{k^2+M_2^2}
\right]
$$
with $C_1+C_2=1$ and $C_1M^2_1+C_2M^2_2=m^2$. All integrals are
finite and we get the following answer
$$
\Gamma_{reg,PV}(p)=\frac{1}{8\pi^2}(A+\frac{B}{2})
(M_1^2M_2^2-m^2(M_1^2+M_2^2-m^2))\times
$$
$$
\times\frac{1}{M_2^2-m^2}\left[ \log\frac{M_1^2+M_2^2-m^2}{M_1^2}
+\frac{M_2^2}{M_1^2-m^2}\log\frac{M_1^2+M_2^2-m^2}{M_2^2}
+\frac{m^2}{M_1^2-m^2}\log\frac{m^2}{M_1^2} \right]
$$
\be +\frac{B}{(4\pi)^2}\left(\frac{m}{\xi|\theta p|}
K_{1}(2m\xi|\theta p|)-C_1\frac{M_1}{\xi|\theta p|}
K_{1}(2M_1\xi|\theta p|)-C_2\frac{M_2}{\xi|\theta p|}
K_{1}(2M_2\xi|\theta p|)\right). \label{nc.uvir.pv}
\end{equation}
Note, that
for fixed values of $M_i$
 there is a cancellation of an IR pole $1/|p\theta|^2$,
i.e. the sum of the last three terms has the following asymptotic
behavior for small $|p\theta|^2$ up to a numeric factor
$$
m^2\log (2m\xi|\theta p|)-C_1 M_1^2 \log (2M_1\xi|\theta p|)- C_2
M_2^2 \log (2M_2\xi|\theta p|).
$$
It is easy to see, that one cannot do this calculation for large
$M_i$.

To remove a regularization (to take limits $M_1\to \infty$ and
$M_2\to \infty$) one has to do the standard UV renormalization which
corresponds to the UV renormalization of the planar part of this
graph. The first two lines in the (\ref{nc.uvir.pv}) contain these
UV divergencies of the tadpole.

The non-planar part, i.e. the third line of (\ref{nc.uvir.pv}),
does not require a renormalization. If one takes limit $M_1\to
\infty$ and $M_2\to \infty$ than two last terms in the third line
of (\ref{nc.uvir.pv}) go to zero since
$$
K_1(z)\sim\sqrt {\frac{\pi}{2z}}e^{-z}~~\mbox{for}~~ z\to \infty.
$$
Therefore, after removing a regularization we are left with the
following term
$$
\frac{B}{(4\pi)^2}\frac{m}{\xi|\theta p|} K_{1}(2m\xi|\theta
p|)
$$
that contains IR pole.

The same result is true for the "cut-off" regularization. We have
(up to numeric factors)
$$
\Gamma_{reg,\Lambda}(p)=\int_0^\infty d\alpha \int d^4k \,\left[
(A+\frac{B}{2})\, e^{- \alpha(k^2 + m^2)} + B\,e^{i k\wedge p -
\alpha(k^2+m^2)} \right] \cdot e^{-\frac{1}{4\alpha \Lambda ^2}}
$$
$$
=\frac{1}{(2\pi)^4} \int_0^\infty d\alpha
\left({\frac{\pi}{\alpha}}\right)^2 \left[(A+\frac{B}{2}) \,
e^{-\alpha m^2} + B\,e^{-\alpha m^2 - \frac{\xi ^2|\theta p|^2}{4
\alpha}} \right] \cdot e^{-\frac{1}{4\alpha \Lambda ^2}}
$$
where $\Lambda$ plays the role of a cut-off. For non-planar part
we have
$$
\frac{B}{(2\pi)^4} \int _0^\infty \frac{d\alpha}{\alpha^2}
e^{-\alpha m^2-\frac{1}{4\alpha \Lambda _{eff}^2}} \sim m^2 \,
\sqrt{\frac{\Lambda _{eff}^2}{m^2}} ~K_1\left( \frac{m}{\Lambda
_{eff}} \right)
$$
where
$$
\frac{1}{\Lambda _{eff}^2}= \xi^2 |p\theta|^2 +\frac{1}{\Lambda^2}.
$$
We see that in the limit $\Lambda \to \infty $ the integral
remains finite and it has infrared pole. This is in an agreement
with an absence of the pole in (\ref{nc.uvir.pv}) for fixed
regularization and a presence of the pole after the regularization
is removed.

The mixing of UV and IR divergencies was noted in \cite{MRS,ABK1}
and as well as consistency of noncommutative quantum field theories
were discussed in many papers \cite{ABK1}-\cite{2loop2} in last two years.

As it is seen from examples, theories without fermions have IR singularities
(usually
poles) at one loop level.
There is a possibility to avoid appearance of IR singularities
for  complex scalar field
in one special case: $B=0$ in our notations. However, it should be
checked that this one loop condition is not violated
at higher loop levels.
In the case of scalar electrodynamics it
is possible to avoid an appearance of IR poles imposing additional
condition
on coupling constant: $\lambda^2 a=-3g^2$.

In \cite{Sus} the $U(1)$ theory with additional fermion degrees of freedom was
analyzed. It was shown that the IR pole modifies the dispersion relation of the
photon in the following way
$$
E^2=p^2-(N_B-N_F)\frac{g^2}{\pi^2}\frac1{(\theta p)^2}
$$
where $N_B$ and $N_F$ are numbers of bosons and fermions in the adjoint
representation. In particular, the energy becomes imaginary for $N_B>N_F$,
indicating that such a theory may suffer from instability.

It is surprising that for the pure gauge theory without fermions we have IR poles
which are not expected from the UV/IR correspondence found for the scalar models
(See \cite{Hay} for detailed calculations).
The most intriguing point is that gauge invariance does not protect from appearence
of the IR singularities. However, such singularities have pure noncommutative nature
becuase of trigonometric factors in the vertices. This phenomenon is clear technically
but not physically.

On the other hand supersymmetric theories do not have pole singularities in the
IR region and this was shown explicitly in the third example. This became
possible thanks to presence of supersymmetry because there is only one
coupling constant which can be tuned and one is unable to remove
singularities using this constant. Although logarithms are still present in the
${\cal N}=2$ NC SYM theory, we emphasize that the logarithm function is
integrable in the origin and cannot produce non-integrable singularities
in multi-loop graphs.

Moreover, it seems likely that the theories with
vanishing $\beta$-function on the ordinary commutative space
are to be free of nonanalytic dependence on
$\theta$ in noncommutative case. In particular, one could expect that
${\cal N}=4$ NC SYM theory should be finite in both UV and IR regions
\cite{Sus}.

Therefore, we see that all theories except ${\cal N}=4$ NC SYM
theory have no a smooth limit $\theta\to 0$ in one-loop
approximation and some of them have IR poles.  This  makes the
issue of self-consistency of such theories rather questionable.

\newpage
\section{Cubic String Field Theory and Matrix Models.}
\label{sec:SFTM}
\setcounter{equation}{0}

\subsection{String Fields as Functionals.}
\subsubsection{Witten's realization of axioms on space of string functionals.}
Witten \cite{Witten} presented realization of the set
(\ref{triplet}-\ref{tr-gr}) in the case when ${\cal A}$ is taken
to be the space of string fields
\begin{equation}
{\cal F} =\{\Psi[X (\sigma); c (\sigma), b
(\sigma)]\}
\label{funct}
\end{equation}
which can be described as functionals of the matter $X (\sigma)$,
ghost $c (\sigma)$ and
antighost fields $b (\sigma)$ corresponding to an open string in 26 dimensions with
$0 \leqslant \sigma \leqslant
\pi$.  For this string field theory $Q_{B}$ is the usual
open string BRST charge of the form
\begin{equation}
Q_B = \int_0^\pi d \sigma \; c (\sigma) \left(T_{B} (\sigma)
+\frac{1}{2}T_{bc} (\sigma) \right)\,.
\label{QB}
\end{equation}
$T_{B}$ and $T_{bc}$ are stress tensors for the matter field and
ghosts (see (\ref{Tx}) and below (\ref{Tbc})).

The star product $\star$ is defined by gluing the right half of one
string to the left half of the other using a delta function
interaction.  The star product factorizes into separate matter and
ghost parts.  For the matter fields the star product is given by
$$
         \left(\Psi \star   \Phi\right) [X(\sigma)]
         \equiv
\int
\prod_{{\pi\over 2} \leqslant \sigma \leqslant \pi} dX'(\sigma) \; dX'' (\pi -\sigma)
\prod_{{\pi\over 2} \leqslant \sigma \leqslant \pi}
\delta[X'(\sigma)-X''(\pi-\sigma)]
\;   \Psi [X'(\sigma)]  \Phi [X''(\sigma)],
$$
\begin{equation}
\label{eq:matter-star}
X(\sigma)  = X'(\sigma) \quad {\rm for} \quad {0}
\leqslant \sigma \leqslant {\pi\over 2}\, ;
 \quad
X(\sigma) = X''(\sigma)\quad {\rm for} \quad   {\pi\over 2} \leqslant \sigma \leqslant \pi\, .
\end{equation}
The integral over a string field factorizes into
matter and ghost parts, and in the matter sector is given by
\begin{equation}
\int \Psi = \int \prod_{0 \leqslant \sigma \leqslant \pi} dX' (\sigma) \;
\prod_{0 \leqslant
\sigma \leqslant \frac{\pi}{2} }
\delta[X'(\sigma)-X'(\pi-\sigma)] \;\Psi[X' (\sigma)]\,.
\label{star-int}
\end{equation}
Performing a Fourier mode expansion of the 26 matter fields $X^\mu$
through
\begin{equation}
X^{\mu}(\sigma) =  x_0^{\mu}+ \sqrt{2\alpha '} \sum_{n=1}^\infty{x_n^{\mu}\cos(n\sigma)}
\label{Xmode}
\end{equation}
one can consider string field $\Phi [X(\sigma)]$
\footnote{Often we shall omit the space-time index $\mu$.}
(we consider there only the $X$-part of string field for
simplicity) as depending on the set $\{x_n\}$, $\Phi =\Phi[\{x_n\}]$.
Then one can regard
 $\Phi[\{x_n\}]$ a wave function in the coordinate representation
of a system with an infinite number of degrees of freedom.

\subsubsection{String functionals as matrices or symbols for operators.}

One can consider  string functionals as vector states in the
coordinate representation
 \begin{equation}
\Psi[X(\sigma)] =
\langle  X(\sigma)|\Psi\rangle .
\label{coor-rep}
\end{equation}
Our goal is  to
obtain  the mapping from string fields $|\Psi\rangle$ to matrices
\begin{equation}
|\Psi\rangle \Longleftrightarrow \Psi_{{\bf n},{\bf m}} \label{state-mat}
\end{equation}
so that
\begin{equation}
|\Psi\rangle\star |\Phi\rangle
\Longleftrightarrow \sum_{{\bf k}}\Psi_{{\bf n},{\bf k}}\Psi_{{\bf k},{\bf
m}},
\end{equation}
\begin{equation}
\int |\Psi\rangle
\Longleftrightarrow Tr \Psi=\sum_{{\bf k}}\Psi_{{\bf k},{\bf k}} .
 \end{equation}

Also we want to find a map from string fields to operators

 \begin{equation}
|\Psi\rangle
\Longleftrightarrow
\hat {\Psi}
\label{state-oper}
\end{equation}
 with
 \begin{equation}
|\Psi\rangle\star |\Phi\rangle
\Longleftrightarrow
\hat {\Psi}\hat {\Phi} .
\label{fun-mult}
\end{equation}
In other words we look for a map between string functionals and operators
 \begin{equation}
\Psi[ X(\sigma)]
\Longleftrightarrow
\hat {\Psi},
\label{func-oper}
\end{equation}
 such that
 \begin{equation}
\Psi[X(\sigma)]\star \Phi[X(\sigma)]
\Longleftrightarrow
\hat {\Psi}\hat {\Phi}.
\label{func-mult}
\end{equation}
There is an analogy between this map and the relation between operators and
their symbols
\cite{Ber}. Using this analogy one can interpret the Witten multiplication
as a Moyal product of two functionals.

Let us make a discretization of the parameter $\sigma$, i.e.
$\sigma_{i}=\frac{i\pi}{2n}$, $i=0,1,...,2n$ and denote
$X_{i}=X(\sigma _{i})$. The string field $\Phi [X(\sigma)]$
becomes a function
 $\Phi (X_{0},...,X_{n-1},X_{n},X_{n+1},...X_{2n})$.
 The product now has the form
\begin {multline} 
                                                          \label {6}
~~~~~~~~~~~~~~~~~~~~~(\Phi \star \Psi )(X_{0},...,X_{n-1},X_{n},X_{n+1},...X_{2n})=\\
\int\Phi(X_{0},...,X_{n-1},X_{n},Y_{n-1},...Y_{0})
\Psi (Y_{0},...,Y_{n-1},X_{n},X_{n+1},...X_{2n})d^{D}Y_{0}...d^{D}Y_{n-1}
.
\end   {multline} 
The rule (\ref {6}) looks like the matrix multiplication, only the
midpoint $X_{n}$ plays a particular role. To obtain a matrix let
us introduce a lattice $\Zh^{D}$ in $\Rh^{D}$ . Then our variables
$x_{i}$ will belong to this lattice. If one restricts himself to a
finite sublattice then one gets a field
$\Phi(\{X_{L,i}\},x,\{X_{R,i}\})$, where $\{
X_{L,i}\}=(X_{0},...,X_{n-1})$ corresponds to $X_{L}(\sigma)$, $\{
X_{R,i}\}=(X_{n+1},...,X_{2n})$ corresponds to $X_{R}(\sigma)$ and
$x=X_{n}$. Denoting multi-indexes $a=\{ X_{L,i}\}$  and $b=\{
X_{R,i}\}$  one has a matrix field $\Phi _{ab}(x)$. $a,b$ can be
enumerated as $a,b=1,...,N$. Therefore, we get a matrix realization
\cite{AV-M} of the Witten algebra where the fields are matrices
$\Phi _{ab}(x)$ depending on parameter $x$, the product $\star$ is
the matrix product
\begin {equation} 
                                                          \label {7}
(\Phi \star \Psi )_{ab}(x)= \Phi _{ac}(x)\Psi _{cb}(x)
\end   {equation} 
and the integral is
\begin {equation} 
                                                          \label {8}
\int \Phi =\sum _{x} \Tr \Phi (x) .
\end   {equation} 

To find a map (\ref{state-mat}) directly in the continuous case it is useful
to consider the half string formalism.

\subsubsection{Half-string functionals.}
In (\ref{eq:matter-star}) and (\ref{star-int})
$\star$ and $\int$ are defined in terms of string overlaps.
Each string has a preferred
point, the "midpoint" $\sigma=\pi/2$. The midpoint divides a string $X$ into left and right halves
$(X_{L},X_{R})$. The product of two string functionals $\Psi[X']$ and $\Psi[X'']$ is zero unless
$X'_{R}$  coincides with $X''_{L}$ in the space-time. So the string field $\Phi [X]$ is considered as a functional
$\Phi [(X_{L},X_{R})]$   and one has
\begin {equation} 
                                                          \label {1}
(\Phi \star \Psi )(X_{L},X_{R})=\int \prod_{\frac{\pi}{2} \leqslant \sigma \leqslant \pi}dY_{R}(\sigma)\;
\Phi [(X_{L},Y_{R})]
\Psi [(r(Y_{R}),X_{R})],
\end   {equation} 
where $$r(Y_{R}(\sigma))=Y_R(\pi-\sigma).$$

It is useful to perform a separate mode expansion of the left and
right pieces of the string. There are different possibilities to
do this \cite{GT01,GT02,rsz-3,Bars} (see also
\cite{bcnt,Abdurrahman-Bordes})). One can define the left
and right pieces of the string with or without a shift on the
midpoint. In the first case one has\footnote{Often we
shall  fix $\alpha'$, here $\alpha '=1$.}
\begin{subequations}
\begin{align}
 X^L(\sigma)\equiv X(\sigma)-X(\tfrac{\pi}{2})&= \sqrt{2\ap}
 \sum_{n=1}^\infty
    x^L_{n}
\cos(2n-1)\sigma \ ,
\label{defl}
\\
   X^R(\sigma)\equiv X( \pi -\sigma)-X(\tfrac{\pi}{2})&= \sqrt{2\ap}
   \sum_{n=1}^\infty
    x^R_{n}
\cos(2n-1)\sigma \ ,
\label{defr}
\end{align}
\end{subequations}
for $0\leqslant \sigma\leqslant \tfrac{\pi}{2}$. Note that $X_L(\sigma)$ and
$X_R(\sigma)$ obey Neumann boundary conditions at $\sigma =0$ and
zero Dirichlet boundary conditions at $\sigma=\tfrac{\pi}{2}$.  These coordinates
are called \cite{Abdurrahman-Bordes} "comma"
coordinates.

An expression for the full open string modes in terms
of the modes of the left-half and modes of the right-half is:
\begin{subequations}
\begin{align}
x_{2n-1}&= \frac12(x^{L}_{n} - x^{R}_{n}), \qquad n\geqslant 1,
\\
x_{2n}&=-\sum_{m=1}B_{2n,2m-1}(\,x^{L}_m+x^{R}_m),\qquad
n\geqslant 1.
\label{x-LReven}
\end{align}
\end{subequations}
Matrices  $B_{2n,2m-1}$  are written explicitly in Appendix 4.1.
For the midpoint coordinate one has
\begin{eqnarray}
X(\tfrac{\pi}{2}) = x_0 +\sqrt{2\ap} \sum_{n\geqslant1} (-1)^n x_{2n}.
\label{midpoint}
\end{eqnarray}
Conversely, the center of mass in the ``comma" representation is:
\begin{eqnarray}
x_0 = X(\tfrac{\pi}{2}) - \frac{\sqrt{2\ap}}{\pi} \sum_{n\geqslant 1}
\frac{(-1)^n}{2n-1} (x^L_n +x^R_n). \label{x0-midpoint}
\end{eqnarray}

Without shift on the midpoint one deals with the following
half string coordinates\footnote{To avoid misunderstanding we use
new letters $l,r$ and $\chi ^L, \chi ^R$ for these half-strings coordinates.}:
\begin{subequations}
\begin{align}
l(\sigma)&\equiv X(\sigma), \label{defl/ws}
\\
r(\sigma)&\equiv X(\pi-\sigma), \label{defr/ws}
\end{align}
\end{subequations}
for $0\leqslant \sigma\leqslant \tfrac{\pi}{2}$. Note that $l(\sigma)$ and
$r(\sigma)$ obey Neumann boundary conditions at $\sigma =0$ and
Dirichlet boundary conditions at $\sigma=\tfrac{\pi}{2}$. One can perform a
separated mode expansion of the left and right half-strings
choosing an odd/even extension to the interval $(\tfrac{\pi}{2},\pi]$. In
the first case one left only with odd modes in the Fourier
expansions
\begin{subequations}
\begin{align}
l(\sigma)&=\sqrt{2\ap}\sum_{n=0}^\infty l_{2n+1}\cos(2n+1)\sigma,
\label{deflw}
\\
r(\sigma)&=\sqrt{2\ap}\sum_{n=1}^\infty r_{2n+1}\cos(2n+1)\sigma.
\label{defrw}
\end{align}
\end{subequations}
Relations between the full-string modes and the half-string modes
$r_{2n+1}$ and $l_{2n+1}$ are
\begin{subequations}
\begin{align}
x_{2n +1}&= \frac12(l_{2n+1} - r_{2n+1}),
\\
x_{2n}& = \frac12\sum _{m=0}X_{2n,2m+1}(\,l_{2m+1} +  r_{2m+1}),
\label{x-lr}
\end{align}
\end{subequations}
where matrices $X_{2n,2m+1}$ are written explicitly in Appendix 4.1.
Choosing an even extension to the interval $(\tfrac{\pi}{2},\pi]$ one left
himself only with even modes in the Fourier expansions of the left
and right half-strings
\begin{subequations}
\begin{align}
\chi^{L}(\sigma)&=\chi^L_0+\sqrt{2\ap}\sum_{n=1}^\infty\chi^L_{2n}\cos(2n\sigma),
\label{l-e}
\\
\chi^{R}(\sigma)&=\chi^R_0+\sqrt{2\ap}\sum_{n=1}^\infty\chi^R_{2n}\cos(2n\sigma).
\label{r-e}
\end{align}
\end{subequations}
Relations between the full-string modes and the half-string modes
$\chi ^L_{2n}$ and $ \chi ^R_{2n}$ are
\begin{subequations}
\begin{align}
x_{2n }&= \frac12(\chi ^L_{2n} + \chi ^R_{2n}),~~n\geqslant 0,
\\
x_{2n-1}&=\frac{\sqrt{2}}{\pi \sqrt{\ap} }\frac{(-1)^n}{2n-1}(\chi^L_{0}-\chi^R_{0})+
\sum_{m=1}B_{2m,2n+1}(\,\chi^L_{2m}-\chi^R_{2m}). \label{x-lr-e}
\end{align}
\end{subequations}

\subsubsection{Half-string functionals and projectors.}

Using for example (\ref{deflw}, \ref{defrw}), one can rewrite a string field
$\Psi[X(\sigma)]$ as a functional of the right and left half-string
degrees of freedom $\Psi[\{l_{2k+1}\};\{r_{2k+1}\}]$.

The star product $\Psi \star \Phi$ in the split string language is
given by
\begin{equation}
(\Psi \star \Phi)[\{l_{2k+1}\};\{r_{2k+1}\}]= \int  \; \prod_{k =
0}^{\infty}d s_{2k+1}\; \Psi[\{l_{2k+1}\};\{s_{2k+1}\}] \;
\Phi[\{s_{2k+1}\};\{r_{2k+1}\}]\,. \label{eq:half-star}
\end{equation}
Since $|\det X| = 1$ (\ref{GoodMatrix}), one can write the string
field integral (\ref{star-int}) as
\begin{equation}
\int \Psi  =
\int \; \prod_{k = 0}^{\infty}
  dl_{2k + 1}
\;\Psi[\{l_{2k + 1}\};\{l_{2k + 1}\}]\,. \label{eq:half-integral}
\end{equation}

One can think that $\{r_{2k+1}\}$ specifies a point $K$ in the
discrete set $K\in {\cal K}$ and consider $\Psi[\{l_{2k +
1}\};\{r_{2k + 1}\}]$ as an infinite dimensional matrix $\Psi
_{R,L}$ so that

\begin{eqnarray}
\int \Psi & \Rightarrow & {\rm Tr}\; \Psi ,\\
(\Psi \star \Phi)[\{l_{2k + 1}\};\{r_{2k + 1}\}] & \Rightarrow & \Psi
_{LK}\Phi_{KR}.
\end{eqnarray}

Since we have divided string into left and right half-strings it
is useful following \cite{GT01},\cite{GT02} to introduce string
fields of the form
\begin{equation}
\Psi[l (\sigma), r (\sigma)] = \psi _L[l (\sigma)] \psi_R[r (\sigma)].
\label{factorized}
\end{equation}
For $ \psi _L=\psi _R$ these string fields are rank one projectors.
One simple class of
them is described by Gaussian functionals
\begin{equation}
\psi[l (\sigma)] \sim \exp\Bigl({-\frac{1}{ 2} l_{2k +
1}M_{kj}l_{2j + 1}} \Bigr).
\label{GaussPr}
\end{equation}
String field (\ref{factorized}) in the full string modes basis
is rewritten as
\begin{equation}
\Psi[\{x_n\}]=\exp \Bigl({-\frac{1}{2}x_n L_{nm} x_m} \Bigr),
\label{factorizedModes}
\end{equation}
where
\begin{align}
L_{2k + 1, 2j + 1}  &=   2M_{kj}, \nonumber\\
L_{2n, 2m} &=   2 X_{2n, 2k + 1} M_{kj} X_{2j + 1, 2m},
\label{projectionForm}\\
L_{2 k + 1, 2m} &= L_{2n, 2j + 1} =  0. \nonumber
\end{align}
This means that string field of the form (\ref{factorizedModes})
 is factorized when the following conditions on the
matrix $L_{nm}$ are imposed\\
\noindent {\bf (a)} The components $L_{nm}$ must vanish when $n+ m$ is odd.\\
\noindent {\bf  (b)} The nonzero components must satisfy
\begin{equation}
L_{2n, 2m} = X_{2n, 2k + 1} L_{2k + 1, 2j + 1} X_{2j + 1, 2m}\,.
\label{projection-conditions}
\end{equation}
Examples of functionals satisfying conditions (a) and (b) are
\begin{equation}
L_{nm}=\delta_{nm}\qquad \text{and}\qquad L_{nm}=\delta_{nm}n^2,
\label{Lexample}
\end{equation}
where identity (\ref{XXiden}) is used.\\

\subsection{$\star$ in Fock Space.}
\subsubsection{String fields as vectors in  Fock space.}

One can regard $\Phi\{x_n\}$  from the previous subsection as a
wave function in the coordinate representation of a system with an
infinite number of degrees of freedom. From the coordinate space
one can make transition to the Fock space. In our case this is
 the Hilbert space ${\cal H}$  of the first-quantized
string theory. To describe ${\cal H}$
let us write
the canonical conjugated momentum
$P_{\nu}(\sigma)$
\begin{equation}
P^{\nu}(\sigma)=\frac{1}{\pi}\left( p^{\nu}+\frac{1}{\sqrt{\alpha
'}} \sum_{n =1}^{\infty}p^{\nu}_n\cos n\sigma\right).
\label{Pmode}
\end{equation}
The canonical commutation relation is
\begin{equation}
[X^{\mu}(\sigma),P^{\nu}(\sigma')]=i\delta^{\mu\nu}\delta(\sigma,\sigma'),
\label{CCR}
\end{equation}
where $\delta(\sigma,\sigma')$ is the $\delta$-function satisfying the
Neumann boundary conditions on $[0,\pi]$:
\begin{equation}
\delta(\sigma,\sigma')=\frac{1}{\pi}\sum_{n=-\infty}^{\infty}\cos n\sigma
\cos n\sigma'
=\delta(\sigma-\sigma')+\delta(\sigma+\sigma').
\label{Ndelta}
\end{equation}

From (\ref{CCR}) the commutation relations for modes are found to
be
\begin{equation}
[x^\mu,p^\nu]=i\eta^{\mu\nu}, \quad [x^\mu_n,p^{\nu}
_{m}]=i\eta^{\mu\nu} \delta_{n,m}. \label{mCCR}
\end{equation}
Now we can introduce the creation and annihilation operators for nonzero modes via
\begin{equation}
x^{\mu}_n=i \frac{\alpha^\mu _n-\alpha^{\mu \dag}_n}{n\sqrt{2}},~~~
p^{\mu}_{n}=\frac{\alpha^\mu _n+\alpha^{\mu \dag}_n}{\sqrt{2}}.
\label{alpha}
\end{equation}
From (\ref{mCCR}) and (\ref{alpha}) we get the following
commutation relations ($\alpha_{-m}=\alpha^{\dag}_m$)
\begin{equation}
[\alpha^\mu  _n,\alpha^\nu _m]=n\eta^{\mu \nu }\delta_{n+m,0}.
\label{alphaCR}
\end{equation}
The state $|0,p\rangle$ is defined as
\begin{equation}
\hat{p}^\mu |0,p\rangle=p^\mu |0,p\rangle ,\quad \alpha^\mu
_n|0,p\rangle=0, \;\;\; n\geqslant 1. \label{xvac}
\end{equation}
A basis for ${\cal H}$ is given by the set of states of the form
\begin{equation}
\alpha^{\mu_1}_{-n_1}\cdots \alpha^{\mu_i}_{-n_i} |0,p\rangle,
\label{xstates}
\end{equation}
where $n>0 $, and $i$ is an  arbitrary positive integer. Then
any state $|\Phi\rangle \in {\cal H}$ can be expanded as
\begin{equation}
|\Phi\rangle =\int dp\left(\phi(p)+A_{\mu}(p)\alpha^{\mu}_{-1}+
B_{\mu\nu}(p)
\alpha^{\mu}_{-1}\alpha^{\nu}_{-1}+\cdots\right)|0,p\rangle,
\label{Fockstates}
\end{equation}
where the coefficients in front of the basis states have the
dependence on the center-of-mass momentum of the string. These
coefficient functions are space-time particle fields.

The eigenstates of the position operators
\begin{equation}
\hat {x}_n|x_n\rangle = x_n|x_n\rangle
\label{coh-state}
\end{equation}
 (here and in eq. (\ref{xvac}) we put hats on $x$ and $p$ despite the fact that we have not put hats in the
 relations
(\ref{mCCR}) and (\ref{alpha})) are given by
 \begin{equation}
|x_n\rangle=\left(\frac{1}{\pi}\right)^{1/4}e^{-\frac{1}{2}nx_n^2-
i\sqrt{2n}x_n\alpha _n^{\dag}+\frac{1}{2n}\alpha _n^{\dag} \alpha
_n^{\dag}}|0\rangle_n, \label{Cstate}
\end{equation}
where $\alpha_n|0\rangle_n=0$.\\
Using the completeness of these states
\begin{equation}
\int \prod_{n=1}^{\infty}|x_n\rangle \langle x_n |~dx_n=1
\label{compl}
\end{equation}
one gets string functional corresponding to a given vector $\Psi$ in the
space ${\cal H}$

\begin{equation}
\Psi [X(\sigma)] \equiv\Psi [\{x_n\}]=
\prod _n  \langle x_n|\Psi\rangle
\label{func-vec}
\end{equation}
and vise versa
\begin{equation}
|\Psi\rangle= \int \Psi [\{x_n\}]\prod _n dx_n |x_n\rangle.
\label{vec-func}
\end{equation}

\subsubsection{Vertices.}
Using (\ref{vec-func}) one can express \cite{GrJe} (see also
\cite{cst}-\cite{Ohta}) the $\star$ operation defined by
(\ref{star-int}) as an operation
\begin{equation}
\star : {\cal H}\otimes {\cal H}~\to ~{\cal H}.
\label{star-vec}
\end{equation}
It is convenient to write the $\star$ product of $|\Psi\rangle$
and $|\Phi\rangle$ using the reflection operator $\langle R|$ and
3-string vertex $\langle V_3|$
\begin{equation}
_{123}\langle V_3|=_{11'}\langle R|_{22'}\langle R|_{33'}\langle
R|V_3 \rangle _{1'2'3'}
\label{<V}
\end{equation}
in the form
 \begin{equation}
|\Psi\rangle\star |\Phi\rangle_3=_{123'}\langle V_3|\Psi\rangle
_1 |\Phi\rangle_2 |R\rangle_{33'} , \label{V3}
\end{equation}
where the subscript 1,2,3 label three interacting strings. One
can rewrite (\ref{V3}) as
 \begin{equation}
|\Psi\rangle\star |\Phi\rangle_3= _1\langle\Psi|_2\langle \Phi|V_3
\rangle _{123}, \label{V3>}
\end{equation}
where
\begin{equation}
_1\langle\Psi|= _{12}\langle R|\Psi\rangle_2 \label{r}.
\end{equation}

\noindent{\textbf{1. Overlaps.}}

\bigskip

$|V_N\rangle$\footnote{We usually use $I$ and $R$ notations for
$V_{1}$ and $V_{2}$ respectively.} solves the overlap conditions
\begin{equation}
X_r(\sigma)-X_{r-1}(\pi-\sigma)=0, \qquad
P_r(\sigma)+P_{r-1}(\pi-\sigma)=0
\label{overlap}
\end{equation}
for $\sigma\in L$, where $L$ denotes the interval $[0, \tfrac{\pi}{2}]$ and $R$
denotes $[\tfrac{\pi}{2}, \pi]$.\\
For the zero mode sector of the Fock space $\cal{H}$ we shall use both oscillator
and momentum representations. We start with the oscillator representation.\\

\noindent{\textbf{2. Vertex in the zero mode oscillator representation.}}

\bigskip

3-vertex with the zero mode in the oscillator representation can be written in the form
\begin{equation}
|V_3\rangle =
\left[
\frac{(2\pi)^{-\frac{1}{4}}}{\sqrt{3}\bigl(\frac{1}{2}+V_{00}\bigr){\ap}^{\frac14}}
\right]^{26}
\exp\Bigl( -\frac12 \sum_{r,s} \sum_{m,n\geqslant 0}
a_m^{r\dag} V'{}^{rs}_{mn} a_n^{s\dag} \Bigr)|0\rangle_{123}.
\label{3vf}
\end{equation}
Here the strange at the first sight normalization factor has
been chosen to cancel unwanted factors in the proceeding expressions.
Note that in (\ref{3vf}) the vacuum $|0\rangle$ is
\begin{equation}
|0\rangle \equiv |0\rangle _{n\geqslant0}~\equiv \bigotimes_{n=0,1,\dots}
|0\rangle _{n}\equiv |0\rangle _{0}\otimes|0\rangle _{n
>0} \label{vac}
\end{equation}
and $a_n$ is related with $\alpha_n$ as
\begin{equation}
a_n=\frac{\alpha_n}{\sqrt{n}},\qquad a^{\dag}_{n}=\frac{\alpha_{-n}}{\sqrt{n}}.
\end{equation}
The vertex functions $V^{'rs}_{mn}$ have  the following properties:
\begin{enumerate}
\item[$1^{\circ}$.] $V^{rs}_{mn}=V^{sr}_{nm}$;
\item[$2^{\circ}$.] cyclic property
\begin{equation}
V^{'rs}=V^{'r+1,s+1}
\label{Vcycle}
\end{equation}
where summation is assumed $\mod 3$;
\end{enumerate}

and can be expressed as \cite{GrJe}
\begin{equation}
 V'{}^{rs}=\frac 13 (C+\alpha^{s-r}U'+\alpha^{r-s}\bar U'),
\label{V'form}
\end{equation}
where $\alpha=\exp(\frac{2\pi i}{3})$, $C_{mn}=(-1)^{n} \delta_{mn}$
is the twist operator and the matrices $U'$ and $\bar
U' \equiv C\,U'\,C$ satisfy the following relations
\begin{eqnarray}
(U')^2=(\bar U')^2=1, \quad (U')^\dag = U'.
\label{U'}
\end{eqnarray}
Equations ($\ref{V'form}$) are equivalent to the fact that the
matrix $V^{'rs}$ can be block-diagonalized in $r$, $s$ indices by
the matrix ${\cal O}$ satisfying ${\cal O}^{-1}={\cal O}^\dag$:
\begin{equation}
V={\cal O}^{-1}V'_D{\cal O} ,
\end{equation}
with
\begin{equation}
V'_D=\left(\begin{array}{ccc} C&0&0\\ 0&U'&0\\ 0&0&{\bar U'
}\end{array}\right), ~~~~~ {\cal O}=\frac{1}{\sqrt3}\left(
     \begin{array}{ccc}1&1&1\\
                       \alpha^*&\alpha&1\\
                       \alpha&\alpha^*&1
     \end{array}
               \right) .
\label{matrO}
\end{equation}

$$~$$

\noindent{\textbf{3. Vertex with zero mode in the momentum representation.}}

\bigskip

Relation between the zero mode vacuum in the momentum and oscillator
representation is
\begin{equation}
|0,p\rangle \equiv |p\rangle _{0} \otimes
|0\rangle _{n>0}= \left[\frac{\ap}{2\pi}\right
]^{\frac{13}{2}}\exp\Bigl(-\frac{\ap}4 p^2+\sqrt{\ap}a_0^\dag p - \frac 12 (a
_0^\dag )^2\Bigr) |0\rangle _{0}\otimes|0\rangle _{n>0}. \label{p-vac}
\end{equation}
One can rewrite (\ref{3vf}) in the momentum representation for the
zero modes
\begin{multline}
|V_3\rangle=\frac{1}{(2\pi)^{13}}\int dp^1dp^2dp^3\,\delta^{(26)}(p^1+p^2+p^3)
\\ \exp\Bigl[ -\frac12 \sum_{r,s}
\sum_{m,n\geqslant 1} a_m^{r\dag} V{}^{rs}_{mn} a_n^{s\dag} -\sqrt{\ap}\sum_{r,s}
\sum_{n\geqslant 1} p^rV{}^{rs}_{0n} a_n^{s\dag} - \frac{\ap}2 \sum_r
p^rV{}^{rr}_{00}p^r \Bigr]|0,p\rangle_{123}. \label{V3imp}
\end{multline}
 The element $V_{00}^{rr}$ is independent of $r$
and we denote it as $V_{00}=-2\log\gamma$, where $\gamma=\frac{4}{3\sqrt{3}}$.
Relations between vertices $V$ and $V'$ are
\begin{subequations}
\begin{align}
V^{\prime rs}_{mn} &= V^{rs}_{mn}-2\sum_t\frac{V_{m0}^{rt}V_{0n}^{ts}}{1+2V_{00}},
\quad\text{for}\quad n,m\geqslant 1;
\\
V^{\prime rs}_{m0} &= V^{\prime sr}_{0m}=\frac{2}{1+2V_{00}}V^{rs}_{m0},
\quad\text{for}\quad m\geqslant 1;
\\
V^{\prime rs}_{00} &= \frac{2}{3}\frac{1}{1+2V_{00}}
+\delta_{rs}\frac{2V_{00}-1}{2V_{00}+1}.
\end{align}
\end{subequations}

The vertex in the zero momentum space is
\begin{equation}
|V_3\rangle = \exp\Bigl( -\frac12 \sum_{r,s} \sum_{m,n\geqslant 1} a_m^{r\dag}
V{}^{rs}_{mn} a_n^{s\dag} \Bigr)|0,0\rangle_{123}, \label{3v-zeromom}
\end{equation}
where $V{}^{rs}_{mn}$ can be expressed in the same form as
$(\ref{V'form})$ \cite{Kostelecky-Potting},\cite{GrJe}.
\be
V{}^{r\,s}=\frac 13 (C+\alpha^{s-r} U+\alpha^{r-s}\bar U) ,
\end{equation}
where matrices $U$ and $\bar U \equiv C\,U\,C$ satisfy the
relations analogous to $(\ref{U'})$
\begin{eqnarray}
(U)^2=(\bar U)^2=1, \quad (U)^\dag = U.
\label{U}
\end{eqnarray}
and can be expressed in terms of the entries of $U'$ as
\be
U_{mn}=U'_{mn}+(V_{00}+\frac{1}{2})U'_{m0}U'_{0n},
\quad m,n\geqslant 1.
\end{equation}

Matrix $V^{rs}$ can be block-diagonalized by the matrix ${\cal O}$
(\ref{matrO}).\\
$|V_3\rangle$ can be explicitly written in terms of the Neumann
coefficients as
\begin{multline}
|V_3\rangle
=\frac{1}{(2\pi)^{13}}\int dp^1 dp^2 dp^3\delta (p^1+p^2+p^3)
\\
\times\exp\left(\frac{1}{2}
\sum_{r,s=1}^3\sum_{n,m\geqslant 1}^{\infty}\alpha_{-n}^{r}N_{nm}^{rs}
\alpha_{-m}^{s}+\sqrt{\ap}\sum_{r,s=1}^3\sum_{m\geqslant 1}p_{r}N^{rs}_{0m}\alpha^{s}_{-m}
+\frac{\ap}{2}N_{00}\sum_{r=1}^{3}p_{i}^{2}
\right)|0,p\rangle_{123}.
 \label{V3os}
\end{multline}
The Neumann coefficients $N_{nm}^{rs}$ represent the effect of
conformal transformations  $f_r^{(3)}$  of the upper half-disks of three
open strings (see Sect.\ref{sec:gluing-and-maps}) to the unit disc
 \begin{equation}
N_{nm}^{rs}=\frac{1}{nm}\oint\frac{dz}{2\pi
i}z^{-n}f_r^{(3)}{}^{\prime}(z) \oint\frac{dw}{2\pi
i}w^{-m}f_s^{(3)}{}^{\prime}(w)\frac{1}{(f_r^{(3)}(z)-f_s^{(3)}(w))^2}. \label{Nc}
\end{equation}

\subsection{$\star$ in  Left-Right Fock Space.}

\subsubsection{Vertices.}
Overlaps for $N$-vertex in terms of half-string coordinates and
momentum have the following forms
\begin{gather}
[\chi^{L}_{i}(\sigma)-\chi^{R}_{i-1}(\sigma)]|V_{N}\rangle=0, \;\;\sigma\in[0,\pi/2),\\
\nonumber
[\wp^{L}_{i}(\sigma)+\wp^{R}_{i-1}(\sigma)]|V_{N}\rangle=0,
\;\;\sigma\in[0,\pi/2),
\end{gather}
where $i=1,2,...,N$.

Using the method developed by Gross and Jevicki \cite{GrJe}
one can solve overlaps
explicitly  \cite{PR99} and get the
following answers for $I$, $V_2$ and $V_3$:
\begin{equation}
|I\rangle=e^{-b^{L\dag}_{1\,n}b^{R\dag}_{1\,n}}|0\rangle^L|0\rangle^R,
\label{1v-C}\end{equation}
\begin{equation}
|V_2\rangle=e^{-\sum_{i=1}^{2}b^{L\dag}_{i\,n}b^{R\dag}_{i-1\,n}}\Pi_{i=1}^{2}|0\rangle^L_i|0\rangle^R_i,
\label{2v-C}\end{equation}
\begin{equation}
|V_3\rangle=e^{-\sum_{i=1}^{3}b^{L\dag}_{i\,n}b^{R\dag}_{i-1\,n}}\Pi_{i=1}^{3}|0\rangle^L_i|0\rangle^R_i.
\label{3v-C}
\end{equation}
Here we use "comma" creation and  annihilation operators introduced
in Appendix 4.2.

In \cite{PR99} the proof of the equivalence of operator
formulation of the Witten vertex and comma theory in the case of $I$ and
$V_2$ is given. This proof is generalized for $3$-vertex in
\cite{9902176}. It is shown that $V_3$ is a solution of comma
overlap equations.
A more direct proof of the equivalence of the two forms of vertices
would be direct calculation of the string vertex operator in
corresponding full string modes and half-string modes starting
from the functional form (\ref{1}).

\subsubsection{$\star$ in half-string local basis.}
\label{sec:123}
From (\ref{1v-C}-\ref{3v-C}) one sees that "comma" vertices can be
written in terms of local variables, for example
\begin{equation}
|V_3\rangle=e^{-\sum_{i=1}^{3}\int d\sigma
b^{L\dag}_{i}(\sigma)b^{R\dag}_{i-1}(\sigma)}
\Pi_{i=1}^{3}|0\rangle^L_i|0\rangle^R_i. \label{3V-L}
\end{equation}
This can be seen in a more direct way. To this purpose one can following \cite{Okuyama}
rewrite conditions (\ref{overlap}) as single condition using
annihilation operator $b(\sigma)$
\footnote{We hope that the use of $b^{\dag},b$
for creation and annihilation operators does not make a confusion with ghost
operators.}
\begin{equation}
b(\sigma)=\sqrt{\pi\alpha'}P(\sigma)-
\frac{i}{2\sqrt{\pi\alpha'}}X(\sigma). \label{NAn}
\end{equation}
The commutator of $b(\sigma)$ and its hermitian conjugate
$b^{\dag}(\sigma)$ is found to be
\begin{equation}
[b(\sigma),b^{\dag}(\sigma')]=\delta(\sigma,\sigma'). \label{NCR}
\end{equation}
The overlap conditions (\ref{overlap}) are
\begin{equation}
b_r(\sigma)=-{b_{r-1}}^{\dag}(\pi-\sigma), ~\text{for}~~\sigma\in
L. \label{newoverlap}
\end{equation}
In terms of $b(\sigma )$ $|V_3\rangle$ has a simple form
\begin{equation}
|V_3\rangle=\exp \left( \sum^{N}_{r=1}\int^{\frac{\pi}{2}}_{0}
d\sigma\, b^{\dag}_r(\sigma)b^{\dag}_{r-1}(\pi-\sigma) \right )|{\cal
O}\rangle _1\otimes|{\cal O}\rangle _2\otimes |{\cal O}\rangle _3
\label{newV3}
\end{equation}
$$=\exp\left(\frac{1}{2}\sum^N_{r,s=1}\int^{\pi}_{0} d\sigma d\sigma'\,
b^{\dag}_r(\sigma)N^{rs}(\sigma,\sigma')b^{\dag}_{s}(\sigma') \right)|{\cal
O}\rangle _1\otimes|{\cal O}\rangle _2\otimes |{\cal O}\rangle _3.
$$
The function $N^{rs}(\sigma,\sigma')$ is defined as
$$
N^{rs}(\sigma,\sigma')=[\delta_{r-1,s}\theta_L(\sigma)
+\delta_{r+1,s}\theta_R(\sigma)]\delta(\pi-\sigma,\sigma'),
$$
where $\theta_L(\sigma)$ and $\theta_R(\sigma)$ are step functions
with supports $[0, \pi/2]$ and $[\pi/2, \pi]$, respectively.
$|{\cal O}\rangle$ is a vacuum with respect to $b(\sigma)$
\begin{equation}
b(\sigma)|{\cal O}\rangle=0, ~~ \text{for}~~ \sigma\in [0,\pi].
\label{a-vac}
\end{equation}

The reflector $_{12}\langle R|$ and the identity $ \langle I|$ are
the subject for the following relations \cite{Okuyama}

\begin{equation}
_{12}\langle R|b_r(\sigma)=-_{12}\langle R|b^{\dag}_{r-1}(\pi-\sigma),
~~ \sigma\in L, ~~ r=1,2, \label{def-refl}
\end{equation}

\begin{equation}
\langle I|b^{\dag}(\sigma)=-\langle I|b(\pi - \sigma).
\label{def-ident}
\end{equation}
In terms of $b(\sigma)$ one has  simple expressions for the
reflector and the identity
\begin{equation}
 _{12}\langle R|=_1\langle {\cal O}|\otimes _2\langle {\cal O}|
\exp [-\sum^{2}_{r=1} \int^{\pi/2}_0d\sigma\,
b_r(\sigma)b_{r-1}(\pi-\sigma)],~~r=1,2, \label{refl}
\end{equation}

\begin{equation}
\langle I|=\langle {\cal O}|\exp [-\frac12 \int^{\pi}_0d\sigma\, b
(\sigma)b(\pi-\sigma) ]. \label{ident}
\end{equation}
$b(\sigma)$ and $b^{\dag}(\sigma)$ can be expanded as
\begin{equation}
b(\sigma) =  b_0\sqrt{\frac{1}{\pi}}+
\sqrt{\frac{2}{\pi}}\sum_{n=1}^\infty{b_n\cos(n\sigma)},
\label{amode}
\end{equation}
\begin{equation}
b^{\dag}(\sigma) =  b^{\dag}_0\sqrt{\frac{1}{\pi}}+
\sqrt{\frac{2}{\pi}}\sum_{n=1}^\infty{b^{\dag}_n\cos(n\sigma)}.
\label{a+mode}
\end{equation}

The pairs $b_n$, $b^{+}_n$ and $a_n$,
 $a^{\dag}_n$ are related by
the Bogoliubov transformation
\begin{equation}
b_n=Ua_nU^{-1},~~b^{+}_n=Ua^{\dag}_nU^{-1},
\label{Bog-tr}
\end{equation}
where $U$ is an operator
\begin{equation}
U=\exp\left(\frac14\sum_{n=1}^{\infty}\log n
(a_n^2-a_{n}^{\dag 2})\right)
\label{Bog-exp}
\end{equation}
or
\begin{equation}
b_n=\frac{n+1}{2\sqrt{n}}a_{n} +\frac{n-1}{2\sqrt{n}}a^{\dag}_{n}.
\label{an-alpha}
\end{equation}
Note that (\ref{Bog-tr}) is not a proper Bogoliubov transformation
from  creation $a^{\dag}_{n}$ and annihilation
$a_{n}$ operators to creation $b^{\dag}_n$ and annihilation
$b_n$ operators since the set $\{(n-1)/(n+1) \}$ is not quadratically
summable.  See \cite{Berezin}, Theorem 1 in Section 2.
The operator $U$
define a map from the Fock space of initial creation and
annihilation operators $a^{\dag}_{n}$,
$a_{n}$ to a new space built on the cyclic
vector $|\cal {O}\rangle$ that is the vacuum for the new
annihilation operators. To give an operator meaning to (\ref{Bog-exp})
one can use a dressing procedure
(see for example \cite{AKulish}).

The zero mode transforms as
\begin{equation}
b_0=\sqrt{\alpha'}p-\frac{i}{2\sqrt{\alpha'}}\,x. \label{a0-alpha}
\end{equation}
The formal relation between $|{\cal O}\rangle$ and the zero momentum
vacuum $|0,0\rangle$ is
\begin{equation}
|{\cal O}\rangle=e^{\frac12 b_0^{\dag 2}}U|0,0\rangle. \label{vac-va}
\end{equation}

\subsubsection{$\star$ in $b$-holomorphic basis.}

A coherent state with respect to
 $b(\sigma)$
\begin{equation}
b(\sigma)|f\rangle =f(\sigma)|f\rangle \label{identa}
\end{equation}
is given by the following formula
\begin{equation}
|f\rangle =\exp [\int^{\pi}_0d\sigma b^{\dag}(\sigma)f(\sigma)]|{\cal
O}\rangle. \label{a+}
\end{equation}
The hermitian conjugate $\langle {f}|$ satisfies
\begin{equation}
\langle  f|b^{\dag}(\sigma )=\langle  f |\bar{f}(\sigma ). \label{hc}
\end{equation}

The completeness condition holds

\begin{equation}
\int [d\bar{f}df]\, |f\rangle e^{-\int^\pi_0d\sigma |f(\sigma
)|^2} \langle  f| =1, \label{complitness}
\end{equation}
where the measure is given by
$$
 [d\bar{f}df]=\prod_{n=0}^{\infty}\frac{d\bar{f}_ndf_n}{2\pi i},
$$
 where
 $f_n$ are
coefficients in the Fourier expansion and
as usual $d\bar{f}_ndf_n=2i dx_ndy_n$.

Extremely simple form takes a product of two coherent states if
the  notation is used for $|f_L, f_R\rangle$ denotes $|f\rangle$
with the functions
$f_{L,R}(\sigma)=\theta_{L,R}(\sigma)f(\sigma)$,
\begin{equation}
|f_L, f_R\rangle\star|g_L, g_R\rangle
=e^{-\int_0 ^{\frac{\pi}{2}}f(\pi-\sigma)g(\sigma)}|f_L, g_R\rangle.
\label{star-coh}
\end{equation}

Let us check that $|{\cal O}\rangle$ satisfies left-right factorization
identities (\ref{projectionForm}). The Bogoliubov transformations
(\ref{a0-alpha}, \ref{an-alpha}) can be rewritten in the form
\begin{gather*}
b_0=\cosh \theta_0 a_0+ \sinh \theta_0 a^{\dag}_0,\\
b_n=\cosh \theta_n a_n+ \sinh \theta_n a^{\dag}_n
\end{gather*}
and
\begin{equation}
|{\cal O}\rangle=\exp(-\frac{1}{2} a^{\dag}_iS_{ij}a^{\dag}_{j})|0,0\rangle,
\label{VacA}
\end{equation}
where the matrix $S_{ij}=\tanh \theta_i \delta_{ij}$ ($i,j=0,1,...$) can be written explicitly
\begin{gather}
S_{00}=\frac{1}{3},\\
S_{nm}=\frac{n-1}{n+1}\delta_{nm}, \quad n,m>0.
\end{gather}
In the coordinate representation (\ref{VacA}) corresponds to
\begin{equation}
\exp\left(-\frac{1}{2}xE^{-1}\frac{1-S}{1+S}E^{-1}x\right),
\end{equation}
where
\begin{equation}
E^{-1}_{ij}=\delta_{ij}\sqrt{i}+\delta_{i0}\delta_{j0}\sqrt{2}.
\end{equation}
Here we have used that
\begin{equation}
|x\rangle=\exp\left( -\frac{1}{2}xE^{-2}x-ia^{\dag}\sqrt{2}E^{-1}x+\frac{1}{2}a^{\dag\,2}\right).
\end{equation}
One gets $L_{ij}$ matrix (\ref{factorizedModes}) of the form
\begin{gather}
L_{00}=(E^{-1}\frac{1-S}{1+S}E^{-1})_{00}=1\\
L_{nm}=(E^{-1}\frac{1-S}{1+S}E^{-1})_{nm}=\delta_{nl}\sqrt{n}\delta_{lk}\frac{1}{l}
\delta_{km}\sqrt{m}=\delta_{nm}.
\label{LO}
\end{gather}
This concludes the proof of the factorization.

\subsubsection{Summary.}
Let us summarize what we presented in this subsection. We have found
a realization of $\star$ in the
string  Fock space  over vacuum ${\cal O}$. This realization has the
following nice properties:
\begin{subequations}
\begin{itemize}
\item
 ${\cal O}$
is a projector with respect to the $\star$-multiplication
\begin{equation}
|{\cal O}\rangle\star|{\cal O}\rangle=|{\cal O}\rangle;
\label{vac-vac}
\end{equation}
\item ${\cal O}$ does not belong the Fock space;
\item ${\cal O}$ satisfies to left-right factorization
identities;
\item There is also the identity  operator $|I\rangle$:
\begin{align}
|I\rangle \star |I\rangle &= |I\rangle ,
\label{vac-id}
\\
|I\rangle\star |{\cal O}\rangle &=|{\cal O}\rangle;
\label{vac-ido}
\end{align}
\item
Vertex has a simple "comma" form (\ref{newV3});
\item
The coherent states over this vacuum are multiplied according to a
simple formula (\ref{star-coh}).
\end{itemize}
\end{subequations}

\subsection{General Properties of Cyclic Vertices in Sliver Basis.}
\subsubsection{Properties of vertex on sliver.}

In this section we will mostly follow the analysis of Kostelecky and
Potting \cite{Kostelecky-Potting} (see also \cite{Japan2}).

Let us suppose that
3-vertex operator (\ref{3v-zeromom}) written in terms of
the creation and annihilation operators
$s^{\dag}$ and $s$ can be presented in the form
\begin{equation}
|V_3\rangle = \exp\Bigl( -\frac12 \sum_{r,s} \sum_{m,n\geqslant 0} s_m^{\dag r} {\hat
V}{}^{rs}_{mn} s_n^{{\dag}s} \Bigr)|\Xi\rangle_{123}, \label{3v-s}
\end{equation}
where $|\Xi\rangle$
is the  s-vacuum
$$
s|\Xi\rangle=0,
$$
${\hat V}{}^{rs}$ obeys the cyclic symmetry (\ref{Vcycle}) and
$({\hat V}{}^{rs})^t={\hat V}{}^{sr}$.

The cyclic symmetry means that the matrix $\hat{V}$ has the
following form
\begin{equation}
\begin{pmatrix}
  \hat{V}^{11} & \hat{V}^{12} & \hat{V}^{21} \\
  \hat{V}^{21} & \hat{V}^{11} & \hat{V}^{12} \\
  \hat{V}^{12} & \hat{V}^{21} & \hat{V}^{11}
\end{pmatrix}
\label{HATV}
\end{equation}

Let us assume that reflector $\langle R|$ has the form
\begin{equation}
\langle R|=_1\!\langle \Xi|\otimes\,  _2\! \langle \Xi| \exp\Bigl(-\sum_{m,n}
s^{1}_n C_{nm}s^{2}_m\Bigr). \label{s-ref}
\end{equation}
Let us also suppose that the following identities (these
identities are the same as identities
(\ref{vac-vac}-\ref{vac-ido}) from the previous section) take
place
\begin{subequations}
\begin{align}
|\Xi\rangle\star|\Xi\rangle&=|\Xi\rangle,
\label{sigma-sigma}
\\
|\tilde{I}\rangle\star |\tilde{I}\rangle&=|\tilde{I}\rangle, \label{sI-sI}
\\
|\Xi\rangle \star |\tilde{I}\rangle&=|\tilde{I}\rangle \star |\Xi\rangle=|\Xi\rangle,
\label{sigma-id}
\end{align}
where
\begin{equation}
|\tilde{I}\rangle=\exp \{-\frac12\sum s^\dag _nC_{nm}s^\dag
_m\}|\Xi\rangle. \label{s-id}
\end{equation}
\end{subequations}

Using the cyclic symmetry one can prove that (\ref{sigma-sigma}) leads
to
\begin{equation}
{\hat V}{}^{rr}=0. \label{diag-s}
\end{equation}
Indeed,
\begin{equation*}
_{1}\!\langle\Xi|\otimes\,_{2}\!\langle\Xi|V\rangle_{123}=
_{12}\langle\Xi|\exp\Bigl(-\frac{1}{2}s_{r}^{\dag}\hat{V}^{rs}s^{\dag}_{s}\Bigr)|\Xi\rangle_{123}
=\exp\Bigl(-\frac{1}{2}s_{3}^{\dag}\hat{V}^{11}s^{\dag}_{3}\Bigr)|\Xi\rangle_{3}.
\end{equation*}
We will widely use this property in the following calculations.
\\
From (\ref{sigma-id}) it follows that
\begin{equation}
\hat{V}^{12}C\hat{V}^{21}=0,\qquad \hat{V}^{21}C\hat{V}^{12}=0.
\label{quad}
\end{equation}
Indeed,
\begin{multline*}
_{2}\langle \tilde{I}|\otimes\,_{1}\langle
\Xi|V\rangle_{123}=_{12}\langle\Xi|\exp\Bigl(-\frac{1}{2}s_2Cs_2\Bigr)
\exp\Bigl(-s_{2}^{\dag}\hat{V}^{12}s^{\dag}_{3}
-s_{1}^{\dag}\hat{V}^{21}s^{\dag}_{3}-s_{1}^{\dag}\hat{V}^{12}s^{\dag}_{2}\Bigr)|\Xi\rangle_{123}\\
=_{2}\langle\Xi|\exp\Bigl(-\frac{1}{2}s_2Cs_2\Bigr)\exp\Bigl(-s^{\dag}_2\hat{V}^{12}s^{\dag}_3\Bigr)|\Xi\rangle_{23}
=\exp\Bigl(-\frac{1}{2}s^{\dag}_{3}\hat{V}^{12}C\hat{V}^{21}s^{\dag}_{3}\Bigr)|\Xi\rangle_3,
\end{multline*}
that leads us to the identity $\hat{V}^{12}C\hat{V}^{21}=0$.
Identity $\hat{V}^{21}C\hat{V}^{12}=0$ can be achieved analogously.

From (\ref{sI-sI}) it follows that
\begin{equation}
\hat{V}^{21}C\hat{V}^{21}C\hat{V}^{21}+\hat{V}^{12}C\hat{V}^{12}C\hat{V}^{12}=C.
\label{lin}
\end{equation}
Indeed,
\begin{align*}
_{2}\langle \tilde{I}|\otimes\,_{1}\langle \tilde{I}|V&\rangle_{123}
=_{12}\langle\Xi|\exp\Bigl(-\frac{1}{2}s_2Cs_2\Bigr)\exp\Bigl(-\frac{1}{2}s_1Cs_1\Bigr)
\exp\Bigl(-s_{2}^{\dag}\hat{V}^{12}s^{\dag}_{3}
-s_{1}^{\dag}\hat{V}^{21}s^{\dag}_{3}-s_{1}^{\dag}\hat{V}^{12}s^{\dag}_{2}\Bigr)|\Xi\rangle_{123}
\\
\propto&\;_{2}\langle\Xi|\exp\Bigl(-\frac{1}{2}s_2Cs_2\Bigr)
\exp\Bigl[-\frac{1}{2}(s^{\dag}_3\hat{V}^{12}+s^{\dag}_2\hat{V}^{21})C
(\hat{V}^{12}s^{\dag}_2+\hat{V}^{21}s^{\dag}_3)\Bigr]
\exp\Bigl(-s_{2}^{\dag}\hat{V}^{12}s^{\dag}_{3}\Bigr)|\Xi\rangle_{23}
\\
\propto&\;_{2}\langle\Xi|\exp\Bigl(-\frac{1}{2}s_2Cs_2\Bigr)\exp\Bigl(-s^{\dag}_2\hat{V}^{21}C\hat{V}^{21}s^{\dag}_3
-s_{2}^{\dag}\hat{V}^{12}s^{\dag}_{3}\Bigr)|\Xi\rangle_{23}
\\
\propto&\;_{2}\langle\Xi|\exp\Bigl(-\frac{1}{2}s_2Cs_2\Bigr)
\exp\Bigl[-s^{\dag}_2(\hat{V}^{21}C\hat{V}^{21}
+\hat{V}^{12})s^{\dag}_{3}\Bigr]|\Xi\rangle_{23}
\\
\propto&\exp\Bigl[-\frac{1}{2}s^{\dag}_3(\hat{V}^{21}+\hat{V}^{12}C\hat{V}^{12})C
(\hat{V}^{21}C\hat{V}^{21}+\hat{V}^{12})s^{\dag}_3\Bigr]|\Xi\rangle_3
\\
\propto&\exp\Bigl[-\frac{1}{2}s^{\dag}_3(\hat{V}^{21}C\hat{V}^{21}C\hat{V}^{21}
+\hat{V}^{12}C\hat{V}^{12}C\hat{V}^{12})s^{\dag}_3\Bigr]|\Xi\rangle_3.
\end{align*}

Introducing notations
\begin{equation}
L=\hat{V}^{12}C,~~~R=\hat{V}^{21}C
\label{prLR}
\end{equation}
one can rewrite (\ref{quad},\ref{lin}) in the form
\begin{equation}
LR=0,~~~ RL=0,~~~ L^3+R^3=1.
\end{equation}
This means that we get the projections operators $R^2=R$ and
$L^2=L$.
Using the notations (\ref{prLR}) and taking into account
(\ref{diag-s}) one can rewrite $C\hat V$
as
\begin{equation}
\label{hV}
C\hat{V}_3=\left(
\begin{array}{ccc}
0&L&R\\
R&0&L\\
L&R&0
\end{array}\right).
\end{equation}

Let us now suppose that creation and annihilation operators $s$
and $s\dag$ are related to $a$ and $a\dag$ via the Bogoliubov
transformation
\begin{equation}
\label{s-a} s=w(a+Sa^\dag),~~~s^\dag=(a^\dag+Sa)w,
\end{equation}
where
\begin{equation}
w=(1-S^2)^{-1/2}.
\end{equation}
Vacua are related via
\begin{eqnarray}
&|\Xi\rangle=&\hbox{Det}(w)^{-1/2} \exp\Bigl(-\frac12a^\dag
Sa^\dag\Bigr)|0\rangle,
\nonumber\\
&|0\rangle=&\hbox{Det}(w)^{-1/2} \exp\Bigl( \frac12 s^\dag
Ss^\dag\Bigr)|\Xi\rangle.
\end{eqnarray}

Let us prove that the vertex given in the $a$ and  $a\dag$ basis
by (\ref{3vf}, \ref{3v-zeromom}) satisfies the cyclic property in the
$s$ and $s\dag$ basis. We must note here that the analysis is the
same for the case of the zero momentum in momentum representation
for the zero mode (\ref{3v-zeromom}) and for the case of oscillator
representation for the zero mode (\ref{3vf}). Reversing the Bogoliubov
transformation
$$a=w(s-Ss^\dag) ,~~~a^{\dag}=(s^\dag-sS)w$$
one can rewrite  the 3-vertex in terms of $s$
\begin{multline}
|V_3\rangle=\exp\Bigl(-\frac{1}{2}a^{r\dag}V^{rs}a^{s\dag}\Bigr)|0\rangle_{123}
\\
\propto\exp\Bigl(-\frac12 a^{r\dag} (V^{rs}-S\delta^{rs})
a^{s\dag}\Bigr)|\Xi\rangle_{123}
\\
\propto\exp\left[-\frac12 (s^\dag+s S)^r\{w(V-S)w \}^{rs}
(s^\dag+Ss)^s\right]|\Xi\rangle_{123}, \label{3vertexs}
\end{multline}
where the symbol $S$ is understood as $S_{mn}\delta^{rs}$.

Using (\ref{bosonicr}) one can rewrite equation (\ref{3vertexs}) as
\begin{equation}
|V_3\rangle\propto\exp \Bigl(-\frac12 s^{r\dag}\hat V{}^{rs}
s^{s\dag}\Bigr)|\Xi\rangle_{123}, \label{3ver-s}
\end{equation}
where
\begin{equation}
\hat V =(1-VS)^{-1}(V-S). \label{3vert-s}
\end{equation}

Since $S$ is diagonal, matrix $\hat V $ given by
(\ref{3vert-s}) can be diagonalized by the same matrix ${\cal O}$
that diagonalizes $V$
\begin{equation}
\hat V ={\cal O}^{-1}(1-V_DS)^{-1}(V_D-S){\cal O},
\end{equation}
where $(1-V_DS)^{-1}(V_D-S)$ is the diagonal matrix with the
diagonal elements $\hat {V}_{D}^{rr}$. Therefore,
\begin{equation}
\hat V=\frac{1}{3}\left(
     \begin{array}{ccc}1&\alpha&\alpha^*\\
                       1&\alpha^*& \alpha\\
                      1&1&1
     \end{array}
               \right)\left(
     \begin{array}{ccc}\hat {V}_{D}^{11}&0&0\\
                       0&\hat {V}_{D}^{22}&0\\
                       0&0&\hat {V}_{D}^{33}
     \end{array}
               \right)
               \left(
     \begin{array}{ccc}1&1&1\\
                       \alpha^*&\alpha&1\\
                       \alpha&\alpha^*&1
     \end{array}
               \right)
\end{equation}
where $\hat {V}_D^{rr}$ are given by
\begin{gather}
\hat {V}_D^{11}=(1-CS)^{-1}(C-S)=C,
\nonumber\\
\hat {V}_D^{22}=(1-US)^{-1}(U-S)\equiv \Lambda,
\nonumber\\
\hat {V}_D^{33}=(1-\bar US)^{-1}(\bar U-S)\equiv {\bar \Lambda}.
\label{transformedev3}
\end{gather}
Note that if $[V,S]=0$ then $\hat V=V$, since $C^2=1$, $U^2=1$ and
$\bar{U}^2=1$ (\ref{U},\ref{U'}).\\
Matrices $\Lambda$ and $\bar\Lambda$ satisfy relations $\Lambda^2=1$ and $\bar\Lambda^2=1$.\\
Indeed,
\begin{multline*}
\Lambda^2=(1-US)^{-1}(U-S)(1-US)^{-1}(U-S)
\\=(U-S)^{-1}U^{-1}(U-S)(U-S)^{-1}U^{-1}(U-S)=1.
\end{multline*}

Let us prove that if $S$ is twist symmetric
$$[C,S]=0$$
then the identity $|I\rangle$ which can be achieved by the
Bogoliubov
transformation is equal to the state introduced by (\ref{s-id}).\\
Indeed, using eq. (\ref{bosonicr})
\begin{multline*}
|I\rangle=\exp\Bigl(-\frac{1}{2}a^{\dag}Ca^{\dag}\Bigr)|0\rangle\\
=\exp\Bigl(-\frac{1}{2}a^{\dag}Ca^{\dag}\Bigr)
\exp\Bigl(\frac{1}{2}a^{\dag}Sa^{\dag}\Bigr)|\Xi\rangle
=\exp\Bigl(-\frac{1}{2}(s^{\dag}-sS)w(C-S)w(-Ss+s^{\dag})\Bigr)|\Xi\rangle\\
\propto\exp\Bigl(-\frac{1}{2}s^{\dag}(1-w(C-S)wS)^{-1}w(C-S)ws^{\dag}\Bigr)|\Xi\rangle=
\exp\Bigl(-\frac{1}{2}s^{\dag}Cs^{\dag}\Bigr)|\Xi\rangle=|\tilde{I}\rangle
\end{multline*}
In the same way it can be shown that $\langle R|$ achieved by the
Bogoliubov transformation is equal to the one
introduced by (\ref{s-ref}).\\
Indeed,
\begin{multline}
_{12}\langle R|=_{12}\langle 0|\exp(-a_{1}Ca_{2})
=_{12}\langle\Xi|\exp\Bigl(\frac{1}{2}a_1Sa_1\Bigr)
\exp\Bigl(\frac{1}{2}a_2Sa_2\Bigr)\exp(-a_{1}Ca_{2})\\
=_{12}\langle\Xi|\exp\Bigl[\frac{1}{2}(s-s^{\dag}S)^s(wTw)^{rs}(-Ss^{\dag}+s)^s\Bigr]
=_{12}\langle\Xi|\exp(-s_1Cs_2),
\end{multline}
where we have used the notation
\begin{equation}
T^{rs}=
\begin{pmatrix}
  S & -C \\
  -C & S
\end{pmatrix}.
\end{equation}
Note that twist symmetry lead us to the relations
$\bar\Lambda=C\Lambda C$ and $L=CRC$.

\subsubsection{Equation for Bogoliubov transformation for sliver basis.}
As one can see from (\ref{diag-s}) the matrix $S$ should satisfy
the following requirement to create the sliver vacuum $\Xi$
\begin{equation}
C+\Lambda+\bar\Lambda=0. \label{SlEq}
\end{equation}
Following \cite{Kostelecky-Potting} we impose the constraints
\begin{equation}
[V,SC]=0.
\label{CommutV}
\end{equation}
Commutation relations (\ref{CommutV}) can
be rewritten in the form
\begin{equation}
CS=SC,\qquad US=S\bar{U},\qquad \bar{U}S=SU. \label{CommutU}
\end{equation}
Let us now solve eq. (\ref{SlEq}):
\begin{equation}
C+(1-US)^{-1}(U-S)+(1-\bar{U}S)^{-1}(\bar{U}-S)=0. \label{SlEqU}
\end{equation}
Multiplying (\ref{SlEqU}) from both sides with $(1-US)$ and using
the identity
\begin{equation}
(\bar{U}-S)(1-US)=(1-\bar{U}S)(\bar{U}-S)
\end{equation}
one gets
\begin{equation}
(C+U+\bar{U})(1+S^2-CS)-3S=0.
\end{equation}
This equation was proposed and solved by Kostelecky and Potting in
\cite{Kostelecky-Potting}.

\subsubsection{Algebra of coherent states on sliver.}
Using the projections $L$ and $R$, following \cite{Japan2} we can
split $s$ into L and R parts
\begin{equation}
 s=s_L+s_R, ~~~s_L=\sum_{n}s_{Ln}e_{n},~~~s_R=\sum_{n}s_{Rn}f_{n},
\end{equation}
where $e_{n}$ and $f_{n}$ are the orthonormal basises of the
eigenspaces of $L$ and $R$:
\begin{equation}
 e_{n}\cdot e_{m}=f_{n}\cdot f_{m}=\delta_{nm},\quad
e_{n}\cdot f_{m}=0,\quad f_{n}=-Ce_{n},
\end{equation}
\begin{equation}
 Le_{n}=e_{n},\quad Re_{n}=0,\quad Lf_{n}=0,\quad Rf_{n}=f_{n},
\end{equation}
\begin{equation}
 L=\sum_{n}e_{n}e_{n}^T,\quad R=\sum_{n}f_{n}f_{n}^T.
\end{equation}
Since $s_L$ and $s_R$ commute,
\begin{equation}
 [s_{Ln},s^{\dag}_{Lm}]=[s_{Rn},s^\dag_{Rm}]=\delta_{nm},\quad
[s_{Ln},s^\dag_{Rm}]=0,
\end{equation}
the Hilbert space is factorized into the Fock spaces of $s_L$ and
$s_R$
\begin{equation}
{\cal H}_{str}={\cal H}_L\otimes {\cal H}_R.
\end{equation}
In terms of $s_{L,R}$, the 3-string vertex takes a ``comma'' form
\begin{equation}
|V_3\rangle=\exp\Big(s^\dag_{R1}s^\dag_{L2}+s^\dag_{R2}s^\dag_{L3}+s^\dag_{R3}s^\dag_{L1}
\Big)|\Xi\rangle_{123} ,
\end{equation}
where
\begin{equation}
 s^\dag_Ls^\dag_R=\sum_{n}s^\dag_{Ln}s^\dag_{Rn}.
\label{Ssum}
\end{equation}
The identity string field and the reflector become
\begin{equation}
|I\rangle=e^{s^\dag_Rs^\dag_L}|\Xi\rangle,\quad _{12}\langle
R|={}_{12}\langle\Xi|e^{s_{R1}s_{L2}+s_{R2}s_{L1}}.
\end{equation}

It is convenient to introduce the coherent states
($f_{L}s^{\dag}_L$, $f_{R}w_{L}$ are defined as in (\ref{Ssum}))
\begin{eqnarray}
 |f_L,f_R\rangle&=&\exp(f_L s^{\dag}_L+f_R s^{\dag}_R)|\Xi\rangle, \\
 I(f_L,f_R)&=& \exp(f_L s^{\dag}_L+f_R s^{\dag}_R)|I\rangle.
\end{eqnarray}
The multiplication rules for these coherent states are
\begin{subequations}
\begin{eqnarray}
 |f_L,f_R\rangle\star|w_L,w_R\rangle&=&e^{f_R w_L}|f_L,w_R\rangle,
 \label{LefRay}\\
 I(f_L,f_R)\star |w_L,w_R\rangle&=&e^{f_R w_L}|f_L+w_L,w_R\rangle,
\label{Icoh}\\
 |f_L,f_R\rangle\star I(w_L,w_R)&=&e^{f_R w_L}|f_L,f_R+w_R\rangle,
\label{cohI}\\
 I(f_L,f_R)\star I(w_L,w_R)&=&e^{f_R w_L}I(f_L+w_L,f_R+w_R).
\label{Izstar}
\end{eqnarray}
\end{subequations}

\subsection{Algebra of String Operators.}

\subsubsection{Operator realization of string functionals.}

There is a general GSM construction which maps
 an algebra of states to an operator algebra.
 Here we show that the comma form of 3-vertex
 permits to write an explicit form of  a map
 of the algebra of states to an operator algebra
 that has the property (\ref{state-oper}, \ref{fun-mult}).
 Following \cite{Japan2}
 we introduce the string fields

 \begin{equation}
|A_{n}\rangle=s^\dag _{Rn}|I\rangle,\quad |A^{\dag}_{n}\rangle=s^\dag
_{Ln}|I\rangle. \label{strosci}
\end{equation}
 They satisfy the canonical commutation relations:
 \begin{equation}
 [|A_{n}\rangle,|A^{\dag}_{m}\rangle]_{\star}=\delta_{nm}|I\rangle,
\label{CCRstar}
\end{equation}
where $[\ ,\ ]_\star$ denotes the commutator with the star string
product:
$$
[A, B]_\star = A \star B-B \star A.
$$


We also have
\begin{equation}
 |A_{n}\rangle\star|\Xi\rangle=|\Xi\rangle\star |A^{\dag}_{n}\rangle=0.
\label{Vac-sl}
\end{equation}
On a linear span of states
\begin{equation}
|\Xi\rangle, \qquad |A^{\dag}_{n_1}\rangle \star |\Xi\rangle, \qquad
|A^{\dag}_{n_1}\rangle \star|A^{\dag}_{n_2}\rangle \star |\Xi\rangle,
\qquad\dots \label{LS2}
\end{equation}
 one can introduce an action of operator
${\bf A}^{\dag}_n=|A^{\dag}_n\rangle \star $
\begin{subequations}
\begin{align}
\nonumber
{\bf A}^{\dag}_n|\Xi\rangle &=
       |A^{\dag}_n\rangle \star |\Xi\rangle ,\\
\nonumber {\bf A}^{\dag}_{n_1}|A^{\dag}_{n_2}\rangle \star |\Xi\rangle &=
|A^{\dag}_{n_1}\rangle \star |A^{\dag}_{n_2}\rangle \star |\Xi\rangle ,\\
\nonumber {\bf A}^{\dag}_{n_3}|A^{\dag}_{n_2}\rangle \star |A^{\dag}_{n_1}\rangle
\star |\Xi\rangle &= |A^{\dag}_{n_3}\rangle \star
|A^{\dag}_{n_2}\rangle \star |A^{\dag}_{n_1}\rangle \star |\Xi\rangle,\\
\nonumber ...&
\end{align}
\end{subequations}
and the operator ${\bf A}_{n}=|A_{n}\rangle \star$
\begin{subequations}
\begin{align}
\nonumber
{\bf A}_{n}|\Xi\rangle &=|A_{n}\rangle \star |\Xi\rangle=0,\\
\nonumber {\bf A}_{n_1}|A_{n_2}\rangle \star |\Xi\rangle &=
|A_{n_1}\rangle\star|A_{n_2}\rangle \star |\Xi\rangle,\\
\nonumber {\bf A}_{n_3}|A_{n_2}\rangle \star |A_{n_1}\rangle \star
|\Xi\rangle &= |A_{n_3}\rangle \star
|A_{n_2}\rangle \star |A_{n_1}\rangle \star |\Xi\rangle,\\
\nonumber ...&
\end{align}
\end{subequations}
We see that due to (\ref{CCRstar}) and the fact that $|I\rangle
\star$ acts on the states (\ref{LS2}) as the identity,
 $ {\bf A}^{\dag}_{n}$
and $ {\bf A}_{n}$ satisfy the commutation relation for creation
and annihilation operators
\begin{equation}
[{\bf A}_{n},{\bf A}^{\dag}_{m}]=\delta_{nm}. \label{CACR2}
\end{equation}
$|\Xi\rangle$ is a vacuum with respect to ${\bf A}_{n}$, i.e.
$$ {\bf A}_n|\Xi\rangle =0. $$
Therefore, the linear span (\ref{LS2}) is naturally embedded in
the Fock space ${\bf{\cal H}}$ representation of  (\ref{CACR2}).
Next, one can perform a  map
 of a linear span
 \begin{equation}
 \label{cLS2}
|\Xi\rangle,~~~
|\Xi\rangle \star |A_n\rangle ,~~~
|\Xi\rangle \star |A_{n_1}\rangle\star |A_{n_2}\rangle ,\qquad \dots
\end{equation}
to the following conjugated states in the space ${\bf{\cal H}}$
  \begin{subequations}
\begin{align}
\nonumber |\Xi\rangle &\Longrightarrow
\langle \Xi|,\\
\nonumber |\Xi\rangle \star |A_n\rangle &
         \Longrightarrow\langle \Xi|{\bf A}_n , \\
\nonumber |\Xi\rangle \star |A_{n_1}\rangle\star |A_{n_2}\rangle &
\Longrightarrow\langle \Xi |{\bf A}_{n_1} {\bf A}_{n_2},\\
\nonumber \cdots &
\end{align}
\end{subequations}

For a general state of the form
\begin{equation}
|N_k,M_l\rangle\equiv |A^{\dag}_{n_{1}}\rangle
\star...|A^{\dag}_{n_k}\rangle \star |\Xi\rangle
 \star|A_{m_1}\rangle \star ...|A_{m_l}\rangle
\label{gen2}
\end{equation}
we can consider a map
\begin{equation}
|N_k,M_l\rangle \Longrightarrow |N_{k}\rangle \langle M_{l}|,
\label{genMap2}
\end{equation}
\begin{equation}
|N_{k}\rangle \langle M_l| = \prod_{i=1}^k{\bf A}_{n_i}|\Xi\rangle
\langle \Xi|\prod _{j=1}^{l}{\bf A}_{m_j}. \label{defNN2}
\end{equation}
This map is consistent with $\star$ multiplication of two states
in the form (\ref{gen2}), i.e it defines a morphism. One can check
this using (\ref{LefRay}--\ref{Izstar}).

It is convenient to introduce  ${\bf A}$-coherent states
$|f\rangle$, and it's conjugate
\begin{equation}
\langle g|=\langle \Xi|e^{ \sum\bar{g}_n{\bf A}_n},~~~ |f\rangle
=e^{\sum f_m{\bf A}^{\dag}_m}|\Xi\rangle. \label{String-coh2}
\end{equation}
Using the morphism (\ref{genMap2}) one can write
\begin{equation}
 |f_L,f_R\rangle= e^{f_Ls^\dag_L}|I\rangle\star|\Xi\rangle\star e^{f_Rs^\dag_R}
 |I\rangle
=e^{f_LA^{\dag}}|\Xi\rangle\langle \Xi|e^{f_RA}\equiv|f_L\rangle\langle f_R|.
\label{cohsA}
\end{equation}
Note that
this relation is consistent with the trace
\begin{equation}
 \Tr \Big(|f_L,f_R\rangle\Big)=\langle I|f_L,f_R\rangle=e^{f_Lf_R}.
\end{equation}

One can write
\begin{equation}
 |N,M\rangle=\prod_{i,j}\frac{s_{L_i}^{\dag n_{i}}s_{Rj}^{\dag m_{j}}}
 {\sqrt{n_{i}!m_{j}!}}|\Xi\rangle \Rightarrow \prod_{i}\frac{A_{i}^{\dag
 n_{i}}}{\sqrt{n_{i}!}}|\Xi\rangle\langle \Xi|\prod_{j}\frac{A_{j}^{m_j}}
{\sqrt{m_{j}!}}=|N\rangle\langle M|.
\end{equation}

The above construction works for half-string representation as
well
 \cite{Okuyama}. In this case using notations of (\ref{sec:123})
we introduce the following string fields
\begin{equation}
|A(\sigma)\rangle=\theta_{R}(\sigma)b^{\dag}(\sigma)\,|I\rangle, \qquad
|A^{\dag}(\sigma)\rangle=-\theta_{R}(\sigma)b^{\dag}(\pi-\sigma)\,|I\rangle.
\qquad
\end{equation}
One can check that
\begin{equation}
[~|A(\sigma)\rangle~,~
|A^{\dag}(\sigma')\rangle ~]_{\star}=\delta (\sigma,\sigma')\,|I\rangle, \qquad
\label{Cr-An}
\end{equation}
We also have
\begin{equation}
|A(\sigma)\rangle \star |{\cal O}\rangle =0, \qquad |{\cal
O}\rangle  \star|A^{\dag}(\sigma)\rangle =0. \label{Vac}
\end{equation}
On a linear span of states
\begin{equation}
|{\cal O}\rangle, \qquad
|A^{\dag}(\sigma)\rangle \star |{\cal O}\rangle, \qquad
|A^{\dag}(\sigma_1)\rangle \star|A^{\dag}(\sigma_2)\rangle \star |{\cal O}\rangle,
\qquad\dots
\label{LS}
\end{equation}
 one can introduce an action of operator
${\bf A}^{\dag}(\sigma)=|A^{\dag}(\sigma)\rangle \star $
\begin{subequations}
\begin{align}
{\bf A}^{\dag}(\sigma)|{\cal O}\rangle &=
       |A^{\dag}(\sigma)\rangle \star |{\cal O}\rangle ,\nonumber\\
{\bf A}^{\dag}(\sigma_1)|A^{\dag}(\sigma_2)\rangle \star |{\cal O}\rangle &=
|A^{\dag}(\sigma_1)\rangle \star |A^{\dag}(\sigma_2)\rangle \star |{\cal O}\rangle ,\nonumber\\
{\bf A}^{\dag}(\sigma_3)|A^{\dag}(\sigma_2)\rangle
\star |A^{\dag}(\sigma_1)\rangle \star |{\cal O}\rangle &=
|A^{\dag}(\sigma_3)\rangle \star
|A^{\dag}(\sigma_2)\rangle \star |A^{\dag}(\sigma_1)\rangle \star |{\cal O}\rangle,\nonumber\\
...&\nonumber
\end{align}
\end{subequations}
and the operator ${\bf A}(\sigma)=|A(\sigma)\rangle \star$
\begin{subequations}
\begin{align}
{\bf A}(\sigma)|{\cal O}\rangle &=|A(\sigma)\rangle \star |{\cal O}\rangle
=0,\nonumber\\
{\bf A}(\sigma_1)|A(\sigma_2)\rangle \star |{\cal O}\rangle &=
|A(\sigma_1)\rangle\star|A(\sigma_2)\rangle \star |{\cal O}\rangle,\nonumber\\
{\bf A}(\sigma_3)|A(\sigma_2)\rangle \star |A(\sigma_1)\rangle
\star |{\cal O}\rangle &= |A(\sigma_3)\rangle \star
|A(\sigma_2)\rangle \star |A(\sigma_1)\rangle \star |{\cal O}\rangle,\nonumber\\
...&\nonumber
\end{align}
\end{subequations}
We see that due to (\ref{Cr-An})
and the fact that $|I\rangle \star$ acts on the states
(\ref{LS}) as the identity,
 operators $ {\bf A}^{\dag}(\sigma)$
and $ {\bf A}(\sigma)$ satisfy the commutation relation for
creation
and annihilation operators
\begin{equation}
[{\bf A}(\sigma),{\bf A}^{\dag}(\sigma')]=\delta(\sigma,\sigma').
\label{CACR}
\end{equation}
$|{\cal O}\rangle$ is a vacuum with respect to ${\bf A}(\sigma)$,
i.e.
$$ {\bf A}(\sigma)|O\rangle =0. $$
Therefore, the linear span (\ref{LS}) is naturally embedded
in  the Fock space ${\bf{\cal H}}$ representation of  (\ref{CACR}).
Next, one can perform a  map
 of a linear span
 \begin{equation}
 \label{cLS}
|{\cal O}\rangle,~~~
|{\cal O}\rangle \star |A(\sigma)\rangle ,~~~
|{\cal O}\rangle \star |A(\sigma_1)\rangle\star |A(\sigma_2)\rangle ,\qquad \dots
\end{equation}
to the following conjugated states in the space
${\bf{\cal H}}$
  \begin{subequations}
\begin{align}
|{\cal O}\rangle &\Longrightarrow
\langle {\cal O}|,\nonumber\\
|{\cal O}\rangle \star |A(\sigma)\rangle &
         \Longrightarrow\langle {\cal O}|{\bf A}(\sigma) , \nonumber\\
|{\cal O}\rangle \star |A(\sigma_1)\rangle\star |A(\sigma_2)\rangle &
\Longrightarrow\langle {\cal O} |{\bf A}(\sigma_1) {\bf A}(\sigma_2),\nonumber\\
\cdots &\nonumber
\end{align}
\end{subequations}

For a general state of the form
\begin{equation}
|N_{n,m}(\{\sigma\},\{\sigma '\})\rangle\equiv
|A^{\dag}(\sigma_1)\rangle \star...|A^{\dag}(\sigma_n)\rangle \star |{\cal O}\rangle
 \star|A(\sigma'_1)\rangle \star ...|A(\sigma'_{m})\rangle
\label{gen}
\end{equation}
we can consider a map
\begin{equation}
|N_{n,m}(\{\sigma\},\{\sigma '\})\rangle
\Longrightarrow
|N_{n}(\{\sigma\}\rangle \langle N_{m}\{\sigma '\})|,
\label{genMap}
\end{equation}
\begin{equation}
|N_{n}(\{\sigma\}\rangle \langle N_{m}\{\sigma '\})|
=
\prod _i^n{\bf A}^{\dag}(\sigma_i)|{\cal O}\rangle
\langle {\cal O}|\prod _i^{n'}{\bf A}(\sigma'_i).
\label{defNN}
\end{equation}
This map is consistent with $\star$ multiplication of two states in
the form (\ref{gen}). Indeed,
\begin{equation}
|N_{n,n'}(\{\sigma\},\{\sigma '\})\rangle \star
|N_{m,m'}(\{\tau\},\{\tau '\})\rangle=\delta _{n',m}\sum _{P}\prod _i
\delta (\sigma _i'-\tau_{P_i})|N_{n,n'}(\{\sigma\},\{\tau '\})\rangle,
\label{mulNN}
\end{equation}
that corresponds to the product
\begin{equation}
|N_{n}(\{\sigma\}\rangle \langle N_{n'}\{\sigma '\})|~
\cdot
~|N_{m}(\{\tau\}\rangle \langle N_{m'}\{\tau '\})|.
\label{prodNN}
\end{equation}

It is convenient to introduce  ${\bf A}$-coherent states $|f\ra $
$$A(\sigma)|f\ra =
\theta_R(\sigma)f(\sigma)|f\ra ,~~~~
\la g|A^{\dag}(\sigma)=
\la g|\theta_R(\sigma)\bar{g}(\sigma).$$
Explicit form of  ${\bf A}$-coherent state and it's conjugate is
\begin{equation}
\la g|=\langle {\cal O}|e^{\int_R \bar{g}\cdot{\bf A}},~~~ |f\ra
=e^{\int_R f\cdot {\bf A}^{\dag}}|{\cal O}\rangle. \label{String-coh}
\end{equation}

The main observation at \cite{Okuyama} is the formula
\begin{equation}
|f_L, f_R\rangle =
e_{\star}^{-\int_R f(\pi-\sigma)|A(\sigma)\rangle}
\star|{\cal O}\rangle
\star e_{\star}^{\int_R f(\sigma)|A(\sigma)\rangle }.
\label{MO}
\end{equation}
Using the map (\ref{genMap}) one can write
\begin{equation}
e_{\star}^{-\int_R f(\pi-\sigma)|A(\sigma)\rangle}
\star|{\cal O}\rangle
\star e_{\star}^{\int_R f(\sigma)|A(\sigma)\rangle }\Longrightarrow
e^{-\int_R f(\pi-\sigma){\bf A}^{\dag}(\sigma)}
|{\cal O}\rangle\langle {\cal O}|
 e^{\int_R f(\sigma){\bf A}(\sigma)}.
 \label{MM}
\end{equation}
Using the notations (\ref{String-coh}) one can rewrite (\ref{MM}) as
\begin{equation}
|f_L, f_R\rangle\Longrightarrow
|-r(f_L)\ra ~
\la \bar{f}_R|.
 \label{MMC}
\end{equation}

One can check directly that if
\begin{subequations}
\begin{align}
|f_L, f_R\rangle &\Longrightarrow |-r(f_L)\ra ~ \la \bar{f}_R|,
\nonumber
\\
 |g_L, g_R\rangle &\Longrightarrow
|-r(g_L)\ra ~ \la \bar{g}_R|, \label{MMCc}
\end{align}
\end{subequations}
then the product of operators in the r.h.s. of eq. (\ref{MMCc}) is
\begin{equation}
|-r(f_L)\ra ~
\la \bar{f}_R|~\cdot ~|-r(g_L)\ra ~
\la \bar{g}_R|=
e^{-\int _R f_R\cdot r(g_L)}~
|-r(f_L)\ra\la \bar{g}_R|,
\label{oprod}
\end{equation}
that in accordance with (\ref{star-coh}) and (\ref{MMC}) corresponds to the product
of states $|f_L, f_R\rangle\star |g_L, g_R\rangle$, that proves the
correspondence (\ref{MMC}).

The integration of the open string field theory
$$\int|f_L,f_R \rangle = \langle I|f_L,f_R \rangle$$
corresponds to
\begin{equation}
\Tr(| -r(f_L)\ra \la f_R|)=\la f_R|
 -r(f_L )\ra =e^{-\int _R \bar{f}_R\cdot r(f_L)}.
\label{tr-int}
\end{equation}

\subsection{Matrix Realization.}

String fields  ${\bf A}(\sigma)$ and ${\bf A}^{\dag}(\sigma)$ can be
expanded as
\begin{equation}
{\bf A}(\sigma)=\sqrt{\frac{2}{\pi}}{\bf A}_0+\frac{2}{\sqrt{\pi}}\sum^{\infty}_{n=1} {\bf A}_n\,
\cos(2n\sigma),
\qquad
{\bf A}^{\dag}(\sigma)=\sqrt{\frac{2}{\pi}}{\bf A}^{\dag}_0+\frac{2}{\sqrt{\pi}}\sum^{\infty}_{n=1} {\bf A}^{\dag}_n\,
\cos(2n\sigma),
\label{A-exp}
\end{equation}
in terms of $\{\cos(2n\sigma)\}_{n=0,1,\dots}$ instead of
$\{\cos(n\sigma)\}_{n=0,1,\dots}$. Although ${\bf
A}(\sigma)$ and ${\bf A}^{\dag}(\sigma)$ are defined on the interval
$[\tfrac{\pi}{2}, \pi]$, the operator ${\bf A}(\sigma)$ can be extended to
an operator ${\bf A}_{ext}$ on the interval $[0, \pi]$ such that
${\bf A}_{ext}(\sigma)=\theta_R(\sigma){\bf A}_{ext}(\sigma)+
\theta_L(\sigma){\bf A}_{ext}(\pi-\sigma)$, and similarly for
${\bf A}_{ext}^{\dag}(\sigma)$. Because these operators ${\bf
A}_{ext}$ and ${\bf A}^{\dag}_{ext}$ satisfy the Neumann boundary
condition, they are expanded in terms of
$\{\cos(n\sigma)\}_{n=0,1,\dots}$. However they also satisfy
${\bf A}_{ext}(\sigma)= {\bf A}_{ext}(\pi-\sigma)$ and are thus
expanded by the even-integer modes
$\{\cos(2n\sigma)\}_{n=0,1,\dots}$. The restriction of ${\bf
A}_{ext}(\sigma)$ and ${\bf A}_{ext}^{\dag}(\sigma)$ to the interval
$[\tfrac{\pi}{2}, \pi]$ gives ${\bf A}$ and ${\bf A}^{\dag}$.

The relation  between the modes $a^{\dag}_n$ and the modes ${\bf A}_n$ and
${\bf A}^{\dag}_n$ can be stated by using the transition matrices collected
in Appendix 4.1.\\
To state a correspondence between operators and matrices one has
to consider the operator $|-r(f_L)\ra ~
\la \bar{f}_R|$ in the occupation numbers basis
\begin{equation}
|{\bf n}\ra=
\sqrt{1\over n_0!\, n_1!\dots}
({\bf A}_0^{\dag})^{n_0}({\bf A}^{\dag}_1)^{n_1}\dots|{\cal O}\rangle,
\label{OB}
\end{equation}
where the infinitely dimensional vector ${\bf n}$ denotes
$(n_0,n_1,\dots)$.  In terms of such states $|{\bf n}\rangle \rangle $,
we have
\begin{equation}
|-r(f)\ra =
\sum^{\infty}_{n_0,n_1,\dots=0}(-1)^{(\sum^{\infty}_{i=0}{n_i})}
{(f_{L,0})^{n_0}(f_{L,1})^{n_1}\dots\over \sqrt{n_0!\, n_1!\dots}}
|{\bf n}\ra,
\label{EOB}
\end{equation}
and we get a map
\begin{equation}
|f_L,f_R\rangle \Longleftrightarrow |-r(f_L)\ra \la
f_R|\Longleftrightarrow f_{\bf n,m}, \label{S-O-M}
\end{equation}
where
\begin{equation}
f_{\bf n,m}=\la {\bf n}|-r(f_L)
\ra \la f_R|{\bf m}
\ra.
\label{con}
\end{equation}

\subsection{Ghost Sector.}
 The ghost sector of the open bosonic string can be described in
terms of fermionic ghost fields $c (\sigma), b (\sigma)$ (about
bosonization  in terms of a single bosonic scalar field $\phi
(\sigma)$ see \ref{sec:supbos}).

To write down the Faddeev-Popov determinant corresponding to the conformal
gauge one has to introduce
two sets of ghosts (see eq. 3.1.35 in \cite{GSW} ) $c^+, b_{++}$
and $c^-, b_{--}$.  They satisfy the first order equations,
\begin{eqnarray}
\partial _-c^+=0,~~\partial _+c^-=0,\\
\partial _-b_{++}=0,~~\partial _+b_{--}=0,
\end{eqnarray}
i.e. $c^+=c^+(\tau+\sigma)$
and $c^-=c^-(\tau-\sigma)$ and the same for $b$.
The Neumann boundary conditions
\begin{equation}
c^+=c^-|_{\sigma=0,\pi},~~~b_{++}=b_{--}|_{\sigma=0,\pi}
\end{equation}
left us with two functions only
$c^+=c^-\equiv c$,  $b_{++}=b_{--}\equiv b$,  one for the
ghost and one for the antighost field, that  satisfy $2\pi $-periodic
boundary conditions
\begin{equation}
c^{\pm} (\sigma + 2 \pi) = c^{\pm} (\sigma), \;\;\;\;\;
b_{\pm\pm} (\sigma + 2 \pi) = b_{\pm\pm} (\sigma)\,.
\label{cb-eq}
\end{equation}
These Grassmann
fields have mode decompositions
\begin{eqnarray}
         c^{\pm}(\sigma,\tau )& = &
\sum_{n } c_ne^{ in(\tau\pm\sigma)}  \, \\
b_{\pm\pm}(\sigma,\tau )& = &
\sum_{n } b_ne^{in(\tau\pm\sigma)}\, .
\label{mod-ghosts}
\end{eqnarray}
The ghost creation and annihilation operators satisfy
\begin{equation}
\{c_n ,b_m\} =\delta_{n+m,0}\, ,\quad  \{c_n ,c_m\}=\{b_n ,b_m\}= 0 \, .
\label{ccr-ghosts}
\end{equation}

The ghost Fock space has a pair of vacua $| \pm \rangle$ annihilated
by $c_n, b_n$ for $n > 0$.  These two vacua
satisfy
\begin{eqnarray*}
c_0 | -\rangle = | + \rangle & \hspace*{0.5in}  &
c_0 | + \rangle = 0  \\
b_0 | + \rangle = | -\rangle &  &  b_0 | -\rangle = 0\, .
\end{eqnarray*}

We define the Fock space so that $b_0, c_0$ are hermitian and
$c^\dagger_n=c_{-n},\,
b^\dagger_n=b_{-n}$.
It follows that $\langle +|+\rangle=
\langle +|c_0|-\rangle =0$, and similarly $\langle -|-\rangle=0.$
We normalize the vacua so that
\begin{equation}
         \langle +|b_0|+\rangle= \langle -|c_0|-\rangle=1\, .
\end{equation}
We will use  the ghost vacuum $| + \rangle\equiv| 0 \rangle $.
The ghost number operator is
\begin{equation}
G=\sum_{n=1 }^\infty\left[ c_{-n}b_n-b_{-n}c_n\right]+\frac{1}{2} \left[
c_{0}b_0-b_{0}c_0\right]  + \frac{1}{2}
\end{equation}
so that the $c_n$ and $b_n$ have ghost number 1 and -1 respectively.
The vacua $|  + \rangle$ and $| -\rangle$ have ghost numbers 1 and 0
respectively.

It is easy to describe the ghost sector in the
bosonized language (see Sect. \ref{sec:supbos}) using the functional
point of view.
In this language the star product in the ghost
sector is given by (\ref{eq:matter-star}) with an extra insertion of $\exp
(3i \phi (\tfrac{\pi}{2})/2)$ inside the integral and  the integration
is given by (\ref{star-int})
with an insertion of $\exp (-3i \phi (\tfrac{\pi}{2})/2)$ inside the integral.

In operator language a precise meaning to these  expressions
(\ref{eq:matter-star},\ref{star-int}) was given  in
\cite{GrJe,cst,Samuel,Ohta}, where ghost part of  the star
product and two- and three-string vertices was written  in terms of
 ghost creation and annihilation  operators $ b_n, c_n$.


\subsubsection{Ghost half-strings.}
In this section we give a construction of the comma ghosts in the
fermionic representation straightly following
\cite{AbdurGhost1},\cite{AbdurGhost2}. Ghost coordinates
(\ref{mod-ghosts}) can be rewritten in form
\begin{gather}
c_{\pm}(\sigma)=\sum_{n=-\infty}^{\infty}c_{n}e^{\pm n
\sigma}=c(\sigma)\pm i\pi_{b}(\sigma),\\
b_{\pm}(\sigma)=\sum_{n=-\infty}^{\infty}b_{n}e^{\pm n
\sigma}=\pi_{c}(\sigma)\pm i b(\sigma).
\end{gather}
The overlap equations for N-strings are
\begin{gather}
c^{s}(\sigma)+c^{s-1}(\sigma)=0,\\
\pi^{s}_{c}(\sigma)-\pi^{s-1}_{c}(\pi-\sigma)=0,
\end{gather}
where $s=1,2,...,N$ and one has similar equations for $b(\sigma)$
and $\pi_{b}(\sigma)$, with the role of coordinates and momenta
exchanged.\\
These overlaps can be explicitly solved to give the following
formula for the identity $|I^{ghost}\rangle$
\begin{equation}
|I^{ghost}\rangle=b_{+}(\tfrac{\pi}{2})b_{-}(\tfrac{\pi}{2})
\exp\Bigl(\sum_{n=1}^{\infty}(-1)^{n}c_{-n}b_{-n}\Bigr) |0\rangle,
\end{equation}
 where the midpoint insertions correspond
to the bosonized insertions mentioned in the previous subsection.\\
The 3-string vertex $|V^{ghost}_3\rangle$ takes the form
\begin{equation}
|V^{ghost}_3\rangle=\exp\Bigl(\sum_{r,s=1}^{3}\sum_{n= 1, m= 0
}^{\infty}c_{-n}^{r}\tilde{V}^{rs}_{nm}b^{s}_{-m}\Bigr)|0\rangle,
\end{equation}
where the matrices $\tilde{V}^{rs}_{mn}$ have cyclic symmetry as
usually.

One can introduce the comma ghost coordinates
$c^{r}_{\pm}(\sigma)=c^{r}(\sigma)\pm i\pi^{r}_{b}$, where
\begin{align}
c^{L}(\sigma)&=c(\sigma),\\
c^{R}(\sigma)&=c(\pi-\sigma),
\end{align}
for $0\leqslant\sigma \leqslant \tfrac{\pi}{2}$. Likewise defined
$\pi_{c}^{r}(\sigma)$, $b^{r}(\sigma)$ and $\pi_{b}^{r}(\sigma)$, where $r=1,2;L,R$.
We choose even extension to the interval $[0,\tfrac{\pi}{2})$ for the ghost
coordinate $c^{r}(\sigma)$ and conjugate momentum
$\pi^{r}_{c}(\sigma)$ and odd extension for the antighost
coordinate $b^{r}(\sigma)$ and conjugate momentum
$\pi^{r}_{b}(\sigma)$:
\begin{subequations}
\begin{align}
c^{r}(\sigma)&=c^{r}_{0}+\sqrt{2}\sum_{n=1}^{\infty}g_{2n}^{r}\cos
2n\sigma,\\
\pi^{r}_{c}(\sigma)&=b_{0}^{r}+\sqrt{2}\sum_{n=1}^{\infty}y_{2n}^{r}\cos
2n\sigma,\\
b^{r}(\sigma)&=\sqrt{2}\sum_{n=1}^{\infty}h_{2n-1}^{r}\sin(2n-1)\sigma,\\
\pi_{b}^{r}(\sigma)&=\sqrt{2}\sum_{n=1}^{\infty}z_{2n-1}^{r}\sin(2n-1)\sigma.
\end{align}
\end{subequations}
Relations between full-string modes and half-string modes for
$c(\sigma)$ and $c^{L,R}(\sigma)$ are
\begin{gather}
c_{0}=\frac{1}{2}(c_{0}^{1}+c_{0}^{2}),\nonumber\\
\frac{1}{\sqrt{2}}(c_{2n}+c_{-2n})=\frac{1}{2}(g_{2n}^{1}+g_{2n}^{2}),\;\;
n\geqslant 1,\label{GHccoor}\\
\frac{1}{\sqrt{2}}(c_{2n-1}+c_{-2n+1})=\frac{\sqrt{2}}{\pi}\;\frac{(-1)^{n}}{2n-1}\sum_{r=1}^{2}(-1)^{r}c_{0}^{r}+\sum_{r=1}^{2}(-1)^{r}\sum_{m=1}^{\infty}B_{2m,2n-1}g_{2m}^{r},\;\;
n\geqslant 1.\nonumber
\end{gather}
Relations between momentum $\pi_{c}(\sigma)$ and $\pi^{L,R}_{c}(\sigma)$
are
\begin{gather}
b_{0}=\frac{1}{2}(b_{0}^{1}+b_{0}^{2}),\nonumber\\
\frac{1}{\sqrt{2}}(b_{2n}+b_{-2n})=\frac{1}{2}(y_{2n}^{1}+y_{2n}^{2}),
\;\; n\geqslant 1,\label{GHpiccoor}\\
\frac{1}{\sqrt{2}}(b_{2n-1}+b_{-2n+1})=\frac{\sqrt{2}}{\pi}\;\frac{(-1)^{n}}{2n-1}\sum_{r=1}^{2}(-1)^{r}b_{0}^{r}+\sum_{r=1}^{2}(-1)^{r}\sum_{m=1}^{\infty}B_{2m,2n-1}y_{2m}^{r},
\;\; n\geqslant 1.\nonumber
\end{gather}
Relations between $b(\sigma)$ and $b^{L,R}(\sigma)$ are
\begin{gather}
\frac{1}{\sqrt{2}}(b_{2n}-b_{-2n})=\sum_{r=1}^{2}\sum_{m=1}^{\infty}(-1)^{r}\frac{2n}{2m-1}B_{2n,2m-1}h^{r}_{2m-1},\nonumber\\
\frac{1}{\sqrt{2}}(b_{2n-1}-b_{-2n+1})=\frac{1}{2}(h^{1}_{2n-1}+h^{2}_{2n-1}).
\label{GHbcoor}
\end{gather}
Relations between momentum $\pi_{b}(\sigma)$ and $\pi^{L,R}_{b}(\sigma)$
are
\begin{gather}
\frac{1}{\sqrt{2}}(c_{2n}-c_{-2n})=\sum_{r=1}^{2}\sum_{m=1}^{\infty}(-1)^{r}\;\frac{2n}{2m-1}B_{2n,2m-1}z^{r}_{2m-1},\nonumber\\
\frac{1}{\sqrt{2}}(c_{2n-1}-c_{-2n+1})=\frac{1}{2}(z^{1}_{2n-1}+z^{2}_{2n-1}).\label{GHpibcoor}
\end{gather}
The inverse relations introduced in Appendix 4.4.\\
Let us introduce the comma ghost modes (see Appendix 4.4)
\begin{gather}
\gamma^{r}_{n}\equiv\frac{1}{2}(y^{r}_{n}+i z^{r}_{n}),
\;\;n\geqslant
1,\\
\beta^{r}_{n}\equiv\frac{1}{2}(g^{r}_{n}+i h^{r}_{n}),
\;\;n\geqslant 1,
\end{gather}
with $\gamma^{r}_{-n}=\gamma^{r\dag}_{n}$ and
$\beta^{r}_{-n}=\beta^{r\dag}_{n}$. The zero modes are defined by
$\gamma_{0}^{r}\equiv \frac{1}{\sqrt{2}}c^{r}_{0}$ and
$\beta_{0}^{r}\equiv \frac{1}{\sqrt{2}}b^{r}_{0}$. These
coordinates satisfy anticommutation relations (see Appendix 4.4)
\begin{equation}
\{\gamma^{r}_{n},\beta^{s}_{m}\}=\delta^{rs}\delta_{n+m,0}.
\end{equation}
Let us now present solutions of overlap equations in the comma
coordinates. The comma overlaps for vertices
$|V^{ghost}_{N}\rangle$ are
\begin{align}
c^{L}_{j}(\sigma)&=-c^{R}_{j-1}(\sigma),\\
\pi^{L}_{c\,j}(\sigma)&=\pi^{R}_{c\,j-1}(\sigma),
\end{align}
where $j=1,2,...,N$. For $b^{r}(\sigma)$ and $\pi^{r}_{b}(\sigma)$
one gets similar equations with the relative sign exchanged. These
overlaps can be explicitly solved in terms of $\gamma^{r}$ and
$\beta^{r}$ using the procedure of \cite{GrJe} and give the
following answer for the identity $|I^{ghost}\rangle$:
\begin{equation}
|I^{ghost}\rangle=\exp\Bigl(\beta^{R}_{-n}\gamma^{L}_{-n}
+\beta^{L}_{-n}\gamma^{R}_{-n}\Bigr)|\Omega\rangle^{L}|\Omega\rangle^{R},\;\;\text{where}
\;\;\gamma^{r}_{0}|\Omega\rangle^{r}=0.
\end{equation}
For $|V^{ghost}_{3}\rangle$ one gets
\begin{multline}
|V^{ghost}_{3}\rangle=\exp\Bigl(\beta^{1,L}_{-n}\gamma^{3,R}_{-n}+\beta^{2,L}_{-n}\gamma^{1,R}_{-n}+\beta^{3,L}_{-n}\gamma^{2,R}_{-n}
\\
+\beta^{1,R}_{-n}\gamma^{2,L}_{-n}+\beta^{2,R}_{-n}\gamma^{3,L}_{-n}+\beta^{3,R}_{-n}\gamma^{1,L}_{-n}\Bigr)\Pi_{i=1}^{3}|\Omega\rangle^{L}_{i}|\Omega\rangle^{R}_{i}.
\end{multline}
In \cite{AbdurGhost2} it is shown that the Witten cubic ghost
vertices solve the comma overlaps. This shows the equivalence of
two theories at the level of vertices.



\subsection{Appendix 4.1. Transition functions between two basis.}
There are two complete sets of orthogonal functions on $[0,\tfrac{\pi}{2}]$
\begin{subequations}
\begin{equation}
 \cos(2m-1)\sigma, ~~m=1,2,...
\label{eq:cos-od}
\end{equation}
or
\begin{equation}
 1,~~\cos2n\sigma , ~~n=1,2,...
\label{eq:cos-odd}
\end{equation}
\end{subequations}
Using explicit formulae for integrals
\begin{equation}
\int_0^{\pi/ 2} d\sigma \cos2n\sigma\cos(2m-1)\sigma=
-\frac{\pi}{2}B_{2n,2m-1},~~~~~~ ({\rm for} \,\, n\geqslant 0)
\label{eq:cosint}
\end{equation}
where
\begin{equation}
\label{B-def} B_{2n,2m-1}=\frac{1}{\pi} (-1)^{n+m}
\left(\frac{1}{2n+2m-1}-\frac{1}{2n-2m+1}\right)
\end{equation}
and
\begin{equation}
\int_0^{\pi/ 2} d\sigma \cos(2n-1)\sigma\cos(2m-1)\sigma= \frac{
\pi}{4}\delta _{n,m},
\label{eq:cosint'}
\end{equation}
\begin{equation}
\int_0^{\pi/ 2} d\sigma \cos 2n\sigma\cos 2m\sigma=\frac{
\pi}{4}\delta _{n,m} \label{eq:cosint''}
\end{equation}
one can rewrite
functions of one set in terms of functions of the other set.
This gives explicit relations between full open string modes
and the left-half and right-half modes.

\subsubsection{$x_n$ and $x^{L,R}_m$.}

We can compare (\ref{defl}), (\ref{defr}) and (\ref{Xmode})
to get an expression for the full open string modes in terms
of the left-half and  right-half modes:

\begin{equation}
x_n = \sum _{m=1}A_{nm}^+ \,x^{L}_m + \sum _{m=1} A^-_{nm} \,
x^{R}_m, \qquad n\geqslant 1,
\label{x-LR}
\end{equation}
where the matrices $A^\pm$ are
\begin{equation}
\label{A+-odd} A_{2n-1,m}^+=-A_{2n-1,m}^-= \frac{1}{2}\delta_{n,m},
\end{equation}
\begin{equation}
\label{A+-even}
A_{2n,m}^+ =A_{2n,m}^- =-B_{2n,2m-1}.
\end{equation}
One can rewrite  (\ref{x-LR}) more explicitly
\begin{equation}
x_{2n-1}= \frac12(x^{L}_{n} - x^{R}_{n}), \qquad n\geqslant 1,
\end{equation}
\begin{equation}
x_{2n}=\sum_{m=1}\frac{1}{\pi}(-1)^{n+m-1}
\left(\frac{1}{2m+2n-1}+\frac{1}{2m-2n-1}\right)(\,x^{L}_m+x^{R}_m),\qquad
n\geqslant 1.
\label{x-LReven-Ap}
\end{equation}
One can invert (\ref{x-LR}) and write the left-half modes and
right-half modes in terms of the full string modes
\begin{equation}
\label{Inv}
x^{L}_m = \sum_{m=1}\tilde A_{mn}^+ \,x_n, ~~~~~
 x^{R}_m = \sum_{m=1}\tilde
A_{mn}^-\, x_n,
\qquad n\ge 1\, ,
\end{equation}
where
\begin{gather}
\nonumber
\tilde A_{m,2n-1}^\pm  =2 A_{2n-1,m}^\pm=\delta_{n,m},\\
 \tilde A_{m,2n}^\pm  =2 A_{2n,m}^\pm -\frac{4}{\pi}
\frac{(-1)^{n+m-1}}{2m-1} = \frac{4}{\pi} (-1)^{n+m-1}
\frac{4n^2}{ \left((2 m-1)^2-4n^2\right)(2m - 1)}. \label{tildeA}
\end{gather}
One can rewrite (\ref{Inv}) as
\begin{equation}
\label{Inv'}
x^{L}_m = x_{2m-1}+\sum _{l=1}\tilde A_{m,2l}^+ \,x_{2l}, ~~~~~
\end{equation}

\begin{equation}
\label{Inv''} x^{R}_m = -x_{2m-1}+\sum _{l=1}\tilde A_{m,2l}^+\,
x_{2l}, \qquad m \geqslant 1 .
\end{equation}

Note that
\begin{equation}
\label{egtwist}
A^+ = C A^-, \quad \tilde A^+ = \tilde A^- C.
\end{equation}

Matrices $\tilde{A}^{\pm}$ and $A^{\pm}$ also satisfy the
relations
\begin{equation}
A^+ \tilde A^+ + A^- \tilde A^- = 1 ,~~
\tilde A^\pm A^\mp = 0 ,~~
\tilde A^\pm A^\pm = 1 .
\end{equation}
To check these properties the following identities are useful

\begin{equation}
\label{ed-B1}
\sum_{k=1}^{\infty}B_{2n,2k-1}B_{2m,2k-1}=\frac{1}{4}\delta_{nm},
\end{equation}
\begin{equation}
\label{ed-B2}
\sum_{k=1}^{\infty}\frac{2k}{2n-1}B_{2n-1,2k}B_{2k,2m-1}=-\frac{1}{4}\delta_{nm}.
\end{equation}

From (\ref{defl}) and (\ref{defr}) one sees that the relationship
between $x_n^\mu$ and $(x_n^{L\mu}$, $x_n^{R\mu})$ does not
involve the zero mode $x_0$ of $X$.

\subsubsection{$x_n$ and $l_{2m+1}, r_{2m+1}$.}

Relations between the full-string modes and the half-string modes
$r_{2n+1}$ and $l_{2n+1}$ are
\begin{equation}
x_{2n +1}= \frac12(l_{2n+1} - r_{2n+1}),
\end{equation}
\begin{equation}
x_{2n} = \frac12\sum _{m=0}X_{2n,2m+1}(\,l_{2m+1} +  r_{2m+1}),
\label{x-llr}
\end{equation}
where
\begin{equation}
X_{2n,2m+1}=X_{2m+1,2n}=-2B_{2n,2m+1}, ~~~n\neq 0, \label{X-A}
\end{equation}
and
\begin{equation}
X_{0,2m+1}=X_{2m+1,0}=-\sqrt{2} B_{0,2m+1}. \label{X0}
\end{equation}
The useful identity for this matrix is
\begin{equation}
\sum_{k=0}^{\infty}X_{2n+1,2k}(2k)^2X_{2k,2m+1}=(2n+1)^2\delta_{nm}\,.
\label{XXiden}
\end{equation}
The matrix
\begin{equation}
\label{GoodMatrix}
X\equiv\begin{pmatrix}
  0 & X_{2k+1,2n} \\
  X_{2n,2k+1} & 0
\end{pmatrix}
\end{equation}
is symmetric and orthogonal: $X=X^{T}=X^{-1}$.\\
One can invert (\ref{x-lr}) and write the left-half modes and
right half modes in terms of the full string modes
\begin{eqnarray}
\label{lr-x-lr}
l_{2k+1}& = & x_{2k+1}+\sum_{n=0}X_{2k+1,2n} \,x_{2n}, \\
r_{2k+1}& = & -x_{2k+1}+\sum_{n=0}X_{2k+1,2n} \,x_{2n}.
\end{eqnarray}

\subsubsection{$x_n$ and $\chi ^{L,R}_{2m}$.}
One can invert (\ref{x-lr-e}) and  write the left-half modes and
right-half modes in terms of the full string modes
\begin{eqnarray}
\label{lr-x-0}
\chi ^L_{0}& = & x_{0}+\frac{2\sqrt{2\ap}}{\pi}\sum_{k=1}
\frac{(-1)^k}{2k-1}\,x_{2k-1}, \\
\chi ^R_{0}& = & x_{0}-\frac{2\sqrt{2\ap}}{\pi}\sum_{k=1}
\frac{(-1)^k}{2k-1}\,x_{2k-1},
\end{eqnarray}
\begin{eqnarray}
\label{lr-x-n}
\chi ^L_{2n}& = & x_{2n}+2\sum_{k=1}B_{2n,2k-1} \,x_{2k-1}, \\
\chi ^R_{2n}& = & x_{2n}-2\sum_{k=1}B_{2n,2k-1} \,x_{2k-1}.
\end{eqnarray}


\subsection{Appendix 4.2. ``Comma'' creation and  annihilation operators.
}

\subsubsection{``Comma'' operators
(with singled out midpoint).}

The quantized momentum conjugate to $x^L_n $, $x^R_n$ and
$X(\pi/2)$ are defined  as usual
\begin{equation}
{\cal P}^{L}_n = -i \frac{\partial}{\partial x^L_n},
\; \; \;
{\cal P}^{R}_n = -i \frac{\partial}{\partial x^R_n},
\; \; \; {\cal P} = -i \frac{\partial}{\partial X(\pi/2)}.
\end{equation}
Relation with the conventional string momentum $p_m$
are given by:
\begin{eqnarray}
{\cal P}^{L}_n & = & \frac {1}{2} p_{2n-1} +
\sum_{m\geqslant1} A^+_{2m,n} p_{2m}
- \frac{\sqrt{2}}{\pi\sqrt{\ap}} \frac{(-1)^n}{2n-1} p_0 ,
\nonumber \\
{\cal P}^{R}_n & = & -\frac {1}{2} p_{2n-1} +
\sum_{m\geqslant1}
 \tilde A^-_{2m,n} p_{2m}
- \frac{\sqrt{2}}{\pi\sqrt{\ap}} \frac{(-1)^n}{2n-1} p_0 ,
\nonumber \\{\cal P} & = & p_0.
\nonumber
\label{LR-mom}
\end{eqnarray}
The inverse relations
read:
\begin{gather}
p_{2n-1}  =  {\cal P}^{L}_n - {\cal P}^{R}_n \\
p_{2n}    =  \sum_{m\geqslant1} \! \tilde A^+_{m,2n} \left( {\cal
P}^{L}_m + {\cal P}^{R}_m \right)+ \sqrt{2} (-1)^n {\cal
P}. \nonumber \label{mom-LR}
\end{gather}

The creation and annihilation operators for the ``comma'' modes are
defined in the following way \footnote{Here we use notations of
\cite{Abdurrahman-Bordes}, where relations between
 creation and annihilation operators for R/L n-modes and corresponding canonical
 operators
are taken to be $a_n=\frac{-i\sqrt{n}}{2}(q_n+i\frac{2}{n}p_n)$}
\footnote{For half-strings we use different notations $r=1,2;$
$L,R;$}:
\begin{eqnarray}
b^{r}_n=\frac{-i}{\sqrt{2}}\left(\frac{2n-1}{2}\right)^{1/2}
\left\{x^{r}_n+i\frac{2}{2n-1}{\cal P}^{r}_n\right\},
\nonumber \\
b^{r\,\dagger}_n=\frac{i}{\sqrt{2}}\left(\frac{2n-1}{2}\right)^{1/2}
\left\{x^{r}_n-i\frac{2}{2n-1}{\cal P}^{r}_n\right\},
\label{LR-cr}
\end{eqnarray}
where $n\geqslant 1$.\\
The vacuum satisfies the relations $$ b^{L}_n \vac_L = 0,
~~~b^{R}_n \vac_R = 0, ~~~n=1,\ldots, \infty .$$ Space of  states
is
$$\hbox{\boldmath $\cal H$} =
{\cal H}_L \otimes {\cal H}_R \otimes {\cal H}_M,$$ where ${\cal
H}_M$ is for the midpoint motion. Relations between $b^{L\#}_n$,
$b^{R\#}_n$ and $a^\#_n$ are
\begin{equation}
b^{L}_n =
\frac{\sqrt{2}}{\pi}
\frac{(-1)^{(n-1)}}{(2n-1)^{3/2}} p_0
+ \frac{1}{\sqrt{2}}
 a_{2n-1}
+\frac{1}{\sqrt{2}}\sum^\infty_{m=1} M_{1\,m,n}a_{2m}-
M_{2\,m,n}a_{2m}^\dagger, \label{b-L}
\end{equation}
\begin{equation}
b^{R}_n =
\frac{\sqrt{2}}{\pi}
\frac{(-1)^{(n-1)}}{(2n-1)^{3/2}} p_0
- \frac{1}{\sqrt{2}}
 a_{2n-1}
+\frac{1}{\sqrt{2}}\sum^\infty_{m=1} M_{1\,m,n}a_{2m}-
M_{2\,m,n}a_{2m}^\dagger, \label{b-R}
\end{equation}
and the corresponding relation for $b^{L\dagger}_n,b^{R\dagger}_n$
can be achieved by changing $a_n$ and $a_n^\dagger$. These
formulae define the Bogoliubov transformation with transformation
matrices
\begin{gather}
M_{1\,n,m}=\frac{1}{2}\sqrt{\frac{2m-1}{2n}}\tilde{A}^{\pm}_{m,2n}+\sqrt{\frac{2n}{2m-1}}A^{\pm}_{2n,m},\\
M_{2\,n,m}=\frac{1}{2}\sqrt{\frac{2m-1}{2n}}\tilde{A}^{\pm}_{m,2n}-\sqrt{\frac{2n}{2m-1}}A^{\pm}_{2n,m}.
\end{gather}

The inverse relations are given by:
\begin{equation*}
a_{2n-1} = b_n^{(-1)},
\end{equation*}
\begin{equation}
a_{2n}  = \frac{(-1)^n}{\sqrt{2n}} {\cal P} + \sum_{m\geqslant
1}M_{1\,n,m} b_m^{(+)}+ M_{2\,n,m} b_m^{(+) \dagger},
\label{ALFA_M}
\end{equation}
where one defines the combinations $$b_m^{(\pm)} = \frac{1}{\sqrt
2} \left( b_m^{L} \pm b_m^{R} \right). $$

Let us write the relation between usual vacuum $a_n |0\rangle =0$
for $ n\geqslant 0$ and vacuum $|0\rangle _L|0\rangle _R$. Since
$a_{2n-1} |0\rangle = b^{(-1)}_n |0\rangle =0$,  only the
combinations $b^{(+)^\dagger}_n$ act on $|0\rangle _L|0\rangle _R$
to get the a-vacuum $|0\rangle$, namely:
\begin{eqnarray}
|0\rangle = \exp \left( -\frac{1}{2} b^{{(+)}\dagger}_n
\phi_{n,m} b^{{(+)}\dagger}_m \right) |0\rangle _L|0\rangle _R
\label{vac-a-lr}
\end{eqnarray}
and the matrix $\phi$ is determined as \cite{Berezin}:
\begin{eqnarray}
\phi = M_1^{-1} M_2.
 \nonumber
\end{eqnarray}
One finds \cite{Abdurrahman-Bordes} $M_1^{-1} M_2$ to be
\begin{eqnarray}
\phi_{n,m} = (2n-1)^{1/2} (2m-1)^{1/2}
\frac{1}{2(n+m-1)}
\left(
\begin{array}{c}
-1/2 \\
n-1
\end{array}
\right)
\left(
\begin{array}{c}
-1/2 \\
m-1
\end{array}
\right).
\label{PHI}
\end{eqnarray}


\subsubsection{``Comma'' operators
(without singled out midpoint).}

The quantized momenta conjugated to $\chi ^L_n $, $\chi^R_n$ are
defined in the usual way
\begin{equation}
\wp^{L}_n = -i \frac{\partial}{\partial \chi ^L_n}, \;\;\;
\wp^{R}_n = -i \frac{\partial}{\partial \chi ^R_n}.
\end{equation}
Relation with the conventional string momentum $p_m$
are given by \cite{PR99}:
\begin{eqnarray}
\wp^{r}_0&=&\frac{1}{2}p_{0}+\sqrt{2}\frac{(-1)^{r}}{\pi}\sum_{n=1}^{\infty}\frac{(-1)^{n}}{2n-1}p_{2n-1},
\\
\wp^{r}_{2n}&=&\frac{1}{2}p_{2n}+(-1)^{r}\sum_{m=1}^{\infty}B_{2n\,2m-1}p_{2m-1},\;
n\geqslant 1.
\nonumber
\end{eqnarray}
The inverse relations are
\begin{gather}
p_{2n}=\wp^{L}_{2n}+\wp^{R}_{2n},\; n\geqslant 0,
\\
p_{2n-1}=\frac{2\sqrt{2}}{\pi}\frac{(-1)^n}{2n-1}(\wp^{L}_{2n}+\wp^{R}_{2n})
+2\sum_{m=1}^{\infty}B_{2m\,2n-1}(\wp^{L}_{2m}-\wp^{R}_{2m}),\;
n\geqslant 1. \nonumber \label{LR-Mom}
\end{gather}

Creation and  annihilation operators satisfying the commutation
relation
\begin{equation}
[b^{r}_{n},b^{s\dag}_{m}]=\delta^{rs}\delta_{nm},\;n\geqslant 1
\end{equation}
can be written in the form
\begin{equation}
b_{0}^{r}=-i\left(\chi_{0}^{r}+\frac{i}{2}\wp_{0}^{r}\right),\;\;\;
b_{n}^{r}=-i\sqrt{\frac{n}{2}}\left(\chi_{2n}^{r}+\frac{i}{n}\wp_{2n}^{r}\right).
\end{equation}
Comma vacuum are introduced as
\begin{equation}
b^{r}_{n}|0\rangle^{r}=0,\;\; n \geqslant 1,
\end{equation}
where we have the following expressions for $b^{r}$
\begin{gather}
b_{0}^{r}=\frac{1}{4}(3a_{0}-a^{\dag}_{0})+
\frac{(-1)^{r}}{2\pi}\sum_{n=1}^{\infty}\frac{(-1)^{n}}{(2n-1)^{3/2}}\{[4+(2n-1)]a_{2n-1}-[4-(2n-1)]a^{\dag}_{2n-1}\},\\
b_{n}^{r}=\frac{1}{\sqrt{2}}a_{2n}+\frac{(-1)^{r+1}}{\sqrt{3}}\sum_{m=1}^{\infty}\left(\frac{2m-1}{2n}\right)^{1/2}[A_{2m-1\,2n}a_{2m-1}-S_{2m-1\,2n}a^{\dag}_{2m-1}],
\end{gather}
where
\begin{equation}
 A_{2m-1\,2n}=B_{2m-1\,2n}- B_{2n\,2m-1},~~~~~
  S_{2m-1\,2n}=B_{2m-1\,2n}+ B_{2n\,2m-1}.
\end{equation}


\subsection{Appendix 4.3. Identities for coherent and squeezed states.}
In \cite{Kostelecky-Potting} the useful identities are introduced.
\begin{multline}
\hbox{Det}(1-S\cdot V)^{1/2}\langle 0|
\exp\Bigl(\lambda\cdot a + \frac{1}{2} a\cdot
S\cdot a\Bigr) \exp\Bigl(\mu\cdot a^\dagger + \frac{1}{2} a^\dagger\cdot V\cdot
a^\dagger\Bigr) |0 \rangle\\
 = \exp\Bigl[\lambda\cdot(1-V\cdot S)^{-1}\cdot\mu +
\frac{1}{2}\lambda\cdot(1-V\cdot S)^{-1} \cdot V\cdot
\lambda+\frac{1}{2}\mu\cdot(1-S\cdot V)^{-1}\cdot S\cdot\mu\Bigr] ,
\label{displacedbosonic}
\end{multline} where the dot indicates
contraction of indices.

If the matrices $A$, $B$, and $C$ satisfy
\begin{equation}
A^T=A, \quad
B^T=B, \quad
AC^T=CA, \quad
BC=C^TB, \quad
C^2=4AB
\label{conditions}
\end{equation}
then the following identity takes place
\begin{multline}
\exp(a^{\dag} Aa^{\dag}+a^{\dag} C a+a B a)\\
=\hbox{Det}\left[(1-C)e^C\right]^{-1/2}
\exp[a^{\dag}(1-C)^{-1}Aa^{\dag}] \exp[-a^{\dag}\ln(1-C) a] \exp[a
B(1-C)^{-1} a]. \label{bosonicr}
\end{multline}


\subsection{Appendix 4.4. Ghost half-string coordinates.}
Here we collect some relations for the half-string and
full-string ghost modes following
\cite{AbdurGhost1},\cite{AbdurGhost2}. The inverse relations to
(\ref{GHccoor}) are
\begin{gather}
c_{0}^{r}=c_{0}+(-1)^{r}\frac{2}{\pi}\sum_{n=1}^{\infty}\frac{(-1)^{n}}{2n-1}(c_{2n-1}+c_{-2n+1}),\\
g_{2n}^{r}=\frac{1}{\sqrt{2}}(c_{2n}+c_{-2n})+\sqrt{2}(-1)^{r}\sum_{m=1}^{\infty}B_{2n,2m-1}(c_{2m-1}+c_{-2m+1}).
\end{gather}
The inverse relations to (\ref{GHpiccoor}) are
\begin{gather}
b_{0}^{r}=b_{0}+(-1)^{r}\frac{2}{\pi}\sum_{n=1}^{\infty}\frac{(-1)^{n}}{2n-1}(b_{2n-1}+b_{-2n+1}),\\
y_{2n}^{r}=\frac{1}{\sqrt{2}}(b_{2n}+b_{-2n})+\sqrt{2}(-1)^{r}\sum_{m=1}^{\infty}B_{2n,2m-1}(b_{2m-1}+b_{-2m+1}).
\end{gather}
The inverse relations to (\ref{GHbcoor}) are
\begin{gather}
h_{2n-1}^{r}=\frac{1}{\sqrt{2}}(b_{2n-1}-b_{-2n+1})-\sqrt{2}(-1)^{r}\sum_{m=1}^{\infty}B_{2n-1,2m}(b_{2m}-b_{-2m}).
\end{gather}
The inverse relations to (\ref{GHpibcoor}) are
\begin{gather}
z^{r}_{2n-1}=\frac{1}{\sqrt{2}}(c_{2n-1}-c_{-2n+1})-\sqrt{2}(-1)^{r}\sum_{m=1}^{\infty}B_{2n-1,2m}(c_{2m}-c_{-2m}).
\end{gather}
We demand the comma modes to satisfy the anticommutation relations
\begin{equation}
\{{\cal O}^{\,r}_{n},\frac{\partial}{\partial {\cal
O}^{\,s}_{m}}\}=\delta^{rs}\delta_{nm}.
\end{equation}
One can check that $\frac{1}{\sqrt{2}}c_{0}^{r}$,
$\frac{1}{\sqrt{2}}g^{r}_{n}$ are conjugate to
$\frac{1}{\sqrt{2}}b_{0}^{r}$, $\frac{1}{\sqrt{2}}y^{r}_{n}$ and
likewise $\frac{\imath}{\sqrt{2}} h^{r}_{2n-1}$ are conjugate to
$\frac{\imath}{\sqrt{2}} z^{r}_{2n-1}$.\\
We now introduce equations relating the full-string ghost modes
$(c_{n},b_{n})$ and the comma ghost modes $(\gamma^{r}_{n},
\beta^{r}_{n})$. For the full-string modes one gets
\begin{gather*}
c_{0}=\frac{1}{\sqrt{2}}(\gamma_{0}^{1}+\gamma_{0}^{2}),\\
c_{2n}=\frac{1}{2}\sum_{r=1}^{2}\frac{1}{\sqrt{2}}(\beta^{r}_{n}+\beta^{r}_{-n})-
\sum_{r=1}^{2}(-1)^{r}\sum_{m=1}^{\infty}\frac{2n}{2m-1}B_{2n,2m-1}\frac{1}{\sqrt{2}}(\gamma^{r}_{m}-\gamma^{r}_{-m}),
\;\;n\geqslant 1,\\
c_{2n-1}=\frac{\sqrt{2}}{\pi}\frac{(-1)^{n}}{2n-1}\sum_{r=1}^{2}(-1)^r\gamma^{r}_{0}
-\frac{1}{2}\sum_{r=1}^{2}\frac{1}{\sqrt{2}}(\gamma^{r}_{n}-\gamma^{r}_{-n})\\
-\sum_{r=1}^{2}(-1)^{r}\sum_{m=1}^{\infty}\frac{2n-1}{2m}B_{2n-1,2m}\frac{1}{\sqrt{2}}(\beta^{r}_{m}+\beta^{r}_{-m}),\;\;
n\geqslant 1,
\end{gather*}
and $c_{-n}\equiv c_{n}^{\dag}$, $n\geqslant 1$. Equations for $b_{n}$ are the same with transposition $\gamma\rightleftharpoons\beta$.\\
The inverse relations are
\begin{gather*}
\gamma^{r}_{0}=c_{0}+\frac{2}{\pi}(-1)^{r}\sum_{n=1}^{\infty}\frac{(-1)^n}{2n-1}(c_{2n-1}+c_{-2n+1}),\\
\gamma^{r}_{n}=\frac{1}{2\sqrt{2}}(b_{2n}+b_{-2n})+\frac{1}{\sqrt{2}}(-1)^{r}\sum_{m=1}^{\infty}B_{2n,2m-1}(b_{2n-1}+b_{-2n+1})\\
-\frac{1}{2\sqrt{2}}(c_{2n-1}-c_{-2n+1})+\frac{1}{\sqrt{2}}(-1)^{r}\sum_{m=1}^{\infty}B_{2n-1,2m}(c_{2n}-c_{-2n}),\;\;
n\geqslant 1,
\end{gather*}
and $\gamma_{-n}\equiv \gamma^{\dag}_{n}$, $n\geqslant 1$. Equations
for $\beta^{r}_{n}$ are the same with transposition
$c\rightleftharpoons b$.

\newpage
\section{Cubic  String Field Theory on Conformal Language. }
\setcounter{equation}{0}
\subsection{Vertex Operators.}
\label{sec:vero}

By using the language of the conformal field theory (CFT) it is possible
to represent each term in the String Field Theory
 action as some correlation function in CFT on a special two dimensional surface.

To specify our notations, let us remind some basic facts concerning two dimensional CFT.
A general solution of the equation
$(\pd^2_{\tau}-\pd^2_{\sigma})X^{\mu}=0$ with Neumann boundary conditions
is
\begin{equation}
\begin{split}
X^{\mu}(\sigma,\tau)&=x^{\mu}+2\alpha^{\,\prime}p^{\mu}\tau
+i\left[\frac{\alpha^{\,\prime}}{2}\right]^{1/2}\sum_{m\ne 0}\frac{\alpha^{\mu}_m}{m}
e^{-im\tau}\left(e^{-im\sigma}+e^{im\sigma}\right),
\\
&\text{where}\quad -\infty<\tau<\infty\quad
0\leqslant\sigma\leqslant \pi.
\end{split}
\label{gs-nbc}
\end{equation}
After Wick rotation we can introduce new complex variables
$z=e^{\tau_E+i\sigma}$, $\oz=e^{\tau_E-i\sigma}$. Note
that this variables are coordinates on the
upper-half complex plane.
The solution \eqref{gs-nbc} in this coordinates
gets the following form
\begin{equation}
X^{\mu}(z,\bar{z})=X^{\mu}_L (z)+X^{\mu}_R (\bar{z}),
\label{x(zz)}
\end{equation}
where
\begin{equation}
X^{\mu}_L(z)=\frac12 x^{\mu}-\frac{i}{2}\alpha^{\,\prime}p^{\mu}\log z^2
+i\left[\frac{\alpha^{\,\prime}}{2}\right]^{1/2}\sum_{m\ne 0}
\frac{\alpha^{\mu}_m}{m z^m}
\quad\text{and}\quad X_R(\bar{z})=X_L(\bar{z}).
\label{x(z)}
\end{equation}
The similar expansion reads for the
 ghost-antighost pair $c(z)$, $b(z)$ as
\begin{equation}
c(z)=\sum_{n}\frac{c_n}{z^{n-1}},\quad 
b(z)=\sum_{n}\frac{b_n}{z^{n+2}}.
\label{c(z)}
\end{equation}

The Neumann correlation function for $X$'s
on the upper half-complex plane is of the from
\begin{equation}
\left\langle X^{\mu}(z,\oz)X^{\nu}(w,\overline{w})\right\rangle
=-\frac{\alpha^{\,\prime}}{2}\eta^{\mu\nu}\,
\left(\log|z-w|^2+\log|\oz-w|^2\right).
\label{Neumann-corf}
\end{equation}
But for computations it is fruitful to
use the traditional shorthand
\begin{equation}
\left\langle X^{\mu}_L(z)X^{\nu}_L(w)\right\rangle
=-\frac{\alpha^{\,\prime}}{2}\eta^{\mu\nu}\,
\log(z-w).
\label{cor}
\end{equation}
Note that this expression is really very formal. For example,
when one wants to compute a correlation function involving
$:e^{ik\cdot X(z,\bar{z})}:$ he encouraged to use
the expression \eqref{Neumann-corf}. But if the point
$(z,\bar{z})$ is on the real axis (i.e. on the boundary of the UHP),
then one can write the equality
$$
\left.:e^{ik\cdot X(z,\bar{z})}:\right|_{z=\bar{z}}
=:e^{2ik\cdot X_L(z)}:
$$
and use the expression \eqref{cor} to compute correlation functions.
Indeed, one can check that these two differently computed correlators
are the same while one assumes $z=\bar{z}$.
In our calculations we will use this trick.

The ground state (\ref{xvac}) is given in terms of conformal fields as
the normal ordered exponent:
\begin{equation}
|0,k\rangle =:e^{ik\cdot X(0,0)}: |0\rangle
\equiv :e^{2ik\cdot X_L(0)}: |0\rangle,
\label{ground-state}
\end{equation}
and any term in the string field expansion (\ref{Fockstates})
corresponds to the conformal field via
\begin{equation}
\left[\prod_i \alpha^{\mu_i}_{m_i}\right]|0,k\rangle
\Longleftrightarrow
:\left[\prod_i \pd^{m_i} X^{\mu_i}(0)\right]
e^{2ik\cdot X_L(0)}:|0\rangle.
\label{state-vertex}
\end{equation}


If it does not
lead to misunderstanding we  also drop out subscript "$L$" and
use  a traditional shorthand $X(z)$
 \begin{equation}
\left\langle X^{\mu}(z)X^{\nu}(w)\right\rangle
=-\frac{\alpha^{\,\prime}}{2}\eta^{\mu\nu}\,
\log(z-w).
\label{cor'}
\end{equation}
Another useful form of relation \eqref{cor} is
\begin{equation}
\label{corfb}
 X^{\mu}(z)X^{\nu}(w)\sim
 -\frac{\alpha^{\,\prime}}{2}\eta^{\mu\nu}\,
\log(z-w).
\end{equation}
The symbol "$\sim$" means that l.h.s and r.h.s are equal up to
regular in $z-w$ terms.
For ghosts we have the following OPE
\begin{equation}
c(z)b(w)\sim\frac 1{z-w}.
\label{corfg}
\end{equation}

The stress tensor for the matter fields
\begin{equation}
\label{Tx}
T_X=-\frac{1}{\alpha^{\,\prime}}\pd X\cdot\pd X
\end{equation}
has the following OPE
\begin{equation}
T_X(z)T_X(w)\sim \frac{26}{2(z-w)^4}+\frac{2}{(z-w)^2}T_X(w)+
\frac{1}{z-w}\pd T_X(w).
\end{equation}

The stress tensor for the ghost fields and the reparametrization algebra are
\begin{align}
&T(z)=-2 b\pd c(z)-1\pd b c(z) \label{Tbc},\\
&T(z)b(w)\sim\frac{2}{(z-w)^2}b(w)+\frac{1}{z-w}\pd b(w),\\
&T(z)c(w)\sim\frac{-1}{(z-w)^2}c(w)+\frac{1}{z-w}\pd c(w),\\
&T(z)T(w)\sim\frac{-13}{(z-w)^4}+\dots .
\end{align}

The system has a $U(1)$ number current
\begin{align}
& j(z)=-bc(z)=\sum_n\frac{j_n}{z^{n+1}},\qquad
j(z)j(w)\sim\frac{1}{(z-w)^2},\\
& j(z)b(w)\sim-\frac{1}{z-w}b(w),\qquad
j(z)c(w)\sim\frac{1}{z-w}c(w),
\end{align}
which charge operator $j_0$ counts $c=+1$, $b=-1$ charge.

The
conformal properties of the current depend on the
vacuum being chosen.
In the $SL_2$ invariant vacuum sector
$\langle c(z)b(w)\rangle=1/(z-w)$,
then the proper relation is
\begin{align}
&T(z)j(w)\sim\frac{-3}{(z-w)^3}+\frac{1}{(z-w)^2}j(z),
\\
&[L_m,j_n]=-nj_{m+n}+\frac{-3}{2}m(m+1)\delta_{m+n,0},
\end{align}
so that $j(z)$ is scale and translation covariant ($m=0,-1$) but
not conformal covariant ($m=1$). Next commutation relations
\begin{equation}
[L_1,j_{-1}]=j_0-3,\qquad [L_1,j_{-1}]^\dag=[L_{-1},j_1]=-j_0
\end{equation}
show the charge asymmetry of the system: $j_0^\dag=-j_0+3$.
Consequently, operator expectation values of charge neutral operators
will vanish
since
\begin{align}
&[j_0,\Oc]=q\Oc,\qquad j_0|0\rangle=
0,\qquad \langle 0|j_0^\dag=-Q\langle 0|,\\
&\langle 0|j_0\Oc|0\rangle=-Q\langle 0|\Oc|0\rangle=
q\langle 0|\Oc|0\rangle.
\end{align}
Only operators which cancel the background charge $-3$ survive.
For example,
\begin{equation}
\langle c(z_1)c(z_2)c(z_3) \rangle=-(z_1-z_2)(z_2-z_3)(z_3-z_1).
\end{equation}

\subsection{Gluing and Conformal Maps.}
\label{sec:gluing-and-maps}
There is a method that allows one
 to represent n-string interactions
via correlation functions on special Riemann surfaces
$R_n$ (the  so-called string configuration).
\begin{equation*}
\int \Phi_{1} \star \dots \star\Phi_{n}=
\langle V_n|\Phi_{1}\rangle _1
\otimes \dots \otimes |\Phi_{n}\rangle _n=
 \langle  \Phi_{1},\dots,\Phi_{n}\rangle _{R_n}
\end{equation*}
Efficiency of this method is a possibility to reduce
calculations of correlation functions on $n$-string configuration to
calculations of correlation functions on the upper half-disk,
or the upper half-plane
using the equality
\begin{equation}
 \langle \Phi_{1},\dots ,\Phi _{n}\rangle _{R_n}=
 \langle F_1^{(n)}\circ\Phi _1\,\dots\,
F_n^{(n)}\circ\Phi_n\rangle .
\label{cm-equality}
\end{equation}
The maps $F_k^{(n)}$ ($k=1\dots n$) are defined as follows:
\begin{subequations}
\begin{align}
&F_k^{(n)}(w)=(P_n\circ f_{k}^{(n)})(w),\\
&f_{k}^{(n)}(w)=e^{\frac{2\pi
i}{n}\,(\mu(n)-k)}\left(\frac{1+iw}{1-iw}\right)^{2/n},
\qquad P_2(z)=i\frac{1-z}{1+z},\qquad
P_3(z)=\frac{i}{\sqrt{3}}\frac{1-z}{1+z}
\label{maps:b}
\\
&\qquad\qquad\qquad\text{where}\quad \mu(n)=
  \begin{cases}
    \frac{n+1}{2} & n=\text{odd}, \\
    \frac{n+2}{2} & n=\text{even}.
  \end{cases}
\end{align}
\label{maps:all}
\end{subequations}
Further we consider only the cases $n=2$ and $n=3$.
The maps $f_{k}^{(3)}$ can be rewritten in the form
$f_{k}^{(3)}=T^{k-1}\circ f_{1}^{(3)}$,
where $T$ is the operator of rotation over $-\frac{2\pi}{3}$.
This property of the maps
as it was shown in \cite{l'Clare}
is very important to have threefold cyclically
symmetric vertex. One can be puzzled why we choose
different maps $P_2$ and $P_3$ and may think
that it gives extra relative factor between
the brackets with $n=2$ and $n=3$. But fortunately
this never happens, because Witten's vertex is
invariant under the conformal group
\cite{l'Clare2},
so one can choose arbitrary functions $P_2$
and $P_3$.
\begin{figure}[!t]
\begin{center}
\includegraphics[width=300pt]{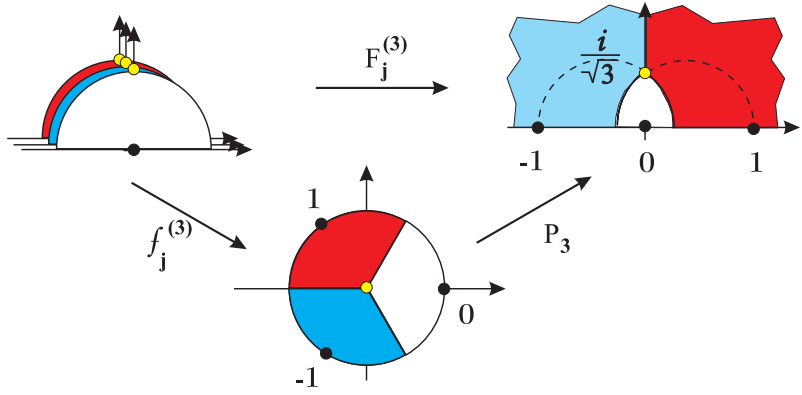}
 \caption{The maps for Witten's vertex, $R_3$.} \label{fig:maps}
\end{center}
\end{figure}

We illustrate the maps \eqref{maps:all} on the
Figure~\ref{fig:maps}. Here we have three half discs
with two marked points. Into the black point we
usually insert the vertex operators corresponding
to the states of the string. The grey  point
is the center of the string (we insert picture changing
operators into this point in the cubic
superstring field theory). Also on the Figure~\ref{fig:maps}
we illustrate where half discs and marked points
maps under the transformations \eqref{maps:all}.

Note that the choice \eqref{maps:b} of the map $f^{(n)}_{k}$ from
the upper half disc to a sector of disc differs from usual one. But
this choice is very convenient for computer calculations, since it
should not evaluate powers of negative numbers.
The maps for $n=2$ have very simple form:
\begin{equation}
f_1^{(2)}(w)=-\frac{1}{w},\qquad
f_2^{(2)}(w)=w.
\label{maps:2}
\end{equation}

$(f\circ\Phi)$ in \eqref{cm-equality} means the conformal transform of
$\Phi$ by $f$. Let us explain this in more details.
If $\Phi$ is a primary field of conformal weight $h$,
then $f\circ\Phi(z)$ is given by
\begin{equation}
(f\circ\Phi)(w)=(f^{\prime}(w))^h\Phi(f(w)).
\label{cm-pf}
\end{equation}
For a derivative of a primary field of conformal weight $h$
one has
\begin{equation}
(f\circ\pd \Phi)(w)=(f^{\prime}(w))^{h+1}\pd \Phi(f(w))+
hf^{\prime\prime}(w)(f^{\prime}(w))^{h-1}\Phi(f(w)).
\label{cm-dpf}
\end{equation}

Since we work with free conformal fields,
all the operators in the theory can be constructed
as normal product of these free fields. Therefore,
we have to know how to find the conformal transformations
of the composite operators.
Let us illustrate this technique on the example of $bc$-current.
The operators $b$ and $c$ are primary ones with
weights $2$ and $-1$ respectively, so
\eqref{cm-pf} yields
\begin{equation}
(f\circ b)(w)=[f^{\prime}(w)]^2\, b(f(w))
\quad\text{and}\quad
(f\circ c)(w)=[f^{\prime}(w)]^{-1}\, c(f(w)).
\label{fbfc}
\end{equation}
The $bc$-current $j_{bc}$ is defined by the following expressions
\begin{equation}
j_{bc}(w)=-:b(w)c(w):=-\lim_{z\to w}:b(z)c(w):
=-\lim_{z\to w}\left(b(z)c(w)-\frac{1}{z-w}\right),
\label{jbc}
\end{equation}
where in the latter equality we have used the OPE
of $b$ with $c$. Now we apply a conformal map $f$ to both
sides of the equality \eqref{jbc}:
\begin{multline}
(f\circ j_{bc})(w)=-\lim_{z\to w}\left(f\circ b(z)
f\circ c(w)-\frac{1}{z-w}\right)
\\
\stackrel{\eqref{fbfc}}{=}
-\lim_{z\to w}\left(\frac{[f^{\prime}(z)]^2}{f^{\prime}(w)}
b(f(z))c(f(w))-\frac{1}{z-w}\right)
\\
\stackrel{\text{OPE}\;bc}{=}
-\lim_{z\to w}\left(\frac{[f^{\prime}(z)]^2}{f^{\prime}(w)}
:b(f(z))c(f(w)):
+\frac{[f^{\prime}(z)]^2 [f^{\prime}(w)]^{-1}}{f(z)-f(w)}
-\frac{1}{z-w}\right)
\\
=f^{\prime}(w)\, j_{bc}(f(w))+\frac{3}{2}
\frac{f^{\prime\prime}(w)}{f^{\prime}(w)}.
\end{multline}
So we get the conformal transformation of the composite
operator $j_{bc}$ using only OPE and conformal
transforms of the operators from which it is built.
This scheme can be applied to find a
conformal transform of any composite operator.

\vspace{0.2cm}
\noindent\textsc{Exercise:\quad}
Show that the stress energy tensor
of the bosonic string $T_X=-\frac{1}{\alpha^{\,\prime}}
\pd X\!\cdot\!\pd X$ changes under conformal transformations
as
\begin{equation}
(f\circ T_X)(w)=\df^2 T_X(f)+\frac{26}{12}\left(
\frac{\dddf}{\df}-\frac{3}{2}\frac{\ddf^2}{\df^2}\right).
\label{cm-TX}
\end{equation}
\vspace{0.2cm}

\vspace{0.2cm}
\noindent\textsc{Exercise:\quad} Check that
\begin{align*}
(f\circ \pd(bc))(w)&=\df^2\pd (bc)(f)
+\ddf b(f)c(f)+\frac{3}{2}\left(\frac{\dddf}{\df}
-\frac{\ddf^2}{\df^2}\right),
\\
(f\circ \pd bc)(w)&=
\df^2\pd b(f)c(f)+2\ddf b(f)c(f)+\frac{5}{6}\frac{\dddf}{\df}
+\frac{1}{4}\frac{\ddf^2}{\df^2},
\\
(f\circ b\pd c)(w)&=
\df^2b(f)\pd c(f)-\ddf b(f)c(f)+\frac{2}{3}\frac{\dddf}{\df}
-\frac{7}{4}\frac{\ddf^2}{\df^2}.
\end{align*}

For future use we list
Taylor's series of maps \eqref{maps:all} in the origin:
\begin{subequations}
\begin{align}
f_1^{(3)}(w)&=1+2\gamma w+3\gamma^2 w^2 +\frac{31}{8}\gamma^3
w^3+\frac{39}{8}\gamma^4 w^4 +\frac{813}{128}\gamma^5 w^5+
O(w^6),
\\
f_2^{(3)}(w)&=\frac{1}{2}\gamma w-\frac{5}{32}\gamma^3 w^3
+\frac{57}{512}\gamma^5 w^5+O(w^7),
\\
f_{3}^{(3)}(w)&=-1+2\gamma w-3\gamma^2 w^2 +\frac{31}{8}\gamma^3
w^3-\frac{39}{8}\gamma^4 w^4 +\frac{813}{128}\gamma^5 w^5+O(w^6).
\end{align}
\label{maps:3}
\end{subequations}
Here $\gamma=\frac{4}{3\sqrt{3}}$.
\subsection{Calculations of the Action for Tachyon and Vector Fields.}
We want to compute Witten's  action
\begin{equation}
S[\Phi]=-\frac{1}{g_o^2}\left[
\frac{1}{2}\la\Phi,\,Q_B\Phi\ra+\frac{1}{3}\la\Phi,\,\Phi,\,\Phi\ra
\right]
\label{w-action}
\end{equation}
for tachyon and vector string fields
\begin{equation}
\Phi(w)=\int \frac{d^{26} k}{(2\pi)^{26}}\, [t(k)\mathrm{V}_t(k,w)+
A_{\mu}(k)\mathrm{V}_v^{\mu}(k,w)
+B(k)\mathrm{V}_B(k,w)],
\label{bsf}
\end{equation}
where vertex operators $\mathrm{V}$ are of the form
\begin{equation}
\begin{split}
\mathrm{V}_t(w,k)&=:\!c(z)e^{2ik\cdot X(w)}\!:,\\
\mathrm{V}_v^{\mu}(w,k)&=i\left[\frac{2}{\alpha^{\,\prime}}\right]^{1/2}
:\!c(w)\pd X^{\mu}(w) e^{2ik\cdot X(w)}\!:,\\
\mathrm{V}_B(w,k)&=:\!\pd c(w)e^{2ik\cdot X (w)}\!:.
\end{split}
\end{equation}
All the coefficients in \eqref{bsf} are chosen in such a way that
the reality condition $\Phi^{\dag}=\Phi$ for string field
produces reality conditions for component fields.

\subsubsection{Action of the BRST charge.}
\noindent First of all we need to calculate the action of the BRST charge
\eqref{QB} on the vertex operators.
The action of the BRST charge on the tachyon vertex is the following:
\begin{multline}
Q_B \mathrm{V}_t(w,k)=\Res_{\zeta=w}\,[c(\zeta)T_B(\zeta)+bc\pd c(\zeta)]
c(w)e^{2ik\cdot X (w)}\\
{=}\Res_{\zeta=w}\,c(\zeta)c(w)
\left[\frac{\alpha^{\,\prime}k^2}{(\zeta-w)^2}+\frac{1}{\zeta-w}\pd_w
\right]e^{2ik\cdot X (w)}
+\frac{1}{\zeta-w}c\pd c(\zeta)e^{2ik\cdot X (w)}\\
=\alpha^{\,\prime}k^2\pd c c e^{2ik\cdot X (w)}+c\pd c e^{2ik\cdot X (w)}
=[-\alpha^{\,\prime}k^2+1]c\pd c e^{2ik\cdot X (w)}.
\end{multline}
The action of the BRST charge on the massless vector vertex is the following:
\begin{multline}
Q_B \mathrm{V}_v^{\mu}(w,k)=\Res_{\zeta=w}\,[c(\zeta)T_B(\zeta)+bc\pd c(\zeta)]
i\left[\frac{2}{\alpha^{\,\prime}}\right]^{\frac12}
c(w)\pd X^{\mu}e^{2ik\cdot X (w)}\\
{=}
i\left[\frac{2}{\alpha^{\,\prime}}\right]^{\frac12}
\Res_{\zeta=w}\,c(\zeta)c(w)
\left[\frac{-i\alpha^{\,\prime}k^{\mu}}{(\zeta-w)^3}e^{2ik\cdot X (w)}
+\left[
\frac{\alpha^{\,\prime}k^2+1}{(\zeta-z)^2}+\frac{1}{\zeta-z}\pd_w
\right]\pd X^{\mu}e^{2ik\cdot X (w)}\right]\\
+\frac{1}{\zeta-w}c\pd c(\zeta)\pd X^{\mu}e^{2ik\cdot X (w)}\\
=i\left[\frac{2}{\alpha^{\,\prime}}\right]^{\frac12}
\left[\frac{-i\alpha^{\,\prime}k^{\mu}}{2}\pd^2 c c
+\pd c c (\alpha^{\,\prime}k^2+1)\pd X^{\mu}+c\pd c\pd X^{\mu}
\right]e^{2ik\cdot X (w)}\\
=i\left[\frac{2}{\alpha^{\,\prime}}\right]^{\frac12}
\left[
\frac{i\alpha^{\,\prime}k^{\mu}}{2}c \pd^2c e^{2ik\cdot X (w)}
+(-\alpha^{\,\prime}k^2) c\pd c\pd X^{\mu}e^{2ik\cdot X (w)}
\right].
\end{multline}
And the action of the BRST charge on the vertex of the auxiliary field
is the following:
\begin{multline}
Q_B \mathrm{V}_B(w,k)=\Res_{\zeta=w}\,[c(\zeta)T_B(\zeta)+bc\pd c(\zeta)]
\pd c e^{2ik\cdot X (w)}\\
{=}
\Res_{\zeta=w}\,c(\zeta)\pd c(w)
\left[\frac{\alpha^{\,\prime}k^2}{(\zeta-w)^2}+\frac{1}{\zeta-w}\pd_w
\right]e^{2ik\cdot X (w)}
+\frac{1}{(\zeta-w)^2}c\pd c(\zeta)e^{2ik\cdot X (w)}\\
=c\pd c\pd e^{2ik\cdot X (w)}+c\pd^2c e^{2ik\cdot X (w)}
=\pd_w\left[c\pd c e^{2ik\cdot X (w)}\right](w).
\end{multline}

It is convenient to rewrite $Q_B\Phi$ in the following way
\begin{multline}
(Q_B \Phi)(0)=\int \frac{d^{26} k}{(2\pi)^{26}}\,
\left[t(k)(Q_B\mathrm{V}_t)(0,k)
+iA_{\mu}(k)(Q_B\mathrm{V}_v^{\mu})(0,k)
+B(k)(Q_B\mathrm{V}_B)(0,k)\right]\\
=\int\left.\frac{d^{26} k}{(2\pi)^{26}}\,
\right[t(k)(-\alpha^{\,\prime}k^2+1)\,c\pd c e^{2ik\cdot X (0)}
+B^{\,\prime}\pd(c\pd c e^{2ik\cdot X (0)})\\
\left.
+\sqrt{2\alpha^{\,\prime}}
A_{\mu}(k)(\eta^{\mu\nu}k^2-k^{\mu}k^{\nu})
c\pd c\pd X^{\nu}e^{2ik\cdot X (0)}
\right],
\label{QPhi}
\end{multline}
where
\begin{equation}
B^{\,\prime}(k)=B(k)
-\left[\frac{\alpha^{\,\prime}}{2}\right]^{\frac12}ik^{\mu}A_{\mu}(k).
\end{equation}

\subsubsection{Conformal transformations.}
The operator for the tachyon field is primary, therefore, its
conformal transformation is the following
\begin{equation}
f\circ \mathrm{V}_t(w,k)=\left[f^{\prime}(w)\right]^{\alpha^{\,\prime}k^2-1}
\mathrm{V}_t(f(w),k).
\label{tr-Vt}
\end{equation}
The operator for the auxiliary field is transformed as follows
\begin{equation}
f\circ \mathrm{V}_B(w,k)=\left[f^{\prime}(w)\right]^{\alpha^{\,\prime}k^2}
\mathrm{V}_B(f(w),k)-\left[f^{\prime}(w)\right]^{\alpha^{\,\prime}k^2}
\frac{f^{\,\prime\prime}}{f^{\,\prime\,2}}
c(f(w))e^{2ik\cdot X (f(w))}.
\label{tr-VB}
\end{equation}
The operator for the vector field is transformed in the following way
\begin{equation}
f\circ \mathrm{V}_v^{\mu}(w,k)
=[f^{\prime}(w)]^{\alpha^{\,\prime}k^2}
\,\mathrm{V}_v^{\mu}(f(w),k)
+\left[\frac{\alpha^{\,\prime}}{2}\right]^{\frac12}
k^{\mu}[f^{\prime}(w)]^{\alpha^{\,\prime}k^2}
\frac{f^{\prime\prime}(w)}{f^{\prime\,2}(w)}
c(f(w))e^{2ik\cdot X (f(w))}
\label{tr-Vv}
\end{equation}
From \eqref{tr-Vv} and \eqref{tr-VB} follows that the operator
\begin{equation}
\mathrm{V}_V^{\mu}(w,k)=\mathrm{V}_{v}^{\mu}(w,k)
+\left[\frac{\alpha^{\,\prime}}{2}\right]^{\frac12}
k^{\mu}\mathrm{V}_B(w,k)
\label{primary-vector}
\end{equation}
is the primary operator of weight $\alpha^{\,\prime}k^2$.

It is also convenient to rewrite $f\circ \Phi$ in the following
form
\begin{multline}
(f\circ \Phi)(w)=\int \frac{d^{26} p}{(2\pi)^{26}}\,
\left[t(p)(f\circ\mathrm{V}_t)(w,p)
+iA_{\mu}(p)(f\circ\mathrm{V}_v^{\mu})(w,p)
+B(p)(f\circ\mathrm{V}_B)(w,p)\right]\\
\stackrel{\ref{primary-vector}}{=}
\int \frac{d^{26} p}{(2\pi)^{26}}\,
[f^{\,\prime}(w)]^{\alpha^{\,\prime}p^2}\left[
t(p)[f^{\,\prime}(w)]^{-1}c(f(w))
-\left[\frac{2}{\alpha^{\,\prime}}\right]^{\frac12}
A_{\mu}(p)[c\pd X^{\mu}
-i\frac{\alpha^{\,\prime}p^{\mu}}{2}\pd c](f(w))
\right.\\
\left.+B^{\prime}\left(\pd c(f(w))
-\frac{f^{\,\prime\prime}}{f^{\,\prime\,2}}c(f(w))\right)
\right]e^{2ip\cdot X (f(w))}.
\label{fPhi}
\end{multline}

\subsubsection{Calculations of kinetic terms.}
Substituting \eqref{fPhi} and \eqref{QPhi} in the quadratic
part of the action \eqref{w-action}
and expanding the obtained expression we get
\begin{align*}
\langle f\circ &\Phi\,Q_B\Phi(0)\rangle=
\int\frac{d^{26}p}{(2\pi)^{26}}\frac{d^{26}k}{(2\pi)^{26}}
f^{\,\prime\,\alpha^{\,\prime}p^2}\left[
t(p)\frac{1}{f^{\,\prime}}t(k)(-\alpha^{\,\prime}k^2+1)\;
\langle c(f)c\pd c(0)\rangle
\langle e^{2ip\cdot X (f)}e^{2ik\cdot X (0)}\rangle
\right.
\\
&+t(p)\frac{1}{f^{\,\prime}}B^{\prime}(k)
\left.\frac{\pd}{\pd w}\right|_{w=0}
\langle c(f)c\pd c(w)\rangle
\langle e^{2ip\cdot X (f)}e^{2ik\cdot X (w)}\rangle
\\
&+t(p)\frac{1}{f^{\,\prime}}\sqrt{2\alpha^{\,\prime}}A_{\mu}(k)
[\eta^{\mu\nu}k^2-k^{\mu}k^{\nu}]\;
\langle c(f)c\pd c(0)\rangle
\langle e^{2ip\cdot X (f)}\pd X_{\nu}(0)e^{2ik\cdot X (0)}\rangle
\\
&-\left.\left[\frac{2}{\alpha^{\,\prime}}\right]^{\frac12}
A_{\beta}(p)t(k)(-\alpha^{\,\prime}k^2+1)
\right\{\langle c(f)c\pd c(0)\rangle
\langle \pd X^{\beta}(f)e^{2ip\cdot X (f)}e^{2ik\cdot X (0)}\rangle
\\
&~~~~~~~~~~~~~~~~~~~~~~~~~~\left.-i\frac{\alpha^{\,\prime}p^{\beta}}{2}
\langle \pd c(f)c\pd c(0)\rangle
\langle e^{2ip\cdot X (f)}e^{2ik\cdot X (0)}\rangle
\right\}
\\
&-\left[\frac{2}{\alpha^{\,\prime}}\right]^{\frac12}
A_{\beta}(p)B^{\prime}(k)\left.\left.\frac{\pd}{\pd w}\right|_{w=0}
\right\{\langle c(f)c\pd c(w)\rangle
\langle \pd X^{\beta}(f)e^{2ip\cdot X (f)}e^{2ik\cdot X (w)}\rangle
\\
&~~~~~~~~~~~~~~~~~~~~~~~~~~\left.-i\frac{\alpha^{\,\prime}p^{\beta}}{2}\langle \pd c(f)c\pd c(w)\rangle
\langle e^{2ip\cdot X (f)}e^{2ik\cdot X (w)}\rangle
\right\}
\\
&-\left.\left[\frac{2}{\alpha^{\,\prime}}\right]^{\frac12}\sqrt{2\alpha^{\,\prime}}
A_{\beta}(p)A_{\mu}(k)[\eta^{\mu\nu}k^2-k^{\mu}k^{\nu}]
\right\{\langle c(f)c\pd c(0)\rangle
\langle \pd X^{\beta}(f)e^{2ip\cdot X (f)}\pd X_{\nu}(0)e^{2ik\cdot X (0)}\rangle
\\
&~~~~~~~~~~~~~~~~~~~~~~~~~~\left.
-i\frac{\alpha^{\,\prime}p^{\beta}}{2}\langle \pd c(f)c\pd c(0)\rangle
\langle e^{2ip\cdot X (f)}\pd X_{\nu}(0)e^{2ik\cdot X (0)}\rangle\right\}
\\
&+B^{\,\prime}(p)
\left.\left[\pd_y-\frac{f^{\,\prime\prime}}{f^{\,\prime}}\right]\right|_{y=f}
t(k)[-\alpha^{\,\prime}k^2+1]\langle c(y)c\pd c(0)\rangle
\langle e^{2ip\cdot X (f)}e^{2ik\cdot X (0)}\rangle
\\
&+B^{\,\prime}(p)B^{\,\prime}(k)
\left.\left[\pd_y-\frac{f^{\,\prime\prime}}{f^{\,\prime}}\right]
\frac{\pd}{\pd w}\right|_{y=f,w=0}\langle c(y)c\pd c(w)\rangle
\langle e^{2ip\cdot X (f)}e^{2ik\cdot X (w)}\rangle
\\
&\left.
+\sqrt{2\alpha^{\,\prime}}B^{\,\prime}(p)A_{\mu}(k)
[\eta^{\mu\nu}k^2-k^{\mu}k^{\nu}]
\left.\left[\pd_y-\frac{f^{\,\prime\prime}}{f^{\,\prime}}\right]\right|_{y=f}
\langle c(y)c\pd c(0)\rangle
\langle e^{2ip\cdot X (f)}\pd X_{\nu}(0)e^{2ik\cdot X (0)}\rangle
\right].
\end{align*}
Using the expressions for OPE \eqref{corfb} and
\eqref{corfg} we
obtain the following result
\begin{multline}
-\int\frac{d^{26}p\,d^{26}k}{(2\pi)^{26}\alpha^{\,\prime\,13}}\delta(p+k)
\frac{f^{\,\prime\,\alpha^{\,\prime}p^2}}{f^{2\alpha^{\,\prime}p^2}}\left[
t(p)t(k)\frac{f^2}{f^{\,\prime}}(-\alpha^{\,\prime}k^2+1)
+t(p)B^{\prime}(k)\frac{2f}{f^{\,\prime}}(\alpha^{\,\prime}k^2-1)
\right.
\\
+\sqrt{2\alpha^{\,\prime}}t(p)A_{\mu}(k)\frac{f}{f^{\,\prime}}
[\eta^{\mu\nu}k^2-k^{\mu}k^{\nu}]i\alpha^{\,\prime}p_{\nu}
-\left[\frac{2}{\alpha^{\,\prime}}\right]^{\frac12}
A_{\beta}(p)t(k
(-\alpha^{\,\prime}k^2+1))if\alpha^{\,\prime}[-k^{\beta}-p^{\beta}]
\\
-\left[\frac{2}{\alpha^{\,\prime}}\right]^{\frac12}
A_{\beta}(p)B^{\prime}(k)
(2\alpha^{\,\prime}k^2-1)i\alpha^{\,\prime}[-k^{\beta}-p^{\beta}]
\\
-2A_{\beta}(p)A_{\mu}(k)[\eta^{\mu\nu}k^2-k^{\mu}k^{\nu}]
\left\{-\frac{\alpha^{\,\prime}}{2}(\eta^{\beta\nu}+2\alpha^{\,\prime}
k^{\beta}k^{\nu})+\alpha^{\prime}p^{\beta}\alpha^{\prime}p^{\nu}\right\}
\\
+B^{\,\prime}(p)t(k)[-\alpha^{\,\prime}k^2+1]
\left[2f-\frac{f^{\,\prime\prime}}{f^{\,\prime}}f^2\right]
+B^{\,\prime}(p)B^{\,\prime}(k)
\left.\left[\pd_y-\frac{f^{\,\prime\prime}}{f^{\,\prime}}\right]
\right|_{y=f}(-2y+2\alpha^{\,\prime}k^2\frac{y^2}{f})
\\
\left.
+\sqrt{2\alpha^{\,\prime}}B^{\,\prime}(p)A_{\mu}(k)
[\eta^{\mu\nu}k^2-k^{\mu}k^{\nu}]
\left[2f-f^2\frac{f^{\,\prime\prime}}{f^{\,\prime}}\right]
i\alpha^{\,\prime}p_{\nu}f^{-1}
\right].
\end{multline}
Using the conservation of momenta and the fact that $\eta^{\mu\nu}k^2-k^{\mu}k^{\nu}$
is orthogonal to $k_{\mu}$ we obtain
\begin{multline}
\int\frac{d^{26}k}{(2\pi)^{26}\alpha^{\,\prime\,13}}
\frac{f^{\,\prime\,\alpha^{\,\prime}p^2}}{f^{2\alpha^{\,\prime}p^2}}\left[
t(-k)t(k)\frac{f^2}{f^{\,\prime}}[-\alpha^{\,\prime}k^2+1]
+t(-k)B^{\prime}(k)(\alpha^{\,\prime}k^2-1)(\frac{2f}{f^{\,\prime}}-2f
+f^2\frac{f^{\,\prime\prime}}{f^{\,\prime\,2}})
\right.
\\
\left.
+2B^{\prime}(-k)B^{\prime}(k)\left[1+
(2-\frac{ff^{\,\prime\prime}}{f^{\,\prime\,2}})(\alpha^{\,\prime}k^2-1)
\right]
+\alpha^{\prime}(\eta^{\mu\nu}k^2-k^{\mu}k^{\nu})A_{\nu}(-k)A_{\mu}(k)
\right].
\label{S2-boson}
\end{multline}
Recall, that the map $f(w)$ (see also \eqref{maps:2}) is defined as follows
\begin{equation}
f(w)=-\frac{1}{w},\qquad
f^{\,\prime}(w)=\frac{1}{w^2},\qquad
f^{\,\prime\prime}(w)=-\frac{2}{w^3}.
\end{equation}
Substituting this map in \eqref{S2-boson} and putting $w=0$
we get the following
expression for the quadratic action
\begin{multline}
S[t,A_{\mu},B]=\frac{1}{g_{o}^2\alpha^{\,\prime\,13}}
\int\frac{d^{26}k}{(2\pi)^{26}}
\left[
\frac12 t(-k)t(k)[-\alpha^{\,\prime}k^2+1]+B^{\,\prime}(-k)B^{\,\prime}(k)
\right.
\\
\left.-
\frac{\alpha^{\,\prime}}{2}(\eta^{\mu\nu}k^2-k^{\mu}k^{\nu})A_{\nu}(-k)
A_{\mu}(k)
\right].
\label{S2boson}
\end{multline}

\subsubsection{Calculation of interaction.}
Here we compute the cubic term of the action \eqref{w-action}.
First, we compute the interaction term for three tachyons. To this end
we substitute only the part of the string field $\Phi$ which
is proportional to $t(k)$:
\begin{multline}
\frac{1}{3}\la\Phi,\,\Phi,\,\Phi\ra=\frac{1}{3}\langle
f_1\circ\Phi\,f_2\circ\Phi\,f_3\circ\Phi\rangle
=\frac13\int\frac{d^{26}k_1}{(2\pi)^{26}}\frac{d^{26}k_2}{(2\pi)^{26}}
\frac{d^{26}k_3}{(2\pi)^{26}}\,f_1^{\prime\,\alpha^{\,\prime}k_1^2-1}
f_2^{\prime\,\alpha^{\,\prime}k_2^2-1}f_3^{\prime\,\alpha^{\,\prime}k_3^2-1}
\\
\times t(k_1)t(k_2)t(k_3)\;\langle c(f_1)c(f_2)c(f_3)\rangle\;
\langle e^{2ik_1\cdot X (f_1)}e^{2ik_2\cdot X (f_2)}e^{2ik_3\cdot X (f_3)}\rangle
\\
=-\frac13\int\frac{d^{26}k_1}{(2\pi)^{26}}\frac{d^{26}k_2}{(2\pi)^{26}}
\frac{d^{26}k_3}{(2\pi)^{26}}\,t(k_1)t(k_2)t(k_3)
(2\pi)^{26}\alpha^{\,\prime\,-13}\delta(p+k+q)
\\
\times f_1^{\prime\,\alpha^{\,\prime}k_1^2-1}
f_2^{\prime\,\alpha^{\,\prime}k_2^2-1}f_3^{\prime\,\alpha^{\,\prime}k_3^2-1}
(f_1-f_2)^{2\alpha^{\,\prime}k_1\cdot k_2+1}
(f_1-f_3)^{2\alpha^{\,\prime}k_1\cdot k_3+1}
(f_2-f_3)^{2\alpha^{\,\prime}k_2\cdot k_3+1}.
\end{multline}
Substituting the maps \eqref{maps:3} into this expression and
performing simplification one gets
\begin{equation}
S_{int}[t]=-\frac{1}{g_o^2\alpha^{\,\prime\,13}}
\int\frac{d^{26}k_1d^{26}k_2d^{26}k_3}{(2\pi)^{26}(2\pi)^{26}}
\delta(k_1+k_2+k_3)
\;\frac{1}{3}t(k_1)t(k_2)t(k_3)\,
\gamma^{\alpha^{\,\prime}k_1^2
+\alpha^{\,\prime}k_2^2+\alpha^{\,\prime}k_3^2-3}.
\label{S3-t-boson}
\end{equation}
Next we compute the term which consists of two vector fields and
one tachyon. So, we compute the following odd bracket
\begin{multline}
\la \mathrm{V}_t(k_1),\,\mathrm{V}_V^{\mu}(k_2),\,
\mathrm{V}_V^{\nu}(k_3)\ra=\langle
f_1\circ\mathrm{V}_t(k_1)\,f_2\circ\mathrm{V}_v^{\mu}(k_2)\,
f_3\circ\mathrm{V}_v^{\nu}(k_3)\rangle
=-\frac{2}{\alpha^{\,\prime}}f_1^{\prime\,\alpha^{\,\prime}k_1^2-1}
f_2^{\prime\,\alpha^{\,\prime}k_2^2}f_3^{\prime\,\alpha^{\,\prime}k_3^2}
\\
\times\left\langle
c(f_1)\left[c\pd X^{\mu}(f_2)-\frac{i\alpha^{\,\prime}k_2^{\mu}}{2}\pd c(f_2)\right]
\left[c\pd X^{\nu}(f_3)-\frac{i\alpha^{\,\prime}k_3^{\nu}}{2}\pd c(f_3)\right]
e^{2ik_1\cdot X (f_1)}e^{2ik_2\cdot X (f_2)}e^{2ik_3\cdot X (f_3)}
\right\rangle
\\
=-\left.\frac{2}{\alpha^{\,\prime}}f_1^{\prime\,\alpha^{\,\prime}k_1^2-1}
f_2^{\prime\,\alpha^{\,\prime}k_2^2}f_3^{\prime\,\alpha^{\,\prime}k_3^2}
\right[\langle c(f_1)c(f_2)c(f_3)\rangle
\langle\pd X^{\mu}(f_2)\pd X_{\nu}(f_3)
e^{2ik_1\cdot X (f_1)}e^{2ik_2\cdot X (f_2)}e^{2ik_3\cdot X (f_3)}\rangle
\\
-\frac{i\alpha^{\,\prime}k_2^{\mu}}{2}\langle c(f_1)\pd c(f_2)c(f_3)\rangle
\langle\pd X_{\nu}(f_3)
e^{2ik_1\cdot X (f_1)}e^{2ik_2\cdot X (f_2)}e^{2ik_3\cdot X (f_3)}\rangle
\\
-\frac{i\alpha^{\,\prime}k_3^{\nu}}{2}\langle c(f_1)c(f_2)\pd c(f_3)\rangle
\langle\pd X_{\mu}(f_2)
e^{2ik_1\cdot X (f_1)}e^{2ik_2\cdot X (f_2)}e^{2ik_3\cdot X (f_3)}\rangle
\\
\left.
-\frac{\alpha^{\,\prime\,2}}{4}k_2^{\mu}k_3^{\nu}
\langle c(f_1)\pd c(f_2)\pd c(f_3)\rangle
\langle e^{2ik_1\cdot X (f_1)}e^{2ik_2\cdot X (f_2)}e^{2ik_3\cdot X (f_3)}\rangle
\right].
\end{multline}
Calculation of the correlation functions leads to the following
expression
\begin{multline}
=-f_1^{\prime\,\alpha^{\,\prime}k_1^2-1}
f_2^{\prime\,\alpha^{\,\prime}k_2^2}f_3^{\prime\,\alpha^{\,\prime}k_3^2}
(2\pi)^{26}\alpha^{\,\prime\,-13}(f_1-f_2)^{2\alpha^{\,\prime}k_1k_2}
(f_1-f_3)^{2\alpha^{\,\prime}k_1k_3}(f_2-f_3)^{2\alpha^{\,\prime}k_2k_3}
\delta(k_1+k_2+k_3)
\\
\times\left[
\eta^{\mu\nu}\frac{(f_1-f_2)(f_3-f_1)}{f_2-f_3}
-2\alpha^{\,\prime}k_1^{\mu}k_1^{\nu}(f_2-f_3)
+2\alpha^{\,\prime}k_1^{\mu}k_2^{\nu}(f_3-f_1)
+2\alpha^{\,\prime}k_3^{\mu}k_1^{\nu}(f_1-f_2)
\right.
\\
-2\alpha^{\,\prime}k_3^{\mu}k_2^{\nu}\frac{(f_1-f_2)(f_3-f_1)}{f_2-f_3}
+\alpha^{\,\prime}k_1^{\nu}k_2^{\mu}(f_1+f_3-2f_2)
+\alpha^{\,\prime}k_2^{\mu}k_2^{\nu}\left[
f_3-f_1+\frac{(f_3-f_1)(f_1-f_2)}{f_3-f_2}\right]
\\
\left.
-\alpha^{\,\prime}k_3^{\nu}k_1^{\mu}(f_1+f_2-2f_3)
+\alpha^{\,\prime}k_3^{\mu}k_3^{\nu}\left[f_1-f_2+\frac{(f_1-f_2)(f_1-f_3)}{f_2-f_3}
\right]
+\alpha^{\,\prime}k_2^{\mu}k_3^{\nu}(f_3-f_2)
\right].
\end{multline}
Performing a simplification of this expression we get the
following formula
\begin{multline}
\la \mathrm{V}_t(k_1),\,\mathrm{V}_V^{\mu}(k_2),\,
\mathrm{V}_V^{\nu}(k_3)\ra=
(2\pi)^{26}\alpha^{\,\prime\,-13}\delta(k_1+k_2+k_3)
[\eta^{\mu\nu}+\alpha^{\,\prime}k_2^{\nu}k_1^{\mu}+\alpha^{\,\prime}k_3^{\mu}k_1^{\nu}]
\\
\times f_1^{\,\prime\,\alpha^{\,\prime}k_1^2-1}
f_2^{\,\prime\,\alpha^{\,\prime}k_2^2}f_3^{\,\prime\,\alpha^{\,\prime}k_3^2}
(f_1-f_2)^{2\alpha^{\,\prime}k_1k_2+1}
(f_1-f_3)^{2\alpha^{\,\prime}k_1k_3+1}
(f_2-f_3)^{2\alpha^{\,\prime}k_2k_3-1}.
\label{TVV}
\end{multline}
To compute the whole interaction term for the tachyon
and two vector fields we need to subtract from \eqref{TVV}
the same terms with permutation $(f_1\leftrightarrow f_2)$
and $(f_1\leftrightarrow f_3)$. After substitution of the maps \eqref{maps:3}
we get the following action
\begin{multline}
S_{int}[t,A]=-\frac{1}{g_o^2\alpha^{\,\prime\,13}}
\int\frac{d^{26}k_1d^{26}k_2d^{26}k_3}{(2\pi)^{26}(2\pi)^{26}}
\delta(k_1+k_2+k_3)
\\
\times
t(k_1)A_{\mu}(k_2)A_{\nu}(k_3)\,
[\eta^{\mu\nu}+\alpha^{\,\prime}k_1^{\mu}k_2^{\nu}+\alpha^{\,\prime}k_3^{\mu}k_1^{\nu}]
\gamma^{\alpha^{\,\prime}k_1^2
+\alpha^{\,\prime}k_2^2+\alpha^{\,\prime}k_3^2}.
\label{S3-TVV}
\end{multline}

The cubic term of the action \eqref{w-action} for whole string
field $\Phi$ \eqref{fPhi} is more complicate. It was computed using the program
on Maple, and the answer is the following
\begin{multline}
S_{int}[t,A_{\mu},B^{\,\prime}]=\frac{1}{g_o^2\alpha^{\,\prime\,13}}
\int\frac{d^{26}k_1 d^{26}k_2 d^{26}k_3}{(2\pi)^{26}(2\pi)^{26}}
\gamma^{\alpha^{\,\prime}k_1^2
+\alpha^{\,\prime}k_2^2+\alpha^{\,\prime}k_3^2}
\left[
\;-\gamma^{-3}\,t(k_1)t(k_2)t(k_3)
\right.
\\
-\gamma^{-1}\;A_{\beta}(k_1)A_{\mu}(k_2)t(k_3)\,
(\eta^{\mu\beta}+\alpha^{\,\prime}k_2^{\beta}k_3^{\mu}
+\alpha^{\,\prime}k_1^{\mu}k_3^{\beta})
+\gamma^{-1}\;B^{\,\prime}(k_1)B^{\,\prime}(k_2)t(k_3)
\\
\left.
+2\gamma^{-1}\left[\frac{\alpha^{\,\prime}}{2}\right]^{\frac12}
\;B^{\,\prime}(k_1)ik_1^{\mu} A_{\mu}(k_2)t(k_3)
\right].
\label{Sint-Boson}
\end{multline}
This expression can also be rewritten in $x$-representation:
\begin{multline}
S_{int}[t,A_{\mu},B^{\,\prime}]=\frac{1}{g_o^2\alpha^{\,\prime\,13}}
\int d^{26}x
\,\left[-\frac{1}{3}\gamma^{-3}\,\tilde{t}\tilde{t}\tilde{t}
-\gamma^{-1}\,\tilde{t}\tilde{A}_{\mu}\tilde{A}^{\mu}
+2\gamma^{-1}\alpha^{\,\prime}\,
\pd^{\mu}\tilde{t}\pd^{\beta}\tilde{A}_{\mu}\tilde{A}_{\beta}
\right.
\\
\left.+\gamma^{-1}\,\tilde{t}\tilde{B}^{\prime}\tilde{B}^{\prime}
+2\gamma^{-1}\left[\frac{\alpha^{\,\prime}}{2}\right]^{\frac12}
\,\tilde{t}\tilde{B}^{\prime}\pd^{\mu}\tilde{A}_{\mu}
\right].
\label{Sint-Boson-x}
\tag{\ref{Sint-Boson}${}^\prime$}
\end{multline}
At all levels the particle fields $\phi$ in the interaction action
are smeared over the distance $\sqrt{\alpha^{\,\prime}}$:
\begin{equation}
\tilde{\phi}(x)=\exp\left[-\alpha^{\,\prime}\log\gamma
\,\pd_{\mu}\pd^{\mu}\right]\;\phi(x).
\end{equation}
\subsection{Koba-Nielson Amplitudes from SFT.
}
\label{sec:kba}
Let us briefly describe how one can obtain Koba-Nielson tree
amplitudes in the Witten
string field theory. One starts with the  N-point diagram:

\begin{equation}
\langle V_{3}\mid \cdots  \langle V_{3}\mid  \Delta ^{(1')}
\mid V_{2}\rangle\cdots
\Delta ^{(N-3')}\mid V_{2}\rangle \mid A_{1}\rangle
\cdots\mid A_{N}\rangle .\label{d.1}
\end{equation}
$\Delta ^{(k)}$ stands for the propagator of the free string of the number $k$.
Usually the propagators are obtained  as a  product of zero-mode
polynomials and the inverse
of $L$, which can be represented as the proper time integral $\int_{0}^{\infty}
e^{-TL}dT$. The vertices  $| V_{3} \rangle$ are three-string Witten vertices and the  vertices
$| V_{2} \rangle$   are specified by the scalar product
(see Sect.4).

\begin{figure}[!h]
\centering
\includegraphics[width=300pt]{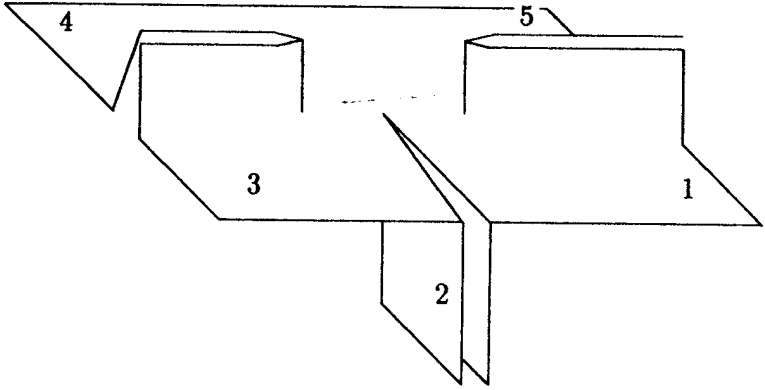}
\caption{The five-point amplitude in the Witten theory.}
\label{fig:five-point-amp}
\end{figure}

It was proved  that  the amplitude (\ref{d.1}) can be presented in
CFT language as
\begin{equation}
\prod_{i=3}^{N-1}
(\int d\tau _{i})\langle
  {\cal O}(z_{1}){\cal O}(z_{2})\oint b(z'_{1})dz'_{1}{\cal O}(z_{3})
\oint b(z'_{2})dz'_{2} \cdots {\cal O}(z_{N})
\rangle_{R}, \label{d.5}
\end{equation}
where the correlation functions must be calculated on a surface  $R$
(the  so-called string scattering configuration).
For Witten's vertex the surface $R$ looks like  it is
sketched in  Figure~\ref{fig:five-point-amp}.

 External strings are semi-infinite rectangular
strips of width $\pi$. The internal propagator is a strip of length
 $T_{i}$ and width $\pi$. Witten's three-point interaction identifies
the first half of one string with the second half of the next string,
for all three strings. Although the world sheet action for
$X^{\mu},c$ and $b$ is quadratic, the complicated
geometry of the string configuration makes the calculations
non-trivial. The key idea of the conformal technique is to
 map conformally from the string configuration to the upper half plane,
where the propagators are known.

The conformal map  $\rho(z)$ which takes the upper-half plane onto the
string configuration for a general N-point tree amplitude
was given by Giddings and Martinec \cite{GM}. The function  $\rho(z)$
satisfies the equation
\begin{equation}
\frac{d\rho}{dz}={\cal N} \frac{\sqrt{\prod_{i=1}^{N-2}
(z-z_{0i})(z- z_{0i}^{\ast})}}{ \prod_{r=1}^{N}(z-z_{r})}, \label{d.6}
\end{equation}
where $z_{0i} $ are complex parameters representing the locations of
the singular intersection points, and $z_{r}$ are the asymptotic
positions on the real axis of the $N$ external states.
The ${\cal  N}$ is a
real normalization constant. If one integrates this equation  one
obtains a map with the following properties:

i)~~logarithmic singularities at the points $z_{r}$;

ii)~~$\sim (z-z_{0i})^{\frac{3}{2}}$ behaviour at the interaction points.

The expression (\ref{d.5}) looks rather similar to the result
known from the covariant  version of the first quantized
approach \cite{fms}
\begin{equation}
\prod_{i=3}^{N-1}(\int dz_{i})   \langle {\cal O}^{1}(z_{1})  {\cal
O}^{2}(z_{2}) {\cal V}^{3}(z_3) \cdots {\cal V}^{n-1}(z_{N-1}){\cal
O}^{N}(z_{N})  \rangle, \label{d.7}
\end{equation}
where  the Koba-Nielsen variables $z_{i}$ for  $i=3, \cdots ,N-1$
are modular parameters and
\begin{equation}
 {\cal O}(z_{i})=c(z_{i}) {\cal V}(z_{i}), \label{d.8}
\end{equation}
with  ${\cal V}(z)$   a conformal  field of dimension one,
 defined solely in terms of the matter  field $x^\mu$. The
amplitude (\ref{d.7}) can  be equivalently rewritten as follows
\begin{equation}   \label{d.9}
\prod ^{N-1}_{i=3} (\int dz_{i} \oint  dz'_{i}) \langle
{\cal O}^{1}(z_{1})  {\cal O}^{2}(z_{2}) b(z'_{3}){\cal O}^{3}(z_3)
 \cdots b(z'_{N-1}) {\cal O}^{N-1}(z_{n-1})
{\cal O}^{N}(z_{N})\rangle,
\end{equation}
where the correlation functions must be
calculated on the entire $z$-plane.

To prove the on-shell agreement of the equations (\ref{d.5})
and (\ref{d.9}) one has to show
that the entire integration region is covered once and the ghost
contribution compensates the Jacobian, which occurs in going from proper space
time variables $\{T_{i} \}$ to the Koba-Nielsen variables $z_{i}$.
The  proof of  this  has been
established by Giddings for  $ N=4  $ \cite{G} and  by Samuel and Bluhm
for $ N>4 $ \cite{SamB}.

\subsection{Sliver.}
Rastelli and Zwiebach \cite{zwiebach} have found a family of string fields
generating a commutative subalgebra of the open string star algebra:
\begin{equation}
| n\rangle \star | m \rangle = | n+ m -1\rangle.
\end{equation}
The state $|n\rangle$ is defined by the following
 requirement for any state $|\phi\rangle$
\begin{equation}
\langle n|\phi\rangle=\langle f_{n}\circ\phi(0)\rangle,
\end{equation}
where \cite{RSZF}
\begin{equation}
f_n(z) = {n\over 2} \tan\left({2\over n} \tan^{-1}(z)\right).
\label{wedgemap}
\end{equation}
This family of states is called wedge states, because
the half-disks representing their local coordinates could be considered
as wedges of the full unit disks.

For $n=1$ the wedge corresponds to the identity state $|I\rangle$
and for $n=2$ the wedge corresponds to the vacuum $|0\rangle$.
In the $n\to\infty$ limit the map \eqref{wedgemap} is
\begin{equation}
f(z) \equiv f_\infty(z) = \tan^{-1}(z)
\end{equation}
and this wedge state has the property
\begin{equation}
|\Xi\rangle\star|\Xi\rangle=|\Xi\rangle.
\label{wedgesquare}
\end{equation}
This wedge state is called sliver. It was found \cite{RSZF} that the sliver
$|\Xi\rangle$ can be written in the factorized form
\begin{equation}
|\Xi\rangle = |\Xi_m\rangle \otimes |
\Xi_{g}\rangle \,.
\label{wedgesliver}
\end{equation}
The matter part of \eqref{wedgesliver} can be defined
 using the CFT technique \cite{l'Clare}
\begin{equation}
|\Xi_m\rangle \propto \exp\Bigl(-
{1\over 2}\eta_{\mu\nu} a^{\mu\dagger}
\cdot S\cdot
a^{\nu\dagger}\Bigr)|0\rangle\, ,
\label{wedgesliveroscil}
\end{equation}
where
\begin{equation}
S_{mn} = -{1\over \sqrt{mn}} \,  \ointop
{d w\over
2\pi i}\, \ointop {dz\over 2\pi i} {1 \over z^n w^m (1 + z^2) (1+w^2)
(\tan^{-1}(z) -
\tan^{-1}(w))^2}\, .
\end{equation}

First Rastelli, Sen and Zwiebach \cite{RSZF} have compared numerically
the sliver \eqref{wedgesliveroscil}
with the matter part of the solution
found by Kostelecky and Potting \cite{Kostelecky-Potting}
(see Sect. \ref{sec:SFTM}) and have got an evidence
that these two solutions are the same.
Later  on these two states  have been identified by  direct
calculations \cite{Japan2}.

\newpage
\section{Cubic Super String Field Theory.}
\label{sec:SFT}
\setcounter{equation}{0}

\subsection{Fermions.}

The action for the open superstring in the conformal gauge has the form
\begin{equation}
S=\frac{1}{2\pi\alpha^{\,\prime}}\int_{-\infty}^{\infty} d\tau
\int_0^{\pi}d\sigma
\left[\frac{1}{2}\pd_a X^{\mu}\pd^a X_{\mu}
-\frac{i}{2}\bar{\Psi}^{\mu}\gamma^a\pd_a\Psi_{\mu}
\right].
\label{string-action}
\end{equation}
Here $\Psi$ is a Majorana spinor and $\bar{\Psi}=\Psi^T\gamma^0$. We
introduce the explicit representation of Euclidean Dirac
$\gamma$-matrices:
\begin{equation}
\gamma^0=
\begin{pmatrix}
0 & 1\\
1 & 0
\end{pmatrix},
\qquad
\gamma^1=
\begin{pmatrix}
0 & -i\\
i & 0
\end{pmatrix}.
\end{equation}
The spinor $\Psi^{\mu}(\sigma,\tau)$ has two components
$$
\Psi^{\mu}(\sigma,\tau)=
\begin{pmatrix}
\psi^{\mu}(\sigma,\tau)\\
\tilde{\psi}^{\mu}(\sigma,\tau)
\end{pmatrix}.
$$
Performing matrix multiplications in \eqref{string-action} and
changing the variables we get the following
action for fermions
\begin{equation}
S=-\frac{1}{2\pi\alpha^{\,\prime}}\int d^2z\,[\psi^{\mu}\opd\psi_{\mu}
+\tilde{\psi}^{\mu}\pd\tilde{\psi}_{\mu}].
\end{equation}
Since $\Psi^{\mu}(\sigma,\tau)$ is spinor, we can impose the following
boundary conditions (the first one can always be reached by redefinition
of the fields)
$$
\psi^{\mu}(\pi,\tau)=\tilde{\psi}^{\mu}(\pi,\tau),
\qquad \psi^{\mu}(0,\tau)=\pm\tilde{\psi}^{\mu}(0,\tau).
$$
For ``$+$" the sector is called Ramond (R), and
for ``$-$" the sector is called Neveu-Schwartz (NS).
Therefore, the solution of the equations of motion (merging $\psi$ and $\tilde{\psi}$)
is of the form
\begin{equation}
\psi^{\mu}(z)=\sum_{n\in \Zh+r}\frac{\psi_n^{\mu}}{z^{n+1/2}},
\end{equation}
where $r=1/2$ for R-sector, and $r=0$ for NS-sector ($z$ is the same as in Sect. \eqref{sec:vero} ).

Following the ordinary quantization
procedure we get the following commutation relations
\begin{equation}
\{\psi_m^{\mu},\psi_n^{\nu}\}
=\eta^{\mu\nu}\delta_{m+n,0}.
\end{equation}
The correlation function is
\begin{equation}
\langle\psi^{\mu}(z)\psi^{\nu}(w)\rangle
=-\frac{\alpha^{\,\prime}}{2}\eta^{\mu\nu}\frac{1}{z-w}.
\end{equation}
\subsection{Superghosts, Bosonization.}
\label{sec:supbos}
Consider a general first order  action \cite{Friedan}
\begin{equation}
S=\frac{1}{2\pi}\int d^2z\,\bb\opd\cb,
\end{equation}
where $\bb$ and $\cb$ denote general conjugate fields of dimension
$\lambda$ and
$1-\lambda$ respectively; they can be either Bose or Fermi
fields\footnote{The formulae in the end of Sec 5.1 are a special case of the general
formulae presented in this section.}.

The $\bb,\cb$ operator product expansion is
\begin{equation}
\cb(z)\bb(w)\sim\frac{1}{z-w},\qquad \bb(z)\cb(w)\sim\frac{\eps}{z-w}.
\label{OPE-bbcc}
\end{equation}
Here and in the sequel, $\eps=+1$ for Fermi statistics and $\eps=-1$ for
Bose statistics. The fields have the following
mode expansions and hermitian properties
\begin{alignat*}{2}
\cb(z)&=\sum_{n}\frac{\cb_n}{z^{n+1-\lambda}},&\quad \cb_n^{\dag}&=\cb_{-n},
\\
\bb(z)&=\sum_{n}\frac{\bb_n}{z^{n+\lambda}},& \bb_n^{\dag}&=\eps \bb_{-n}.
\end{alignat*}
The OPE \eqref{OPE-bbcc}
determines the (anti)commutation relations
\begin{equation}
\cb_m\bb_n+\eps \bb_n\cb_m=\delta_{m+n,0}.
\end{equation}
There are NS and R sectors of the theory, specified by

\smallskip
\begin{center}
\renewcommand{\arraystretch}{1.5}
\begin{tabular}{||c|c||}
\hline
NS & R\\
\hline
$\bb_n,\,\,n\in \Zh-\lambda~~~~$ & $\bb_n,\,\,n\in \frac12+\Zh-\lambda$\\
$\cb_n,\,\,n\in \Zh+\lambda~~~~$ & $\cb_n,\,\,n\in \frac12+\Zh+\lambda$\\
\hline
\end{tabular}
\end{center}
The stress tensor and reparametrization algebra are
\begin{align}
&T(z)=-\lambda \bb\pd \cb(z)+(1-\lambda)\pd\bb \cb(z),\\
&T(z)\bb(w)\sim\frac{\lambda}{(z-w)^2}\bb(w)+\frac{1}{z-w}\pd\bb(w),\\
&T(z)\cb(w)\sim\frac{1-\lambda}{(z-w)^2}\cb(w)+\frac{1}{z-w}\pd\cb(w),\\
&T(z)T(w)\sim\frac{c/2}{(z-w)^4}+\dots,\qquad c=\eps(1-3Q^2),\qquad
Q=\eps(1-2\lambda).
\end{align}

Special cases are the
reparametrization ghost algebra with $\eps=1$,
$\lambda=2$,
$Q=-3$ and $c=-26$; and the local superconformal ghost algebra
with
$\eps=-1$, $\lambda=3/2$, $Q=2$ and $c=11$.

The system has a global $U(1)$ current
\begin{alignat}{2}
j(z)&=-\bb\cb(z)=\sum_n\frac{j_n}{z^{n+1}},&\qquad
j(z)j(w)&\sim\frac{\eps}{(z-w)^2},\\
j(z)\bb(w)&\sim-\frac{1}{z-w}\bb(w),&
j(z)\cb(w)&\sim\frac{1}{z-w}\cb(w),
\end{alignat}
which charge operator $j_0$ counts $\cb=+1$, $\bb=-1$ charge. The
conformal
properties of the current depend on the vacuum being chosen.
In the $SL_2$ invariant vacuum
$\langle\cb(z)\bb(w)\rangle=1/(z-w)$,
the proper relation is
\begin{subequations}
\begin{align}
&T(z)j(w)\sim\frac{Q}{(z-w)^3}+\frac{1}{(z-w)^2}j(z),\label{eq1}\\
&[L_m,j_n]=-nj_{m+n}+\frac{Q}{2}m(m+1)\delta_{m+n,0}.
\end{align}
\end{subequations}
So $j(z)$ is scale and translation covariant ($m=0,-1$), but
not conformal covariant ($m=1$). Next commutation relations
\begin{equation}
[L_1,j_{-1}]=j_0+Q,\qquad [L_1,j_{-1}]^\dag=[L_{-1},j_1]=-j_0,
\end{equation}
show the charge asymmetry of the system: $j_0^\dag=-j_0-Q$.
Consequently, operator expectation values of charge neutral operators
will vanish
since
\begin{align*}
&[j_0,\Oc]=q\Oc,\qquad j_0|0\rangle=0,\qquad \langle0|j_0^\dag=-Q\langle 0|,\\
&\langle0|j_0\Oc|0\rangle=-Q\langle0|\Oc|0\rangle=q\langle0|\Oc|0\rangle.
\end{align*}
Only operators which cancel the background charge $Q$ survive.

The $U(1)$ number current may be used to build a stress energy tensor that
reproduces
the relation (\ref{eq1})
\begin{subequations}
\begin{align}
&T_j(z)=-\frac{\eps}{2}(-j^2(z)+Q\pd j(z))
\label{Tj-def},\\
&T_j(z)j(w)\sim\frac{Q}{(z-w)^3}+\frac{j(z)}{(z-w)^2},\\
&T_j(z)T_j(w)\sim\frac{1-3\eps Q^2}{2(z-w)^4}+\dots
\end{align}
\end{subequations}
Thus the new stress tensor has $c_j=1-2\eps Q^2$; the original $\bb$,
$\cb$ stress
tensor has
\smallskip
\begin{center}
\renewcommand{\arraystretch}{1}
\begin{tabular}{lll}
$c=c_j$, & $\eps=1$, & (Fermi)\\
$c=c_j-2$, & $\eps=-1$, & (Bose)
\end{tabular}
\end{center}
To reproduce correct central charge in the Bose case
one has to add to \eqref{Tj-def} a proper Fermi system with central
charge $-2$ such that
\begin{equation}
T=T_j+T_{-2}.
\end{equation}
More precisely, this has to be
Fermi system with $\lambda_{-2}=1$ and $Q_{-2}=-1$ composed of a dimension
$1$ field $\eta(z)$ and dimension $0$ field $\xi(z)$.

Now let us discuss bosonization of $\bb,\cb$-system.
To this end, we introduce bosonic field $\phi$ such that
\begin{equation}
j(z)=\eps\pd\phi(z),\qquad \phi(z)\phi(w)\sim\eps\log(z-w).
\end{equation}
This means that the stress energy tensor of the field $\phi$
is of the form
\begin{equation*}
T_{\phi}=-\frac{\eps}{2}\pd\phi\pd\phi-\frac{Q}{2}\pd^2\phi.
\end{equation*}
The main object in bosonization construction is a
normal ordered exponent $e^{q\phi(z)}$.
It has the following OPEs with stress-energy tensor $T_{\phi}$
and $U(1)$-current $j$
\begin{subequations}
\begin{align}
T_{\phi}(z)e^{q\phi(w)}&=\left[\frac{\frac{\eps}{2}q(q+Q)}{(z-w)^2}+
\frac{1}{z-w}\pd_w\right] e^{q\phi(w)},\\
j(z)e^{q\phi(w)}&\sim \frac{q}{z-w}e^{q\phi(w)}\quad
\Rightarrow\quad [j_0,e^{q\phi(w)}]=qe^{q\phi(w)},
\\
e^{q\phi(z)}e^{q'\phi(w)}&\sim (z-w)^{\eps qq'}e^{q\phi(z)+q'\phi(w)}.
\label{ferm-like}
\end{align}
\end{subequations}
So $e^{q\phi(w)}$ has $j_0$-charge equal to $q$ and conformal weight
$\frac{\eps}{2}q(q+Q)$.
From \eqref{ferm-like} it follows that $e^{q\phi(z)}$
for odd $q$ behaves like anticommuting operator.
So the main bosonization formula are
\smallskip
\begin{center}
\renewcommand{\arraystretch}{1.5}
\begin{tabular}{||c|c||}
\hline
Fermi & Bose\\
\hline
$\cb(z)=e^{\phi(z)}\equiv e^{\sigma(z)}~~$ &
$\cb(z)\equiv\gamma=\eta(z)e^{\phi(z)}$ \\
$\bb(z)=e^{-\phi(z)}\equiv e^{-\sigma(z)}$ &
$\bb(z)\equiv\beta=e^{-\phi(z)}\pd\xi(z)$ \\
\hline
\end{tabular}
\end{center}
The $(\eta,\xi)$-system has its own conserved current $\xi\eta$.
Using this current we can construct a new one, which is
called the picture current and is of the form
\begin{equation*}
j_p(z)=-\pd\phi-\xi\eta.
\end{equation*}
In Table \ref{tab:gh} we collect the assignments
of the ghost number $n_g$ and the picture number $n_p$ for
the conformal operators described above.
Let us note that $SL(2,\Rh)$ invariant vacuum has
zero ghost and picture numbers.
\begin{table}[!t]
\centering
\renewcommand{\arraystretch}{1.2}
\begin{tabular}{||c|c|c|c||}
\hline
field &  $n_g $ & $n_p$ & $h$\\
\hline
$b$ & $-1 $ & $0$ &$2$ \\
\hline
$c$&$ 1 $&$0$&$-1 $\\
\hline
$ e^{q\phi}$&$0$&$ q$&$-\frac{1}{2}q(q+2)$ \\
\hline
$\xi $ & $-1 $ & $1$ & $0$\\
\hline
$\eta $ & $1$  &  $-1 $& $1$\\
\hline
$\gamma $&1&0&-1/2\\
\hline
$\beta $&-1&0&3/2\\
\hline
\end{tabular}
\caption{Ghost number $n_g$, picture number $n_p$
and conformal weight $h$ assignments}
\label{tab:gh}
\end{table}

Correlation functions are to be computed in the following normalization
\begin{equation*}
{}^*\langle\xi(z) c\partial c\partial^2 c(w) e^{-2\phi(y)}\rangle = -2.
\end{equation*}
It is a good point to note here, that the above expression
is written in the so-called large Hilbert space.
The problem is that in the bosonization formula we have used
only $\pd\xi$ rather then $\xi$, therefore, the zero mode $\xi_0$
is not important for bosonization. Moreover the
Hilbert space, which can be associated in a standard
way with the bosonized superghosts, does not
form an irreducible representation of the superghost algebra.
One can make it irreducible by excluding the mode $\xi_0$.
In the succeeding sections we will assume this restriction
and correlation functions will be normalized as follows
\begin{equation*}
 \label{eb1}
\langle c\partial c\partial^2 c(w) e^{-2\phi(y)}\rangle = -2,
\quad\text{where}\quad\langle 0|={}^*\langle 0|\xi_0.
\end{equation*}

\subsection{Problems with Action in -1 Picture.}

The original Witten's proposal \cite{w2} for NSR superstring
field theory action reads:
\begin{equation}
S_{W}\cong \int \mathcal{A}\star Q_{NS} \mathcal{A} + \frac23\int
\mathrm{X}\mathcal{A} \star \mathcal{A}\star \mathcal{A}
+\int Y(i)\Psi \star Q_{R} \Psi +2\int \Psi \star \mathcal{A}\star \Psi .
\label{w}
\end{equation}
Here $Q_{NS}$ and $Q_R$ are the BRST charges
in NS and R sectors, $\int$ and $\star$ are Witten's
string integral and star product to be specified below.
States in the extended Fock space $\mathcal{H}$ are
created by the modes of the
matter fields $X^{\mu}$ and $\psi^{\mu}$, conformal ghosts $b,\,c$
and superghosts $\beta,\,\gamma$:
\begin{equation}
\mathcal{A}=\sum _{m,j,i \in \Zh +\frac12}A_{i\dots}(x)\,\beta_i
...\gamma_j...b_k...c_l...\alpha^{\mu}_n... \psi^{\nu}_m
|0\rangle_{-1}, \label{f}
\end{equation}
\begin{equation}
\Psi=\sum _{m,j,i \in \Zh }\Psi^{A}{}_{i\dots}(x)\,\beta_i
...\gamma_j...b_k...c_l...\alpha^{\mu}_n... \psi^{\nu}_m
|A\rangle_{-\frac12}. \label{fr}
\end{equation}

The characteristic feature of the action (\ref{w}) is the choice
of the $-1$ picture for the string field $\mathcal{A}$. The vacuum
$|0\rangle _{-1}$  in the NS sector is defined as
\begin{equation}
|0\rangle_{-1}=c(0)e^{-\phi(0)}|0\rangle,
\end{equation}
where $|0\rangle$ stands for $SL(2,\mathbb{R})$-invariant vacuum.
In the description of the open NSR superstring the string field
$\mathcal{A}$ is
subjected to be GSO$+$.

 The vacuum $|A\rangle_{-\frac12}$ in the R sector is defined as
\begin{equation}
|A\rangle_{-\frac12}=c(0)e^{-\frac12\phi(0)}S_A(0)|0\rangle,
\end{equation}
where $S_A(z)$ is a spin operator of weight $\frac58$.

The
insertion of the picture-changing operator \cite{fms}
\begin{equation}
\mathrm{X}=\frac1{\alpha^{\,\prime}} e^{\phi}\psi\cdot\partial
X+c\partial\xi +\frac14 b\partial\eta
e^{2\phi}+\frac14\partial(b\eta e^{2\phi}) \label{picX}
\end{equation}
in the cubic term is aimed to absorb the unwanted unit of the
$\phi$ charge as only $\langle 0|e^{-2\phi} |0\rangle \neq 0$.

The action (\ref{w}) suffers from the contact term divergencies
\cite{GS,GK,AM,wendt} which arise when a pair of $\mathrm{X}$-s
collides in a point.
This sort of singularities appears already at the tree level.

Really, the tree
level graphs are generated by solving the classical equation of
motion by perturbation theory. For Witten's action \eqref{w} the
equation reads
$$
Q_B\mathcal{A}+\mathrm{X}\mathcal{A}\star \mathcal{A} =0.
$$
Here $Q_B=Q_{NS}$. The first nontrivial iteration ($4$-point function)
involving the
pair of $\mathrm{X} A\star A$ vertices produces the contact term
singularity when two of $\mathrm{X}$-s collide in a point.

\subsection{Action in  $0$ Picture and Double
Step Inverse Picture Changing Operator.}
\label{sec:act0pic}

To overcome the troubles that we have discussed in the previous
subsection  it was proposed to change the picture of NS
string fields from $-1$ to $0$, i.e. to replace $|0\rangle_{-1}$
in (\ref{f}) by $|0\rangle$ \cite{AMZ1,PTY}. States in the $-1$
picture can be obtained from the states in the $0$ picture by the
action of the inverse picture-changing operator $Y$ \cite{fms}
$$
Y=4c\partial\xi e^{-2\phi}(w)
$$
with $\mathrm{X}Y=Y\mathrm{X}=1$. This identity holds outside the ranges
$\ker \mathrm{X}(w)$ and $\ker Y(w)$ \cite{AM2}.
These kernels are:
\begin{equation*}
\ker X(w)=\{\Oc(w)\;|\;\lim_{z\to w}X(w)\Oc(z)=0\}
\quad\text{and}\quad
\ker Y(w)=\{\Oc(w)\;|\;\lim_{z\to w}Y(w)\Oc(z)=0\}.
\end{equation*}
This means that $\ker X(w)$ and $\ker Y(w)$ depend
on a point $w$. In particular, the following statement
is true: for all operators $\Oc(z)$, $z\ne w$
the product $X(w)\Oc(z)\ne 0$.

The existence of these local kernels leads
to the fact that at the $0$ picture
there are states that can not
be obtained by applying the picture changing operators
$\mathrm{X}$ and $Y$ to the states at the $-1$ picture.

 The action for the NS string field in the $0$ picture has the
cubic form with the insertion of a double-step inverse
picture-changing operator $Y_{-2}$ \cite{AMZ1,PTY}\footnote{One
can cast the
 action into the same form as the action (\ref{action}) for the bosonic string if one
 modifies the NS string integral accounting the "measure" $Y_{-2}$:
 $\int^{\prime}=\int Y_{-2}$.}:
\begin{equation}
\label{AMZPTY} S\cong \int Y_{-2}\mathcal{A}\star Q_B\mathcal{A}
 + \frac23\int Y_{-2}\mathcal{A}\star \mathcal{A}\star \mathcal{A}.
\end{equation}
We discuss $Y_{-2}$ in the next subsection.

 In the $0$ picture there is a variety
of auxiliary fields as compared with the $-1$ picture. These fields become
zero due to the free equation of motion: $Q_B
\mathcal{A}=0$, but they play a significant role in the
off-shell calculations. For instance, a low level
off-shell NS string field expands as
\begin{align}
\mathcal{A}\cong\int dk\; \{& u(k) c_1-\frac{1}{2}A_{\mu}(k)
ic_1\alpha _{-1}^{\mu}-\frac{1}{4}B_{\mu}(k) \gamma
_{\frac{1}{2}}\psi_{-\frac{1}{2}}^{\mu}
\notag
\\
&+\frac{1}{2}F_{\mu  \nu }(k)c_1\psi _{-\frac{1}{2}}^{\mu} \psi
_{-\frac{1}{2}}^{\nu} +B(k)c_0 +r(k)c_1\gamma _{\frac{1}{2}}\beta
_{-\frac{3}{2}}+ \dots\}e^{ik\cdot X(0,0)}|0\rangle.
\label{4.10}
\end{align}
Here $u$ and $r$ are just the auxiliary fields mentioned above.

The SSFT based on the action \eqref{AMZPTY} is free from the
drawbacks of the Witten's action  \eqref{w}. The absence of
contact singularities can be explained shortly.  The
action \eqref{AMZPTY} yields the following equation
\begin{equation}
Y_{-2}(Q_B\mathcal{A}+\mathcal{A}\star \mathcal{A}) =0.
\label{eq-with-Y}
\end{equation}
Since operator $Y_{-2}$ is inserted in the mid point in the bulk
 and $\Ac$ is an operator
in a point on the boundary we can drop out $Y_{-2}$ and
write the following equation
\begin{equation}
Q_B\mathcal{A}+\mathcal{A}\star \mathcal{A} =0.
\label{eq-without-Y}
\end{equation}
So the interaction vertex does not contain any
insertion leading to singularity. The complete proofs of this fact
can be found in \cite{AMZ1,PTY,AMZ2}.

To have well defined SSFT \eqref{AMZPTY}
the double step inverse picture changing
operator must be restricted to be

a) in accord with the identity\footnote{We assume that this
equation is true up to BRST exact operators.}:
\begin{equation}
Y_{-2}\mathrm{X} =\mathrm{X}Y_{-2}=Y,\label{condition}
\end{equation}

b) BRST invariant: $[Q_B,Y_{-2}]=0,$

c) scale invariant conformal field, i.e. conformal weight of
$Y_{-2}$ is $0$,

d) Lorentz invariant conformal field, i.e. $Y_{-2}$ does not
depend on momentum.

\noindent The point a) provides the formal equivalence between the
improved \eqref{AMZPTY} and the original Witten \eqref{w}
actions. The point b) allows us ``to move'' BRST
charge from one string field to another.
The point c) is
necessary to make the insertion of $Y_{-2}$ compatible with the
$\star$-product. And the point d) is necessary to preserve
unbroken Lorentz symmetry.

As it was shown in paper \cite{PTY}, there are two (up to BRST
equivalence) possible choices for the operator $Y_{-2}$.

The first operator, the chiral one \cite{AMZ1}, is built
from holomorphic fields in the upper half plane and is given by
\begin{equation}
Y_{-2}(z)=-4e^{-2\phi(z)}-\frac{16}{5\alpha^{\,\prime}}
e^{-3\phi}c\partial\xi\psi\cdot\partial X(z). \label{chiral}
\end{equation}
The identity \eqref{condition} for this operator reads
$$
Y_{-2}(z)\mathrm{X}(z)= \mathrm{X}(z)Y_{-2}(z) =Y(z).
$$
This $Y_{-2}$ is uniquely defined by the constraints a) -- d).

The second operator, the nonchiral one \cite{PTY}, is built from
both holomorphic and antiholomorphic fields in the upper half
plane
 and is of the form:
\begin{equation}
Y_{-2}(z,\overline{z})= Y(z) Y(\overline{z})
\label{nonchiral}.
\end{equation}
Here by $Y(\overline{z})$ we denote the
antiholomorphic field $4c(\overline{z})
\overline{\partial}\xi (\overline{z})e^{-2\phi (\overline{z})}$.
For this choice of $Y_{-2}$ the identity \eqref{condition} takes
the form:
$$
Y_{-2}(z,\overline{z})\mathrm{X}(z)= \mathrm{X}(z)Y_{-2}(z,\overline{z})
=Y(\overline{z}).
$$

One can find the inverse operators $W$ to $Y_{-2}(z,\overline{z})$ and
$Y_{-2}(z)$
\begin{subequations}
\begin{alignat}{2}
&W(z,\overline{z})Y_{-2}(z,\overline{z})=1,&\qquad
&W(z,\overline{z})=X(z) X(\overline{z});
\label{invnonchiral}
\\
&W(z)Y_{-2}(z)=1,& &W(z)= X^2(z)+ \alpha _{1}W_{1}(z)+\alpha _{2}
W_{2}(z)+\alpha _{3} W_{3}(z),
\label{invchiral}
\end{alignat}
\end{subequations}
where $ W_i(z)=\{Q,\Omega _i(z)\}$,  $\Omega_i$ is given by
\begin{equation}
\Omega_1=(\partial b)\partial e^{2\phi},~~
\Omega_{2}=(\partial  ^2b)e^{2\phi},~~
\Omega_{3}=b\partial ^2e^{2\phi}
\end{equation}
and $\alpha_i$ are the following numbers
\begin{equation}
\alpha _{1}=-\frac{19}{192},\;\alpha _{2}=-\frac{5}{48},
\;\alpha_{3}=-\frac{1}{96}.
\label{e.25}
\end{equation}

The issue of equivalence between the theories based on chiral or
nonchiral insertions still remains open. The first touch to the
problem was performed in \cite{UZ}. It was shown that the actions
for low-level space-time fields are different depending on
insertion being chosen.

\subsection{SSFT in  Conformal Language.}
As for the bosonic string (see Sec.~\ref{sec:gluing-and-maps})
for SSFT calculations it is convenient to employ the tools
of CFT \cite{l'Clare}.
In the conformal language $\int$ and
$\star$-product mentioned above are replaced by the odd bracket
$\langle\!\langle...|\dots\rangle\!\rangle$, defined as follows
\begin{equation}
\begin{split}
\langle\!\langle Y_{-2}|A_1,\dots,A_n\rangle\!\rangle&=
\left\langle\, P_n\circ Y_{-2}(0,0)\,f_1^{(n)}\circ A_1(0)\dots
f_n^{(n)}\circ A_n(0)\right\rangle_{R_n}\\
&=\left\langle\, f_j^{(n)}\circ Y_{-2}(i,\overline{i})\,f_1^{(n)}\circ A_1(0)\dots
f_n^{(n)}\circ A_n(0)\right\rangle_{R_n}
, \quad n=2,3.
\end{split}
\label{oddbracket}
\end{equation}
Here r.h.s. contains $SL(2,\mathbb{R})$\,-invariant correlation
function in CFT on
string configuration surface $R_n$ described
in Section~\ref{sec:gluing-and-maps}.
$A_j$~$(j=1,\dots,n)$  are vertex operators and
$\{f^{(n)}_j\}$ are given by eq.(\ref{maps:b})

$Y_{-2}$ is the double-step inverse picture changing operator
\eqref{nonchiral} inserted in the center of the unit disc. This
choice of the insertion point is very important, since all the
functions $f_j^{(n)}$ maps the points $i$ (the middle points of
the individual strings) to the same point that is the origin.
In other words, the origin is a unique common point for
all strings (see Figure \ref{fig:maps}). The next important
fact is the zero weight of the operator $Y_{-2}$, so its conformal
transformation is very simple $f\circ
Y_{-2}(z,\overline{z})=Y_{-2}(f(z),\overline{f}(\overline{z}))$.
Due to this property it can be inserted in any string.
This note shows that the definition \eqref{oddbracket} is
self-consistent and does not depend on a choice of a string on
which we insert $Y_{-2}$.

Due to the Neumann boundary conditions, there is a relation between
holomorphic and antiholomorphic fields. So it is convenient to
employ a doubling trick (see details in \cite{polchinski}).
Therefore, $Y_{-2}(z,\overline{z})$ can be rewritten in the
following form\footnote{Because of this formula, the operator
$Y_{-2}(z,\overline{z})$ is sometimes called bilocal.}
$$
Y_{-2}(z,\overline{z})=Y(z)Y(z^*).
$$
Here $Y(z)$ is the holomorphic field and $z^*$ denotes the conjugated
point of $z$ with respect to a boundary, i.e. for the unit disc
$z^*=1/\overline{z}$ and $z^*=\overline{z}$ for the upper half
plane. From now on  we work only on the whole complex
plane. Hence the odd bracket takes the form
\begin{equation} \langle\!\langle
Y_{-2}|A_1,\dots,A_n\rangle\!\rangle=\left\langle\,
Y(P_n(0))Y(P_n(\infty))\,f_1^{(n)}\circ A_1(0)\dots f_n^{(n)}\circ
A_n(0)\right\rangle. \label{oddbracket2}
\end{equation}

To summarize, the action we start with reads
\begin{equation}
S[\mathcal{A}]=\frac{1}{g_{o}^2}\left[\frac12\langle\!\langle
Y_{-2}|\mathcal{A},Q_B\mathcal{A}\rangle\!\rangle +\frac13\langle\!\langle
Y_{-2}|\mathcal{A},\mathcal{A},\mathcal{A}\rangle\!\rangle\right],
\label{action0}
\end{equation}
where $g_o$ is a dimensionless coupling constant.

\subsection{Free Equation of Motion.}
The aim of this section is to illustrate that the free
equation of motion \eqref{eq-with-Y} obtained from the action \eqref{action0}
and the free equation of motion \eqref{eq-without-Y} are the
same\footnote{In this section
we use the convention $\ap=2$}.

To this purpose we consider the following GSO$+$ string field
(in picture zero and ghost number one)
\begin{multline}
\Ac=\int\frac{d^{26}k}{(2\pi)^{26}}\,\{u(k)ce^{2ik\cdot X_L(w)}+
\frac{1}{2}A_{\mu}(k)c\partial X^{\mu}e^{2ik\cdot X_L(w)}-
\frac{1}{4}B_{\mu}(k)\eta e^{\phi}\psi^{\mu}e^{2ik\cdot X_L(w)}\\
+\frac{1}{2}F_{\mu\nu}(k)c\psi^{\mu}\psi^{\nu}e^{2ik\cdot X_L(w)}+
r_{1}(k)\partial c e^{2ik\cdot X_L(w)}+
r_{2}(k)c\partial \phi e^{2ik\cdot X_L(w)} \}
\label{ex-filed}
\end{multline}
and show that free equations of motion for the component fields
are usual Maxwell equations for the massless
vector field plus some relations between
the other fields.

The \eqref{ex-filed} contains all possible excitations
in picture $0$ on levels $L_0=-1$ and $L_0=0$.
There are several unexpected properties of this formula:
\begin{itemize}
\item First, we find
that the level $L_0=-1$ is not empty --- it contains
one field $u(k)$;
\item Second, we find
that  there are two candidates for the Maxwell field: $A_\mu$
and $B_\mu$;
\item Third, we find
that  there are too many fields as compared with the
massless spectrum of the first quantized string.
\end{itemize}
As it will be shown in Section~\ref{sec:rest-on-fields}
the field $B_{\mu}$ is physical one,
while all other fields are auxiliary ones. 

Let us first examine the equations that follow from
$ Q_B|\Ac\rangle=0$. We have
\begin{align*}
\{Q_B,\Ac\}=\int&\frac{d^{26}k}{(2\pi)^{26}}\,
\Bigl[u(k)[(2 k^{2}-1)\partial c c
+c\eta e^{\phi}ik\cdot \psi+\frac{1}{4}\partial \eta \eta e^{2\phi}]e^{2ik\cdot X_L(w)}
\\
&+A_{\mu}(k)[k^{2}\partial c c \partial X^{\mu} -\frac{1}{2} {i} k^{\mu} \partial^{2}c c
+\frac{1}{4}c\partial(\eta e^{\phi}\psi^{\mu})
\\
&~~~~~~~~~~+\frac{1}{2}c\eta e^{\phi}{i} k \cdot \psi\partial X^{\mu}+
\frac{1}{8}\partial \eta \eta e^{2\phi}\partial X^{\mu}]e^{2ik\cdot X_L(w)}
\\
&+B_{\mu}(k)[-\frac{1}{2}k^{2}\partial c\eta e^{\phi}\psi^{\mu}-\frac{1}{4}c\partial(\eta e^{\phi} \psi^{\mu})
-\frac{1}{2}c\eta e^{\phi}\psi^{\mu}{i} k \cdot \partial X
\\
&~~~~~~~~~~+\frac{1}{8}{i} k^{\mu}\partial(\partial\eta\eta e^{2\phi})-
\frac{1}{8}\partial\eta\eta e^{2\phi}\partial X^{\mu}-
\frac{1}{4}\partial\eta\eta e^{2\phi}{i} k \cdot \psi\psi^{\mu}]e^{2ik\cdot X_L(w)}
\\
&+F_{\mu\nu}[k^{2}\partial c c \psi^{\mu}\psi^{\nu}-
\frac{1}{2}c\partial(\eta e^{\phi}){i} k_{[\mu}\psi_{\nu]}+
\frac{1}{2}c\eta e^{\phi}{i} k \cdot \psi\psi^{\mu}\psi^{\nu}
\\
&~~~~~~~~~~-\frac{1}{4}c\eta e^{\phi}\psi_{[\mu}\partial X_{\nu]}+
\frac{1}{8}\partial\eta\eta e^{2\phi}\psi^{\mu}\psi^{\nu}]e^{2ik\cdot X_L(w)}
\\
&+r_{1}(k)[-\partial^{2}c c -\partial c c\, 2{i} k \cdot \partial X
+\partial c\eta e^{\phi}{i} k \cdot \psi+
\frac{1}{4}\partial(\partial\eta\eta e^{2\phi})]e^{2ik\cdot X_L(w)}
\\
&+r_{2}(k)[-\partial^{2}c c+2 k^{2}\partial c c\partial\phi-
\frac{1}{2}c\eta e^{\phi}\psi\cdot\partial X-
c e^{\phi}\partial(\eta\,{i} k\cdot\psi)\\
&~~~~~~~~~~+c\eta\partial e^{\phi}2{i} k\cdot\psi+\frac{5}{8}\partial\eta\eta\partial e^{2\phi}
+\frac{1}{2}bc\partial\eta\eta e^{2\phi}+\frac{1}{2}\partial^{2}\eta\eta e^{2\phi}]e^{2ik\cdot X_L(w)}
\Bigr].
\end{align*}
From $\{Q_B,\,\Ac\}=0$ one gets the following
equations for component fields:
\begin{subequations}
\label{equat-QA}
\begin{align}
u=0,~~~r_2=0,
\label{FP-eq}
\\
B_{\mu}=A_{\mu},~~~F_{\mu\nu}=\frac12 ik_{[\mu}A_{\nu ]},~~~
r_1=-\frac14 ik_{\mu} A_{\mu},
\label{AB-eq}
\\
k^2A_{\mu}=4ik_{\mu} r_1.
\label{M-eq}
\end{align}
\end{subequations}
Note that we get Maxwell equation for $B_{\mu}$
after excluding $r_1$ in (\ref{M-eq}) via (\ref{AB-eq})
\begin{equation}
k^2B_\mu=k_\mu k_\nu B_\nu.
\end{equation}
Eqs. \eqref{FP-eq} exclude fields $u$ and $r_2$
and (\ref{AB-eq}) leave only one vector field.

Now let us present the result of computations of
the action \eqref{action0} on the string field  \eqref{ex-filed}:
\begin{multline*}
S_{2}=\int\frac{d^{26}k}{(2\pi)^{26}}\,\Bigl[
u(-k)u(k)-\frac{1}{4}A_{\mu}(-k)A_{\mu}(k)
-\frac{1}{4}B_{\mu}(-k)B_{\mu}(k)
+\frac{1}{2}A_{\mu}(-k)B_{\mu}(k)
\\
-\frac{1}{2}F_{\mu\nu}(-k)F_{\mu\nu}(k)+i k_{\mu}F_{\mu\nu}(-k)B_{\nu}(k)
-2i r_{2}(-k)k_{\mu}B_{\mu}(k)
-10 r_{2}(-k)r_{2}(k)-4r_{1}(-k)r_{2}(k)
\Bigr].
\end{multline*}
One can check that the equations of motion following from this action
are precisely the eqs. \eqref{equat-QA}.

Moreover, free equations of motion following
from $Q_B|\Ac\rangle=0$ and from $Y_{-2}Q_B|\Ac\rangle=0$
are the same for any string field $\Ac$.
\subsection{Scattering Amplitudes.}
\subsubsection{Gauge fixing.}
 Before  proceeding with  the investigation  of tree  graphs in the theory with
the  action (\ref{action0}), we should  specify a   gauge fixing procedure. It  is convenient
to choose the
Siegel  gauge \cite {48}
\begin{equation}
b_0 |A\rangle  =0.
\label{4.13}
\end{equation}
To find the propagator we have to invert the operator
$  Y_{-2}Q $ on the space of states satisfying (\ref{4.13}).
 The propagator can be presented in a number of forms:
\begin{equation}
\Delta =\frac{b_{0}}{L}WQ\frac{b_{0}}{L}
=\frac{b_{0}}{L}W-\frac{b_{0}}{L}W\frac{b_{0}}{L}Q
=W\frac{b_{0}}{L}+Q\frac{b_{0}}{L}W\frac{b_{0}}{L} .
\label{4.15}
\end{equation}
One can prove that $ \Delta Y_{-2}Q =1$ in the  space  (\ref {4.13})
by a simple calculation based on the
following set of equalities:
\begin{equation}
[L,Q]=0,~~~\{b_{0},Q\}=L,~~~[W,Q]=0,~~~[Y_{-2},Q]=0,~~~Y_{-2}W=1.
\label{4.16}
\end{equation}


\subsubsection{4-point on-shell amplitude.}
\setlength{\unitlength}{0.5in}

Now  we turn to  the  calculation  of the four-boson scattering  amplitude in
the
cubic theory.
We shall proceed  as  close as  possible to  the analogous calculation
in the usual  Witten theory.  In the bracket representation for two- and
three-point vertices one can write

\begin{equation}
{\cal A}_{4} ={}_{41i}\langle V_3|  ~_{j23}\langle V_3|  Y_{-2}^{(i)} Y_{-2}^{(j)}
\frac{b^{(i)}_o}{L^{(i)}} WQ \frac{b^{(i)}_o}{L^{(i)}}  |V_2\rangle_{ij}
|\tilde{A} _{4} \rangle_{4}
|\tilde{A} _{1} \rangle_{1}
|\tilde{A} _{2} \rangle_{2}
|\tilde{A} _{3} \rangle_{3}.
\label{5.1}
\end{equation}

In  comparison with the usual  formula (\ref{d.1}),  we  evidently  find in
(\ref{5.1}) the additional  midpoint insertions of $Y_{-2}^{(i)}$ ,
$Y_{-2}^{(j)}$ and the  new  propagator (\ref{4.15}) . The  string fields
$|\tilde{A}_r\rangle =X|A\rangle$ is a field in the -1 picture  defined
 by the expression  like (\ref{4.10}).

Taking into account the BRST invariance of the vertices and the
on-shell condition for
$|\tilde{A}_{r}\rangle$, we can  rewrite  eq. (\ref{5.1}) in the form

\begin{equation}
{\cal A}_{4} =~ _{41i}\langle V_{3}|_{j23}\langle V_3|  Y_{-2}^{(i)}
\frac{b^{(i)}_{o}}{L^{(i)}}  |V_2\rangle_{ij}
|\tilde{A} _{4} \rangle_{4}
|\tilde{A} _{1} \rangle_{1}
|\tilde{A} _{2} \rangle_{2}
|\tilde{A} _{3} \rangle_{3},
\label{5.2}
\end{equation}
where one  of the two potentially dangerous $ Y_{-2}$'s   is cancelled by the
operator $W$ in the propagator.

In the conformal formulation this expression corresponds to
\begin{equation}
{\cal A}_{4}=\int^{\infty}_{0} dT  \oint \frac {dz}{2\pi i} \frac {dz}{d\rho}
\langle \tilde{{\cal O}}_1 \tilde{{\cal O}}_2 \tilde{{\cal O}}_3 \tilde{{\cal O}}_4
  b Y_{-2} \rangle.\label{5.3}
\end{equation}
where $Z(\rho )$ is the Giddings map for $N=4$.
 To be more rigorous we can consider  first the regularized expression
\begin{equation}
{\cal A}_{4}=\int  ^{\infty}_{\tau} dT  \oint \frac {dz}{2\pi i}\frac {dz}{d\rho }
\langle \tilde{{\cal O}}_1 \tilde{{\cal O}}_2 \tilde{{\cal O}}_3 \tilde{{\cal O}}_4
  b Y_{-2} \rangle,\label{5.4}
\end{equation}
with the regularization  parameter $\tau$ .

  We will prove that:

i)~the correlation function (\ref{5.4}) is well-defined in
the limit $\tau \to 0$;

ii) in the limit  $\tau \to 0$  the correlation function (\ref{5.3})
after summation
over all  channels is equal to the Koba-Nielsen amplitude.

We will proceed in the same way as described for Witten's diagram
in  Section \ref{sec:kba} above. The first step   is to  move two of the $X(0)$, say $X^{(1)}$
and $X^{(4)}$ ,
to the midpoint via the famous rule
\begin{equation}
X^{(k)}(0)  - X^{(k)}(i) =\{ Q,\delta \xi ^{(k)}\} ,\label{5.5}
\end{equation}
where $$\delta \xi ^{(k)}=\xi ^{(k)}(i) - \xi ^{(k)}(0) $$
and   we find  $A_{4}$ in the form

  \begin{eqnarray}\label{5.6}
{\cal A}_{4}= {\cal A}_{4}^{0}+ {\cal A}_{4}^{1}+
{\cal A}_{4}^{2}~~~~~~~~~~~~~~~~~~~~~~~~~~~~\nonumber\\
\nonumber\\
 \begin{picture}(3.5,2.3)
\put(0,0.5){\line(1,1){0.5}}
\put(0.5,1){\line(-1,1){0.5}}
\put(0.5,1){\line(1,0){2}}
\put(2.5,1){\line(1,1){0.5}}
\put(3,0,5){\line(-1,1){0.5}}
\put(0.5,1){\circle*{0.1}}                                    
\put(0,0.2){$\tilde  {4}$}
\put(0,1.7){$\tilde{1}$}
\put(2.7,0.2){$\tilde{3}$}
\put(2.7,1.7){$\tilde{2}$}
\put(0.5,0.6){$Y_{-2}$}
\put(3,1){=}
\end{picture}
 \begin{picture}(3.5,2.3)
\put(0,0.5){\line(1,1){0.5}}
\put(0.5,1){\line(-1,1){0.5}}
\put(0.5,1){\line(1,0){2}}
\put(2.5,1){\line(1,1){0.5}}
\put(3,0,5){\line(-1,1){0.5}}
\put(0.5,1){\circle*{0.1}}                                    
\put(0,0.2){$\tilde  {4}$}
\put(0,1.7){$1$}
\put(2.7,0.2){$\tilde{3}$}
\put(2.7,1.7){$\tilde{2}$}
\put(0.5,0.6){$Y$}
\put(3,1){+}
\end{picture}
~~~~~~~\begin{picture}(3,2.3)
\put(0,0.5){\line(1,1){1}}
\put(0,1.5){\line(1,-1){1}}                  
\put(0,0.2){$\tilde {4}$}
\put(0,1.7){$\xi  1$}
\put(0.5,1){\circle*{0.1}}
\put(1.1,1.7){$\tilde {2}$}
\put(1.1,0.2){$\tilde {3}$}
\put(0.8,0.8){$\xi Y_{-2}$}
\put(3,1){=}
\end{picture}    \nonumber\\
\nonumber\\
 \begin{picture}(3.5,2.3)
\put(0,0.5){\line(1,1){0.5}}
\put(0.5,1){\line(-1,1){0.5}}
\put(0.5,1){\line(1,0){2}}
\put(2.5,1){\line(1,1){0.5}}
\put(3,0,5){\line(-1,1){0.5}}          
\put(0,0.2){$4$}
\put(0,1.7){$1$}
\put(2.7,0.2){$\tilde{3}$}
\put(2.7,1.7){$\tilde{2}$}
\put(3,1){$+$}
\end{picture}
\begin{picture}(2,2.3)
\put(0,0.5){\line(1,1){1}}
\put(0,1.5){\line(1,-1){1}}                  
\put(0,0.2){$4 \xi $}
\put(0,1.7){$1$}
\put(1.1,1.7){$\tilde {2}$}
\put(0.5,1){\circle*{0.1}}
\put(1.1,0.2){$\tilde {3}$}
\put(0.8,0.8){$\xi$}
\put(1.5,1){$+$}
\end{picture}
\begin{picture}(3,2.3)
\put(0,0.5){\line(1,1){1}}
\put(0,1.5){\line(1,-1){1}}                  
\put(0,0.2){$\tilde  {4}$}
\put(0,1.7){$\xi  1$}
\put(0.5,1){\circle*{0.1}}
\put(1.1,1.7){$\tilde {2}$}
\put(1.1,0.2){$\tilde {3}$}
\put(0.8,0.8){$\xi Y_{-2}$}
\end{picture}~~~~~~~~~~    \\ \nonumber
\end{eqnarray}

In  the RHS of eq. (\ref{5.6}) the first term contributes to the Koba-Nielsen amplitude
\begin{equation}
{\cal A}_{4}^{0}=\int ^{\infty}_{\tau}  dT \oint \frac {dz}{2\pi i}
\frac {dz}{d\rho}
\langle {\cal O}_1 \tilde{{\cal O}}_2 \tilde{{\cal O}}_3 {\cal O}_4
  b \rangle \label{5.7}.
\end{equation}
The last two terms in (\ref{5.6}), $ A_{4}^{1}$ and $ A_{4}^{2}$ ,
are contact terms with cancelled  propagator
\begin{equation}
{\cal A}_{4}^{1}=  \langle   Y \delta  \xi ^{(4)} {\cal O}_1 \tilde{{\cal O}}_2
\tilde{{\cal O}}_3 {\cal O}_4
  b \rangle \label{5.8},
\end{equation}
and
\begin{equation}
{\cal A}_{4}^{2}=  \langle   Y_{-2} \delta  \xi ^{(1)}
 {\cal O}_1 \tilde{{\cal O}}_2 \tilde{{\cal O}}_3\tilde{{\cal O}}_4
  b \rangle. \label{5.9}
\end{equation}

In  contrast to the usual  calculation (see  Appendix A in \cite{AMZ2}
for details) we have  not obtained
any singularity in contact terms (see Appendix \ref{sec:pcl})
so  there is no need to look for a
regularization  to define  $A_{4}$  correctly. Therefore, the limit
$\tau \to 0$ does exist . Finally note that the  amplitudes ${\cal A}^1_4$
and ${\cal A}^2_4$ after summation over all channels are identically
zero.
For illustration we will
 prove  this for ${\cal A}^1_4$. The
 explicit
formula for ${\cal A}^1_4$ in the $s$-channel reads
\begin{equation}
{\cal A}^{1(s)}_4=\int ^{\infty}_{0} dT\oint \frac {dz}{2\pi i}
\frac{dz}{d\rho}\langle {\cal O}_1(z_1)\tilde{{\cal O}}_2(z_2)
\tilde{{\cal O}}_3(z_3){\cal O}_4(z_4)b(z)((\xi (z_4)-\xi (z_0))
Y(z_0)\rangle .
\label{5.10}
\end{equation}
In the $t$-channel we get
\begin{equation}
{\cal A}^{1(t)}_4=\int ^{\infty}_{0} dT\oint \frac {dz}{2\pi i}
\frac{dz}{d\rho}\langle\tilde{{\cal O}}_2(z_1)\tilde{{\cal O}}_3(z_2)
{\cal O}_4(z_3){\cal O}_1(z_4)b(z)((\xi (z_3)-\xi (z_0))
Y(z_0)\rangle .
\label{5.11}
\end{equation}
Making a $SL(2,C)$ transformation of the form
\begin{equation}
z \rightarrow z^{\prime} = \frac {z-1}{z+1}
\label{5.12},
\end{equation}
which gives
$$(z_1,z_2,z_3,z_4) \rightarrow (z_2,z_3,z_4,z_1), $$
and using the fact that ${\cal O}_r $ are anticommuting
operators we find that
$$ {\cal A}^{1(t)}_4 \rightarrow  -{\cal A}^{1(s)}_4,  $$
and so the sum of (\ref{5.10}) and (\ref{5.11}) is equal to zero.
In the oscillator language this corresponds to the cancellation of 4-point
contact terms by cyclic symmetry arguments.
   The term (\ref{5.7}) is also well-defined in the limit $\tau \rightarrow 0$
and  gives the $s$-channel contribution in the total Koba-Nielson amplitude
\cite{42,41,60}.

As in a usual gauge theory the issue of finiteness of 4-point off-shell
amplitudes crucially depends on the choice of a gauge. One can use the  alternative
gauge $b_0(Y_{-2}\tilde{A})=0$, which leads to the propagator
\begin{equation}
W(i)\frac{b_0}{L}Y_{-2}Q\frac{b_0}{L}W(i)= W(i)\frac{b_0}{L}-
W(i)\frac{b_0}{L}Y_{-2}\frac{b_0}{L}W(i)Q. \label{5.13}
\end{equation}
In this gauge the NS 4-point function (off-shell NS amplitude) is obviously finite.
\subsubsection{N-point on-shell amplitudes.}

Let us start with the example of the 5-point amplitude. From now on we
 choose the propagator in the form (see eq. (\ref{4.15}))
\begin{equation}
\Delta=\frac{b_0}{L}W(i)-\frac{b_0}{L}W(i)\frac{b_0}{L}Q,
\label{7.1}
\end{equation}
then the 5-point function is  given by the sum of two Feynman graphs
\begin{equation}
  \begin{picture}(3.5,2.3)
\put(0,0.5){\line(1,1){0.5}}
\put(0.5,1){\line(-1,1){0.5}}
\put(0.5,1){\line(1,0){2}}
\put(2.5,1){\line(1,1){0.5}}
\put(3,0,5){\line(-1,1){0.5}}        
\put(1.5,1){\line(0,1){0.5}}
\put(0,0.2){$\tilde{1}$}
\put(0,1.7){$\tilde{2}$}
\put(2.7,0.2){$\tilde{5}$}
\put(1.7,1.7){$\tilde{3}$}
\put(2.7,1.7){$\tilde{4}$}
\put(3,1){~+}
\end{picture}
   \begin{picture}(5.5,1.8)
\put(0,0.5){\line(1,1){0.5}}
\put(0.5,1){\line(-1,1){0.5}}
\put(0.5,1){\line(2,0){3}}
\put(3.5,1){\line(1,1){0.5}}
\put(4,0.5){\line(-1,1){0.5}}
\put(2.5,1){\line(0,1){0.5}}
\put(0.5,1){\circle*{0.1}}                                    
\put(1.5,1){\circle*{0.1}}
\put(2.5,1){\circle*{0.1}}
\put(0,0.2){$\tilde{1}$}
\put(0,1.7){$\tilde{2}$}
\put(4,0.2){$\tilde{5}$}
\put(4,1.7){$\tilde{4}$}
\put(2.4,1.7){$\tilde{3}$}
\put(0.5,0.6){$Y_{-2}$}
\put(2.5,0.6){$ZQ$}
\put(1.5,0.6){$W$}
\end{picture}  ~~~~~~
\label{7.2}
\end{equation}
where the line with a dot stands for the second term in eq. (\ref{7.1}).
To analyze the graphs of this type, which also appear in N-point
amplitudes, we adopt the following regularization
\begin{eqnarray} \label{7.3}
\Delta ~~\rightarrow~~\Delta _{\tau _1, \tau _2}=\frac{b_0}{L}
e^{-\tau _1L}W(i)-
\frac{b_0}{L}e^{-\tau _2L}W(i)e^{-\tau _2L}\frac{b_0}{L} Q \nonumber\\
  \nonumber\\
=b_0 \int^{\infty}_{\tau _{1}}d\tau e^{-\tau L}W(i)-
b_0 \int^{\infty}_{\tau _{2}}d\tau e^{-\tau L}W(i)
b_0 \int^{\infty}_{\tau _{2}}d\tau e^{-\tau L}Q. \\ \nonumber \end{eqnarray}
This regularization is a lagrangian one and it's special case was
used previously for the 4-point function.

Adopting the regularization eq. (\ref{7.3}) we find that in the limit
$\tau_1 \to 0$ the operator $Q$ in the second term of  (\ref{7.2})
directly cancels the right-standing propagator thus leading to the
diagram with the contact 4-vertex.
\begin{equation}
 \begin{picture}(5.5,1.8)
\put(0,0.5){\line(1,1){0.5}}
\put(0.5,1){\line(-1,1){0.5}}
\put(0.5,1){\line(2,0){2}}
\put(2.5,1){\line(1,1){0.5}}
\put(3,0.5){\line(-1,1){0.5}}
\put(2.5,1){\line(0,1){0.5}}
\put(0.5,1){\circle*{0.1}}                                    
\put(1.5,1){\circle*{0.1}}
\put(2.5,1){\circle*{0.1}}
\put(0,0.2){$\tilde{1}$}
\put(0,1.7){$\tilde{2}$}
\put(3.2,0.2){$\tilde{5}$}
\put(3.2,1.7){$\tilde{4}$}
\put(2.7,1.7){$\tilde{3}$}
\put(0.5,0.6){$Y_{-2}$}
\put(2.5,0.6){$Y_{-2}$}
\put(1.5,0.6){$W$}
\end{picture}
  \end{equation}
According to the Lemma in  Appendix \ref{sec:pcl} we cannot remove the regularization parameter
$\tau _{2}$ in this diagram. However, literally reproducing the arguments of the previous
subsection, we conclude
that after summation over all the channels these contact terms give zero.
Hence we find that only the first graph in (\ref{7.2}) contributes
on-shell. Reproducing once again the arguments of the previous subsection, we
finally conclude    that the   first graph in (\ref{7.2})
gives the Koba-Nielsen formula.

Following just the same device one can analyze any N-point on-shell amplitude
and demonstrate the total agreement with the Koba-Nielsen formulae.

The off-shell analysis of N-point functions is much more subtle. The key
point is the behavior of multi-factor OPEs of the form
$Y_{-2}\cdot W\cdot Y_{-2}\cdots Y_{-2}$.
\subsection{Restriction on  String Fields.}
\label{sec:rest-on-fields}
The aim of this section is to try to reduce the number
of excitations appearing in the zero picture string field.
Such reductions can be made if we have degenerate quadratic
part of the action. But the reduction we describe
is not of this sort. In this construction we
use only the fact that the expectation value of the operator
on the complex plane in super conformal two dimensional field theory
is non-vanishing only if it compensates all background
charges of the theory.

Let us
decompose the string field $\Ac$
according to the $\phi$-charge $q$:
$$
\Ac=\sum_{q\in\Zh}\Ac_q,\qquad \Ac_q\in V_q,
$$
where
\begin{equation}
[j_0,\,\Ac_q]=q\Ac_q \quad\text{with}\quad j_0=\frac{1}{2\pi i
}\oint d\zeta\,\pd\phi(\zeta).
\label{phich}
\end{equation}

The BRST charge $Q_B$ has also  the natural decomposition
over $\phi$-charge:
\begin{equation}
Q_B=Q_0+Q_1+Q_2.
\end{equation}
Since $Q_B{}^2=0$ we get the identities:
\begin{equation}
Q_0{}^2=0,\quad\{Q_0,\,Q_1\}=0,\quad\{Q_1,\,Q_2\}=0,
\quad Q_2{}^2=0 \quad\text{and}\quad
\{Q_0,\,Q_2\}+Q_1{}^2=0.
\end{equation}
The non-chiral inverse double step picture changing operator
$Y_{-2}$ has $\phi$-charge equal to $-4$.
Therefore to be non zero the expression
in the brackets $\la Y_{-2}|\dots\ra$  must have
$\phi$-charge equal to $+2$. Hence the quadratic $S_2$ and cubic $S_3$
terms of the  action (\ref{action}) read:
\begin{subequations}
\begin{align}
S_2&=-\frac{1}{2}\sum_{q\in\Zh}\la Y_{-2}|\Ac_{2-q},\,Q_0\Ac_q\ra
-\frac{1}{2}\sum_{q\in\Zh}\la Y_{-2}|\Ac_{1-q},\,Q_1\Ac_q\ra
-\frac{1}{2}\sum_{q\in\Zh}\la Y_{-2}|\Ac_{-q},\,Q_2\Ac_q\ra.
\label{resA}
\\
S_3&=-\frac{1}{3}\sum_{q,\,q^{\prime}\in\Zh}
\la Y_{-2}|\Ac_{2-q-q^{\prime}},\Ac_{q^{\prime}},\Ac_q\ra.
\end{align}
\label{resQact}
\end{subequations}
 We see that all the fields $\Ac_q$, $q\ne 0,1$, give only linear
contribution to the quadratic action \eqref{resA}.
We propose to exclude such fields.  We will
consider the action
\eqref{resQact}  with fields that belong to the space
$V_0\oplus V_1$ only.
To make this prescription meaningful we have to check that
the restricted action is gauge invariant.

The action restricted to  subspaces $V_0$ and $V_1$
takes the form
\begin{subequations}
\begin{align}
S_{2,\,\text{restricted}}&=-\frac{1}{2}\la Y_{-2}|\Ac_{0},\,Q_2\Ac_0\ra
-\la Y_{-2}|\Ac_{0},\,Q_1\Ac_1\ra
-\frac{1}{2}\la Y_{-2}|\Ac_{1},\,Q_0\Ac_1\ra,
\label{act-2}
\\
S_{3,\,\text{restricted}}&=
-\la Y_{-2}|\Ac_{0},\Ac_{1},\Ac_1\ra.
\end{align}
\label{act}
\end{subequations}
The action \eqref{act} has a nice structure.
One sees that since the charge $Q_2$ does not contain
zero modes of stress energy tensor
the fields $\Ac_0$ play a role of auxiliary fields.
On the contrary,
all  fields $\Ac_1$ are physical ones, i.e. they
have non-zero kinetic terms.

Let us now check that the action \eqref{act} has
gauge invariance. The action (\ref{action0}) is gauge invariant under
\begin{equation}
\delta\Ac=Q_B\Lambda+[\Ac,\,\Lambda].
\label{gaugeP}
\end{equation}
But after the restriction to the space $V_0\oplus V_1$ it might be lost.
Decomposing the gauge parameter $\Lambda$ over
$\phi$-charge $\Lambda=\sum\limits_q\Lambda_q$ we rewrite the
gauge transformation \eqref{gaugeP} as:
\begin{equation}
\delta\Ac_q=Q_0\Lambda_q+Q_1\Lambda_{q-1}+Q_2\Lambda_{q-2}
+\sum_{q^{\prime}\in\Zh}[A_{q-q^{\prime}},\,\Lambda_{q^{\prime}}].
\label{gaugeR}
\end{equation}
Assuming that $\Ac_q=0$ for $q\neq0,1$
from \eqref{gaugeR} we get
\begin{subequations}
\begin{align}
\delta\Ac_{-2}&=Q_0\Lambda_{-2};
\\
\delta\Ac_{-1}&=Q_0\Lambda_{-1}+[\Ac_0,\,\Lambda_{-1}]
+Q_1\Lambda_{-2}+[\Ac_1,\,\Lambda_{-2}];
\\
\delta\Ac_{0}&=Q_0\Lambda_{0}+[\Ac_0,\,\Lambda_{0}]
+Q_1\Lambda_{-1}+[\Ac_1,\,\Lambda_{-1}]
+Q_2\Lambda_{-2};
\\
\delta\Ac_{1}&=Q_0\Lambda_{1}+[\Ac_0,\,\Lambda_{1}]
+Q_1\Lambda_{0}+[\Ac_1,\,\Lambda_{0}]
+Q_2\Lambda_{-1};
\\
\delta\Ac_{2}&=Q_1\Lambda_{1}+[\Ac_1,\,\Lambda_{1}]
+Q_2\Lambda_{0};\label{da2}
\\
\delta\Ac_{3}&=Q_2\Lambda_{1}.
\end{align}
\label{restr}
\end{subequations}
To make the restriction consistent with transformations
(\ref{restr}) the variations of the fields
$\Ac_{-2},\,\Ac_{-1},\,\Ac_{2}$ and $\Ac_3$ must be zero. Since
the gauge parameters cannot depend on string fields we must put
$\Lambda_{-2}=\Lambda_{-1}=\Lambda_{1}=0$. So we are left with the
single parameter $\Lambda_0$, but to have zero variation of $\Ac_2$ we
must require in addition $Q_2\Lambda_0=0$. Therefore the gauge
transformations take the form
\begin{equation}
\begin{split}
\label{resgt} \delta\Ac_0&=Q_0\Lambda_0+[\Ac_0,\,\Lambda_0],
\\
\delta\Ac_1&=Q_1\Lambda_0+[\Ac_1,\,\Lambda_0],
\quad\text{with}\quad Q_2\Lambda_0=0.
\end{split}
\end{equation}

It is easy to check that (\ref{resgt}) form a closed algebra. It
is also worth to note that  the restriction $Q_2\Lambda_0=0$
leaves the gauge transformation of the massless vector field
unchanged.

\subsection{Appendix.}

\subsubsection{Cyclicity property.}\label{app:cyclic}
The proof of the cyclicity property is very similar to the
one given in \cite{0002211}.
But there is one specific point --- insertion of the
double step inverse picture changing operator $Y_{-2}$
\eqref{nonchiral}.
So we repeat the proof with all necessary modifications.

Let $T_n$ and $R$ denotes rotation
by $-\frac{2\pi}{n}$ and $-2\pi$ respectively:
$$
T_n(w)=e^{-\frac{2\pi i}{n}}w,
\qquad R(w)=e^{-2\pi i}w.
$$
These transformations have two fixed points namely
$0$ and $\infty$. Let us apply the transformation
$T_n$ to the maps $f^{(n)}_k$ \eqref{maps:b} and
we get the identities:
\begin{equation}
T_n\circ f_{k}^{(n)}=f_{k+1}^{(n)},\quad k<n,\qquad
T_n\circ f_{n}^{(n)}=R\circ f_{1}^{(n)},
\qquad n=2,3.
\end{equation}
Since the weight of the operator $Y_{-2}$
is zero and $0$ and $\infty$ are fixed points of
$T_n$ and $R$,
the operator $Y_{-2}$ remains unchanged.
Due to $SL(2,\mathbb R)$-invariance
of the correlation function
we can write down a chain of equalities
\begin{multline}
\langle Y_{-2}\,F_1^{(n)}\circ A_1\dots
F_{n-1}^{(n)}\circ A_{n-1}F_{n}^{(n)}\circ
\mathcal{O}_h\rangle
=\langle Y_{-2}\,f_1^{(n)}\circ A_1\dots
f_{n-1}^{(n)}\circ A_{n-1}f_{n}^{(n)}\circ
\mathcal{O}_h\rangle
\\
=\langle Y_{-2}\,T_n\circ f_1^{(n)}\circ A_1\dots
T_n\circ f_{n-1}^{(n)}\circ A_{n-1}T_n\circ f_{n}^{(n)}
\circ \mathcal{O}_h\rangle
\\
=e^{-2\pi i h}\langle Y_{-2}\,F_{1}^{(n)}\circ \mathcal{O}_h\,
F_2^{(n)}\circ A_1\dots F_{n}^{(n)}\circ A_{n-1}\rangle.
\end{multline}
In the last line we assume that $\mathcal{O}_h$ is a
primary field of weight $h$
and use the transformation law of primary
fields under rotation:
$$
(R\circ\mathcal{O}_h)(w)=e^{-2\pi ih}\mathcal{O}_h(e^{-2\pi i}w).
$$
Also we change the order of operators in correlation
function
without change of a sign, because
the expression inside the brackets should be odd (otherwise it will be equal to zero)
and therefore no matter whether $\Phi$ odd or even. So the cyclicity property
reads
\begin{equation}
\langle\!\langle Y_{-2}|A_1,\dots,A_n\rangle\!\rangle=e^{-2\pi i h_{n}}\langle\!\langle Y_{-2}|A_n,A_1,\dots,A_{n-1}\rangle\!\rangle
\label{cyclic}
\end{equation}

\medskip
\textsc{Examples.} Now we consider some applications of the cyclicity property (\ref{cyclic}).
GSO$+$ sector consists of the fields with integer weights
and therefore their
exponential factor is equal to $1$,
while GSO$-$ sector consists of the fields with half integer
weights and therefore their
exponential factor is $-1$. Now we give few examples
\begin{subequations}
\label{cyclicGSO-}
\begin{align}
\langle\!\langle Y_{-2}|\mathcal{A}_+,Q_B\mathcal{A}_+\rangle\!\rangle&
=\langle\!\langle Y_{-2}|Q_B\mathcal{A}_+,\mathcal{A}_+\rangle\!\rangle,
\\
\langle\!\langle Y_{-2}|\mathcal{A}_-,Q_B\mathcal{A}_-\rangle\!\rangle&
=-\langle\!\langle Y_{-2}|Q_B\mathcal{A}_-,\mathcal{A}_-\rangle\!\rangle,
\\
\langle\!\langle Y_{-2}|\mathcal{A}_+,\mathcal{A}_-,\mathcal{A}_-\rangle\!\rangle&
=-\langle\!\langle Y_{-2}|\mathcal{A}_-,\mathcal{A}_+,\mathcal{A}_-\rangle\!\rangle
=\langle\!\langle Y_{-2}|\mathcal{A}_-,\mathcal{A}_-,\mathcal{A}_+\rangle\!\rangle.
\end{align}
\end{subequations}

\subsubsection{Twist symmetry.}\label{app:twist}
The proof of the twist symmetry is similar to the one given in \cite{0002211}.
But there is one specific point --- insertion of the operator $Y_{-2}$.
So we repeat this proof here with all necessary modifications.

A twist symmetry is a relation between correlation functions of operators written in
one order and in the inverse one:
\begin{equation}
\langle\!\langle Y_{-2}|\mathcal{O}_1,\dots,\mathcal{O}_{n}\rangle\!\rangle=(-1)^{?}\langle\!\langle Y_{-2}|\mathcal{O}_n,\dots,\mathcal{O}_{1}\rangle\!\rangle.
\label{twist-problem}
\end{equation}
We are interesting in this relation for $n=3$.

\noindent 1) Let us consider the following transformations
$M(w)=e^{-i\pi}w$ and  $\tilde{I}(w)=e^{i0}/w$.
The transformation $\tilde{I}$ has the following properties:
$$
\tilde{I}(z_1)\tilde{I}(z_2)=\tilde{I}(z_1z_2)\quad
\text{and}\quad
\tilde{I}(z^{2/3})=(\tilde{I}(z))^{2/3}.
$$
The pair of points $0$ and $\infty$ is not affected by
$M$ and $\tilde{I}$, therefore the
double-step inverse picture changing operator $Y_{-2}$ remains unchanged.
For the maps \eqref{maps:b} we have got the following composition laws
\begin{equation}
f_1^{(3)}\circ M=\tilde{I}\circ f_{3}^{(3)},\quad
f_2^{(3)}\circ M=\tilde{I}\circ f_{2}^{(3)}
\quad\text{and}\quad
f_3^{(3)}\circ M=\tilde{I}\circ f_{1}^{(3)}.
\end{equation}

\noindent 2) Since there is an identity $M\circ\mathcal{O}(0)=e^{-i\pi h}\mathcal{O}(0)$ we can apply
it to (\ref{twist-problem})
\begin{multline}
\langle\!\langle Y_{-2}|\mathcal{O}_1,\mathcal{O}_2,\mathcal{O}_{3}\rangle\!\rangle
=e^{i\pi\sum h_j}\langle Y_{-2}\,f_{1}^{(3)}\circ M\circ\mathcal{O}_1\,
f_{2}^{(3)}\circ M\circ\mathcal{O}_2\,
f_{3}^{(3)}\circ M\circ\mathcal{O}_{3}\rangle
\\
=e^{i\pi\sum h_j}
\langle Y_{-2}\,\tilde{I}\circ f_{3}^{(3)}\circ\mathcal{O}_1
\,\tilde{I}\circ f_{2}^{(3)}\circ\mathcal{O}_2
\,\tilde{I}\circ f_{1}^{(3)}\circ\mathcal{O}_{3}\rangle
\\
=e^{i\pi\sum h_j}
\langle Y_{-2}\,f_{3}^{(3)}\circ\mathcal{O}_1
\,f_{2}^{(3)}\circ\mathcal{O}_2
\,f_{1}^{(3)}\circ\mathcal{O}_{3}\rangle
\end{multline}
in the last line we use the invariance with respect to $SL(2,\mathbb R)$.
Let $N_{odd}$ and $N_{even}$
be a number of \textit{odd} or \textit{even}
respectively fields in the set $\{\mathcal{O}_1,\mathcal{O}_2,\mathcal{O}_3\}$.
After rearranging the fields one gets
\begin{equation}
\langle\!\langle Y_{-2}|\mathcal{O}_1,
\mathcal{O}_2,\mathcal{O}_{3}\rangle\!\rangle
=e^{i\pi\sum h_j}
(-1)^{\frac{N_{odd}\left(N_{odd}-1\right)}{2}}
\langle\!\langle Y_{-2}|\mathcal{O}_3,
\mathcal{O}_2,\mathcal{O}_{1}\rangle\!\rangle.
\label{pretwist1}
\end{equation}

\noindent 3) Since the correlation function
is non zero only for odd expression, number
$N_{odd}$ is odd and
$N_{odd}=2m+1$ for some integer $m$.
Also we have an identity $N_{even}+N_{odd}=3$.
It's not difficult to check that
\begin{equation}
(-1)^{\frac{N_{odd}\left(N_{odd}-1\right)}{2}}=(-1)^m
=(-1)^{\frac{N_{odd}-1}{2}+
\frac{N_{odd}+N_{even}-3}{2}}
=(-1)^{N_{odd}+\frac{N_{even}}{2}}.
\label{pretwist2}
\end{equation}
Combining \eqref{pretwist1} and \eqref{pretwist2} we get the twist property
\begin{equation}
\begin{split}
\langle\!\langle Y_{-2}|\mathcal{O}_1,
\mathcal{O}_2,\mathcal{O}_3\rangle\!\rangle&
=\Omega_1\Omega_2\Omega_3\,
\langle\!\langle Y_{-2}|\mathcal{O}_3,
\mathcal{O}_2,\mathcal{O}_1\rangle\!\rangle,
\\
\mbox{where}\quad\Omega_j&=
\left\{
\begin{tabular}{LL}
(-1)^{h_j+1}, & h_j\in\mathbb Z\quad \mbox{ (i.e. GSO$+$)}\\
(-1)^{h_j+\frac{1}{2}}, & h_j\in \mathbb Z  +\frac12
\quad \mbox{ (i.e. GSO$-$)}.
\end{tabular}
\right.
\end{split}
\label{twist-gen}
\end{equation}

\medskip
\noindent\textsc{Examples.}

\noindent $i)$ Let $A_+=A_1+A_2$. Each term in
$$
A_+^3=A_1^3+(A_1A_2^2+A_2^2A_1)+A_2A_1A_2+(A_2^2A_1+A_1A_2^2)+A_1A_2A_1+A_2^3
$$
should be twist invariant to be nonzero. Therefore we get
$\Omega_1=1$ and $\Omega_2=1$.

\noindent $ii)$ Let we have fields $A_+$ and $A_-=a_1+a_2$.
Using cyclicity property (\ref{cyclicGSO-})
one gets
$$
2(A_+,a_1+a_2,a_1+a_2)=(A_+a_1^2+a_1^2A_+)+(A_+a_2^2+a_2^2A_+)+
(A_+a_1a_2+a_2a_1A_+)+(A_+a_2a_1+a_1a_2A_+).
$$
So we get $\Omega_+=1$ and $\Omega_+\Omega_1\Omega_2=1$. If $\Omega_1=1$ (tachyon's sector)
then $\Omega_2=1$ too. Therefore we
can consider a sector with $\Omega=1$ only.


\subsubsection{Power-counting lemma.}
\label{sec:pcl}
\newtheorem{Lemma}{Lemma }
\begin{Lemma}
Consider an off-shell N-point tree graph with $V=N-2$ three-string
vertices, and with $L_0$ internal lines corresponding to the
usual propagator $\Delta _{0}=
\frac {b_{0}}{L}W$ and $L_{1}$ internal lines
corresponding to $W$-propagator term $\frac {b_{0}}{L}W\frac {b_{0}}{L}$.
In such a graph
there is $L=L_0+2L_1$, $V-L_0$ and $L_1$ operators $b(z)$, $Z(z)$ and
$W(z)$
respectively. Let us denote the power of the leading singularity in
the product $(b(z))^L (Z(z))^{V-L_0} (W(z))^{L_1}$ by $s$, i.e.
$$\prod _{n=1}^{L}b(z_{0}+a_{i}\epsilon )\prod _{j=1}^{V-L_0}
Z(z_{0}+a'_{j}\epsilon )
\prod _{k=1}^{L_1}W(z_{0}+a''_{k}\epsilon)=\frac{1}{\epsilon ^{s}}
[~{\cal O}'(z_{0})+o(\epsilon)~],$$
where $a_{i},a'_{j},a''_{k}$ are some constants.

The  amplitude is non-singular if $s<2L$, it can contain a logarithmic
singularity if $s=2L$ and it is divergent in the case $s>2L$.
\end{Lemma}
\vspace{10pt}
{\bf Proof}.

Note that we consider only the case when there are no cancelled
propagators in the tree or,
in other words, all the $Q$-s from the
propagators $ \frac {b_{0}}{L}W\frac {b_{0}}{L}Q$ sit on external legs.

Going to the z-plane one can represent the amplitude in the form
\begin{equation}
{\cal A}=\prod_{n=1}^{L}(\int_0^{\infty}dT_n){\cal B}(T_1,\ldots,T_L),
\label{x.1}
\end{equation}
where
\begin{equation}
{\cal B}(T_1,\ldots,T_L)=\prod_{n=1}^{L}
(\oint \frac {dw_n}{2\pi i}\frac {dw_n}{d\rho})
\langle \prod _{r=1}^{N}{\cal O}_r(z_r)
\prod _{n=1}^{L}b(w_n)\prod _{i=1}^{V-L_0}Z(z_{0i})\prod _{i=1}^{L_1}W(z'_{0i})\rangle.
\label{x.2}
\end{equation}
In the space of modular variables $\{T_n\}$ we choose the spherical
coordinates:  $$T_n=\tau \alpha _n(\Omega),$$

where $\tau $ is the radius and $\Omega $ is the set of $n-1$ angular variables.
 Then
\begin{equation}
{\cal A}=\int_0^{\infty}\tau  ^{L-1}d\tau d\Omega {\cal B}(\tau ,\Omega ),
\label{x.3}
\end{equation}

where ${\cal B}(\tau ,\Omega)\equiv {\cal B}(\tau \alpha _1,\ldots ,\tau
\alpha _n)$.

To analyze the question of finiteness of the amplitude ${\cal A}$
let us estimate
the behavior of the integrand $ \tau^{L-1}{\cal B}(\tau  ,\Omega  )$
when $\tau $ tends to zero.

Let us denote by $z_0$ the common limit of the points $z_{0i}$ and $z'_{0i}$
when $\tau \to  0 $.
If we put $\epsilon  =z_{0i}-z_{0} $ for some $i$ then for every $j$ one has:
$z_{0,j}=z_{0}+a_{j}\epsilon $.
     In this notation it is possible to write the OPE for the product of
the operators $b$, $Z$, and $W$ in eq. (\ref{x.2}) as
\begin{equation}
\prod _{n=1}^{L}b(w_n)\prod _{i=1}^{V-L_0}Z(z_{0i})\prod _{i=1}^{L_1}W(z'_{0i})
=\sum _{\alpha}O'_{\alpha}(z_0)\epsilon ^{k_{\alpha}}\prod_{n=1}^{L}(w_n-z_0)
^{j_{\alpha,n}}.
\label{x.4}
\end{equation}

By using eq. (\ref{x.4}) one can write eq.(\ref{x.2}) in the form
\begin{equation}
{\cal B}(\epsilon)=\langle \prod _{r=1}^{N}{\cal O}_r(z_r)
\sum _{\alpha}O'_{\alpha}(z_0)\epsilon ^{k_{\alpha}}\prod _{n=1}^{L}
\oint \frac {dw_n}{2\pi i}\frac {w_n}{d\rho}(w_n-z_0)^{j_{\alpha,n}}\rangle.
\label{x.6}
\end{equation}

Because  the conformal map defined by equation (\ref{d.6}) we have
\begin{equation}
\oint \frac {dw_n}{2\pi i}\frac {dw_n}{d\rho}(w_n-z_0)^{j_{\alpha,n}}
~\sim~\epsilon^{-\frac{V}{2}+j_{\alpha ,n}+1}.
\label{x.8}
\end{equation}
In summary  we conclude  that when $\epsilon \rightarrow 0$, $\cal {B}(\epsilon)$
has a  behavior:

\begin{eqnarray}\label{x.9}
{\cal B}(\epsilon)&~\sim~&\langle \prod _{r=1}^{N}{\cal O}_r(z_r)
\sum _{\alpha}{\cal O}'_{\alpha}(z_0)\epsilon ^{k_{\alpha}}\prod _{n=1}^{L}
\epsilon^{-\frac{V}{2}+j_{\alpha ,n}+1}\rangle \nonumber\\
\nonumber\\
&~\sim~& \epsilon ^{-(\frac{V}{2}-1)L}
\sum _{\alpha}{\cal B}_{\alpha}
\epsilon ^{k_{\alpha}}\prod  _{n=1}^{L}\epsilon  ^{j_{\alpha ,n}}.
\\ \nonumber\end{eqnarray}

If $s$ is the maximal power of singularity for the OPE (\ref{x.4})
near the point
$z_0$ (one must also put $w_{n}-z_{0}~\sim~\epsilon $ in eq. (\ref{x.4}) then
$$\sum _{\alpha}{\cal B}_{\alpha}
\epsilon ^{k_{\alpha}}\prod  _{n=1}^{L}\epsilon  ^{j_{\alpha ,n}}
~\sim~\epsilon^{-s},$$
and eq.(\ref{x.9}) takes the form
\begin{equation}
{\cal  B}(\epsilon)~\sim~\epsilon ^{-(\frac{V}{2}-1)L-s}.
\label{x.10}
\end{equation}

Next we change the variables in (\ref{x.10}) from $\epsilon$ to $\tau $.
As it is obvious from
(\ref{d.6}) when $\tau \rightarrow 0$ :
\begin{equation}
\tau~\sim~\epsilon ^{\frac{V}{2}+1}.
\label{x.11}
\end{equation}

In terms of $\tau $ eq. (\ref{x.10}) looks like
\begin{equation}
{\cal B}~\sim ~\tau ^{-L+\frac{2}{V+2}(2L-s)}.
\label{x.12}
\end{equation}

That is why the entire integrand in (\ref{x.3}) behaves as
\begin{equation}
\tau^{L-1}\tau ^{-L+\frac{2}{V+2}(2L-s)}=
\tau ^{-1+\frac{2}{V+2}(2L-s)}.
\label{x.13}
\end{equation}

when $\tau \rightarrow 0$. From (\ref{x.13}) we can derive
the statement of our Lemma.

\newpage
\subsubsection{Table of notations, correlation functions and OPE.}
\renewcommand{\arraystretch}{1.5}
\begin{longtable}[h]{||LL||}
\hline
X_L^{\mu}(z)X_L^{\nu}(w)\sim-\frac{\alpha^{\prime}}{2}\eta^{\mu\nu}
\log(z-w)
& \pd X^{\mu}(z)\pd X^{\nu}(w)\sim-\frac{\alpha^{\,\prime}}{2}\eta^{\mu\nu}
\frac{1}{(z-w)^2}
\\
\psi^{\mu}(z)\psi^{\nu}(w)\sim-\frac{\alpha^{\prime}}{2}\eta^{\mu\nu}
\frac{1}{z-w}
&
\\
\hline
\hline
c(z)b(w)\sim b(z)c(w)\sim\frac{1}{z-w}
&
\gamma(z)\beta(w)\sim -\beta(z)\gamma(w)\sim\frac{1}{z-w}
\\
\phi(z)\phi(w)\sim-\log(z-w)
& \xi(z)\eta(w)\sim \eta(z)\xi(w)\sim\frac{1}{z-w}
\\
\langle c(z_1)c(z_2)c(z_3)\rangle=-(z_1-z_2)(z_2-z_3)(z_3-z_1)~~~~
&\langle e^{-2\phi(y)}\rangle=1
\\
\multicolumn{2}{||L||}{
\left\langle e^{i2p_{\alpha}X_L^{\alpha}(y)-i2b_{i}X_L^{i}(y)}
e^{i2k_{\beta}X_L^{\beta}(z)+i2b_{j}X_L^{j}(z)}\right\rangle=
(2\pi)^{p+1}\alpha^{\,\prime-\frac{p+1}{2}}\delta(p+k)(y-z)^{-2\alpha^{\,\prime}(k^2+b^2)}
}\\
\hline
\hline
T_B=-\frac{1}{\alpha^{\,\prime}}\pd X\cdot\pd X-\frac{1}{\alpha^{\,\prime}}\pd \psi\cdot\psi
& T_F=-\frac{1}{\alpha^{\,\prime}}\pd X\cdot\psi
\\
T_{bc}=-2b\pd c-\pd bc
& T_{\beta\gamma}=-\frac32\beta\pd\gamma-\frac12\pd\beta\gamma
\\
T_{\phi}=-\frac12\pd\phi\pd\phi-\pd^2\phi
& T_{\eta\xi}=\pd\xi\eta
\\
\hline
\hline
\multicolumn{2}{||L||}{
T_B(z)T_B(w)\sim \frac{15}{2(z-w)^4}+\frac{2}{(z-w)^2}T_B(w)+\frac{1}{z-w}\pd T_B(w)}
\\
\multicolumn{2}{||L||}{
T_B(z)T_F(w)\sim\frac{3/2}{(z-w)^2}T_F(w)+\frac{1}{z-w}\pd T_F(w)}
\\
\multicolumn{2}{||L||}{
T_F(z)T_F(w)\sim\frac{5/2}{(z-w)^3}+\frac{1/2}{z-w}T_B(w)}
\\
\hline
\hline
\gamma=\eta e^{\phi}
& \beta=e^{-\phi}\pd\xi
\\
T_{\beta\gamma}=T_{\phi}+T_{\eta\xi}
& \gamma^2=\eta\pd\eta e^{2\phi}
\\
\hline
\hline
\phi-\text{charge:}\quad
\left[\frac{1}{2\pi i}\oint d\zeta\partial\phi(\zeta),
\,\mathcal{A}_q\right]=q\mathcal{A}_q
&
\\
\hline
\hline
\multicolumn{2}{||L||}{Q_B=\frac{1}{2\pi i}\oint d\zeta\,\left[
c(T_B+T_{\beta\gamma}+\frac12 T_{bc})+\frac{1}{\alpha^{\,\prime}}\gamma\psi\cdot\pd X
-\frac14 b\gamma^2\right]}
\\
\multicolumn{2}{||L||}{Q_B=\frac{1}{2\pi i}\oint d\zeta\,\left[
c(T_B+T_{\phi}+T_{\eta\xi}+\frac12 T_{bc})+\frac{1}{\alpha^{\,\prime}}\eta e^{\phi}\psi\cdot\pd X
+\frac14 b\pd\eta\eta e^{2\phi}\right]}
\\
\hline
\hline
\multicolumn{2}{||L||}{
\mathrm{X}=\frac1{\alpha^{\,\prime}} e^{\phi}\psi\cdot\pd X+c\pd\xi
+\frac14 b\pd\eta e^{2\phi}+\frac14\pd(b\eta e^{2\phi})}
\\
Y=4c\pd\xi e^{-2\phi} &\\
\hline
\end{longtable}
\renewcommand{\arraystretch}{1}

\newpage
\section{Cubic (Super)String Field Theory on Branes
and Sen Conjectures.}
\label{sec:SSFT}
\setcounter{equation}{0}

\subsection{Sen's Conjecture.}

As we have seen in Section~5 there  is a tachyon mode in the spectrum of
bosonic open string. There is also a  tachyon mode  in GSO$-$ sector
of fermionic
string.
Normally we impose GSO$+$ projection on superstring spectrum
and obtain tachyon free spectrum. But there are objects
in the string theory such as non-BPS branes or brane-anti-brane pairs,
which necessarily contain both GSO$+$ and GSO$-$ sector \cite{9904207}.
Therefore, it becomes important to explore the tachyon phenomenon.
The existence of the tachyon mode does not necessarily signify a
sickness of the theory,
but may simply indicate an existence of a ground state with the energy
density lower than in the starting configuration.
One can compare this situation
with Higgs phenomenon. The Higgs field has a potential looking like
a Mexican hat. On the top of the potential the Higgs
field has negative mass square,
while at the bottom it has positive mass square.

To find a new vacuum in the theory of strings one has to be able
to calculate the effective tachyon potential, that is a functional on
a constant field configuration.
Since a non zero constant  tachyon field is far
away from its on-shell value one has to use an off-shell formulation of
string theory to do this. As we know from sections 4-6
string filed theories give such off-shell formulations of the string
theory. There are also so-called background independent string filed theories
(BSFT) which are also used to calculate  the effective tachyon
potential \cite{0009103,additive3,0009148}.

First attempts to find a stable vacuum in the bosonic string using SFT was
made by Kostelecky and Samuel \cite{KS}. They have computed tachyon potential
using level truncation scheme and shown that there is another vacuum,
in which there is no tachyonic state.

About two years ago A.~Sen \cite{9805170} suggested to interpret tachyon condensation as a
decay of unstable D-branes.
Let us consider one non-BPS D$9$-brane. As it is mentioned above
the string attached to this brane has a tachyon in the spectrum.
For this case, the Sen's conjecture says \cite{9904207}:

\begin{enumerate}
\item The tachyon potential $\mathcal{V}_{c}(t)$
looks like Mexican hat with minimum at the point $t_c$.
The difference $\mathcal{V}(0)-\mathcal{V}(t_c)$
is precisely the tension $\tilde{\tau}_9$ of the non-BPS D$9$-brane.

\item The  theory around the new vacuum does not contain any
open string excitation.
Therefore, this new vacuum can be also considered as a vacuum of a closed string.

\item All D-branes in type IIA (IIB) string theory can be regarded
as classical solutions in the open string field theory living on a
system of non-BPS D$9$-branes (or D$9$-anti-D$9$-brane pair).
\end{enumerate}

In this section we will show how one can use SFT
to check these conjectures.
We will mainly concentrate on the first conjecture
and will say only few words about others.
 Before proceed let us summarize what was obtained in the literature.

The first Sen's conjectures was numerically verified almost in all SFTs
listed in the Introduction.
In Table below we collect results of the computations
of the ratio $(\mathcal{V}(0)-\mathcal{V}(t_c))/\tau$performed in different
SFTs using the level truncation scheme.
\renewcommand{\arraystretch}{1.4}
\begin{longtable}[!h]{||l|C|C|C||}
\hline
SFT Type& \text{level} & \text{number of the fields} &
\frac{\mathcal{V}(0)-\mathcal{V}(t_c)}{\tau}
\\
\hline
\hline
Cubic SFT & 10 & 102 & 0.9991
\\
BI SFT & \text{---} & 1 & 1\text{ (exact)}
\\
\hline
\hline
Nonpolynomial SSFT & 4 & 70 & 0.944
\\
Modified Cubic SSFT & 2 & 10 & 1.058
\\
BI SSFT & \text{---} & 1 & 1\text{ (exact)}
\\
\hline
\end{longtable}
\renewcommand{\arraystretch}{1}
These results were presented in \cite{0002237}, \cite{0009103}, \cite{0109182}, \cite{ABKM}
and \cite{0009148} correspondingly. See also \cite{0002211}, \cite{0003220},
\cite{0109182} about calculations of the tachyon potential in
nonpolynomial SSFT \cite{9503099}, \cite{0001084} and \cite{0004112}
about calculations in the Witten cubic SSFT.
One sees that Background Independent SFTs give exact
results for the minimum of the tachyon potential.
The formulation of Background Independent SFT can
be found in \cite{BCFT,0009103}, and for further developments related to
tachyon see
\cite{0009191,0010108,0012198,0101200,0103089,0104099,0105076,
0105098,0105115,0105238,0106107,0107098}.

The second Sen's conjecture was also verified almost in all SFTs
\cite{ZwiebachToy,Dasgupta,0106068,0005036,
lump-mogila,0008101,KochRod,KochJev,HarveyKraus,
jatkVat}
and \cite{kink-mogila}. For superstrings
the solution representing, for example, a decay non-BPS D$9$ $\rightarrow$
BPS D$8$ ($\mathrm{\bar{D}}8$) is a
kink (anti-kink) solution. There are also solutions which represent
a decay D$p$ $\rightarrow$ D$q$. Such solutions can be obtained
by combining several kink solutions representing the decay D$p$ $\rightarrow$ D$(p-1)$.
This chain relations between the solutions are called the brane descend
relations \cite{9904207,9905006}.
In the case of the bosonic strings the solution representing
a decay D$p$ $\rightarrow$ D$q$ is called lump solution of
codimension $(p-q)$. Below we summarize some numeric
results concerning lump solutions.

The third conjecture  implies that there are different solutions of the
equations of motion of the SFT which can be identified with
different objects appearing in the string theory.
For example, the vortex and kink solutions in the Super SFT represent lower
dimensional D-branes (of various types) in IIA(B) theory.
Another example is
lump solutions in the Bosonic SFT, which also represent lower dimensional
D-branes.

\subsection{CFT on Branes.}

D$p$-brane is incorporated into the string field theory by considering
boundary CFT. This boundary CFT is constructed by $p+1$ bosons
$X^{\alpha}(z,\oz)$ ($\alpha=0,\dots,p$) with Neumann boundary
conditions and $25-p$ bosons
$X^{\,\prime\,i}(z,\oz)$ ($i=p+1,\dots,25$) with Dirichlet
boundary conditions. But to be able to use
vertex operators we need to consider T-dual CFT. This CFT consist of
$26$ bosons $X^{\mu}(z,\oz)$ with Neumann boundary conditions, but part of them
is related to $X^{\,\prime\,i}(z,\oz)$ by T-duality transformation:
\begin{equation}
X^i(z,\oz)=X_L^i(z)+X_R^i(\oz)
\quad\stackrel{T}{\Longrightarrow}\quad
X^{\,\prime\,i}(z,\oz)=X_L^i(z)-X_R^i(\oz).
\end{equation}

\subsection{String Field Theory on Pair of Branes.}

To incorporate more than one brane into string field theory one
can use Chan-Paton factors. We will consider the string field
theory for a pair of D$p$-branes. To this end we need to introduce
$2\times 2$ Chan-Paton factors:
\begin{align*}
  \begin{pmatrix}
  1 & 0\\
  0 & 0
  \end{pmatrix} & \qquad\text{denotes a string that begins
     and ends on the \textit{first} brane;}
\\
  \begin{pmatrix}
  0 & 1\\
  0 & 0
  \end{pmatrix} & \qquad\text{denotes a string that begins on the
    \textit{first} and ends on the \textit{second} brane;}
\\
  \begin{pmatrix}
  0 & 0\\
  0 & 1
  \end{pmatrix} & \qquad\text{denotes a string that
      begins and ends on the \textit{second} brane;}
\\
  \begin{pmatrix}
  0 & 0\\
  1 & 0
  \end{pmatrix} & \qquad\text{denotes a string that begins
        on the \textit{second} and ends on the \textit{first} brane.}
\end{align*}
For these matrices the following compositions are true:
\begin{subequations}
\begin{align}
\begin{pmatrix}
0 & 1\\
0 & 0
\end{pmatrix}
\begin{pmatrix}
0 & 0\\
1 & 0
\end{pmatrix}
&=
\begin{pmatrix}
1 & 0\\
0 & 0
\end{pmatrix},
\label{CPmul:a}
\\
\begin{pmatrix}
0 & 0\\
1 & 0
\end{pmatrix}
\begin{pmatrix}
0 & 1\\
0 & 0
\end{pmatrix}
&=
\begin{pmatrix}
0 & 0\\
0 & 1
\end{pmatrix}.
\label{CPmul:b}
\end{align}
\label{CPmul}
\end{subequations}

The string field is modified in the following  way:
\begin{equation}
\hat{\Phi}=\Phi^{(1)}\otimes\begin{pmatrix}1 & 0\\0 & 0\end{pmatrix}
+\phi\otimes\begin{pmatrix}0 & 0\\1 & 0\end{pmatrix}
+\phi^*\otimes\begin{pmatrix}0 & 1\\0 & 0\end{pmatrix}
+\Phi^{(2)}\otimes\begin{pmatrix}0 & 0\\0 & 1\end{pmatrix}.
\label{PhiHat}
\end{equation}
The action is of the form
\begin{multline}
S[\hat{\Phi}]=-\frac{1}{g_o^2}\Tr\left[
\frac{1}{2}\la\hat{\Phi},\,\hat{Q}_B\hat{\Phi}\ra
+\frac{1}{3}\la\hat{\Phi},\,\hat{\Phi},\,\hat{\Phi}\ra
\right]=
\\
-\frac{1}{g_o^2}\left[
\frac{1}{2}\la\Phi^{(1)},Q_B\Phi^{(1)}\ra
+\frac{1}{2}\la\Phi^{(1)},Q_B\Phi^{(3)}\ra
+\frac{1}{2}\la\phi^*,Q_B\phi\ra
+\frac{1}{2}\la\phi,Q_B\phi^*\ra
\right.
\\
+\frac{1}{3}\la\Phi^{(1)},\Phi^{(1)},\Phi^{(1)}\ra
+\frac{1}{3}\la\Phi^{(2)},\Phi^{(2)},\Phi^{(2)}\ra
\\
\left.
+\frac{1}{3}(\la\Phi^{(1)},\phi^{*},\phi\ra+c.p.)
+\frac{1}{3}(\la\Phi^{(2)},\phi,\phi^{*}\ra+c.p.)
\right].
\label{action2x2b}
\end{multline}
The dimensionless coupling constant $g_o^2$ is the
same as in the action \eqref{w-action}.
The expression \eqref{action2x2b} can be simplified if we
employ cyclic symmetry of the odd bracket:
\begin{multline}
S[\Phi^{(1)},\Phi^{(2)},\phi,\phi^{*}]=
-\frac{1}{g_o^2}\left[
\frac{1}{2}\la\Phi^{(1)},Q_B\Phi^{(1)}\ra
+\frac{1}{2}\la\Phi^{(2)},Q_B\Phi^{(2)}\ra
+\la\phi^*,Q_B\phi\ra
\right.
\\
\left.
+\frac{1}{3}\la\Phi^{(1)},\Phi^{(1)},\Phi^{(1)}\ra
+\frac{1}{3}\la\Phi^{(2)},\Phi^{(2)},\Phi^{(2)}\ra
+\la\Phi^{(1)},\phi^{*},\phi\ra
+\la\Phi^{(2)},\phi,\phi^{*}\ra
\right].
\tag{\ref{action2x2b}${}^{\prime}$}
\label{action2x2B}
\end{multline}

Let us consider the following string fields
\begin{subequations}
\begin{align}
\Phi^{(1)}(w)&=\int\frac{d^{p+1}k}{(2\pi)^{p+1}}\;i\delta\chi_j(k_{\alpha})
\mathrm{V}_V^{j}(w,k_{\alpha},0),
\label{Vchi}
\\
\phi(w)&=\int\frac{d^{p+1}k}{(2\pi)^{p+1}}\;t(k_{\alpha})
\mathrm{V}_t(w,k_{\alpha},\tfrac{b^j}{2\pi\alpha^{\,\prime}})
+iu_i(k_{\alpha})\mathrm{V}_V^{i}(w,k_{\alpha},\tfrac{b^j}{2\pi\alpha^{\,\prime}}),
\\
\phi^{*}(w)&=\int\frac{d^{p+1}k}{(2\pi)^{p+1}}\;t^*(k_{\alpha})
\mathrm{V}_t(w,k_{\alpha},-\tfrac{b^j}{2\pi\alpha^{\,\prime}})
+iu_i^*(k_{\alpha})\mathrm{V}_V^{i}(w,k_{\alpha},-\tfrac{b^j}{2\pi\alpha^{\,\prime}},)
\end{align}
\end{subequations}
where $\mathrm{V}_V^{i}$ is a \textit{primary} operator
representing the vector field \eqref{primary-vector}
and $\mathrm{V}_t$ is the tachyon vertex operator.
And we introduce
additional restrictions $b^iu_i(k_{\alpha})=0$ and $b^iu^*_i(k_{\alpha})=0$.
We are interested in the following action
\begin{equation}
\tilde{S}[\delta\chi_i,t,u_i]=-\frac{1}{g_o^2}\left[
\frac{1}{2}\la\Phi^{(1)},Q_B\Phi^{(1)}\ra
+\la\phi^*,Q_B\phi\ra
+\la\Phi^{(1)},\phi^{*},\phi\ra
\right].
\label{2D}
\end{equation}
The quadratic action can be easily read from \eqref{S2boson} and it
is of the form
\begin{multline}
\tilde{S}_2[\delta\chi_i,t,u_i]=\frac{1}{g_o^2\alpha^{\,\prime\,\frac{p+1}{2}}}
\int\frac{d^{p+1}k}{(2\pi)^{p+1}}\;\left[
\frac{\alpha^{\,\prime}k_{\alpha}^2}{2}\delta\chi_i(-k)\delta\chi^i(k)
\right.
\\
\left.
-\left[\alpha^{\,\prime}k_{\alpha}^2
+\alpha^{\,\prime}m^2_b-1\right]
t^*(-k)t(k)
+\left[\alpha^{\,\prime}k_{\alpha}^2
+\alpha^{\,\prime}m^2_b\right]
u^*_i(-k)u^i(k)
\right],
\label{tildeS2}
\end{multline}
where $m_b^2=\left(\frac{b_i}{2\pi\alpha^{\,\prime}}\right)^2$.

To compute the interaction terms we need to know the following
correlation functions:
\begin{multline}
\la i\delta\chi_j(k_1)\mathrm{V}_V^{j}(k_1,0),\,
t^*(k_2)\mathrm{V}_t(k_2,-\tfrac{b_i}{2\pi\alpha^{\,\prime}}),\,
t(k_3)\mathrm{V}_t(k_3,\tfrac{b_i}{2\pi\alpha^{\,\prime}})\ra
\\
=i\delta\chi_j(k_1)t^*(k_2)t(k_3)
i\left[\frac{2}{\alpha^{\,\prime}}\right]^{\frac12}
f_1^{\,\prime\,\alpha^{\,\prime}k_1^2}
f_2^{\,\prime\,\alpha^{\,\prime}k_2^2-1}f_3^{\,\prime\,\alpha^{\,\prime}k_3^2-1}
\\
\times
\left\langle c\pd X^{j}(f_1)
c(f_2)c(f_3)e^{2ik_1\cdot X_L(f_1)}e^{2ik_2\cdot X_L(f_2)}e^{2ik_3\cdot X_L(f_3)}
\right\rangle
\\
=\delta\chi^j(k_1)t^*(k_2)t(k_3)
\left[\frac{2}{\alpha^{\,\prime}}\right]^{\frac12}
f_1^{\,\prime\,\alpha^{\,\prime}k_1^2}
f_2^{\,\prime\,\alpha^{\,\prime}k_2^2-1}f_3^{\,\prime\,\alpha^{\,\prime}k_3^2-1}
(2\pi)^{p+1}\alpha^{\,\prime\,-\frac{p+1}{2}}\delta(k_1+k_2+k_3)
\\
\times
(f_1-f_2)^{2\alpha^{\,\prime}k_1k_2}
(f_1-f_3)^{2\alpha^{\,\prime}k_1k_3}
(f_2-f_3)^{2\alpha^{\,\prime}k_2k_3}
(f_1-f_2)(f_2-f_3)(f_3-f_1)
\left[\frac{-i\alpha^{\,\prime}\tfrac{-b_j}{2\pi\alpha^{\,\prime}}}{f_1-f_2}
+\frac{-i\alpha^{\,\prime}\tfrac{b_j}{2\pi\alpha^{\,\prime}}}{f_1-f_3}\right]
\\
=-\frac{ib_j\delta\chi^j(k_1)}{2\pi}
t^*(k_2)t(k_3)\left[\frac{2}{\alpha^{\,\prime}}\right]^{\frac12}
(2\pi)^{p+1}\alpha^{\,\prime\,-\frac{p+1}{2}}\delta(k_1+k_2+k_3)
\\
\times
f_1^{\,\prime\,\alpha^{\,\prime}k_1^2}
f_2^{\,\prime\,\alpha^{\,\prime}k_2^2-1}
f_3^{\,\prime\,\alpha^{\,\prime}k_3^2-1}
(f_1-f_2)^{2\alpha^{\,\prime}k_1k_2}
(f_1-f_3)^{2\alpha^{\,\prime}k_1k_3}
(f_2-f_3)^{2\alpha^{\,\prime}k_2k_3+2}
.
\label{PhiTTf}
\end{multline}
Substitution of the maps \eqref{maps:3} into \eqref{PhiTTf}
leads to the following expression
\begin{equation}
-\left[\frac{2}{\alpha^{\,\prime}}\right]^{\frac12}
\frac{ib_j\delta\chi^j(k_1)}{2\pi}
t^*(k_2)t(k_3)
(2\pi)^{p+1}\alpha^{\,\prime\,-\frac{p+1}{2}}\delta(k_1+k_2+k_3)
\gamma^{\alpha^{\,\prime}k_1^2+\alpha^{\,\prime}k_2^2+\alpha^{\,\prime}k_3^2-2}.
\label{PhiTT}
\end{equation}
 Now we consider the correlation functions
\begin{multline}
\la i\delta\chi_i(k_1)\mathrm{V}_V^i(k_1,0),\,
iu^*_j(k_2)\mathrm{V}_V^j(k_2,-\tfrac{b_i}{2\pi\alpha^{\,\prime}}),\,
iu_k(k_3)\mathrm{V}_V^k(k_3,\tfrac{b_i}{2\pi\alpha^{\,\prime}})
\ra
\\
=-\delta\chi_i(k_1)u^*_j(k_2)u_k(k_3)
f_1^{\,\prime\,\alpha^{\,\prime}k_1^2}
f_2^{\,\prime\,\alpha^{\,\prime}k_2^2}
f_3^{\,\prime\,\alpha^{\,\prime}k_3^2}
\left[\frac{2}{\alpha^{\,\prime}}\right]^{\frac32}
\\
\times\left\langle
c\pd X^i e^{2ik_1\cdot X_L(f_1)}
c\pd X^j e^{2ik_2\cdot X_L(f_2)}
c\pd X^k e^{2ik_3\cdot X_L(f_3)}
\right\rangle.
\end{multline}
Using the fact that $u_i(k_{\alpha})b^i=u^*_i(k_{\alpha})b^i=0$
we get the following
expression
\begin{multline}
=\frac{2}{\alpha^{\,\prime}}\left[\frac{2}{\alpha^{\,\prime}}\right]^{\frac12}
\delta\chi_i(k_1)u_j^*(k_2)u_k(k_3)
f_1^{\,\prime\,\alpha^{\,\prime}k_1^2}
f_2^{\,\prime\,\alpha^{\,\prime}k_2^2}
f_3^{\,\prime\,\alpha^{\,\prime}k_3^2}
(f_1-f_2)(f_2-f_3)(f_3-f_1)
\\
\times
\left[-\frac{2}{\alpha^{\,\prime}}\right]
\frac{\eta^{jk}}{(f_2-f_3)^2}(-i\alpha^{\,\prime})
\left[
\frac{-\tfrac{b_i}{2\pi\alpha^{\,\prime}}}{f_1-f_2}
+\frac{\tfrac{b_i}{2\pi\alpha^{\,\prime}}}{f_1-f_3}
\right]
(f_1-f_2)^{2\alpha^{\,\prime}k_1k_2}
(f_1-f_3)^{2\alpha^{\,\prime}k_1k_3}
(f_2-f_3)^{2\alpha^{\,\prime}k_2k_3}
\\
=-\left[\frac{2}{\alpha^{\,\prime}}\right]^{\frac12}
\frac{ib^i\delta\chi_i(k_1)}{2\pi}u_j^*(k_2)u^j(k_3)
f_1^{\,\prime\,\alpha^{\,\prime}k_1^2}
f_2^{\,\prime\,\alpha^{\,\prime}k_2^2}
f_3^{\,\prime\,\alpha^{\,\prime}k_3^2}
(f_1-f_2)^{2\alpha^{\,\prime}k_1k_2}
(f_1-f_3)^{2\alpha^{\,\prime}k_1k_3}
(f_2-f_3)^{2\alpha^{\,\prime}k_2k_3}.
\label{VVVf}
\end{multline}
Substitution of the maps \eqref{maps:3} into \eqref{VVVf}
leads to the following expression
\begin{equation}
\left[\frac{2}{\alpha^{\,\prime}}\right]^{\frac12}
\frac{ib^i\delta\chi_i(k_1)}{2\pi}u_j^*(k_2)u^j(k_3)
(2\pi)^{p+1}\alpha^{\,\prime\,-\frac{p+1}{2}}\delta(k_1+k_2+k_3)
\gamma^{\alpha^{\,\prime}k_1^2+\alpha^{\,\prime}k_2^2+\alpha^{\,\prime}k_3^2}.
\label{VVV}
\end{equation}
Combining \eqref{PhiTT} and \eqref{VVV} we get
the interaction term of the action \eqref{2D}
\begin{multline}
\tilde{S}_3[\delta\chi_i,t,u_i]=
\frac{1}{g_o^2\alpha^{\,\prime\,\frac{p+1}{2}}}
\left[\frac{2}{\alpha^{\,\prime}}\right]^{\frac12}
\int\frac{d^{p+1}k_1d^{p+1}k_2d^{p+1}k_3}{
(2\pi)^{p+1}(2\pi)^{p+1}}\;\delta(k_1+k_2+k_3)
\frac{ib_i\delta\chi^i(k_1)}{2\pi}
\\
\times
\gamma^{\alpha^{\,\prime}k_1^2
+\alpha^{\,\prime}k_2^2+\alpha^{\,\prime}k_3^2}
\left[
\gamma^{-2}t^*(k_2)t(k_3)
-u_j^*(k_2)u^j(k_3)
\right],
\label{tildeS3}
\end{multline}
where $\gamma=\frac{4}{3\sqrt{3}}$.
Let us choose $\delta\chi^i(x)$ to be a constant
$$
\delta\chi^i(k)=(2\pi)^{p+1}\delta\chi^i\delta(k)
$$
and let $t,\,t^*$ and $u_i^*,\,u_i$ be on mass shell. This
simplifies the action \eqref{tildeS3}. The whole
action $\tilde{S}$ \eqref{2D} is of the form
\begin{multline}
\tilde{S}[t,u_i]=\frac{1}{g_o^2\alpha^{\,\prime\,\frac{p+1}{2}}}
\int\frac{d^{p+1}k}{(2\pi)^{p+1}}\;
\left[-\alpha^{\,\prime}k_{\alpha}^2
-\alpha^{\,\prime}\left[\frac{b_i}{2\pi\alpha^{\,\prime}}\right]^2+1+
\left[\frac{2}{\alpha^{\,\prime}}\right]^{\frac12}
\frac{ib_i\delta\chi^i}{2\pi}
\right]
t^*(-k)t(k)
\\
-\left[\alpha^{\,\prime}k_{\alpha}^2
+\alpha^{\,\prime}\left[\frac{b_i}{2\pi\alpha^{\,\prime}}\right]^2
-\left[\frac{2}{\alpha^{\,\prime}}\right]^{\frac12}
\frac{ib_i\delta\chi^i}{2\pi}\right]
u^*_i(-k)u^i(k).
\end{multline}
One sees that the term proportional to $\delta\chi_i$ has
natural interpretation as a shift of mass of the fields $t$ and $u_i$.
But we want to interpret the term with $\delta\chi_i$ as a shift
of $b_i$. Therefore, we have to compare two shifts of the mass: one produced
by $\delta\chi_i$ and another produced by a shift of $b_i$:
\begin{equation}
\alpha^{\,\prime}\frac{2b_i\delta b^i}{4\pi^2\alpha^{\,\prime\,2}}
=-\left[\frac{2}{\alpha^{\,\prime}}\right]^{\frac12}
\frac{ib_i\delta\chi^i}{2\pi}.
\end{equation}
So we get
\begin{equation}
\delta\chi^i=i\left[\frac{\alpha^{\,\prime}}{2}\right]^{\frac12}
\frac{\delta b^i}{\pi\alpha^{\,\prime}}.
\label{chi-b}
\end{equation}
This expression can be obtained in a more simple way.
Let us consider the following operator product expansion:
\begin{multline}
e^{2i\frac{\delta b_i}{2\pi}X^i_L(z)}
e^{2i\frac{b_j}{2\pi}X^j_L(w)}=
(z-w)^{2\alpha^{\,\prime}\frac{\delta b_i b^i}{(2\pi)^2}}\,
e^{2i\frac{\delta b_i}{2\pi}X^i_L(z)+2i\frac{b_j}{2\pi}X^j_L(w)}
\\
=(z-w)^{2\alpha^{\,\prime}\frac{\delta b_i b^i}{(2\pi)^2}}\,
e^{2i\frac{\delta b_i+b_i}{2\pi}X_L^i(w)+2i\frac{\delta b_i}{2\pi}\pd X^i(w)(z-w)+O(z-w)^2}
\\
=(z-w)^{2\alpha^{\,\prime}\frac{\delta b_i b^i}{(2\pi)^2}}\,
e^{2i\frac{\delta b_j+b_j}{2\pi}X_L^i(w)}\left[
1+(z-w)\frac{i\delta b_i}{\pi}\pd X^i(w)+O(z-w)^2
\right].
\end{multline}

The
state $|0,b\rangle$ is generated by the vertex operator
$V_t(0,\tfrac{b_i}{2\pi\alpha^{\,\prime}})$, therefore, the change
of this operator under a small shift of $b_i$ must be
equal to $\delta\chi_iV_V^i(0,\tfrac{b_i}{2\pi\alpha^{\,\prime}})$.
So we get
\begin{multline}
\delta\left[ c(w)e^{\tfrac{ib_j}{\pi\alpha^{\,\prime}}X_L^j(w)}\right]
=
c(w)\frac{i\delta b_j}{\pi\alpha^{\,\prime}}\pd X^j(w)
\,e^{\tfrac{ib_j}{\pi\alpha^{\,\prime}}X_L^j(w)}
\\
=-i\left[\frac{\alpha^{\,\prime}}{2}\right]^{\frac12}
\frac{i\delta b_j}{\pi\alpha^{\,\prime}}\;
i\left[\frac{2}{\alpha^{\,\prime}}\right]^{\frac12}
c(w)\pd X^j(w)
\,e^{\tfrac{ib_j}{\pi\alpha^{\,\prime}}X_L^j(w)}
=\left[\frac{\alpha^{\,\prime}}{2}\right]^{\frac12}
\frac{\delta b_j}{\pi\alpha^{\,\prime}}
\;V_v(0,\tfrac{b_i}{2\pi\alpha^{\,\prime}}),
\label{simpleB}
\end{multline}
which is equal to the previously obtained expression up to the vertex operator
of the auxiliary field. So we can now identify $\delta\chi^i$ by using
\begin{equation}
i\delta\chi^i=\left[\frac{\alpha^{\,\prime}}{2}\right]^{\frac12}
\frac{\delta b_j}{\pi\alpha^{\,\prime}}.
\label{7.19'}
\end{equation}
One sees that up to the sign (that is easily explained) we get the
same result as in \eqref{chi-b}.

Now we substitute \eqref{chi-b} or \eqref{7.19'} into \eqref{tildeS2} and
get for the field $\delta b^i$ the following action
\begin{equation}
\tilde{S}_2[\delta b_i]=\frac{1}{2\pi^2g_o^2\alpha^{\,\prime\,\frac{p+1}{2}}}
\int\frac{d^{p+1}k}{(2\pi)^{p+1}}\;
\frac{k_{\alpha}^2}{2}\delta b_i(-k)\delta b^i(k)
\end{equation}
As a consequence we get the brane's tension:
\begin{equation}
\tau_p=\frac{1}{2\pi^2g_o^2\alpha^{\,\prime\,\frac{p+1}{2}}}.
\end{equation}
\subsection{Superstring Field Theory on non-BPS $\mathrm{D}$-brane.}

To describe the open string states living on a
single non-BPS $\mathrm{D}$-brane one has to add
GSO$-$ states \cite{9904207}.
GSO$-$ states are Grassmann even, while
GSO$+$ states are Grassmann odd (see Table \ref{tab:1}).
\begin{table}[!h]
\begin{center}
\renewcommand{\arraystretch}{1.4}
\begin{tabular}[h]{||C|c|C|C|C|c||}
\hline \textrm{Name}& Parity & \textrm{GSO} & \stackrel{\text{\small Superghost}}{\text{\small number}}
 & \textrm{Weight}\, (h)& Comments\\ \hline
\hline \mathcal{A}_+ & odd & + & 1 & h\in\Zh,\,h\geqslant -1
& string \\
\cline{1-5} \mathcal{A}_- & even & - & 1& h\in\Zh+\frac12,\,h\geqslant -\frac12 & fields\\
\hline \Lambda_+ & even & + & 0& h\in\Zh,\,h\geqslant 0 & gauge\\
\cline{1-5} \Lambda_- & odd & - & 0& h\in\Zh+\frac12,\,h\geqslant \frac{1}{2} & parameters\\ \hline
\end{tabular}
\end{center}
\vspace{-0.5cm}\caption{Parity of string fields and gauge
parameters in the 0 picture.}\label{tab:1}
\end{table}

The unique (up to rescaling of the fields)
gauge invariant action unifying GSO$+$
and GSO$-$ sectors is found to be
\begin{equation}
\begin{split}
S[\mathcal{A}_+,\mathcal{A}_-]&=\frac{1}{g^2_o}\left[
\frac{1}{2}\langle\!\langle Y_{-2}|\mathcal{A}_+,Q_B\mathcal{A}_+
\rangle\!\rangle+\frac{1}{3}\langle\!\langle
Y_{-2}|\mathcal{A}_+,\mathcal{A}_+,\mathcal{A}_+\rangle\!\rangle
\right.\\
&~~~~~~~~~\left.+\frac{1}{2}\langle\!\langle
Y_{-2}|\mathcal{A}_-,Q_B\mathcal{A}_-\rangle\!\rangle
-\langle\!\langle
Y_{-2}|\mathcal{A}_+,\mathcal{A}_-,\mathcal{A}_-\rangle\!\rangle\right].
\end{split}
\label{action7}
\end{equation}
Here the factors before the odd brackets are fixed by the
constraint of gauge invariance, that is specified below, and
reality of the string fields $\mathcal{A}_{\pm}$. Variation of
this action with respect to $\mathcal{A}_+$, $\mathcal{A}_-$
yields the following equations of motion\footnote{We assume that
r.h.s. is zero modulo $\ker Y_{-2}$.}
\begin{equation}
\begin{split}
Q_B\mathcal{A}_++\mathcal{A}_+\star \mathcal{A}_+
-\mathcal{A}_-\star \mathcal{A}_-&=0,\\
Q_B\mathcal{A}_-+\mathcal{A}_+\star \mathcal{A}_-
-\mathcal{A}_-\star \mathcal{A}_+&=0.\\
\end{split}
\label{eqmotion}
\end{equation}
To derive these equations we used the
cyclicity property of the odd
bracket \eqref{cyclicGSO-}
(see Section~\ref{app:cyclic}). The action \eqref{action7} is invariant under the
gauge transformations
\begin{equation}
\begin{split}
\delta \mathcal{A}_+&=Q_B\Lambda_++[\mathcal{A}_+,\Lambda_+]
+\{\mathcal{A}_-,\Lambda_-\},
\\
\delta\mathcal{A}_-&=Q_B\Lambda_-+[\mathcal{A}_-,\Lambda_+]
+\{\mathcal{A}_+,\Lambda_-\},
\end{split}
\label{gauge7}
\end{equation}
where $[\,,]$ ($\{\,,\}$) denotes $\star$-commutator
(-anticommutator). To prove the gauge invariance, it is sufficient to
check the covariance of the equations of motion \eqref{eqmotion}
under the gauge
transformations \eqref{gauge7}. A simple calculation leads to
\begin{equation*}
\begin{split}
\delta(Q_B\mathcal{A}_+&+\mathcal{A}_+\star \mathcal{A}_+
-\mathcal{A}_-\star \mathcal{A}_-)\\
&=[Q_B\mathcal{A}_++\mathcal{A}_+\star \mathcal{A}_+
-\mathcal{A}_-\star \mathcal{A}_-,\,\Lambda_+]-
[Q_B\mathcal{A}_-+\mathcal{A}_+\star \mathcal{A}_-
-\mathcal{A}_-\star \mathcal{A}_+,\,\Lambda_-],\\
\delta(Q_B\mathcal{A}_-&+\mathcal{A}_+\star \mathcal{A}_-
-\mathcal{A}_-\star \mathcal{A}_+)\\
&=[Q_B\mathcal{A}_-+\mathcal{A}_+\star \mathcal{A}_-
-\mathcal{A}_-\star \mathcal{A}_+,\,\Lambda_+]+
[Q_B\mathcal{A}_++\mathcal{A}_+\star \mathcal{A}_+
-\mathcal{A}_-\star \mathcal{A}_-,\,\Lambda_-].
\end{split}
\end{equation*}
Note that to obtain this result the associativity of
$\star$-product and Leibnitz rule for $Q_B$ must be employed. These
properties follow from the cyclicity property of the odd bracket.
The formulae above show that the gauge transformations define a Lie algebra.

\subsection{Computation of Tachyon Potential in Cubic SSFT.}
\label{sec:calc}

Here we explore the
tachyon condensation on the non-BPS $\mathrm{D}$-brane. In the
first subsection, we describe the expansion of the
string field relevant to the tachyon condensation and the level
expansion of the action. In the second subsection we calculate
the tachyon potential up to levels 1 and 4, and find its minimum.

\subsubsection{Tachyon string field in cubic SSFT.}\label{sec:t-field}
The useful devices for computation of the tachyon potential were elaborated in
\cite{9912249,9911116,0001084}. We employ these devices without additional references.

Denote by $\mathcal{H}_1$ the subset of vertex operators of ghost
number $1$ and picture $0$, created by the matter stress tensor
$T_B$, matter supercurrent $T_F$ and the ghost fields $b$, $c$,
$\pd\xi$, $\eta$ and $\phi$. We restrict the string fields
$\mathcal{A}_+$ and $\mathcal{A}_-$ to be in this subspace
$\mathcal{H}_1$. We also restrict ourselves by Lorentz scalars and
put the momentum in vertex operators equal to zero.

Next we expand $\mathcal{A}_{\pm}$ in a basis of $L_0$ eigenstates,
and write the action (\ref{action}) in terms of space-time component
fields. The string field is now a series with each term being
a vertex operator from $\mathcal{H}_1$ multiplied by
a space-time component field. We define
the level $K$ of string field's component $A_i$ to be $h+1$, where
$h$ is the conformal dimension of the vertex operator multiplied by $A_i$,
i.e. by convention the tachyon is taken to have
level $1/2$.
To compose the action truncated at level $(K,\,L)$
we select all the quadratic
and cubic terms  of total level not more than $L$
for the space-time fields of levels not more than $K$.
Since our action is cubic,
$L \le 3K$.

To calculate the action up to level $(2,6)$ we have
a collection of vertex operators listed in
Table~\ref{tab:2}.
\begin{table}[!t]
\centering
\renewcommand{\arraystretch}{1.4}
\begin{tabular}{||C|C|C|C|C|C||C||}
\hline
\textrm{Level}&\textrm{Weight}&\textrm{GSO}&\textrm{Twist}&\textrm{Name}&
\mathrm{Picture}\;$-1$&\mathrm{Picture}\;$0$\\ \cline{6-7}
L_0+1&L_0&(-1)^{F}&\Omega&&\multicolumn{2}{c||}{\textrm{Vertex operators}}\\
\hline
\hline
0&-1&+&\mathrm{even}&u&\textrm{---}&c\\
\hline
1/2&-1/2&-&\mathrm{even}&t&ce^{-\phi}&e^{\phi}\eta\\
\hline
1&0&+&\mathrm{odd}&r_i&c\partial c\partial\xi e^{-2\phi}&\partial c,\;\;c\partial\phi\\
\hline
3/2&1/2&-&\mathrm{odd}&s_i&c\partial\phi e^{-\phi}&cT_F ,\quad \partial(\eta e^{\phi})\\
&&&&&&bc\eta e^{\phi},\quad \eta\partial e^{\phi}\\
\hline
2&1&+&\mathrm{even}&v_i&\eta,\;\;T_Fce^{-\phi}&\partial^2c ,\;\; cT_B,\;\; cT_{\xi\eta}\\
&&&&&\partial\xi c\partial^2c e^{-2\phi}&cT_{\phi},\;\; c\partial^2\phi ,\;\; T_F\eta e^{\phi}\\
&&&&&\partial^2\xi c\partial c e^{-2\phi}& bc\pd c,\;\; \pd c\pd\phi\\
&&&&&\partial\xi c\partial c \partial e^{-2\phi}& \\
\hline
\end{tabular}
\caption{Vertex operators in pictures $-1$ and $0$.}
\label{tab:2}
\end{table}
Note that there are extra fields in the $0$ picture as compared
with the picture $-1$ (see Section \ref{sec:act0pic}).
Surprisingly the level $L_0=-1$ is not empty, it contains the field $u$.
One can check that this field is auxiliary. In the following analysis it plays
a significant role. Only due to this field in the next subsection we get
a nontrivial tachyon potential
(as compared with one given in \cite{0004112}) already at level
$(1/2,\,1)$.

As it is shown in Appendix~\ref{app:twist} the string
field theory action in the restricted
subspace $\mathcal{H}_1$ has $\mathbb{Z}_2$ twist symmetry.
Since the tachyon vertex operator
has even twist we can consider a further truncation  of the string field
 by restricting $\mathcal{A}_{\pm}$ to be twist even. Therefore,
the fields $r_i ,s_i$ can be dropped out.
Moreover, we impose one more restriction and require our
fields to have the $\phi$-charge (see \ref{phich}) equal to
$0$ and $1$.
String fields (in GSO$\pm$ sectors) up to level $2$ take the
form\footnote{The string fields are presented
without any gauge fixing conditions.}
\begin{equation}
\begin{split}
\mathcal{A}_+(z)&=u\,c(z)+v_1\,\partial^2c(z)+v_2\,cT_B(z)+v_3\,cT_{\eta\xi}(z)
+v_4\,cT_{\phi}(z)\\
&~~~+v_5\,c\partial^2\phi(z)+v_6\,T_F\eta e^{\phi}(z)
+v_7\,bc\pd c(z)+v_8\,\pd c\pd\phi(z),\\
\mathcal{A}_-(z)&=\frac t4\,e^{\phi(z)}\eta(z).
\end{split}
\label{str-f}
\end{equation}

\subsubsection{Tachyon potential.}
\label{sec:t-potential}
Here we give expressions for the action and the potential by truncating them
up to level $(2,\,6)$.
Since the field (\ref{str-f}) expands over the levels $0$,
$\frac12$, and $2$
we can truncate the action at levels $(1/2,\,1)$ and $(2,\,6)$ only.
All the calculations have been performed on a specially
written program on Maple. All we need is to give to the program the
string fields (\ref{str-f}) and we get the following
lagrangians
\begin{align}
\mathcal{L}^{(\frac12,\,1)}&=\frac1{g_o^2{\alpha^{\prime}}^{\frac{p+1}2}}
\left[u^2+\frac{1}{4}t^2+\frac{1}{3\gamma^2}ut^2\right],
\label{level1}
\\\mathcal{L}^{(2,\,6)}&=\frac1{g_o^2{\alpha^{\prime}}^{\frac{p+1}2}}
\left[u^2+\frac{1}{4}t^2+(4v_1-2v_3-8v_4+8v_5+2v_7)u\right.
\nonumber
\\&+4v_1^2+\frac{15}{2}v_2^2+v_3^2+\frac{77}{2}v_4^2+22v_5^2+10v_6^2
+8v_1v_3-32v_1v_4+24v_1v_5+4v_1v_7 \nonumber
\\&-16v_3v_4+4v_3v_5-2v_3v_7+12v_3v_8-52v_4v_5-8v_4v_7-20v_4v_8
+8v_5v_7+8v_5v_8 \nonumber
\\&+(-30v_4+20v_5+30v_2)v_6+4v_7v_8\nonumber
\\&+\left(\frac{1}{3\gamma
^2}u+\frac{9}{8}v_1-\frac{25}{32}v_2-\frac{9}{16}v_3-\frac{59}{32}v_4
+\frac{43}{24}v_5+\frac{2}{3}v_7 \right)t^2 \nonumber
\\&\left.
-\left(\frac{40\gamma }{3}u+45\gamma ^3v_1 -\frac{45\gamma^3}{4}v_2
-\frac{45\gamma ^3}{2}v_3-\frac{295\gamma ^3}{4}v_4
+\frac{215\gamma ^3}{3}v_5+\frac{80\gamma^3}{3}v_7\right)v_6^2\right],
\label{level4}
\end{align}
where $\gamma=\frac4{3\sqrt{3}}$.
To simplify the succeeding analysis we use a special gauge choice
\begin{equation}
3v_2-3v_4+2v_5=0.
\label{gauge}
\end{equation}
This gauge eliminates the terms linear in $v_6$ and drastically
simplifies the calculation of the effective potential for the tachyon field.
We will discuss an issue of validity of this gauge
in Section \ref{sec:glts}.
The effective tachyon potential is
defined as $\mathcal{V}(t)=-\mathcal{L}(t,u(t),v_i(t))$,
where $u(t)$ and $v_i(t)$ are solution to equations of motion
$\partial_{u}\mathcal{L}=0$ and $\partial_{v_i}\mathcal{L}=0$.
In our gauge the equation $\partial_{v_6}\mathcal{L}=0$
admits a solution $v_6=0$ and, therefore, the tachyon potential
computed at levels $(2,\,4)$ and $(2,\,6)$ is the same.
The potential at levels $(1/2,\,1)$ and $(2,\,6)$
has the following form:
\begin{equation}
\begin{split}
\mathcal{V}_{\mathrm{eff}}^{(\frac12,\,1)}(t)&=\frac{1}{g_o^2\alpha^{\,\prime\,\frac{p+1}{2}}}
\left[ \frac{81}{1024}t^4-\frac{1}{4}t^2
\right],\\
\mathcal{V}_{\mathrm{eff}}^{(2,\,6)}(t)&=\frac{1}{g_o^2\alpha^{\,\prime\,\frac{p+1}{2}}}
\left[ \frac{5053}{69120}t^4-\frac{1}{4}t^2 \right]
\end{split}
\label{potential-g}
\end{equation}
One sees that the potential has two global minima, which are
reached at points $t_c=\pm1.257$ at level $(1/2,\,1)$ and at points
$t_c=\pm 1.308 $ at level $(2,\,6)$ (see also Figure \ref{Fig:potential}
and Table \ref{tab:3}).

\subsection{Tension of Non-BPS $\mathrm{D}p$-brane in Cubic SSFT
 and Sen's Conjecture.
}
\label{sec:tension}

To find a tension of  $\mathrm{D}p$-brane following \cite{0002211}
one considers the
SFT describing a pair of $\mathrm{D}p$-branes and calculates the
string field action on a special string field. This string field
\begin{wrapfigure}[11]{l}{55mm}
\vspace{-6mm}
\begin{center}
\includegraphics[width=130pt]{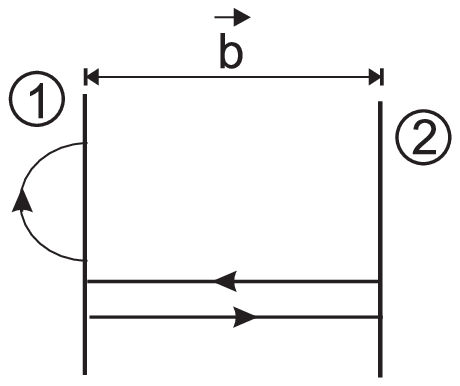}
\end{center}
\vspace{-2mm}
\caption{The system of two non BPS D$p$-branes and
strings attached to them.}
\label{fig:branes}
\end{wrapfigure}
contains a field describing a displacement of one of the branes
and a field describing arbitrary excitations of the strings
stretched between the two branes. For simplicity one can use
low-energy
excitations of the strings stretched between
the branes.

The cubic SSFT describing a pair of non-BPS $\mathrm{D}p$-branes
includes
 $2\times 2$ Chan-Paton (CP)
factors \cite{0002211,polchinski}
and has the following form
\begin{equation}
\begin{split}
S[\mathcal{\hat{A}}_+,\mathcal{\hat{A}}_-]&=\frac{1}{g^2_{o}}\left[
\frac{1}{2}\langle\!\langle \hat {Y}_{-2}| \hat {\mathcal{A}}_+,
\hat {Q}_B\hat {\mathcal{A}}_+
\rangle\!\rangle+\frac{1}{3}\langle\!\langle
\hat {Y}_{-2}|\hat {\mathcal{A}}_+,\hat {\mathcal{A}}_+,
\hat {\mathcal{A}}_+\rangle\!\rangle
\right.\\
&~~~~~~~~~\left.+\frac{1}{2}\langle\!\langle
\hat {Y}_{-2}|\hat {\mathcal{A}}_-,\hat {Q}_B\hat {\mathcal{A}}_
-\rangle\!\rangle
-\langle\!\langle
\hat {Y}_{-2}|\hat {\mathcal{A}}_+,\hat {\mathcal{A}}_-,
\hat {\mathcal{A}}_-\rangle\!\rangle\right].
\end{split}
\label{mat-action}
\end{equation}
Here $g_o$ is a dimensionless coupling constant.
The hatted BRST charge $\hat {Q}_B$ and double step inverse picture changing
operator $\hat {Y}_{-2}$ are $Q_B$ and $Y_{-2}$ tensored
by $2\times 2$ unit
matrix.
The string fields are also $2\times2$ matrices
\begin{equation}
\hat{\mathcal{A}}_{\pm}=\mathcal{A}_{\pm}^{(1)}\otimes
\begin{pmatrix}
1 & 0\\
0 & 0
\end{pmatrix}
+
\mathcal{A}_{\pm}^{(2)}\otimes
\begin{pmatrix}
0 & 0\\
0 & 1
\end{pmatrix}
+
\mathcal{B}_{\pm}^*\otimes
\begin{pmatrix}
0 & 1\\
0 & 0
\end{pmatrix}
+
\mathcal{B}_{\pm}\otimes
\begin{pmatrix}
0 & 0\\
1 & 0
\end{pmatrix}
\label{string-fields}
\end{equation}
and the odd bracket includes the trace over matrices.

The action is invariant
under the following gauge transformations:
\begin{equation}
\begin{split}
\delta \hat {\mathcal{A}}_+&=\hat {Q}_B\hat {\Lambda}_+
+[\hat {\mathcal{A}}_+,\hat {\Lambda}_+]
+\{\hat {\mathcal{A}}_-,\hat {\Lambda}_-\},\\ \delta
\hat {\mathcal{A}}_-&=\hat {Q}_B\hat {\Lambda}_-+[\hat {\mathcal{A}}_
-,\Lambda_+]
+\{\hat {\mathcal{A}}_+,\hat {\Lambda}_-\}.
\end{split}
\label{tgauge}
\end{equation}

The fields ${\mathcal A}^{(1)}_{\pm}$
 describe excitations of the string attached to the
first brane, while
$\mathcal{A}^{(2)}_{\pm}$ describe excitations of the string
attached to the second one.
Excitations of the stretched strings are represented by
the fields $\mathcal{B}_{\pm}$ and
$\mathcal{B}_{\pm}^*$ (see Figure~\ref{fig:branes}).
The action for a single non-BPS D-brane (\ref{action}) that we have used above is
derived from the universal action (\ref{mat-action})
by setting
$\mathcal{A}^{(2)}_{\pm}$, $\mathcal{B}_{\pm}$ and $\mathcal{B}^*_{\pm}$
to zero.
Note also that we have not changed the value of the coupling constant $g_o$.
The "$\pm$" subscript specify the GSO sector.

Let us take the following  string fields $\hat{\mathcal{A}}_{\pm}$:
\begin{equation}
\hat{\mathcal{A}}_+=\hat{A}^{(1)}_+ +\hat{B}^*_++\hat{B}_+,
\qquad
\hat{\mathcal{A}}_-=\hat{B}^*_-+\hat{B}_-,
\label{fie}
\end{equation}
where
\begin{align*}
\hat{A}^{(1)}_+&=\int
\frac{d^{p+1}k}{(2\pi)^{p+1}}\;A_i(k_{\alpha})
\mathrm{V}_v^i(k_{\alpha},0)
\otimes
\begin{pmatrix}
1 & 0\\
0 & 0
\end{pmatrix},
\\
\hat{B}_+&=\int\frac{d^{p+1}k}{(2\pi)^{p+1}}\;
iB_{i}(k_{\alpha})
\mathrm{V}_v^{i}(k_{\alpha},
\tfrac{\overrightarrow{b}}{2\pi\alpha^{\prime}})
\otimes
\begin{pmatrix}
0 & 0\\
1 & 0
\end{pmatrix},
\\
\hat{B}^*_+&=\int\frac{d^{p+1}k}{(2\pi)^{p+1}}\;
iB_{i}^* (k_{\alpha})
\mathrm{V}_v^{i}(k_{\alpha},
-\tfrac{\overrightarrow{b}}{2\pi\alpha^{\prime}})
\otimes
\begin{pmatrix}
0 & 1\\
0 & 0
\end{pmatrix}
\\
\hat{B}_-&=\int\frac{d^{p+1}k}{(2\pi)^{p+1}}\;
t(k_{\alpha})
\mathrm{V}_t(k_{\alpha},
\tfrac{\overrightarrow{b}}{2\pi\alpha^{\prime}})
\otimes
\begin{pmatrix}
0 & 0\\
1 & 0
\end{pmatrix},
\\
\hat{B}^*_-&=\int\frac{d^{p+1}k}{(2\pi)^{p+1}}\;
t^*(k_{\alpha})
\mathrm{V}_t(k_{\alpha},
-\tfrac{\overrightarrow{b}}{2\pi\alpha^{\prime}})
\otimes
\begin{pmatrix}
0 & 1\\
0 & 0
\end{pmatrix}
.
\end{align*}
Here $b_i$ is a distance between the branes,
$\alpha=0,\dots,p$ and
$i=p+1,\dots,9$ and
$\mathrm{V}_v^i$ and $\mathrm{V}_t$
are vertex operators of a massless vector and tachyon
fields respectively defined by
\begin{equation}
\begin{split}
\mathrm{V}_v^{\mu}(k_{\alpha},k_i)&=\frac{i}{2}\left[\frac{2}
{\alpha^{\,\prime}}\right]^{1/2}
\left[c\partial X^{\mu}+c2ik\cdot\psi\psi^{\mu}- \frac12\eta
e^{\phi}\psi^{\mu}\right]e^{2ik\cdot X_L(0)},
\\
\mathrm{V}_t(k_{\alpha},k_i)&=\frac{1}{2}
\left[c2ik\cdot\psi- \frac12\eta
e^{\phi}\right]e^{2ik\cdot X_L(0)}.
\end{split}
\label{vec-ver}
\end{equation}
These vertex operators are written in the $0$ picture and
can be obtained
by applying the picture changing operator \eqref{picX} to
the corresponding operators in picture $-1$.
The Fourier transform of $A_i(k_{\alpha})$ has an interpretation
of the $\mathrm{D}p$-brane's coordinate up to an overall
normalization factor \cite{polchinski}. Further we will assume that
$b^iB_{i}(k_{\alpha})=0$.

The action for the field \eqref{fie}
depending on the local fields
$B_i(k_{\alpha})$,
$B^*_i(k_{\alpha})$, $t(k_{\alpha})$,
$t^*(k_{\alpha})$ and $A_i(k_{\alpha})$
is given by
\begin{multline}
S[A_i,B_i,t]=\frac{1}{g^2_o}\left[
\frac{1}{2}\langle\!\langle Y_{-2}|\hat{A}^{(1)}_+,\hat{A}^{(1)}_+\rangle\!\rangle
+\langle\!\langle
Y_{-2}|\hat{B}^*_+,\hat{B}_+\rangle\!\rangle +
\langle\!\langle
Y_{-2}|\hat{B}^*_-,\hat{B}_-\rangle\!\rangle
\right.
\\
\left.+\langle\!\langle Y_{-2}|\hat{A}^{(1)}_+,\hat{B}^*_+,
\hat{B}_+\rangle\!\rangle
-\langle\!\langle Y_{-2}|\hat{A}^{(1)}_+,\hat{B}^*_-,\hat{B}_-
\rangle\!\rangle\right]
=\frac{\alpha^{\,\prime}}{g^2_o
\alpha^{\,\prime\frac{p+1}{2}}}
\int \frac{d^{p+1}k}{(2\pi)^{p+1}}\;
\Biggl\{\frac{k_{\alpha}^2}{2}
A^i(-k)A^i(k)
\\
+B^*_i(-k)B^i(k)\left[k_{\alpha}^2
+\tfrac{b_i^2}{(2\pi\alpha^{\,\prime})^2}\right]
+t^*(-k)t(k)\left[k_{\alpha}^2
+\tfrac{b_i^2}{(2\pi\alpha^{\,\prime})^2}-\tfrac{1}{2\alpha^{\,\prime}}\right]
\\
+\int \frac{d^{p+1}p}{(2\pi)^{p+1}}\;
\gamma^{2\alpha^{\,\prime}(k^2+p^2+p\cdot k+\tfrac{b_i^2}{(2\pi\alpha^{\,\prime})^2})}
\left[B^*_i(-p-k)B^i(k)+\gamma^{-1}t^*(-p-k)t(k)\right]
\frac{b_jA^j(p)}{2\pi\alpha^{\prime}\sqrt{2\alpha^{\prime}}} \Biggr\},
\label{tensc}
\end{multline}
where $\gamma=\frac{4}{3\sqrt{3}}$.
Let us now consider the constant field
$A_i(\xi)=\mathrm{const}$, where $\xi^{\alpha}$ are coordinates
on the brane. Its Fourier transform is of the form
$A_i(p)=(2\pi)^{p+1}A_i\delta(p)$.
Let also $B_i(k)$, $B_i^*(k)$ and
$t(p)$, $t^*(p)$ be on-shell,
i.e.
$$
k_{\alpha}^2+\frac{b_i^2}{(2\pi\alpha^{\,\prime})^2}=0
$$
and
$$
p_{\alpha}^2+\frac{b_i^2}{(2\pi\alpha^{\,\prime})^2}-
\frac{1}{2\alpha^{\,\prime}}=0.
$$
In this case the action \eqref{tensc} is simplified and takes the form:
\begin{multline}
\tilde{S}[t,B_i]=\frac{\alpha^{\,\prime}}{g_o^2\alpha^{\,\prime\,\frac{p+1}{2}}}
\int_{\text{mass shell}}\frac{d^{p+1}k}{(2\pi)^{p+1}}\;
\left[k_{\alpha}^2
+\frac{b_i^2}{(2\pi\alpha^{\,\prime})^2}-\frac{1}{2\alpha^{\,\prime}}+
\frac{b_i A^i}{2\pi\alpha^{\,\prime}\sqrt{2\alpha^{\,\prime}}}
\right]
t^*(-k)t(k)
\\
+\frac{\alpha^{\,\prime}}{g_o^2\alpha^{\,\prime\,\frac{p+1}{2}}}
\int_{\text{mass shell}}\frac{d^{p+1}k}{(2\pi)^{p+1}}\;
\left[k_{\alpha}^2
+\frac{b_i^2}{(2\pi\alpha^{\,\prime})^2}
+\frac{b_iA^i}{2\pi\alpha^{\,\prime}\sqrt{2\alpha^{\,\prime}}}\right]
B^*_j(-k)B^j(k).
\end{multline}
The terms proportional to $A_i$ have natural interpretation as shifts of
masses of the fields $t$ and $B_i$. We want to interpret the term with $A_i$
 as a shift of brane coordinate $b_i$. Therefore, we have to compare two shifts
 of the mass: one produced by $A_i$ and another produced by a shift of $b_i$

\begin{equation}
\delta(m^2)=\frac{2b_i\delta b^i}{4\pi^2\alpha^{\,\prime\,2}}
=\frac{b_iA^i}{2\pi\alpha^{\,\prime}\sqrt{2\alpha^{\,\prime}}}.
\end{equation}
So one gets
\begin{equation}
A^i=\frac{1}{\pi}\left[\frac{2}{\alpha^{\,\prime}}\right]^{\frac12}
\delta b^i.
\label{chi-b7}
\end{equation}
This formula determines the normalization of the field $A_i$.
Therefore, we can introduce the profile of the first non-BPS D$p$-brane as
$x_i(\xi)=\pi\left[\frac{\alpha^{\,\prime}}{2}\right]^{1/2}A_i(\xi)$.
Substitution of the $A_i(\xi)$ into the first term of the action
(\ref{tensc}) yields
\begin{equation}
S_0=-\frac{2}{g^2_o\pi^2\alpha^{\,\prime\frac{p+1}{2}}}
\int d^{p+1}\xi\;\frac{1}{2}
\partial_{\alpha}x^i(\xi)\partial^{\alpha}x_i(\xi).
\label{S0X}
\end{equation}
The coefficient before the integral is the non-BPS
D$p$-brane tension $\tilde{\tau}_p$:
\begin{equation}
\tilde{\tau}_p=\frac{2}{g^2_o\pi^2\alpha^{\,\prime\frac{p+1}{2}}}.
\label{tauBPS}
\end{equation}
As compared with the expression for the tension given in \cite{0002211}
we have the addition factor $4$. The origin of this factor is the difference in
the normalization of the superghosts $\beta,\gamma$.

\begin{figure}[t]
\centering
\includegraphics[width=320pt]{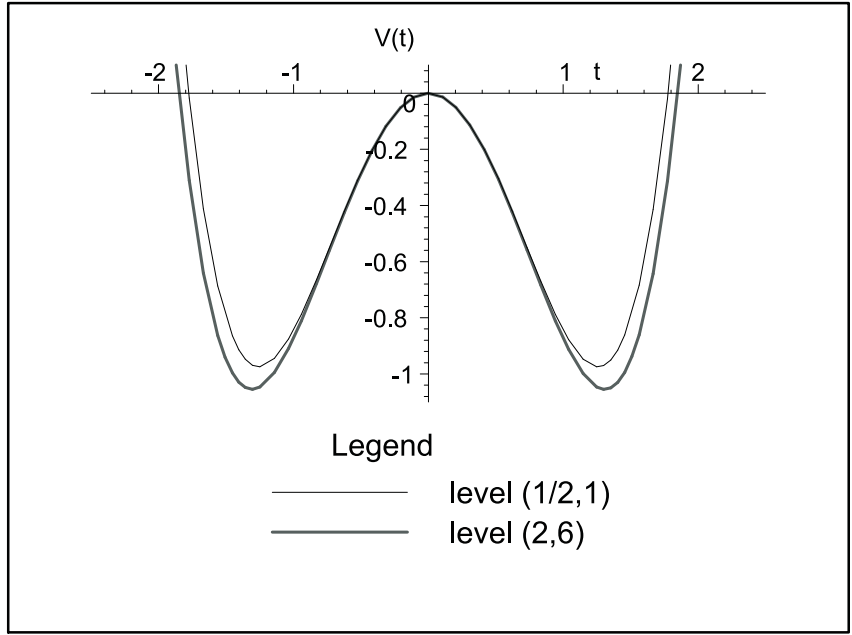}
\caption{Graphics of the tachyon potential at the levels
$(1/2,\,1)$ and $(2,\,6)$.
``$-1$" is equal to the minus tension of non BPS D$p$-brane
$\tilde{\tau}_p=\frac{2}{\pi^2 g^2_o\alpha^{\,\prime\,\frac{p+1}{2}}}$.}
\label{Fig:potential}
\end{figure}

Now we can express the coupling constant $g_o^2$ in terms of the
tension $\tilde{\tau}_p$.
Hence the potential \eqref{potential-g} at levels $(1/2\,1)$ and $(2\,6)$
takes the form (see also Figure \ref{Fig:potential})

\begin{equation}
\begin{split}
\mathcal{V}_{\mathrm{eff}}^{(\frac12,\,1)}(t)&=\frac{\pi^2\tilde{\tau}_p}{2}
\left[ \frac{81}{1024}t^4-\frac{1}{4}t^2
\right],\\
\mathcal{V}_{\mathrm{eff}}^{(2,\,6)}(t)&=\frac{\pi^2\tilde{\tau}_p}{2}
\left[ \frac{5053}{69120}t^4-\frac{1}{4}t^2 \right].
\end{split}
\end{equation}

The critical points of these functions are collected in Table
\ref{tab:3}.
\begin{table}[!thb]
\centering
\renewcommand{\arraystretch}{1.4}
\begin{tabular}{||C|L|L||}
\hline
\textrm{Potential}& \textrm{Critical points} & \textrm{Critical values}\\
\hline
\hline
\mathcal{V}_{\mathrm{eff}}^{(\frac1 2,\,1)} & t_c=0& \mathcal{V}_c=0\\
& t_c=\pm\frac{8\sqrt{2}}{9}\approx \pm1.257&
\mathcal{V}_c\approx-0.975\tilde{\tau}_p\\
\hline
\mathcal{V}_{\mathrm{eff}}^{(2,\,6)} & t_c=0& \mathcal{V}_c=0\\
& t_c=\pm\frac{24}{5053}\sqrt{75795}\approx\pm 1.308&
\mathcal{V}_c\approx-1.058\tilde{\tau}_p\\
\hline
\end{tabular}
\caption{The critical points of the tachyon potential at levels
$(1/2,\,1)$ and $(2,\,6)$.}
\label{tab:3}
\end{table}
One sees that the potential has a global minimum and the value of this
minimum is $97.5\%$ at level $(1/2,\,1)$
and $105.8\%$ at level $(2,\,6)$
of the tension $\tilde{\tau}_p$ of the non-BPS $\mathrm{D}p$-brane.

\subsection{Test of Absence of Kinetic Terms Around Tachyon
Vacuum.}
\label{sec:calc1}
In this section we perform calculations of the effective potential at level 0,
i.e. we take only low levels fields:
\begin{equation}
\begin{split}
\Ac_+&=\int\frac{d^{p+1}k}{(2\pi)^{p+1}}\,u(k) \left.\mathrm{U}(w,k)\right|_{w=0},
\quad
\mathrm{U}(w,k)=c(w)e^{2ik\cdot X_L(w)},\\
\Ac_-&=\int\frac{d^{p+1}k}{(2\pi)^{p+1}}\,t(k) \left.\mathrm{T}(w,k)\right|_{w=0},
\quad
\mathrm{T}(w,k)=\frac{1}{2}\left[c(w)2ik\cdot\psi(w)-\frac{1}{2}\eta(w)e^{\phi(w)}\right]
e^{2ik\cdot X_L(w)}.
\end{split}
\end{equation}
Note that the expression for the tachyon field can be obtained by applying
rasing picture changing operator $\mathrm{X}$ to the tachyon vertex operator
in picture $-1$.
The action in the momentum representation is of the form
\begin{equation}
\label{action-moment}
S[u,t]=
\frac{1}{g_o^2\alpha^{\,\prime \frac{(p+1)}{2}}}\int\frac{d^{p+1}k}{(2\pi)^{p+1}}\left[
u(-k)u(k)-\frac12 t(-k)t(k)\,\left[\alpha^{\,\prime}k_{\alpha}^2-1/2
\right]\right.
\end{equation}
$$
\left.-\frac{1}{3\gamma ^2}\int\frac{d^{p+1}p}{(2\pi)^{p+1}}\, t(-k-p)t(p)u(k)
\left(\gamma \right)^{2\alpha^{\,\prime}(p^2
+k^2+kp)}
\right].
$$

This expression can be shortly rewritten in the $x$-representation
\begin{equation}
S[u,t]=\frac{1}{g_o^2\alpha^{\,\prime \frac{(p+1)}{2}}}
\int\; d^{p+1}x\left[
u(x)u(x)-\frac12 t(x)\,\left(-\alpha^{\,\prime}\pd_{\alpha}\pd^{\alpha}-1/2\right)t(x)
-\frac{1}{3\gamma^2} \tilde{u}(x)\tilde{t}(x)\tilde{t}(x)
\right],
\label{action-x}
\end{equation}
where
\begin{equation}
\tilde{\phi}(x)=\exp(-\alpha^{\,\prime}\log\gamma\,\pd_{\alpha}\pd^{\alpha})\phi(x).
\label{tilde}
\end{equation}

The operation $\tilde{}$ has the following properties
\begin{itemize}
\item
$\tilde{t}_0=t_0$ if $t_0=\textrm{const}$,
\item
$\int dx\;\tilde{t}(x)=\int dx\;t(x)$,
\item
$\int dx\;\tilde{a}(x)b(x)=\int dx\;a(x)\tilde{b}(x)$.
\end{itemize}
Using the second property we can move $\tilde{}$ from $u(x)$ on the
term $\tilde{t}(x)\tilde{t}(x)$. After this one can integrate over $u(x)$
and gets an effective action
\begin{equation}
S_{\textrm{eff}}[t]=\frac{1}{g_o^2\alpha^{\,\prime \frac{(p+1)}{2}}}
\int\; d^{p+1}x\left[
-\frac12 t(x)\,\left(-\alpha^{\,\prime}\pd_{\alpha}\pd^{\alpha}-1/2\right)t(x)
-\frac{1}{4\cdot 9\gamma^4} \left(\widetilde{\tilde{t}\tilde{t}}\right)^2(x)
\right].
\label{action-eff}
\end{equation}
Expanding
$t(x)=t_0+t_1(x)$
and keeping only terms that are quadratic in $t_1$ and linear in
$\alpha '$ we get
\begin{equation}
S^{(2)}_{2,shifted}[t]=\frac {a}{2}
\int\; d^{p+1}x\left[
1+\frac{10t_0^2}{9\gamma^4}\log\gamma \right]
t_1(x)\,\alpha^{\,\prime}\pd_{\alpha}\pd^{\alpha} t_1(x).
\label{action-eff1}
\end{equation}

So to vanish the kinetic terms for $t_1$ we have
to have $t_0=1.099$, which is not far from
the critical value $t_0=1.257$ for the minimum
of the tachyon potential at level $(\frac12,1)$.

\subsection{Lump Solutions.}
Here we give a brief review of the numeric computations of lump solutions
(see review \cite{0106068} and \cite{0005036}-\cite{kink-mogila}
for more details\footnote{See \cite{GhoshalPADIC,MinahanPADIC}
for relations with p-adic strings
\cite{VVZ,freund,padic-str,Frampton,padic-VS}}).

First, let us define what is the lump solution. Let us denote
by $\Phi_0$ the string field representing the solution
of momentum independent equations of motion. In other words,
$\Phi_0$ is a tachyon condensate. We want to find a lump solution $\Phi$
of codimension one, which represents a decay D$25$ $\rightarrow$ D$24$.
This means that string field $\Phi$ is a static configuration depending on
one space coordinate, say $x^{25}\equiv x$. This configuration has nontrivial boundary
conditions
$$
\Phi(x)\rightarrow \Phi_0\quad\text{as}\quad x\rightarrow\pm\infty.
$$
Physically this condition means that we want to have true string vacuum
at plus and minus infinity. We expect that the solution is
concentrated near the origin and, therefore, it distinctly differs
from $\Phi_0$ only in small domain around the origin.

Second, to be able to find a solution numerically we have to have finite number of fields.
To this purpose we use level truncation scheme for the computations of the tachyon potential.
To find the lump solution authors of the paper \cite{0005036} proposed to use a modified
level truncation scheme. Their idea consists of two steps:
\begin{enumerate}
\item We compactify the direction $x$ on a circle of radius $R\sqrt{\ap}$.
This allows us to represent component fields $\phi_i(x)$ of
the string field $\Phi(x)$ by
Fourier series rather than Fourier integral:
\begin{equation}
\phi_i(x)\sim\sum_{n}\phi_{i,n}e^{\frac{inx}{R\sqrt{\ap}}}.
\end{equation}
\item We modify the level grading:
\begin{equation}
\mathrm{lvl}_R(\phi_{i,n})=\frac{n^2}{R^2}+N_i+1,
\label{grading}
\end{equation}
where $N_i$ is oscillator number operator evaluated on the vertex operator
corresponding to the field $\phi_i$ and the normalization of the grading
was chosen in such a way that it is equal to $1$ on the zero momentum tachyon field.
Using this level grading we can define $(M,N)$-level approximation of the action.
This means that we get all fields with $\mathrm{lvl}_R\leqslant M$ from the string field $\Phi$
and keep only the terms in the action for which $\mathrm{lvl}_{R}\leqslant N$.
\end{enumerate}

Third, we have to choose the Hilbert space $\Hc$ in which we are going to find our
solution. The analysis performed in \cite{0005036}
shows that without loss of generality we can consider the following reduced space:
\begin{equation}
\Hc=\left.\Hc^{\text{univ}'}_{\text{matter}}\oplus\Hc_{\text{ghost}}\oplus\Hc_{\text{matter}}(x)
\right|_{even},
\end{equation}
where $\Hc^{\text{univ}'}_{\text{matter}}$ is a universal (background independent)
Hilbert space \cite{0005036} generated by matter Virasoro generators
restricted to the directions $x^1,\dots,x^{24}$,
$\Hc_{\text{ghost}}$ is a Hilbert space spanned by all ghost operators
and $\Hc_{\text{matter}}(x)$ is a Hilbert space generated by all oscillators
$\alpha_n^{25}$.

\begin{figure}[t]
\centering
\includegraphics[width=450pt]{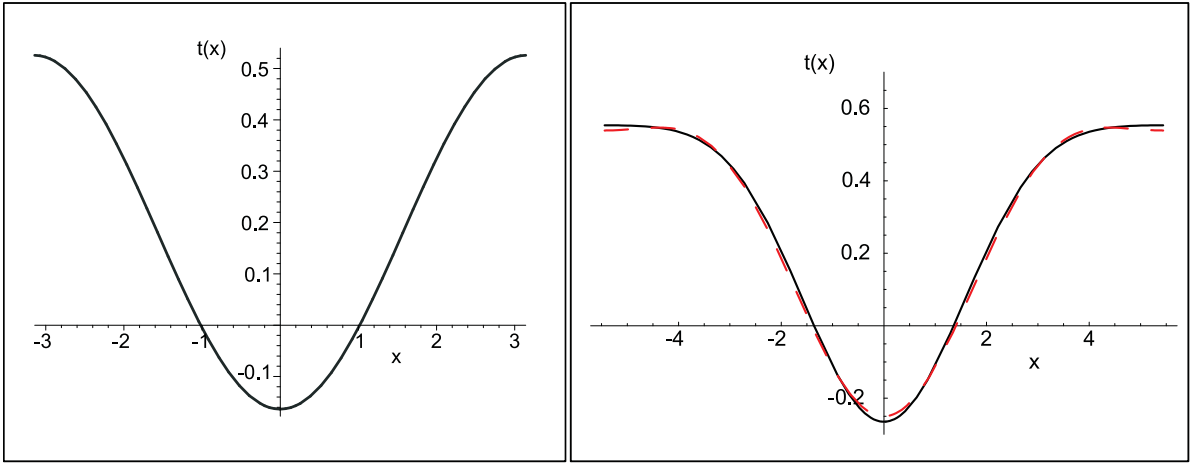}
\caption{Plot of $t(x)$ for $R=\sqrt{3}$ on different levels.
On the left pane the solid line represents $t(x)$ at level $(\frac{1}{3},\frac{2}{3})$.
On the right pane the dashed line shows a plot of $t(x)$ at level
$(\frac{7}{3},\frac{14}{3})$ and
solid line shows a plot of $t(x)$ at level
$(3,6)$.}\label{Fig:lumppict}
\end{figure}

Now consider a simple example of the computations.
Let us choose $R=\sqrt{3}$ and assume that we have only
tachyon field $t(x)$. Since it has to be even we can write
\begin{equation}
t(x)=\sum_{n=0}^{\infty}t_n \cos\Bigl(\frac{nx}{R\sqrt{\ap}}\Bigr).
\end{equation}
The terms $t_n$ have the following grading:
$$
\mathrm{lvl}_{\sqrt{3}}\,{t_n}=1+\frac{n^2}{3}.
$$
Now substitution into the cubic action \eqref{S2boson}, \eqref{Sint-Boson-x}
yields the following expressions at levels $(0,0)$ and $(\frac13,\frac23)$:
\begin{subequations}
\begin{align}
\mathcal{V}^{(0,0)}&=-\frac12 t_0^2+\frac{1}{3\gamma^3}t_0^3,
\\
\mathcal{V}^{(\frac13,\frac23)}&=-\frac16 t_1^2+\frac{1}{2\gamma^{\frac{11}{3}}}t_0t_1^2,
\end{align}
\label{zerolvl}
\end{subequations}
where $\gamma=\frac{4}{3\sqrt{3}}$ and $\mathcal{V}$ is related to $S$ by formula
\begin{equation}
S=\frac{2\pi R\sqrt{\ap}\, \mathrm{Vol}_{24}}{g_o^2\alpha^{\,\prime 13}}\,\mathcal{V}.
\end{equation}
The solution of the equations of motion obtained from \eqref{zerolvl} is
presented on the left pane on Figure~\ref{Fig:lumppict}.

One can also compute the ratio of tension of the lump solution and
the tension of original D$25$-brane. One expect that the tension
of the lump solution of codimension $1$ is equal to the tension of
D$24$-brane. Therefore, we expect that
\begin{equation}
\frac{\tau(\text{lump})}{\tau_{\text{D}24}}=1,
\quad\text{where}\quad
\tau(\text{lump})=\frac{2\pi R\sqrt{\ap}}{g_o^2\alpha^{\,\prime 13}}\,
(\mathcal{V}(\Phi)-\mathcal{V}(\Phi_0)).
\label{estimation}
\end{equation}
There are several ways to check this equality numerically.
We can express the tension of D$25$ brane through the coupling
constant $g_o$ as \cite{9912249}
$$
\tau_{\text{D}25}=\frac{1}{2\pi^2 g_o^2\alpha^{\prime\,13}}.
$$
Substitution in \eqref{estimation} yields
\begin{equation}
r=\frac{\tau(\text{lump})}{\tau_{\text{D}24}}=
\frac{\tau_{\text{D}25}}{\tau_{\text{D}24}}2\pi R\sqrt{\ap}2\pi^2(\mathcal{V}(\Phi)-\mathcal{V}(\Phi_0))
=R(2\pi^2\mathcal{V}(\Phi)-2\pi^2\mathcal{V}(\Phi_0)).
\end{equation}
So, now we can compute the value of $r$ for our approximate solution and
check how accurately it approaches unity.
Since we know that $2\pi^2\mathcal{V}(\Phi_0)=-1$, we can write
two expressions for $r$:
\begin{subequations}
\begin{align}
r^{(1)}&=R\Bigl[2\pi^2\mathcal{V}^{(M,N)}(\Phi)+1\Bigr],
\\
r^{(2)}&=2\pi^2R\Bigl[\mathcal{V}^{(M,N)}(\Phi)-\mathcal{V}^{(M,N)}(\Phi_0)\Bigr].
\end{align}
\end{subequations}

By adding more fields one can increase the level and make computations
of the solution more and more accurate. Now we present the results
of such computations, which were obtained in \cite{0005036} for
radius $R=\sqrt{3}$. On the right pane of Figure~\ref{Fig:lumppict}
we present a plot of $t(x)$ at levels $(\frac{7}{3},\frac{14}{3})$ and
$(3,6)$. In the Table below one can find the values of $r^{(1)}$ and
$r^{(2)}$ depending on the level.
\renewcommand{\arraystretch}{1.4}
\begin{longtable}[!h]{||L|L|L||}
\hline
\text{Level} & r^{(1)} & r^{(2)} \\
\hline
(\frac13,\frac23) & 1.32002 & 0.77377 \\
\hline
(\frac43,\frac83) & 1.25373 & 0.707471 \\
\hline
(2,4) & 1.11278 & 1.02368 \\
\hline
(\frac73,\frac{14}{3}) & 1.07358 & 0.984467 \\
\hline
(3,6) & 1.06421 & 0.993855 \\
\hline
\end{longtable}
\renewcommand{\arraystretch}{1}
So, it seems that $r^{(1)}$ and $r^{(2)}$ converge to unity from
top and bottom correspondingly.

\newpage
\section{Level Truncation and Gauge Invariance.}
\label{sec:glts}
\setcounter{equation}{0}

\subsection{Gauge Symmetry on Constant Fields.}
\label{sec:gscf}
In this section we restrict our attention to scalar fields at zero momentum,
which are relevant for calculations of a Lorentz-invariant vacuum.
The zero-momentum scalar string fields $\Ac_+$ and $\Ac_-$
can be expanded as
\begin{equation}
\Ac_+ = \sum_{i = 0}^{\infty}\phi^i\Phi_{i}\qquad\text{and}\qquad
\Ac_-= \sum_{a = 0}^{\infty}t^a\mathrm{T}_{a},
\end{equation}
where conformal operators $\Phi_{i}$ and $\mathrm{T}_{a}$ are taken at zero
momentum
and $\phi^i$ and $t^a$ are  constant  scalar fields.
The action \eqref{action} for the component
fields $\phi^i$, $t^a$ is a  cubic polynomial
of the following form
\begin{equation}
S = -\frac{1}{2}\sum _{i,j} \mathscr{M}_{ij}\phi^i\phi^j
-\frac{1}{2}\sum_{a,b} \mathscr{F}_{ab}t^at^b
-\frac{1}{3}\sum _{i,j,k} \mathscr{G}_{ijk}\phi^i\phi^j\phi^k
+\sum \mathscr{G}_{i,a,b}\phi^it^at^b,
\label{action-mat}
\end{equation}
where
\begin{subequations}
\begin{alignat}{2}
\mathscr{M}_{ij} &= \la Y_{-2}| \Phi_i,Q_B\Phi_j\ra,
&\qquad\qquad
\mathscr{G}_{ijk} &= \la Y_{-2}| \Phi_i,\Phi_j, \Phi_k\ra,
\\
\mathscr{F}_{ab} &= \la Y_{-2}| \mathrm{T}_a,Q_B\mathrm{T}_b\ra, &
\mathscr{G}_{iab} &= \la Y_{-2}| \Phi_i,\mathrm{T}_a,
\mathrm{T}_b\ra.
\end{alignat}
\end{subequations}

For the sake of simplicity we consider the gauge transformations
with GSO$-$ parameter $\Lambda_-$ equal to zero.
The scalar constant
gauge parameters $\{\delta\lambda^{\alpha}\}$ are the components of a ghost number zero
GSO$+$ string field
\begin{equation}
\Lambda_{+} =
\sum_{\alpha} \delta\lambda^{\alpha}\Lambda_{+,\alpha} .
\end{equation}
Assuming that the basis $\{\Phi_j,\,\mathrm{T}_b\}$ is complete
we write the following identities:
\begin{subequations}
\label{def-J}
\begin{align}
Q_B\Lambda_{+,\alpha}
=\sum \mathscr{V}^i_{\alpha}\Phi _i,\\
\Phi_j\star\Lambda_{+,\alpha}-\Lambda_{+,\alpha}\star\Phi_j&=
\sum_{i}\mathscr{J}^i{}_{j\alpha}\Phi_i,
\\
\mathrm{T}_b\star\Lambda_{+,\alpha}-\Lambda_{+,\alpha}\star\mathrm{T}_b&=
\sum_{a}\mathscr{J}^a{}_{b\alpha}\mathrm{T}_a.
\end{align}
\end{subequations}
The variations of the component fields $\phi^i$ and $t^a$
with respect to the gauge transformations \eqref{gauge7}
generated by  $\delta\lambda^{\alpha}$ can be
expressed in terms of the ``structure constants''
\begin{subequations}
\begin{align}
\delta \phi^i &\equiv
\delta_0\phi^i +
\delta _1\phi^i =
(\mathscr{V}^i_{\alpha} +  \mathscr{J}^i{}_{j\alpha} \phi^j)
\delta\lambda^{\alpha},
\label{ptrans}
\\
\delta t^a &\equiv
\delta_1 t^a =\mathscr{J}^a{}_{b\alpha} t^b\delta\lambda^{\alpha}.
\label{strans}
\end{align}
\label{trans}
\end{subequations}
The constants $\mathscr{V}^i_{\alpha}$ solve the zero vector
equation for the matrix $\mathscr{M}_{ij}$:
\begin{equation}
\mathscr{M}_{ij}\mathscr{V}^j_{\alpha}=0
\label{zero}
\end{equation}
and therefore the quadratic action is always invariant with respect to free
gauge transformations.

In the bosonic case one deals only with the gauge transformations
of the form \eqref{ptrans}
and finds $\mathscr{J}^i{}_{j\alpha}$ \cite{Taylor} using
an explicit form of $\star$-product
in terms of the Neumann functions
\cite{GrJe}. In our case it is more suitable to employ the
conformal field theory calculations by using
the following identity:
\begin{equation}
\la Y_{-2}| \Phi_1,\,\Phi_2\star\Phi_3\ra=\la Y_{-2}| \Phi_1,\,\Phi_2,\,\Phi_3\ra.
\label{star}
\end{equation}
To this end it is helpful to
use a notion of dual conformal operator.
Conformal operators $\{\tilde{\Phi}^i,\,\tilde{\mathrm{T}}^a\}$ are
called dual to the operators $\{\Phi_j,\,\mathrm{T}_b\}$ if the following
equalities hold
\begin{equation}
\la Y_{-2}| \tilde {\Phi}^i,\,\Phi_j\ra=\delta^i{}_{j}
\qquad\text{and}\qquad
\la Y_{-2}| \tilde {\mathrm{T}}^a,\,\mathrm{T}_b\ra=\delta^a{}_{b}.
\label{Def_d}
\end{equation}

Using \eqref{star}, \eqref{Def_d} and \eqref{def-J} we can express the
structure constants $\mathscr{J}^i{}_{j\alpha}$
and $\mathscr{J}^a{}_{b\alpha}$ in terms of the correlation functions:
\begin{subequations}
\begin{align}
\mathscr{J}^i{}_{j\alpha}&=
\la Y_{-2}| \tilde {\Phi}^i,\,\Phi_j,\,\Lambda_{+,\alpha}\ra
-\la Y_{-2}| \tilde {\Phi}^i,\,\Lambda_{+,\alpha},\,\Phi_j\ra,
\\
\mathscr{J}^a{}_{b\alpha}&=
\la Y_{-2}| \tilde {\mathrm{T}}^a,\,\mathrm{T}_b,\,\Lambda_{+,\alpha}\ra
-\la Y_{-2}| \tilde {\mathrm{T}}^a,\,\Lambda_{+,\alpha},\,\mathrm{T}_b\ra.
\end{align}
\label{str-const}
\end{subequations}
In the next section these formulae are used to write down
gauge transformations explicitly.
\subsection{Calculations of Structure Constants.}
\label{sec:strcon}

In Section~\ref{sec:t-potential} we have computed \cite{ABKM2}
the restricted action \eqref{act-2} up to
level $(2,6)$. The relevant conformal fields
with $\phi$-charge $1$ and $0$ are
\begin{subequations}
\begin{alignat}{3}
\Phi_0\equiv U&=\,c
&\qquad \Phi_3\equiv V_3&=\,cT_{\eta\xi}
&\qquad \Phi_6\equiv V_6&=\,T_F\eta e^{\phi}
\\
\Phi_1\equiv V_1&=\,\partial^2c
& \Phi_4\equiv V_4&=\,cT_{\phi}
& \Phi_7\equiv V_7&=\,bc\pd c
\\
\Phi_2\equiv V_2&=\,cT_B
& \Phi_5\equiv V_5&=\,c\partial^2\phi
& \Phi_8\equiv V_8&=\,\pd c\pd\phi
\\
&&\mathrm{T}_0&=\frac 14\,e^{\phi}\eta
\end{alignat}
\label{operators}
\end{subequations}
with $\phi^i=\{u,v_1,\dots,v_8\}$ and $t^a=\{t\}$.
For this set of fields we have got
\begin{align}
S_2^{(2,4)}&=u^2+\frac{1}{4}t^2+(4v_1-2v_3-8v_4+8v_5+2v_7)u
\nonumber
\\
&+4v_1^2+\frac{15}{2}v_2^2+v_3^2+\frac{77}{2}v_4^2+22v_5^2+10v_6^2
+8v_1v_3-32v_1v_4+24v_1v_5+4v_1v_7
\nonumber
\\
&-16v_3v_4+4v_3v_5-2v_3v_7+12v_3v_8-52v_4v_5-8v_4v_7-20v_4v_8
+8v_5v_7+8v_5v_8
\nonumber
\\
&+(-30v_4+20v_5+30v_2)v_6+4v_7v_8,
\label{S24}
\\
S_3^{(2,6)}&=\left(\frac{1}{3\gamma ^2}u+\frac{9}{8}v_1-\frac{25}{32}v_2-\frac{9}{16}v_3-\frac{59}{32}v_4
+\frac{43}{24}v_5+\frac{2}{3}v_7
\right)t^2
\nonumber
\\
&+\left(-\frac{40\gamma }{3}u-45\gamma ^3v_1
+\frac{45\gamma ^3}{4}v_2+\frac{45\gamma ^3}{2}v_3+\frac{295\gamma ^3}{4}v_4
-\frac{215\gamma ^3}{3}v_5-\frac{80\gamma ^3}{3}v_7\right)v_6^2.
\label{S34}
\end{align}

Here $\gamma=\frac4{3\sqrt{3}}$. There is no gauge transformation at level zero. At level 2 the gauge
parameters are  zero picture conformal fields
with ghost number 0 and the weight $h=1$, see Table \ref{tab:2}.
 There are two such conformal
fields with $0$ $\phi$-charge
$bc$ and $\pd\phi$, i.e.
on the conformal language the gauge parameter $\Lambda_+$
with  the weight $1$ is of the form
\begin{equation}
\Lambda_+=\delta\lambda_1\,bc+\delta\lambda_2\pd\phi.
\label{lh1CFT}
\end{equation}
The zero order gauge transformation (\ref{gaugeP})
of level $2$ fields has the form
\begin{multline}
\delta_0\Ac_+(w)\equiv Q_B\Lambda_+(w)=\\
(-\delta\lambda_2+\frac{3}{2}\delta\lambda_1)\pd^2 c(w)
+\delta\lambda_1\,cT_B(w)
+\delta\lambda_1\,cT_{\xi\eta}(w)
+\delta\lambda_1\,cT_{\phi}(w)
+\delta\lambda_2\,c\pd^2\phi(w)
\\
-\delta\lambda_2\,\eta e^{\phi}T_F(w)
+\delta\lambda_1\,bc\pd c(w)
+\delta\lambda_2\,\pd c\pd\phi(w)
+\frac{1}{4}(\delta\lambda_1-2\delta\lambda_2)\,b\eta\pd\eta e^{2\phi(w)}.
\end{multline}
We see that in accordance with (\ref{da2}) one gets the field $\Phi_9 =
b\eta\pd\eta e^{2\phi}$ from the sector with $q=2$.
To exclude this field from the consideration we
impose the condition $Q_2 \Lambda_+ =0$, which links
parameters $\delta\lambda_1$
and $\delta\lambda_2$ appearing in \eqref{lh1CFT}:
\begin{equation*}
Q_2\Lambda_+
=\frac{1}{4}(\delta\lambda_1-2\delta\lambda_2)
b\eta\pd\eta e^{2\phi}(w)=0,
\end{equation*}
i.e.
\begin{equation}
\delta\lambda_1=2\delta\lambda_2\equiv 2\delta\lambda.
\label{lambda}
\end{equation}
We are left with
the following zero order gauge transformations
of the restricted  action on level $2$:
\begin{alignat}{3}
\delta _0v_1&=2\delta\lambda,
&\qquad\qquad
\delta _0v_4&=2\delta\lambda,
&\qquad\qquad
\delta _0v_7&=2\delta\lambda,
\nonumber
\\
\delta _0v_2&=2\delta\lambda,&
\delta _0v_5&=\delta\lambda,&
\delta _0v_8&=\delta\lambda,
\label{rd2}
\\
\delta _0v_3&=2\delta\lambda,&
\delta _0v_6&=-\delta\lambda,&
\delta _0u&=0.
\nonumber
\end{alignat}
Transformations \eqref{rd2} give
the vector
$\mathscr{V}^{i}_1\equiv \mathscr{V}^{i}$ in \eqref{ptrans}
in the form
\begin{equation}
\mathscr{V}^{i}=\{0,2,2,2,2,1,-1,2,1\}.
\label{V9}
\end{equation}

One can check that the quadratic action
at level $(2,4)$ (\ref{S24}) is invariant
with respect to this transformation,
\begin{equation}
\delta _0 S_2=\delta\lambda \sum _{i=1}^{9}
\frac{\partial S_2}{\partial \phi^{i}}\mathscr{V}^i=0,
\end{equation}
or in other words
$9$-component vector $\mathscr{V}^i$ \eqref{V9}
is the zero vector of the matrix $\mathscr{M}_{ij}$ defined
by \eqref{S24}.

Now we would like to find the nonlinear terms in the transformations
(\ref{trans}). At  level $2$ we have
$\mathscr{J}^i_{j1}=\mathscr{J}^i_{j}$.The dual operators
\eqref{Def_d} to the operators \eqref{operators} are the following
\begin{subequations}
\begin{alignat}{2}
\tilde{\Phi}^1&=\frac{1}{16}\,\eta\partial\eta\bigl[ 1
+\partial b c\bigr]e^{2\phi},
&\qquad
\tilde{\Phi}^5&=-\frac{1}{16}\,\eta\partial\eta\bigl[4-\partial^2\phi
+2\,\partial\phi\partial\phi\bigr] e^{2\phi},
\\
\tilde{\Phi}^2&=\frac{1}{60}\,\eta\partial\eta e^{2\phi}T_B,
&\qquad
\tilde{\Phi}^6&=-\frac{1}{20}\,cT_F\partial\eta e^{\phi},
\\
\tilde{\Phi}^3&=\frac{1}{48}\,\bigl[\partial\eta\partial^2\eta
-6\,\eta\partial\eta \bigr]e^{2\phi},
&\qquad
\tilde{\Phi}^7&=\frac{1}{8}\,\eta\partial\eta\bigl[1
+b\partial c\bigr] e^{2\phi},
\\
\tilde{\Phi}^4&=-\frac{1}{8}\,\eta\partial\eta\partial\phi\partial\phi e^{2\phi},
&\qquad
\tilde{\Phi}^8&=\frac{1}{8}\,\eta\partial\eta bc\partial\phi e^{2\phi},
\end{alignat}
\vspace{-8mm}
\begin{alignat}{2}
\tilde{\Phi}^0&=\frac{1}{8}\,\eta\partial\eta\bigl[
6-\partial (b c)-2\partial^2\phi\bigr] e^{2\phi}
+\frac{1}{48}\,\partial\eta\partial^2\eta e^{2\phi},
&\qquad
\tilde{\mathrm{T}}^0&=\frac{1}{2}\,c e^{\phi}\partial\eta.~~~~
\end{alignat}
\label{dual}
\end{subequations}
It is straightforward to check that
\begin{equation}
\la Y_{-2}|\tilde{\Phi}^{j},\,\Phi_i\ra=\delta^j{}_{i}
\qquad\text{and}\qquad
\la Y_{-2}| \tilde{\mathrm{T}}^{0},\,\mathrm{T}_{0}\ra=1.
\end{equation}

We find the coefficients $\mathscr{J}^i_j$ in (\ref{trans}) up to
level $(2, 4)$ and this gives the following gauge transformations
\begin{subequations}\label{delta1S}
\begin{align}
\delta_1 u&=[(-\frac{82}{3}\gamma^3+32\gamma)v_1-\frac{16}{3}\gamma^3v_4
+(-19\gamma^3+16\gamma)v_5
+(-\frac{73}{3}\gamma^3+16\gamma)v_7
\nonumber
\\
&~~~~~~~~~~~~~+(\frac{154}{3}\gamma^3-32\gamma)v_8]~\delta\lambda,
\\
\delta_1 t&=\frac{4}{3}t~\delta\lambda,
\\
\delta_1 v_1&=[(-\frac{27}{2}\gamma^3+\frac{8}{3}\gamma)v_1
+(-\frac{11}{4}\gamma^3+\frac{4}{3}\gamma)v_5
+(-\frac{17}{12}\gamma^3+\frac{4}{3}\gamma)v_7
+(\frac{3}{2}\gamma^3-\frac{8}{3}\gamma)v_8]~\delta\lambda,
\\
\delta_1 v_2&=[-\frac{5}{3}\gamma^3v_1-\frac{5}{6}\gamma^3v_5-\frac{5}{6}\gamma^3v_7
+\frac{5}{3}\gamma^3v_8]~\delta\lambda,
\\
\delta_1 v_3&=[(\frac{17}{3}\gamma^3-\frac{16}{3}\gamma)v_1
+(\frac{17}{6}\gamma^3-\frac{8}{3}\gamma)v_5
+(\frac{17}{6}\gamma^3-\frac{8}{3}\gamma)v_7+(-\frac{17}{3}\gamma^3+
\frac{16}{3}\gamma)v_8]~\delta\lambda,
\\
\delta_1 v_4&=[-\frac{5}{3}\gamma^3v_1+\frac{32}{3}\gamma^3v_4-\frac{37}{6}\gamma^3v_5
-\frac{5}{6}\gamma^3v_7+\frac{5}{3}\gamma^3v_8]~\delta\lambda,
\\
\delta_1 v_5&=[-\frac{4}{3}\gamma u+(\frac{61}{6}\gamma^3-\frac{32}{3}\gamma)v_1
+\frac{25}{8}\gamma^3v_2
-\frac{5}{12}\gamma^3v_3+\frac{481}{24}\gamma^3v_4
\nonumber
\\
&~~~~~~~~~~~
+(-\frac{31}{6}\gamma^3-\frac{16}{3}\gamma)v_5
+(\frac{14}{3}\gamma^3-\frac{16}{3}\gamma)v_7+(-\frac{52}{3}\gamma^3+
\frac{32}{3}\gamma)v_8]~\delta\lambda,
\\
\delta_1 v_6&=
-\frac{4}{3}\gamma^3v_6~\delta\lambda,
\\
\delta_1 v_7&=[\frac{16}{3}\gamma u+(\frac{38}{3}\gamma^3+\frac{16}{3}\gamma)v_1
-\frac{25}{2}\gamma^3v_2+\frac{5}{3}\gamma^3v_3
-\frac{193}{6}\gamma^3v_4
\nonumber
\\
&~~~~~~~~~~~~+(\frac{70}{3}\gamma^3+\frac{8}{3}\gamma)v_5
+(8\gamma^3+\frac{8}{3}\gamma)v_7
+(16\gamma^3-\frac{16}{3}\gamma)v_8]~\delta\lambda,
\\
\delta_1 v_8&=[\frac{4}{3}\gamma u+\frac{43}{6}\gamma^3v_1-\frac{25}{8}\gamma^3v_2
+\frac{5}{12}\gamma^3v_3+\frac{95}{24}\gamma^3v_4
+\frac{11}{6}\gamma^3v_5+4\gamma^3v_7]~\delta\lambda,
\end{align}
\end{subequations}
where $\gamma=\frac{4}{3\sqrt{3}}\approx 0.770$.
\subsection{Breaking of Gauge Invariance by Level Truncation.}

As it has been mentioned above (see \eqref{zero}) the
quadratic restricted
action (\ref{act-2})
 is invariant with respect to the free gauge transformation
\eqref{rd2}.

In contrast to the bosonic case \cite{Taylor}
already the first order gauge
invariance is broken by the level truncation scheme.
Explicit calculation shows that
\begin{equation}
\delta S|_{\text{first order}}\equiv
\delta_1S_2^{(2,4)}+\delta_0S_3^{(2,6)}=-\frac13t^2\delta\lambda+
\{\text{quadratic terms in } v_i\}\delta\lambda.
\end{equation}
Note that the terms in the braces belong to level $6$
and therefore we neglect them.

The origin of this breaking is in the presence of non-diagonal
terms in the quadratic action \eqref{S24}. More precisely, in the bosonic
case the operators with different weights are orthogonal
to each other,
while in the fermionic string due
to the presence of
$Y_{-2}$ this orthogonality is broken.
Indeed, the substitution of $\{\phi^i\}=\{u,v_i,\,w^I\}$
and $\{t^a\}=\{t,\,\tau^A\}$ (here by $w^I$ and $\tau^A$
we denote higher level fields) into the action \eqref{action-mat}
yields
\begin{subequations}
\begin{align}
S_2 &= \frac14t^2 -\sum_A \mathscr{F}_{tA}t\tau^A
-\frac12\sum_{A,B} \mathscr{F}_{AB}\tau^A\tau^B
-\frac12\sum_{i,j} \mathscr{M}_{ij}\phi^i\phi^j,
\\
S_3&=-\frac13\sum _{i,j,k}\mathscr{G}_{ijk}\phi^i\phi^j\phi^k
+\sum_{i} \mathscr{G}_{itt}\phi^it^2
+\sum_{i,A} (\mathscr{G}_{itA}+\mathscr{G}_{iAt})\phi^i t\tau^A
+\sum_{i,A,B} \mathscr{G}_{iAB}\phi^i\tau^A\tau^B.
\end{align}
\end{subequations}
For the gauge transformations (\ref{resgt}) with $\Lambda_+$
in the form (\ref{lh1CFT}), (\ref{lambda}) we have
\begin{subequations}
\begin{align}
\delta_0 t&=0,\quad\delta_0\tau^A=0,\quad\text{and}\quad \delta_0w^I=0,
\label{0tr}
\\
\delta_1t&=\frac43t\delta\lambda
\quad\text{and}\quad
\delta_1\tau^A= (\mathscr{J}^A{}_{t} t+\mathscr{J}^A{}_{B} \tau^B)
\delta\lambda.
\label{1tr}
\end{align}
\end{subequations}

The first order
gauge transformation of the action produces the following quadratic in $t^a$ terms
\begin{multline}
\label{exact}
\left.(\delta_1 S_2+\delta_0S_3)\right|_{t^at^b-\text{terms}}=
\left.\left(\frac{\pd S_2}{\pd t}\delta_1 t
+\frac{\pd S_2}{\pd \tau^A}\delta_1\tau^A
+\frac{\pd S_3}{\pd u}\delta_0 u
+\frac{\pd S_3}{\pd v_i}\delta_0v_i\right)\right|_{t^at^b-\text{terms}}.
\end{multline}
Here we take into account that due to (\ref{0tr})
\begin{equation}
\delta_0 S_3|_{\text{contributions from higher levels}}=0.
\end{equation}
The exact gauge invariance means that (\ref{exact})
 equals to zero. In the presence of non-diagonal terms in $S_2$ we have
\begin{equation}
\frac{\partial S_2}{\partial \tau^A}\delta_1 \tau^A
=-\mathscr{F}_{At}\mathscr{J}^A{}_{t}t^2
-(\mathscr{F}_{tA}\mathscr{J}^A{}_{B}
+\mathscr{F}_{AB}\mathscr{J}^A{}_t)t\tau^B-
\mathscr{F}_{AB}\mathscr{J}^B{}_{C}\tau^A\tau^C.
\label{next}
\end{equation}
Generally speaking, $\mathscr{F}_{At}\neq 0$
and  (\ref{next}) contains $t^2$ term.
Therefore, if we exclude fields $\tau ^A$  from $S_2$
we break
the first order gauge invariance.

We can estimate
the contribution of higher level fields $\{\tau^A\}$
to \eqref{next}. Let us consider only $t^2$ terms in
\eqref{exact}:
\begin{equation}
\left(\frac23 t^2-\mathscr{F}_{tA}\mathscr{J}^A{}_t t^2\right)\delta\lambda
+\mathscr{G}_{itt}\delta_0 v_i=0.
\label{EQSD}
\end{equation}
One can check that $\mathscr{G}_{itt}\delta_0 v_i=-1$ and hence
$\mathscr{F}_{tA}\mathscr{J}^A{}_t=-\frac13$.
Therefore, we see that the contribution of the higher levels
into equality \eqref{EQSD} is only $33\%$. This gives us a hope that the
gauge invariance rapidly restores as level grows. For the bosonic case
this restoration was advocated at \cite{Taylor}.

\subsection{The Orbits.}
\label{sec:orbits}
\setcounter{equation}{0}
\renewcommand{\arraystretch}{1}

In this section we discuss the method of checking
the validity of gauge fixing condition.

Let us consider the simplest case then we have only one
gauge degree of freedom and therefore only one gauge
fixing condition. For this case one can find orbits of gauge transformations.
Using this technique we will analyze the validity
of the Feynman-Siegel gauge\footnote{Note that the analysis
of the validity of FS gauge was performed in \cite{Taylor},
where it was shown that FS gauge affects the region
of validity of Taylor series used to compute tachyon potential
at higher levels.}
 used in bosonic SFT computations
\cite{zwiebach}
and the validity of the gauge
\begin{equation}
G(\phi^i)\equiv 3v_2-3v_4+2v_5=0
\label{gaugecond}
\end{equation}
used in Section~\ref{sec:calc} on level $2$ in computations in Super SFT.

Since all our computations are based on level truncation
scheme our study of the validity of a gauge will be also
based on level truncation scheme.
The statement that gauge is valid means
that for any field configuration $\{\phi^i_0\}$ one
can find the gauge equivalent configuration $\{\phi'{}^i\}$
such that $G(\phi'{}^i)=0$.
In general a given field configuration $\{\phi^i_0\}$
defines gauge orbit $\{\phi^i(\lambda;\phi_0)\}$ with
condition $\phi^i(0;\phi_0)=\phi^i_0$.
So the previous statement can be reformulated as follows:
for any field configuration $\{\phi^i_0\}$
there exists $\lambda=\lambda_0$ such that
$G(\phi^i(\lambda_0;\phi_0))=0$.

Therefore, to prove gauge validity one has to find
gauge orbit $\phi^i(\lambda;\phi_0)$ and show that
equation $G(\phi^i(\lambda;\phi_0))=0$ has solution
for any $\phi_0^i$. In the case of one parametric gauge
transformations the orbit can be found as a solution
of the following differential equations:
\begin{subequations}
\begin{align}
\frac{d\phi^i(\lambda)}{d\lambda}&=
\mathscr{V}^i+ \mathscr{J}^i{}_{j}\phi^j(\lambda)
\quad\text{with}\quad\phi^i(0)=\phi^i_0,
\label{eqphi}
\\
\frac{dt^a(\lambda)}{d\lambda}&=
\mathscr{J}^a{}_{b}t^b(\lambda)
\quad\text{with}\quad t^a(0)=t^a_0.
\label{eqt}
\end{align}
\label{equations}
\end{subequations}
Here we use notations from \eqref{def-J}.
To write down explicit solutions of \eqref{eqphi} and \eqref{eqt}
we will use a basis for $\phi_i$ in which the matrices
$\mathscr{J}$ have the canonical Jordan form.

\subsubsection{Orbits in bosonic string field theory.}
As a simple example of the gauge fixing in the level truncation scheme
let us consider the Feynman-Siegel gauge
at  level $(2,6)$ in bosonic open string field theory.
Here we use notations of \cite{zwiebach}.
Up to level $2$ the string field has the expansion:
\begin{equation}
\Phi =\sum _1^4 \phi^i\Phi_i\qquad\text{with}\quad
\phi=\{t,\,v,\,u,\,w\}
\label{bas_f}
\end{equation}
and
\begin{equation}
\Phi_1=c,~\Phi_2=cT_B,~\Phi_3=\frac12\partial^2c,~\Phi_4=bc\partial c.
\label{oper-b}
\end{equation}
The Feynman-Siegel gauge $b_0\Phi=0$ restricted to level $2$ gives
the condition
\begin{equation}
G_{FS}(\phi^i)\equiv \phi^4=0.
\label{FZgauge}
\end{equation}
To compute structure constants $\mathscr{J}^i{}_{j1}$
using method described in Section~\ref{sec:gscf} one has
to know the dual operators to \eqref{oper-b}.
One can check that the set
\begin{equation}
\tilde{\Phi}^1=c\partial c,~\tilde{\Phi}^2
=\frac{1}{3}c\partial cT_B,
~\tilde{\Phi}^3=\frac12\partial c\partial^2c,
~\tilde{\Phi}^4=\frac16\partial^3cc.
\end{equation}
satisfies the required properties \eqref{Def_d}
\begin{equation}
\la\tilde{\Phi}^i,\Phi_j\ra=\delta^i{}_{j}, \quad i,j=1,\dots,4.
\end{equation}

The gauge parameter at level $2$  is:
$$
\Lambda=\delta\lambda_1\Lambda_1,
\quad\text{where}\quad\Lambda_1=bc\quad\text{and}\quad\delta\lambda_1\equiv
\delta\lambda.
$$
In this case the vector $\mathscr{V}^i$ defined by \eqref{def-J} is of the form
\begin{equation}
\mathscr{V}^i=
\begin{bmatrix}
0\\
1/2\\
-3\\
-1
\end{bmatrix}.
\end{equation}
The structure constants of the gauge transformation are given by the
matrix $\mathscr{J}^i{}_{j1}\equiv \mathscr{J}^i{}_{j}$
\begin{equation}
\mathscr{J}^i{}_{j}
=\la\tilde{\Phi}^i,\Phi_j,\Lambda_1\ra
-\la\tilde{\Phi}^i,\Lambda_1,\Phi_j\ra.
\end{equation}
The matrix $\mathscr{J}^i{}_j$ has the following entries
\begin{equation}
[\mathscr{J}^i{}_{j}]=
\begin{bmatrix}
-\frac{1}{\gamma}&\frac{65}{16}\gamma
&\frac{29}{16}\gamma&-\frac{3}{2}\gamma\\
\\
\frac{5}{16}\gamma&-\frac{581}{256}\gamma^3&-\frac{145}{256}\gamma^3
&\frac{15}{32}\gamma^3\\
\\
\frac{11}{16}\gamma&-\frac{715}{256}\gamma^3&-\frac{703}{256}\gamma^3
&-\frac{47}{32}\gamma^3\\
\\
\frac{21}{8}\gamma&-\frac{1356}{128}\gamma^3&\frac{31}{128}\gamma^3&
\frac{63}{16}\gamma^3
\end{bmatrix},
\label{J-bos}
\end{equation}
where $
\gamma=\frac4{3\sqrt{3}}$.
This result coincides with the one obtained in
eq. (9) of \cite{Taylor} with obvious redefinition
of the fields $\phi^i$.

To solve equation \eqref{eqphi} it is convenient to rewrite it
in eigenvector basis of matrix $\mathscr{J}^i{}_j$.
The characteristic polynomial $\mathscr{P}$ of the matrix
$\mathscr{J}^i{}_j$
$$
\mathscr{P}(\mathscr{J},\omega)=
\omega^4+\frac{335}{324}\sqrt{3}\,\omega^3
-\frac{3584}{6561}\,\omega^2 +\frac{11869696}{4782969}\sqrt{3}\,\omega
+\frac{819200}{531441}.
$$
has the following roots
\begin{align}
\{\omega\}=\{-2.565,\,-0.332,\,0.553\pm i1.226\}.
\end{align}
The corresponding four eigenvectors are
\begin{equation}
\nu_{\omega}=
\begin{pmatrix}
0.131\omega^3+1.661\omega^2-0.804\omega+4.655
\\
0.464\omega^3+0.687\omega^2-0.532\omega+2.276
\\
-0.249\omega^3-0.127\omega^2+0.298\omega-1.189
\\
1
\end{pmatrix}.
\label{eigenstate-b}
\end{equation}

\begin{figure}[t]
\centering
\includegraphics[width=380pt]{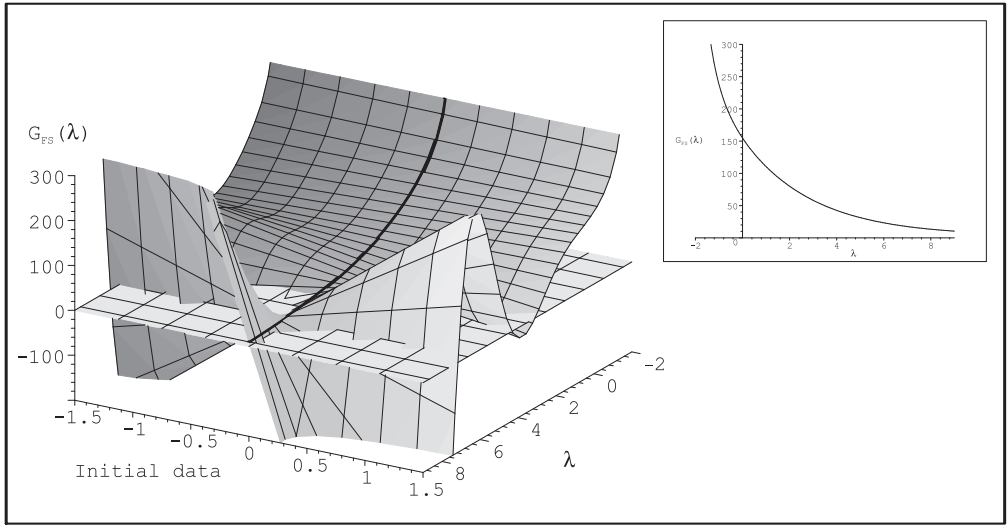}
\caption{$G_{FS}(\phi^i(\lambda;\phi_0))$ restricted to level $2$
in bosonic string field theory.}
\label{pic:boson}
\end{figure}

We solve  system \eqref{eqphi} in the basis of these eigenvectors and get
the following dependence of $G_{FS}$ \eqref{FZgauge}
on $\lambda$ (see Figure~\ref{pic:boson}):
\begin{equation}
G_{FS}(\lambda)\equiv G_{FS}(\phi^i(\lambda))=
[a\sin{(1.23\lambda)}+b\cos{(1.23\lambda)}]e^{0.553\lambda}
+c e^{-0.332\lambda}
+d e^{-2.57\lambda}+2.89,
\label{FSorbit}
\end{equation}
where
\begin{subequations}
\begin{align}
a&=1.13\,t_0-2.13\,v_0+0.997\,u_0+0.921\,w_0-0.0861,
\\
b&=0.172\,t_0-0.939\,v_0-0.346\,u_0+1.04\,w_0-1.49,
\\
c&=0.0466\,t_0+0.300\,v_0-0.0149\,u_0-0.00958\,w_0-0.741,
\\
d&=-0.218\,t_0+0.639\,v_0+0.360\,u_0-0.0337\,w_0-0.657.
\end{align}
\end{subequations}
It is obvious that if one takes an initial data $\phi_0\equiv\{t_0,\,v_0,\,u_0,\,w_0\}$
such  that $a=b=0$ and $c,d\geqslant 0$
then the function $G_{FS}(\phi^i(\lambda;\phi_0))$ is strongly positive,
and therefore there is no representative element in FS gauge
for this set of field configurations.
This situation is depicted in Figure~\ref{pic:boson} by the thick line
on 3-dimensional plot and by 2-dimensional plot.

\subsubsection{Orbits in superstring field theory.}
The logic of previous subsection can be simply applied
to analysis of the validity of the gauge \eqref{gaugecond}
in cubic Super SFT.

The vector $\mathscr{V}^i$ has been already computed in \eqref{V9}.
The matrix $\mathscr{J}^i{}_j$ on the level $2$
can be obtained from \eqref{delta1S}.
The characteristic polynomial $\mathscr{P}$ of the
matrix $\mathscr{J}^i{}_j$
\begin{equation}
\mathscr{P}(\mathscr{J},\omega)=
\bigl(\omega^4-\frac{1187840}{59049}\omega^2+\frac{451911090176}{3486784401}\bigr)
\bigl(\omega+\frac{256}{729}\sqrt{3}\bigr)
\omega^4
\end{equation}
has the following eigenvalues
\begin{align}
\{\omega\}=\{
0,\, 0,\, 0,\, 0,\, -0.608,\,
\eta
\}\qquad\text{where}\qquad \eta=\pm3.274\pm i0.814.
\label{eigenvalues-s}
\end{align}
The the corresponding eigenvectors are
\begin{align}
\nu_{0}^{(1)}=
\begin{bmatrix}
1\\ 0\\ 0\\ -5.4\\ 0\\ 0\\ 0\\ 0\\ 0
\end{bmatrix},
\qquad
\nu_{0}^{(2)}=
\begin{bmatrix}
0\\ 0\\ 0\\ 39.3\\ 1\\ 2\\ 0\\ -6\\ -2
\end{bmatrix},
\qquad
\nu_{0}^{(3)}=
\begin{bmatrix}
0\\ 0\\ 1\\ 7.5\\ 0\\ 0\\ 0\\ 0\\ 0
\end{bmatrix},
\qquad
\nu_{0}^{(4)}=
\begin{bmatrix}
0\\ 1\\ 0\\ 155.6\\ 0\\ 0\\ 0\\ -18\\ -8
\end{bmatrix},
\\
\nonumber
\\
\nu_{-0.608}=
\begin{bmatrix}
0\\ 0\\ 0\\ 0\\ 0\\ 0\\ 1\\ 0\\ 0
\end{bmatrix},\qquad
\nu_{\eta}=
\begin{bmatrix}
0.011\eta^3-0.17\eta^2+0.12\eta-1.77
\\
-0.015\eta^3+0.022\eta^2-0.16\eta-0.28
\\
0.5
\\
1
\\
0.038\eta^3+0.26\eta^2+0.39\eta-0.044
\\
-0.03\eta^3+0.044\eta^2+\eta+1.95
\\
0
\\
0.29\eta^3+0.995\eta^2-2.27\eta-7.33
\\
0.12\eta^3+0.54\eta^2-0.13\eta-2.97
\end{bmatrix}.
\label{eigenstate-s}
\end{align}

The solution of \eqref{equations}
yields the following dependence on $\lambda$ of the gauge fixing
condition $G$ \eqref{gaugecond}
\begin{multline}
G(\lambda)\equiv G(\phi(\lambda;\phi_0))
=[a\sin(0.814\lambda)+b\cos(0.814\lambda)]e^{3.27\lambda}
\\
+[c\sin(0.814\lambda)+d\cos(0.814\lambda)]e^{-3.27\lambda}+
4.15\lambda +f,
\end{multline}
where
\begin{figure}[t]
\centering
\includegraphics[width=380pt]{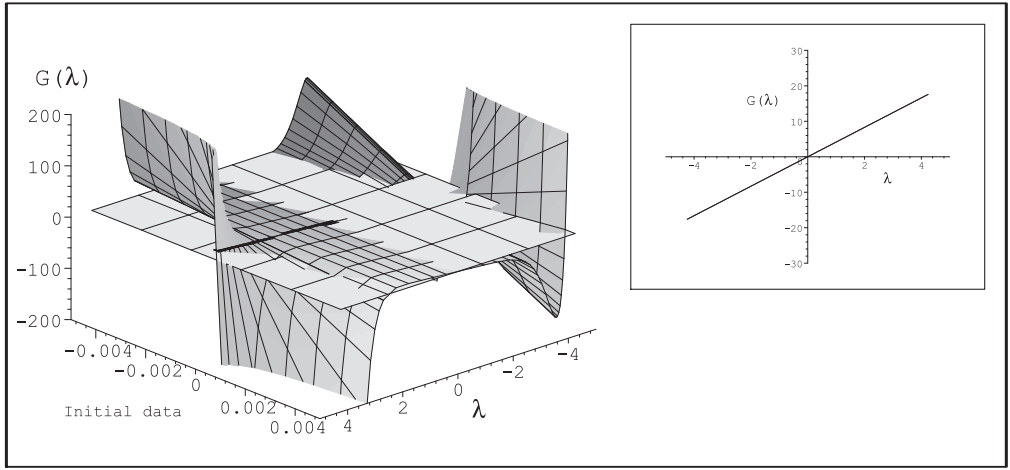}
\caption{$G(\phi^i(\lambda;\phi_0))$ on level $2$
in NS string field theory.}
\label{pic:super}
\end{figure}
\begin{subequations}
\begin{align}
a&=-1.05 u_0 -2.81 v_{1,0} +1.46 v_{2,0} -0.195 v_{3,0}\notag\\
&~~~~~~~~~~+5.82 v_{4,0} -4.81 v_{5,0} -2.13 v_{7,0} +0.614 v_{8,0} -0.177\\
b&=-0.406 u_0 -1.04 v_{1,0} +0.559 v_{2,0} -0.0754 v_{3,0}\notag\\
&~~~~~~~~~~-0.03 v_{4,0} -0.59 v_{5,0} -0.64 v_{7,0} -0.155 v_{8,0} -0.924\\
c&=-0.0933 u_0 -1.38 v_{1,0} +0.130 v_{2,0} -0.0174 v_{3,0}\notag\\
&~~~~~~~~~~+1.06 v_{4,0} -0.572 v_{5,0} + 0.166 v_{7,0} -0.885 v_{8,0} +0.516\\
d&=-0.0636 u_0+ 0.104 v_{1,0} +0.088 v_{2,0} -0.0118 v_{3,0}\notag\\
&~~~~~~~~~~+0.552 v_{4,0} -0.195 v_{5,0} +0.085 v_{7,0} -0.407 v_{8,0} -0.186\\
f&=0.465 u_0+ 0.939 v_{1,0} +2.35 v_{2,0} +0.0867 v_{3,0}\notag\\
&~~~~~~~~~~-3.51 v_{4,0} +2.77 v_{5,0} +0.56 v_{7,0} +0.563 v_{8,0} +1.12
\end{align}
\end{subequations}
and $u_0, v_{i,0}$ are initial data for the corresponding differential
equations \eqref{equations}.

A simple analysis shows that there are no restrictions
on the range of validity of the gauge (\ref{gaugecond}).
So the gauge condition \eqref{gaugecond} is a valid
choice for the computations of the tachyon potential.
The 2-dimensional plot on the Figure~\ref{pic:super}
corresponds to the special initial data $a=b=c=d=0$.

It is interesting to note that there is another gauge which strongly
simplifies the effective potential. Namely, this is the gauge $v_6=0$.
The orbits of this gauge condition have the form
$$
v_6(\lambda)=(1.64+v_{6,0})e^{-0.608\lambda}-1.64.
$$
It is evident that this gauge condition is not always reachable and
cannot be used in the calculation of the tachyon potential.

\section*{Acknowledgments.}

We would like to thank
B. Dragovic, J. Lukierski and I. Volovich for discussions.\\
I.A. thanks the organizing committee of the Swieca Summer School,
Summer School in  Sokobanja, Max Born Symposium in Karpacz
and Workshop "Noncommutative Geometry,
Strings and Renormalization" in Leipzig
for warmest hospitality.
A.K. would like to thank E. Kiritsis and the HEP theory group of the
University of Crete for the warmest hospitality where the final stage of this work
has been done.

This work was supported in part
by RFBR grant 99-01-00166 and by RFBR grant for leading scientific
schools. I.A., D.B., A.K. and P.M. were supported in part by INTAS grant
99-0590.
A.K. was supported in part by the NATO fellowship program.



\end{document}